\documentstyle[12pt,clpsref,extras,epsfig]{article}

\newcommand{\ipar}{\hspace*{6mm}}
\newcommand{\ilskp}{\hspace*{27mm}}
\renewcommand{\thefigure}{\thesection.\arabic{figure}}

\newcommand{\BD}{\mbox{\boldmath $D$}}

\newcommand{\Real}{\mbox{$\rule{1pt}{0.7em}{\rm R}$}}
\newcommand{\sfrac}[2]{\mbox{\footnotesize $\frac{#1}{#2}$}}
\newcommand{\vp}{{\vec p}}
\newcommand{\wvp}{{\vec P}}
\newcommand{\sect}[1]{\section{{\bf\sc #1}}
       \setcounter{table}{0}\setcounter{figure}{0}
       \setcounter{equation}{0}\vspace*{-1.0\baselineskip}}
\newcommand{\subsect}[1]{\subsection{{\it #1}}
        \setcounter{equation}{0}\vspace*{-1.0\baselineskip}} 
\newcommand{\subsubsect}[1]{\bigskip\underline{#1}\ipar}
\newcommand{\lsim}{\mathrel{\rlap{\lower4pt\hbox{\hskip0pt$\sim$}}
\raise1pt\hbox{$<$}}}           
\newcommand{\gsim}{\mathrel{\rlap{\lower4pt\hbox{\hskip0pt$\sim$}}
\raise1pt\hbox{$>$}}}           
\begin{document}
\setcounter{page}{1}
\rightline{ANL-PHY-9530-TH-2000}
\rightline{nucl-th/0005064}
\begin{center}
\parbox{30em}{
DYSON-SCHWINGER EQUATIONS:\\DENSITY, TEMPERATURE AND CONTINUUM STRONG QCD}
\vspace*{5mm}

Craig D. Roberts and Sebastian M. Schmidt
\vspace*{1em}

Physics Division, Building 203, Argonne National Laboratory, Argonne, IL
60439-4843, USA

\vspace*{1em}

{\it e-mail: cdroberts@anl.gov \hspace*{\parskip} basti@theory.phy.anl.gov}
\end{center}
%

\ilskp ABSTRACT\\[1em]
%
Continuum strong QCD is the application of models and continuum quantum field
theory to the study of phenomena in hadronic physics, which includes; e.g.,
the spectrum of QCD bound states and their interactions; and the transition
to, and properties of, a quark gluon plasma.  We provide a contemporary
perspective, couched primarily in terms of the Dyson-Schwinger equations but
also making comparisons with other approaches and models.  Our discourse
provides a practitioners' guide to features of the Dyson-Schwinger equations
[such as confinement and dynamical chiral symmetry breaking] and canvasses
phenomenological applications to light meson and baryon properties in cold,
sparse QCD.  These provide the foundation for an extension to hot, dense QCD,
which is probed via the introduction of the intensive thermodynamic
variables: chemical potential and temperature.  We describe order parameters
whose evolution signals deconfinement and chiral symmetry restoration, and
chronicle their use in demarcating the quark gluon plasma phase boundary and
characterising the plasma's properties.  Hadron traits change in an
equilibrated plasma.  We exemplify this and discuss putative signals of the
effects.  Finally, since plasma formation is not an equilibrium process, we
discuss recent developments in kinetic theory and its application to
describing the evolution from a relativistic heavy ion collision to an
equilibrated quark gluon plasma.

\bigskip

\ipar PACS: \hspace*{4.0em}
05.20.Dd, 
11.10.Wx, 
11.15.Tk, 
12.38.Mh, 
24.85.+p, 
25.75.Dw  

\ipar KEYWORDS:  \vspace*{-2.3\baselineskip}

\hspace*{9.0em}\parbox{140mm}{\flushleft Dyson-Schwinger Equations, Hadron
Physics, Kinetic Theory, Nonperturbative QCD Modelling, Particle Production,
Quark-Gluon Plasma}

\vspace*{20mm}

{\it To appear in:\\ 
\underline{Progress in Particle and Nuclear Physics {\bf 45} (2000) 2nd issue}}

\vfill\eject
\parbox{185mm}{\tableofcontents}
\pagebreak
\sect{Prologue}
Continuum strong QCD.  The phrase comes to us via Ref.~\cite{mikeCsQCD},
although that is likely not its debut.  In our usage it embraces all
continuum nonperturbative methods and models that seek to provide an
intuitive understanding of strong interaction phenomena, and particularly
those where a direct connection with QCD can be established, in one true
limit or another.  The community of continuum strong QCD practitioners is a
large one and that is helped by the appealing simplicity of models, which
facilitates insightful contributions that are not labour intensive.

Our goal is to present an image of the gamut of strong QCD phenomena
currently at the forefront of the interaction between experiment and theory,
and of contemporary methods and their application.  Our portrayal will not be
complete but ought to serve as a conduit into this field.  We judge that the
best means to pursue our aim is to focus on a particular approach, making
connections and comparisons with others when possible and helpful.  All
methods have strengths and weaknesses, and, when modelling is involved, they
are not always obviously systematic.  That is the cost of leaving
perturbation theory behind.  They can also yield internally consistent
results that nevertheless are without fidelity to QCD.  Vigilance is
therefore necessary but herein lies another benefit of a leagued community.

Everyone has a favourite tool and the Dyson-Schwinger equations [DSEs] are
ours: they provide the primary medium for this discourse.  The framework is
appropriate here because the last decade has seen something of a renaissance
in its phenomenological application and we chronicle that herein.
Additionally, the DSEs have been applied simultaneously to phenomena as
apparently unconnected as low-energy $\pi \pi$ scattering, $B \to D^\ast$
decays and the equation of state for a quark gluon plasma, and hence they
provide a single framework that serves to conduct us through a wide range of
hadron physics.  Focusing on one approach is no impediment to a broad
perspective: the continuum, nonperturbative studies complement each other,
with agreement between them being a strong signal of a real effect.  In this
and their flexibility they provide a foil to the phlegmatic progress of
numerical simulations of lattice-QCD, and a fleet guide to unanticipated
phenomena.

To close this short introduction we present a list of abbreviations.  They
are defined in the text when first introduced, however, that point can be
difficult to locate.
\begin{tabbing}
DCSB \= \ldots\ \= dynamical chiral symmetry breaking \hspace*{1em}\= 
        DCSB \= \ldots\ \= dynamical chiral symmetry breaking \kill
BSA \> \ldots\ \> Bethe-Salpeter amplitude
        \> BSE \> \ldots\ \> Bethe-Salpeter equation\\
DCSB \> \ldots\ \> dynamical chiral symmetry breaking
        \> DSE \> \ldots\ \> Dyson-Schwinger equation \\
EOS \> \ldots\ \> equation of state 
        \> LHC \> \ldots\ \> large hadron collider \\
QCD \> \ldots\ \> quantum chromodynamics 
        \> QC$_2$D \> \ldots \> two-colour QCD\\
QED \> \ldots\ \> quantum electrodynamics 
        \> QED$_3$ \> \ldots\ \> three-dimensional QED \\
QGP \> \ldots\ \> quark gluon plasma 
        \> RHIC \> \ldots \> relativistic heavy ion collider
\end{tabbing}

\sect{Dyson-Schwinger Equations} 
\label{sectionDSEs}
In introducing the DSEs we find useful the brief text book discussions in
Chap.~10 of Ref.~\cite{iz80} and Chap.~2 of Ref.~\cite{rivers}, and
particularly their use in proving the renormalisability of quantum
electrodynamics [QED] in Ref.~\cite{bd65}, Chap.~19.  However, these sources
do not cover the use of DSEs in contemporary nuclear and high-energy physics.
Reference~\cite{cdragw} ameliorates that with a wide ranging review of the
theoretical and phenomenological applications extant when written, and it
provides a good foundation for us.

The DSEs are a nonperturbative means of analysing a quantum field theory.
Derived from a theory's Euclidean space generating functional, they are an
enumerable infinity of coupled integral equations whose solutions are the
$n$-point Schwinger functions [Euclidean Green functions], which are the same
matrix elements estimated in numerical simulations of lattice-QCD.  In
theories with elementary fermions, the simplest of the DSEs is the {\it gap}
equation, which is basic to studying dynamical symmetry breaking in systems
as disparate as ferromagnets, superconductors and QCD.  The gap equation is a
good example because it is familiar and has all the properties that
characterise each DSE: its solution is a $2$-point function [the fermion
propagator] while its kernel involves higher $n$-point functions; e.g., in a
gauge theory, the kernel is constructed from the gauge-boson $2$-point
function and fermion--gauge-boson vertex, a $3$-point function; a
weak-coupling expansion yields all the diagrams of perturbation theory; and
solved self-consistently, the solution of the gap equation exhibits
nonperturbative effects unobtainable at any finite order in perturbation
theory; e.g, dynamical symmetry breaking.

The coupling between equations; i.e., the fact that the equation for a given
$m$-point function always involves at least one $n>m$-point function,
necessitates a truncation of the tower of DSEs in order to define a tractable
problem.  One systematic and familiar truncation is a weak coupling expansion
to reproduce perturbation theory.  However, that precludes the study of
nonperturbative phenomena and hence something else is needed for the
investigation of strongly interacting systems, bound state phenomena and
phase transitions.

In analysing the ferromagnetic transition, the Hartree-Fock approximation
yields qualitatively reliable information and in QED and QCD its analogue:
the rainbow truncation, has proven efficacious.  However, {\it a priori} it
can be difficult to judge whether a given truncation will yield reliable
results and a systematic improvement is not always obvious.  It is here that
some model-dependence enters but that is not new, being typical in the study
of strongly-interacting few- and many-body systems.  To proceed with the DSEs
one just employs a truncation and explores its consequences, applying it to
different systems and constraining it, where possible, by comparisons with
experimental data and other theoretical approaches on their common domain.
In this way a reliable truncation can be identified and then attention paid
to understanding the keystone of its success and improving its foundation.
This pragmatic approach has proven rewarding in strong QCD, as we shall
describe.

\subsect{DSE Primer}
Lattice-QCD is defined in Euclidean space because the zero chemical potential
[$\mu=0$] Euclidean QCD action defines a probability measure, for which many
numerical simulation algorithms are available.  The Gaussian distribution:
\begin{equation}
{\cal K}_t(q,q^\prime)= (2\pi t)^{-3/2}\,\exp[-\left(q-q^\prime\right)^2/(2
t)]\,,
\end{equation}
defines the simplest probability measure: $dq^\prime \, {\cal
K}_t(q,q^\prime)$.  ${\cal K}_t$ is positive and normalisable, which allows
its interpretation as a probability density.  An heuristic exposition of
probability measures in quantum field theory can be found in
Ref.~\cite{rivers}, Chap.~6, while Ref.~\protect\cite{gj81}, Chaps.~3 and 6,
provides a more rigorous discussion in the context of quantum mechanics and
quantum field theory.  

Working in Euclidean space, however, is more than simply pragmatic: Euclidean
lattice field theory is currently a primary candidate for the rigorous
definition of an interacting quantum field theory~\cite{gj81,seiler} and that
relies on it being possible to {\it define} the generating functional via a
proper limiting procedure.  The moments of the measure; i.e., vacuum
expectation values of the fields, are the $n$-point Schwinger functions and
the quantum field theory is completely determined once all its Schwinger
functions are known.  The time-ordered Green functions of the associated
Minkowski space theory can be obtained in a well-defined fashion from the
Schwinger functions.  This is one reason why we employ a Euclidean
formulation.  Another is a desire to maintain contact with perturbation
theory where the renormalisation group equations for QCD and their solutions
are best understood~\cite{gross75}.

To make clear our conventions: for $4$-vectors $a$, $b$:
\begin{equation}
a\cdot b := a_\mu\,b_\nu\,\delta_{\mu\nu} := \sum_{i=1}^4\,a_i\,b_i\,,
\end{equation}
so that a spacelike vector, $Q_\mu$, has $Q^2>0$; our Dirac matrices are
Hermitian and defined by the algebra
\begin{equation}
\{\gamma_\mu,\gamma_\nu\} = 2\,\delta_{\mu\nu}\,;
\end{equation}
and we use
\begin{equation}
\gamma_5 := -\,\gamma_1\gamma_2\gamma_3\gamma_4
\end{equation}
so that
\begin{equation}
{\rm tr}\left[ \gamma_5 \gamma_\mu\gamma_\nu\gamma_\rho\gamma_\sigma \right] 
= - 4 \,\varepsilon_{\mu\nu\rho\sigma}\,,\;
\varepsilon_{1234}= 1\,.
\end{equation}
The Dirac-like representation of these matrices is:
\begin{equation}
\vec{\gamma}=\left(
\begin{array}{cc}
0 & -i\vec{\tau}  \\
i\vec{\tau} & 0
\end{array}
\right),\;
\gamma_4=\left(
\begin{array}{cc}
\tau^0 & 0 \\
0 & -\tau^0
\end{array}
\right),
\end{equation}
where the $2\times 2$ Pauli matrices are:
\begin{equation}
\label{PauliMs}
\tau^0 = \left(
\begin{array}{cc}
1 & 0 \\
0 & 1
\end{array}\right),\;
\tau^1 = \left(
\begin{array}{cc}
0 & 1 \\
1 & 0
\end{array}\right),\;
\tau^2 = \left(
\begin{array}{cc}
0 & -i \\
i & 0
\end{array}\right),\;
\tau^3 = \left(
\begin{array}{cc}
1 & 0 \\
0 & -1
\end{array}\right).
\end{equation}

Using these conventions the [unrenormalised] Euclidean QCD action is
\begin{equation}
\label{ActionQCD}
S[\bar q, q, A] = \
\int d^4x\,\left[\sfrac{1}{4}\,F_{\mu\nu}^a F_{\mu\nu}^a
+ \sfrac{1}{2\xi}\,\partial\cdot A^a\,\partial\cdot A^a
+ \sum_{f=1}^{N_f}\,\bar q_f\left(
        \gamma\cdot \partial  + m_f 
        + i g\, \sfrac{1}{2}\lambda^a\,\gamma\cdot A^a\right) q_f
 \right],
\end{equation}
where: $F_{\mu\nu}^a= \partial_\mu A_\nu^a - \partial_\nu A_\mu^a - g f^{abc}
A_\mu^b A_\nu^s$; $N_f$ is the number of quark flavours; $m_f$ are the
current-quark masses; $\{\lambda^a: a=1,\ldots,8 \}$ with
$[\lambda^a,\lambda^b] = 2 i f^{abc}\lambda^c$ are the Gell-Mann matrices for
$SU(3)$ colour; and $\xi$ is the covariant gauge fixing parameter.  The
generating functional follows:
\begin{equation}
\label{partitionZ}
{\cal Z}[\bar\eta,\eta,J]= \int\,d\mu(\bar q,q,A,\bar\omega,\omega)\,\exp\int
d^4x \left[ \bar q \,\eta + \bar \eta \,q + J_\mu^a \,A_\mu^a\right] ,
\end{equation}
with sources: $\bar \eta$, $\eta$, $J$, and a functional integral measure
\begin{eqnarray}
\lefteqn{d\mu(\bar q,q,A,\bar\omega,\omega) := }\\
&& \nonumber
\prod_x \prod_{\phi} {\cal D} \bar q_\phi(x) {\cal D} q_\phi(x) \prod_{a}
{\cal D} \bar \omega^a(x) {\cal D} \omega^a(x) \prod_{\mu} {\cal D}A_\mu^a(x)
\exp(-S[\bar q,q,A]-S^g[\bar\omega,\omega,A])\,,
\end{eqnarray}
where $\phi$ represents both the flavour and colour index of the quark field,
and $\bar\omega$ and $\omega$ are scalar, Grassmann [ghost] fields that are a
necessary addition in covariant gauges, and most other gauges too.  [Without
gauge fixing the action is constant along trajectories of gauge-equivalent
gluon field configurations, which leads to a gauge-orbit volume-divergence in
the continuum generating functional.]  The normalisation
\begin{equation}
{\cal Z}[\bar\eta=0,\eta=0,J=0] = 1
\end{equation}
is implicit in the measure.  

The ghosts only couple directly to the gauge field:
\begin{equation}
\label{ghostaction}
S_g[\bar\omega,\omega,A] = \int d^4x\,\left[
-\partial_\mu \bar\omega^a\, \partial_\mu \omega^a - g
f^{abc}\,\partial_\mu\bar\omega^a \,\omega^b\,A_\mu^c
\right]\,,
\end{equation}
and restore unitarity in the subspace of transverse [physical] gauge fields.
Practically, the normalisation means that ghost fields are unnecessary in the
calculation of gauge invariant observables using lattice-regularised QCD
because the gauge-orbit volume-divergence in the generating functional,
associated with the uncountable infinity of gauge-equivalent gluon field
configurations in the continuum, is rendered finite by the simple expedient
of only summing over a finite number of configurations.

It is not necessary to employ the covariant gauge fixing condition in
constructing the measure although it does yield DSEs with a simple form.
Indeed a rigorous definition of the measure may require a gauge fixing
functional that either completely eliminates Gribov copies or restricts the
functional integration domain to a subspace without them.  Concerned, as they
are, with unobservable degrees of freedom, such modifications of the measure
cannot directly affect the colour-singlet $n$-point functions that describe
physical observables.  However, they have the capacity to modify the infrared
behaviour of coloured $n$-point functions~\cite{zwanzigerold,zwanziger} and
thereby influence the manner in which we intuitively understand observable
effects.

The DSEs are derived from the generating functional using the elementary
observation that, with sensible Euclidean space boundary conditions, the
integral of a total derivative is zero:
\begin{equation}
0 = \int\,\frac{\delta}{\delta A}\,d\mu(\bar q,q,A,\bar\omega,\omega)\,
\exp\int d^4x \left[ \bar q \,\eta + \bar \eta \,q + J_\mu^a
\,A_\mu^a\right].
\end{equation}
Examples of such derivations: the QED gap equation, the equation for the
photon vacuum polarisation and that for the fermion-fermion scattering matrix
[a $4$-point function], are presented in Ref.~\cite{iz80}, Chap.~10.  The
scattering matrix is important because it lies at the heart of two-body bound
state studies in quantum field theory, being a keystone of the Bethe-Salpeter
equation.  Section~2.1 of Ref.~\cite{cdragw} repeats the first two
derivations and also makes explicit the effects of renormalisation.  Herein
we simply begin with the relevant DSE leaving its derivation as an exercise.
That too can be side-stepped if a Minkowski space version of the desired DSE
is at hand.  Then one need merely employ the {\it transcription} {\it rules}:
\begin{center}
\parbox{132mm}{
\parbox{60mm}{Configuration Space
\begin{enumerate}
\item $\displaystyle \int^M\,d^4x^M \, \rightarrow \,-i \int^E\,d^4x^E$
\item $\slash\!\!\! \partial \,\rightarrow \, i\gamma^E\cdot \partial^E $
\item $\slash \!\!\!\! A \, \rightarrow\, -i\gamma^E\cdot A^E$
\item $A_\mu B^\mu\,\rightarrow\,-A^E\cdot B^E$
\item $x^\mu\partial_\mu \to x^E\cdot \partial^E$
\end{enumerate}}\hspace*{10mm}
\parbox{60mm}{Momentum Space
\begin{enumerate}
\item $\displaystyle \int^M\,d^4k^M \, \rightarrow \,i \int^E\,d^4k^E$
\item $\slash\!\!\! k \,\rightarrow \, -i\gamma^E\cdot k^E $
\item $\slash \!\!\!\! A \, \rightarrow\, -i\gamma^E\cdot A^E$
\item $k_\mu q^\mu \, \rightarrow\, - k^E\cdot q^E$
\item $k_\mu x^\mu\,\rightarrow\,-k^E\cdot x^E$
\end{enumerate}}}
\end{center}
These rules are valid in perturbation theory; i.e., the correct Minkowski
space integral for a given diagram will be obtained by applying these rules
to the Euclidean integral: they take account of the change of variables and
rotation of the contour.  However, for the diagrams that represent DSEs,
which involve dressed $n$-point functions whose analytic structure is not
known {\it a priori}, the Minkowski space equation obtained using this
prescription will have the right appearance but it's solutions may bear no
relation to the analytic continuation of the solution of the Euclidean
equation.

\subsect{Core Issues}
The use of DSEs as a unifying, phenomenological tool in QCD has grown much
since the publication of Ref.~\cite{cdragw}, and Ref.~\cite{revpeter} reviews
applications to electromagnetic and hadronic interactions of light mesons and
the connection to the successful Nambu-Jona-Lasinio~\cite{vw91,njltwo,sk92}
and Global-Colour~\cite{gcm98} models.  In most of their widespread
applications these two models yield results kindred to those obtained with
the rainbow-ladder DSE truncation.

\addcontentsline{toc}{subsubsection}{\protect\numberline{ } {Quark DSE [Gap
Equation]}}
\subsubsect{Quark DSE [Gap Equation]} 
\label{mr97sect}
The idea that the fermion DSE, or gap equation, can be used to study
dynamical chiral symmetry breaking [DCSB] perhaps began with a study of the
electron propagator~\cite{bjw64}.  Since that analysis there have been many
studies of dynamical mass generation in QED, and those reviewed in
Ref.~\cite{cdragw} establish the context: there is some simplicity in dealing
with an Abelian theory.  Continuing research has led to an incipient
understanding of the role played by multiplicative renormalisability and
gauge covariance in constraining the dressed-fermion-photon
vertex~\cite{mrpAdelaide}, and progress in this direction provides intuitive
guidance for moving beyond the rainbow truncation.  However, the studies have
also made plain the difficulty in defining the chiral limit of a theory
without asymptotic freedom.  Renormalised, strong coupling, quenched QED
yields a scalar self energy for the electron that is not positive-definite:
damped oscillations \label{`oscillations'} appear after the renormalisation
point~\cite{fred94}.  [Quenched $=$ bare photon propagator in the gap
equation.  It is the simplest and most widely explored truncation of the
gauge sector.]  This pathology can be interpreted as a signal that four
fermion operators $\sim [\bar\psi(x)\psi(x)]^2$ have acquired a large
anomalous dimension and have thus become relevant operators for the range of
gauge couplings that support DCSB.  That hypothesis has implications for the
triviality of the theory, and is reviewed and explored in
Refs.~\cite{cdragw,manuel}.  Nonperturbative QED remains an instructive
challenge.

The study of DCSB in QCD is much simpler primarily because the chiral limit
is well-defined.  We discuss that now using the renormalised quark-DSE:
\begin{eqnarray}
\label{gendse}
S(p)^{-1} & = & Z_2 \,(i\gamma\cdot p + m_{\rm bare})
+\, Z_1 \int^\Lambda_q \,
g^2 D_{\mu\nu}(p-q) \frac{\lambda^a}{2}\gamma_\mu S(q)
\Gamma^a_\nu(q,p) \,,
\end{eqnarray}
where $D_{\mu\nu}(k)$ is the renormalised dressed-gluon propagator,
$\Gamma^a_\nu(q;p)$ is the renormalised dressed-quark-gluon vertex, $m_{\rm
bare}$ is the $\Lambda$-dependent current-quark bare mass that appears in the
Lagrangian and $\int^\Lambda_q := \int^\Lambda d^4 q/(2\pi)^4$ represents
mnemonically a {\em translationally-invariant} regularisation of the
integral, with $\Lambda$ the regularisation mass-scale.  The final stage of
any calculation is to remove the regularisation by taking the limit $\Lambda
\to \infty$.  Using a translationally invariant regularisation makes possible
the preservation of Ward-Takahashi identities, which is crucial in studying
DCSB, for example.  One implementation well-suited to a nonperturbative
solution of the DSE is Pauli-Villars regularisation, which has the quark
interacting with an additional massive gluon-like vector boson:
\mbox{mass$\,\sim \Lambda$}, that decouples as $\Lambda\to
\infty$~\cite{mr97}.  An alternative is a numerical implementation of
dimensional regularisation, which, although more cumbersome, can provide the
necessary check of scheme-independence~\cite{sizer}.

In Eq.~(\ref{gendse}), $Z_1(\zeta^2,\Lambda^2)$ and $Z_2(\zeta^2,\Lambda^2)$
are the quark-gluon-vertex and quark wave function renormalisation constants,
which depend on the renormalisation point, $\zeta$, and the regularisation
mass-scale, as does the mass renormalisation constant
\begin{equation}
\label{Zmass}
Z_m(\zeta^2,\Lambda^2) = Z_4(\zeta^2,\Lambda^2)/Z_2(\zeta^2,\Lambda^2) ,
\end{equation}
with the renormalised mass given by
\begin{equation}
\label{mzeta}
m(\zeta) := m_{\rm bare}(\Lambda)/Z_m(\zeta^2,\Lambda^2).
\end{equation}
Although we have suppressed the flavour label, $S$, $\Gamma^a_\mu$ and
$m_{\rm bare}$ depend on it.  However, one can always use a
flavour-independent renormalisation scheme, which we assume herein, and hence
all the renormalisation constants are flavour-independent~\cite{mr97,kusaka}.

The solution of Eq.~(\ref{gendse}) has the form
\begin{equation}
\label{sinvp}
S(p)^{-1} = i \gamma\cdot p \,A(p^2,\zeta^2) + B(p^2,\zeta^2)
        = \frac{1}{Z(p^2,\zeta^2)}\left[ i\gamma\cdot p +
        M(p^2,\zeta^2)\right]\,.
\end{equation}
The functions $A(p^2,\zeta^2)$, $B(p^2,\zeta^2)$ embody the effects of vector
and scalar quark-dressing induced by the quark's interaction with its own
gluon field.  The ratio: $M(p^2,\zeta^2)$, is the quark mass function and a
pole mass would be the solution of
\begin{equation}
m_{\rm pole}^2 - M^2(p^2=-m_{\rm pole}^2,\zeta^2) = 0.
\end{equation}
A widely posed conjecture is that confinement rules out a solution of this
equation~\cite{gastao}.  We discuss this further below.

Equation~(\ref{gendse}) must be solved subject to a renormalisation
[boundary] condition, and because the theory is asymptotically free it is
practical and useful to impose the requirement that at a large spacelike
$\zeta^2$
\begin{equation}
\label{renormS}
\left.S(p)^{-1}\right|_{p^2=\zeta^2} = i\gamma\cdot p + m(\zeta)\,,
\end{equation}
where $m(\zeta)$ is the renormalised current-quark mass at the scale $\zeta$.
By ``large'' here we mean $\zeta^2 \gg \Lambda_{\rm QCD}^2$ so that in
quantitative, model studies extensive use can be made of matching with the
results of perturbation theory.  It is the ultraviolet stability of QCD;
i.e., the fact that perturbation theory is valid at large spacelike momenta,
that makes possible a straightforward definition of the chiral limit.  It
also provides the starkest contrast to strong coupling QED.

Multiplicative renormalisability in gauge theories entails that
\begin{equation}
\frac{A(p^2,\zeta^2)}{A(p^2,\bar\zeta^2)\rule{0ex}{2ex}}
= \frac{Z_2(\zeta^2,\Lambda^2)}{Z_2(\bar\zeta^2,\Lambda^2)\rule{0ex}{2ex}}
= A(\bar\zeta^2,\zeta^2) 
= \frac{1}{A(\zeta^2,\bar\zeta^2)\rule{0ex}{2ex}}
\end{equation}
and beginning with Ref.~\cite{cp90} this relation has been used efficaciously
to build realistic {\it Ans\"atze} for the fermion--photon vertex in quenched
QED.  A systematic approach to such nonperturbative improvements is
developing~\cite{dong,bashir} and these improvements continue to provide
intuitive guidance in QED, where they complement the perturbative calculation
of the vertex~\cite{ayse}.  They are also useful in exploring model
dependence in QCD studies.

At one loop in QCD perturbation theory
\begin{equation}
\label{z2mu}
Z_2(\zeta^2,\Lambda^2) 
= \left[\frac{ \alpha(\Lambda^2) }{\alpha(\zeta^2)}
        \right]^{-\gamma_F/\beta_1}\!,\;
\gamma_F= \sfrac{2}{3}\xi,\;
\beta_1= \sfrac{1}{3}N_f - \sfrac{11}{2},
\end{equation}
and at this order the running strong coupling is
\begin{equation}
\label{alphaq2}
\alpha(\zeta^2) = \frac{\pi} {\rule{0mm}{1.2\baselineskip} -\sfrac{1}{2}
             \beta_1 \ln\left[\zeta^2/\Lambda_{\rm QCD}^2\right]}\,.
\end{equation}
In Landau gauge: $\xi = 0$, so $Z_2 \equiv 1$ at one loop order.  This, plus
the fact that Landau gauge is a fixed point of the renormalisation group
[Eq.~(\ref{landau})], makes it the most useful covariant gauge for model
studies.  It also underlies the quantitative accuracy of Landau gauge,
rainbow truncation estimates of the critical coupling in strong
QED~\cite{mrpAdelaide,adnan}.  In a self consistent solution of
Eq.~(\ref{gendse}), $Z_2 \neq 1$ even in Landau gauge but, at large
$\zeta^2$, the $\zeta$-dependence is very weak.  However, as will become
evident, in studies of realistic QCD models this dependence becomes
significant for $\zeta^2 \lsim 1$--$2\,$GeV$^2$, and is driven by the same
effect that causes DCSB.

The dressed-quark mass function:
$M(p^2,\zeta^2)=B(p^2,\zeta^2)/A(p^2,\zeta^2)$, is independent of the
renormalisation point; i.e., with $\zeta\neq \bar\zeta$
\begin{equation}
\label{Mrpi}
M(p^2,\zeta^2) = M(p^2,\bar\zeta^2):= M(p^2)\,, \; \forall\, p^2\,:
\end{equation}
it is a function only of $p^2/\Lambda_{\rm QCD}^2$, which is another
constraint on models.  At one loop order the running [or renormalised] mass
\begin{equation}
\label{masanom}
m(\zeta) = M(\zeta^2) = \frac{\hat m} {\left(\rule{0mm}{1.2\baselineskip}
\sfrac{1}{2}\ln\left[\zeta^2/\Lambda_{\rm QCD}^2
\right]\right)^{\gamma_m}}\,,\; \gamma_m= 12/(33-2 N_f)\,,
\end{equation}
where $\hat m$ is the renormalisation point independent current-quark mass,
and the mass renormalisation constant is, Eq.~(\ref{Zmass}),
\begin{equation}
\label{zmdef}
Z_m(\zeta^2,\Lambda^2) 
= \left[\frac{ \alpha(\Lambda^2) }{\alpha(\zeta^2)}\right]^{\gamma_m}\,.
\end{equation}
The mass anomalous dimension, $\gamma_m$, is independent of the gauge
parameter to all orders perturbation theory and for two different quark
flavours the ratio: $m_{f_1}(\zeta) / m_{f_2}(\zeta) = \hat m_{f_1}/\hat
m_{f_2}$, which is independent of the renormalisation point and of the
renormalisation scheme.  The chiral limit is unambiguously defined by
\begin{equation}
\label{chirallimit}
{\bf chiral~limit}:\; \hat m = 0\,.
\end{equation}
In this case there is no perturbative contribution to the scalar piece of the
quark self energy; i.e., $B(p^2,\zeta^2)\equiv 0$ at every order in
perturbation theory and in fact there is no scalar mass-like divergence in
the calculation of the self energy.  This is manifest in the quark DSE, with
Eq.~(\ref{gendse}) yielding, in addition to the perturbative result:
$B(p^2,\zeta^2)\equiv 0$, a solution
$M(p^2)=B(p^2,\zeta^2)/A(p^2,\zeta^2)\neq 0$ that is power-law suppressed in
the ultraviolet: $M(p^2) \sim 1/p^2$, guaranteeing convergence of the
associated integral without subtraction\label{`without subtraction'}.
Dynamical chiral symmetry breaking is
\begin{equation}
{\bf DCSB:}\;M(p^2)\neq 0\,\; {\rm when} \;\hat m = 0\,.
\end{equation}
As we shall see, in QCD this is possible if and only if the quark condensate
is nonzero.  The criteria are equivalent.

The solution of the quark DSE depends on the anatomy of the dressed-gluon
propagator, which in concert with the dressed-quark-gluon vertex encodes in
Eq.~(\ref{gendse}) all effects of the quark-quark interaction.  In a
covariant gauge the renormalised propagator is
\begin{equation}
\label{gluoncovariant}
D_{\mu\nu}(k) = \left( \delta_{\mu\nu} - \frac{k_\mu k_\nu}{k^2}\right)
                \frac{d(k^2,\zeta^2)}{k^2} + \xi\,\frac{k_\mu k_\nu}{k^4}\,,
\end{equation}
where $d(k^2,\zeta^2) = 1/[1+\Pi(k^2,\zeta^2)]$, with $\Pi(k^2,\zeta^2)$ the
renormalised gluon vacuum polarisation for which the conventional
renormalisation condition is 
\begin{equation}
\label{renormPi}
\Pi(\zeta^2,\zeta^2) = 0\,;\; {\rm i.e.,}\; d(\zeta^2,\zeta^2) =1\,.
\end{equation}
For the dressed gluon propagator, multiplicative renormalisability entails
\begin{equation}
\label{mrPi}
\frac{d(k^2,\zeta^2)}{d(k^2,\bar\zeta^2)\rule{0ex}{2ex}}
= \frac{Z_3(\bar\zeta^2,\Lambda^2)}{Z_3(\zeta^2,\Lambda^2)\rule{0ex}{2ex}}
= d(\zeta^2,\bar\zeta^2)
= \frac{1}{d(\bar\zeta^2,\zeta^2)\rule{0ex}{2ex}}\,,
\end{equation}
and at one loop in perturbation theory
\begin{equation}
\label{z3mu}
Z_3(\zeta^2,\Lambda^2) 
= \left[\frac{ \alpha(\Lambda^2) }{\alpha(\zeta^2)}
        \right]^{-\gamma_1/\beta_1},\;
\gamma_1 = \sfrac{1}{3}N_f - \sfrac{1}{4}(13 - 3\,\xi)\,.
\end{equation}
The gauge parameter is also renormalisation point dependent; i.e., the
renormalised theory has a running gauge parameter.  However, because of
Becchi-Rouet-Stora [BRST or gauge] invariance, there is no new dynamical
information in that: its evolution is completely determined by the gluon wave
function renormalisation constant
\begin{equation}
\xi(\zeta^2) = Z_3^{-1}(\zeta^2,\Lambda^2)\,\xi_{\rm bare}(\Lambda)\,.
\end{equation}
One can express $\xi(\zeta^2)$ in terms of a renormalisation point invariant
gauge parameter: $\hat \xi$, which is an overall multiplicative factor in the
formula and hence
\begin{equation}
\label{landau}
{\bf Landau~Gauge:}\; \hat \xi = 0 \Rightarrow \xi(\zeta^2)\equiv 0
\end{equation}
at all orders in perturbation theory; i.e., Landau gauge is a fixed point of
the renormalisation group.

The renormalised dressed-quark-gluon vertex has the form
\begin{equation}
\Gamma^a_\nu(k,p) = \frac{\lambda^a}{2} \Gamma_\nu(k,p)\,;
\end{equation}
i.e., the colour matrix structure factorises.  It is a fully amputated
vertex, which means all the analytic structure associated with elementary
excitations has been eliminated.  To discuss this further we introduce the
notion of a particle-like singularity.  It is one of the form: $P= k-p$,
\begin{equation}
\frac{1}{(P^2+b^2)^{x}},\; x\in(0,1]\,.
\end{equation}
If the vertex possesses such a singularity then it can be expressed in terms
of non-negative spectral densities, which is impossible if $x>1$.  $x=1$ is
the ideal case of an isolated $\delta$-function distribution in the spectral
densities and hence an isolated free-particle pole.  $x\in(0,1)$ corresponds
to an accumulation at the particle pole of branch points associated with
multiparticle production, as occurs with the electron propagator in QED
because of photon dressing.  

The vertex is a fully amputated $3$-point function.  Hence, the presence of
such a singularity entails the existence of a flavour singlet composite
(quark-antiquark bound state) with colour octet quantum numbers and mass
$m=b$.  The bound state amplitude follows immediately from the associated
homogeneous Bethe-Salpeter equation, which the singularity allows one to
derive.  Such an excitation must not exist as an asymptotic state, that would
violate the observational evidence of confinement, and any modelling of
$\Gamma_\mu^a(k,p)$ ought to be consistent with this.

Expressing the Dirac structure of $\Gamma_\nu(k,p)$ requires $12$ independent
scalar functions:
\begin{equation}
\Gamma_\nu(k,p) = \gamma_\nu \,F_1(k,p,\zeta) + \ldots\,,
\end{equation}
which for our purposes it is not necessary to reproduce fully.  A pedagogical
discussion of the perturbative calculation of $\Gamma_\nu(k,p)$ can be found
in Ref.~\cite{pt84} while Refs.~\cite{bc80,vertex} explore its
nonperturbative structure and properties.  We only make $F_1(k,p,\zeta)$
explicit because the renormalisability of QCD entails that it alone is
ultraviolet divergent.  Defining
\begin{equation}
f_1(k^2,\zeta^2):= F_1(k,-k,\zeta)\,,
\end{equation}
the conventional renormalisation boundary condition is
\begin{equation}
f_1(\zeta^2,\zeta^2) = 1\,,
\end{equation}
which is practical because QCD is asymptotically free.  Multiplicative
renormalisability entails
\begin{equation}
\frac{f_1(k^2,\zeta^2)}{f_1(k^2,\bar\zeta^2)\rule{0ex}{2ex}}
= \frac{Z_1(\zeta^2,\Lambda^2)}{Z_1(\bar\zeta^2,\Lambda^2)\rule{0ex}{2ex}}
= f_1(\bar\zeta^2,\zeta^2)
= \frac{1}{f_1(\zeta^2,\bar\zeta^2)\rule{0ex}{2ex}}\,,
\end{equation}
and at one loop order
\begin{equation}
\label{z1mu}
Z_1(\mu^2,\Lambda^2) = \left[
                \frac{\alpha(\Lambda^2)}{\alpha(\mu^2)}
                        \right]^{-\gamma_\Gamma/\beta_1}\,,\;
\gamma_\Gamma = \sfrac{1}{2}[ \sfrac{3}{4} (3+\xi) + \sfrac{4}{3}\xi]\,.
\end{equation}

\addcontentsline{toc}{subsubsection}{\protect\numberline{ } {Dynamical Chiral
Symmetry Breaking}} \subsubsect{Dynamical Chiral Symmetry Breaking}
At this point each element in the quark DSE, Eq.~(\ref{gendse}), is defined,
with some of their perturbative properties elucidated, and the question is
how does that provide an understanding of DCSB?  It is best answered using an
example, in which the model-independent aspects are made clear.

The quark DSE is an integral equation and hence its elements must be known at
all values of their momentum arguments, not just in the perturbative domain
but also in the infrared.  While the gluon propagator and quark-gluon vertex
each satisfy their own DSE, that couples the quark DSE to other members of
the tower of equations and hinders rather than helps in solving the gap
equation.  Therefore here, as with all applications of the gap equation, one
employs {\it Ans\"atze} for the interaction elements [$D_{\mu\nu}(k)$ and
$\Gamma_\nu(k,p)$], constrained as much and on as large a domain as possible.
This approach has a long history in exploring QCD and we illustrate it using
the model of Ref.~\cite{mr97}.

The renormalised dressed-ladder truncation of the quark-antiquark scattering
kernel [$4$-point function] is
\begin{equation}
\bar K(p,q;P)_{tu}^{rs} = 
g^2(\zeta^2)\, D_{\mu\nu}(p-q) \,
\left[\rule{0mm}{0.7\baselineskip} \Gamma^a_\mu(p_+,q_+)\,S(q_+) \right]_{tr} 
\,
\left[ \rule{0mm}{0.7\baselineskip}S(q_-)\,\Gamma^a_\nu(q_-,p_-) \right]_{su},
\end{equation}
where $p_\pm=p \,\pm \,P/2$, $q_\pm=q \,\pm \,P/2$, with $P$ the total
momentum of the quark-antiquark pair, and although we use it now we have
suppressed the $\zeta$-dependence of the Schwinger functions.  From
Eqs.~(\ref{renormPi}-\ref{z3mu}) it follows that for $Q^2:=(p-q)^2$ large and
spacelike
\begin{equation}
d(Q^2,\zeta^2)= \frac{Z_3(\zeta^2,\Lambda^2)}{Z_3(Q^2,\Lambda^2)}
\,d(\zeta^2,\zeta^2) 
= \left[\frac{\alpha(Q^2)}{\alpha(\zeta^2)}\right]^{\gamma_1/\beta_1}
\Rightarrow D_{\mu\nu}(p-q) = 
\left[\frac{\alpha(Q^2)}{\alpha(\zeta^2)}\right]^{\gamma_1/\beta_1}\,
\!\!\!D_{\mu\nu}^{\rm free}(p-q)\,.
\end{equation}
Using this and analogous results for the other Schwinger functions then on
the kinematic domain for which $Q^2 \sim p^2\sim q^2$ is large and spacelike
[$g^2(\zeta^2):= 4\pi\alpha(\zeta^2)$]
\begin{equation}
\bar K(p,q;P)_{tu}^{rs} \approx 
4\pi\alpha(Q^2)\, D_{\mu\nu}^{\rm free}(p-q) \,
\left[\rule{0mm}{0.7\baselineskip} \sfrac{1}{2}\lambda^a\gamma_\mu\, S^{\rm
free}(q_+) \right]_{tr}
\, \left[ \rule{0mm}{0.7\baselineskip}S^{\rm
free}(q_-)\,\sfrac{1}{2}\lambda^a\gamma_\nu \right]_{su},
\end{equation}
because Eqs.~(\ref{z2mu}), (\ref{z3mu}) and (\ref{z1mu}) yield
\begin{equation}
\label{sumanom}
\frac{2\, \gamma_F}{\beta_1} + \frac{\gamma_1}{\beta_1}  -  \frac{2\,
\gamma_\Gamma}{\beta_1} = 1\,. 
\end{equation}

This is one way of understanding the origin of an often used {\it Ansatz} in
studies of the gap equation; i.e., 
making the replacement
\begin{equation}
\label{abapprox}
g^2 D_{\mu\nu}(k) \to 4\pi\,\alpha(k^2) \,D_{\mu\nu}^{\rm free}(k)
\end{equation}
in Eq.~(\ref{gendse}), and using the ``rainbow truncation:''
\begin{equation}
\label{rainbow}
\Gamma_\nu(q,p)=\gamma_\nu \, .
\end{equation}
Equation~(\ref{abapprox}) is often described as the ``Abelian approximation''
because the left- and right-hand-sides [r.h.s.] are {\it equal} in QED.  In
QCD, equality between the two sides cannot be obtained easily by a selective
resummation of diagrams.  As reviewed in Ref.~\cite{cdragw}, Eqs.~(5.1-5.8),
it can only be achieved by enforcing equality between the renormalisation
constants for the ghost-gluon vertex and ghost wave function: $\tilde
Z_1=\tilde Z_3$.  A mutually consistent constraint, which follows formally
from $\tilde Z_1=\tilde Z_3$, is to enforce the Abelian Ward identity: $Z_1 =
Z_2$.  At one-loop this corresponds to neglecting the contribution of the
3-gluon vertex to $\Gamma_\nu$, in which case $\gamma_\Gamma \to
\sfrac{2}{3}\xi = \gamma_F$.  This additional constraint provides the basis
for extensions of Eq.~(\ref{rainbow}); i.e., using {\it Ans\"atze} for
$\Gamma_\nu$ that are consistent with the QED vector Ward-Takahashi identity;
e.g., Ref.~\cite{fredthesis}.

Arguments such as these inspire the following {\it Ansatz} for the kernel in
Eq.~(\ref{gendse})~\cite{mr97}:
\begin{equation}
\label{ouransatz}
Z_1\, \int^\Lambda_q \, g^2 D_{\mu\nu}(p-q) \frac{\lambda^a}{2}\gamma_\mu
S(q) \Gamma^a_\nu(q,p) \to \int^\Lambda_q \, {\cal G}((p-q)^2)\,
D_{\mu\nu}^{\rm free}(p-q) \frac{\lambda^a}{2}\gamma_\mu S(q)
\frac{\lambda^a}{2}\gamma_\nu \,,
\end{equation}
with the ultraviolet behaviour of the ``effective coupling:'' ${\cal
G}(k^2)$, fixed by that of the strong running coupling.  Since it is not
possible to calculate $Z_1$ nonperturbatively without analysing the DSE for
the dressed-quark-gluon vertex, this {\it Ansatz} absorbs it in the model
effective coupling.

Equation~(\ref{ouransatz}) is a model for the product of the
dressed-propagator and dressed-vertex and its definition is complete once the
behaviour of ${\cal G}(k^2)$ in the infrared is specified; i.e., for $k^2
\lsim 1$-$2\,$GeV$^2$.  Reference~\cite{mr97} used
\begin{equation}
\label{gk2}
\frac{{\cal G}(k^2)}{k^2} = 8\pi^4 D \delta^4(k) + \frac{4\pi^2}{\omega^6} D
k^2 {\rm e}^{-k^2/\omega^2} + 4\pi\,\frac{ \gamma_m \pi} {\sfrac{1}{2}
\ln\left[\tau + \left(1 + k^2/\Lambda_{\rm QCD}^2\right)^2\right]} {\cal
F}(k^2) \,,
\end{equation}
with ${\cal F}(k^2)= [1 - \exp(-k^2/[4 m_t^2])]/k^2$ and $\tau={\rm e}^2-1$.
For $N_f=4$, $\Lambda_{\rm QCD}^{N_f=4}= 0.234\,{\rm GeV}$.  The qualitative
features of Eq.~(\ref{gk2}) are plain.  The first term is an integrable
infrared singularity~\cite{mn83} and the second is a finite-width
approximation to $\delta^4(k)$, normalised such that it has the same $\int
d^4k$ as the first term.  In this way the infrared strength is split into the
sum of a zero-width and a finite-width piece.  The last term in
Eq.~(\ref{gk2}) is proportional to $\alpha(k^2)/k^2$ at large spacelike-$k^2$
and has no singularity on the real-$k^2$ axis.

There are ostensibly three parameters in Eq.~(\ref{gk2}): $D$, $\omega$ and
$m_t$.  However, in Ref.~\cite{mr97} the authors fixed
$\omega=0.3\,$GeV$(=1/[.66\,{\rm fm}])$ and $m_t=0.5\,$GeV$(=1/[.39\,{\rm
fm}])$, and only varied $D$ and the renormalised $u=d$- and $s$-current-quark
masses in an attempt to obtain a good description of low-energy $\pi$- and
$K$-meson properties, using a renormalisation point $\zeta=19\,$GeV that is
large enough to be in the perturbative domain.  [The numerical values of
$\omega$ and $m_t$ are chosen so as to ensure that ${\cal G}(k^2)\approx 4\pi
\alpha(k^2)$ for $k^2> 2\,$GeV$^2$.  Minor variations in $\omega$ and
$m_t$ can be compensated by small changes in $D$.]  Such a procedure could
self-consistently yield $D=0$, which would indicate that agreement with
observable phenomena precludes an infrared enhancement in the effective
interaction.  However, that was not the case and a good fit required
\begin{equation}
\label{Dvalue}
D= (0.884\,{\rm GeV})^2\,,
\end{equation}
with renormalised current-quark masses
\begin{equation}
\label{params}
\begin{array}{cc}
m_{u,d}(\zeta) = 3.74\,{\rm MeV}\,,\; &
m_s(\zeta) = 82.5\,{\rm MeV}\,,
\end{array}
\end{equation}
which are in the ratio $1\,$:$\,22$, and yielded, in MeV,
\begin{equation}
\begin{array}{l|cccc}
         & m_\pi & m_K & f_\pi & f_K \\\hline
{\rm Calc.\,~\protect\cite{mr97}} & 139 & 497 & 131 & 154 \\
{\rm Expt.~\protect\cite{pdg98}} & 139 & 496 & 131 & 160
\end{array}
\end{equation}
and other quantities to be described below.  An explanation of how this fit
was accomplished requires a discussion of the homogeneous Bethe-Salpeter
equation, which we postpone.  Here we instead focus on describing the
properties of the DSE solution obtained with these parameter values.

Using Eqs.~(\ref{gendse}-\ref{mzeta}) and (\ref{ouransatz}) the gap equation
can be written
\begin{eqnarray}
\label{dsemod}
S(p,\zeta)^{-1} & = & Z_2\, i\gamma\cdot p + Z_4\, m(\zeta) + \Sigma^\prime
(p,\Lambda)\,,
\end{eqnarray}
with the regularised quark self energy
\begin{eqnarray}
\label{sigmod}
\Sigma^\prime(p,\Lambda) & := & \int^\Lambda_q \, {\cal G}((p-q)^2)\,
D_{\mu\nu}^{\rm free}(p-q) \frac{\lambda^a}{2}\gamma_\mu S(q)
\frac{\lambda^a}{2}\gamma_\nu \,.
\end{eqnarray}
When $\hat m \neq 0$ the renormalisation condition, Eq.~(\ref{renormS}), is
straightforward to implement.  Writing
\begin{equation}
\Sigma^\prime(p,\Lambda) := i \gamma\cdot p \, \left(
                A^\prime(p^2,\Lambda^2) - 1\right) +
                B^\prime(p^2,\Lambda^2)\,,
\end{equation}
which emphasises that these functions depend on the regularisation
mass-scale, $\Lambda$, Eq.~(\ref{renormS}) entails
\begin{equation}
\label{z2def}
Z_2(\zeta^2,\Lambda^2) = 2 - A^\prime(\zeta^2,\Lambda^2)
\;\; {\rm and} \;\;
m(\zeta) = Z_2(\zeta^2,\Lambda^2)\,m_{\rm bm}(\Lambda^2) + 
        B^\prime(\zeta^2,\Lambda^2)
\end{equation}
so that
\begin{equation}
\label{arenbren}
A(p^2,\zeta^2)  =  1 
        +  A^\prime(p^2,\Lambda^2) 
        - A^\prime(\zeta^2,\Lambda^2)\,,\;\;
B(p^2,\zeta^2)  =  m(\zeta) 
        +  B^\prime(p^2,\Lambda^2) 
        - B^\prime(\zeta^2,\Lambda^2)\,.
\end{equation}
Multiplicative renormalisability requires that having fixed the solutions at
a single renormalisation point, $\zeta$, their form at another point,
$\bar\zeta$, is given by
\begin{equation}
S^{-1}(p,\bar\zeta) = i \gamma\cdot p\, A(p^2,\bar\zeta^2) + B(p^2,\bar\zeta^2)
        = \frac{Z_2(\bar\zeta^2,\Lambda^2)}{Z_2(\zeta^2,\Lambda^2)}
                S^{-1}(p,\zeta)\,.
\end{equation}
This feature is evident in the solutions obtained in Ref.~\cite{mr97}.  It
means that, in evolving the renormalisation point to $\bar\zeta$, the ``1''
in Eqs.~(\ref{arenbren}) is replaced by
$Z_2(\bar\zeta^2,\Lambda^2)/Z_2(\zeta^2,\Lambda^2)$, and the ``$m(\zeta)$''
by $m(\bar\zeta)$; i.e., the ``seeds'' in the integral equation evolve
according to the QCD renormalisation group. This is why Eq.~(\ref{ouransatz})
is called a ``renormalisation-group-improved rainbow truncation.''

Turning to the chiral limit, it follows from Eqs.~(\ref{Zmass}),
(\ref{mzeta}), (\ref{masanom}) and (\ref{chirallimit}) that for $\hat m = 0$
\begin{equation}
\label{chiralA}
Z_2(\zeta^2,\Lambda^2) \,m_{\rm bare}(\Lambda^2)=0\,,\;\forall \Lambda\,.
\end{equation}
Hence, as remarked on page~\pageref{`without subtraction'}, there is no
subtraction in the equation for $B(p^2,\zeta^2)$; i.e., Eq.~(\ref{arenbren})
becomes
\begin{equation}
B(p^2,\zeta^2)  =  B^\prime(p^2,\Lambda^2) \,,\;
\lim_{\Lambda\to \infty} B^\prime(p^2,\Lambda^2) < \infty\,,
\end{equation}
which is only possible if the mass function is at least $1/p^2$-suppressed.
This is not the case in quenched strong coupling QED, where the mass function
behaves as $\sim \cos({\rm const.}\,\ln
[p^2/\zeta^2])/(p^2/\zeta^2)^{1/2}$~\cite{fukuda,cahillQED}, and that is the
origin of the complications indicated on
page~\pageref{`oscillations'}~\cite{cdragw,manuel,sizer}.

\begin{figure}[t]
\centering{\ \epsfig{figure=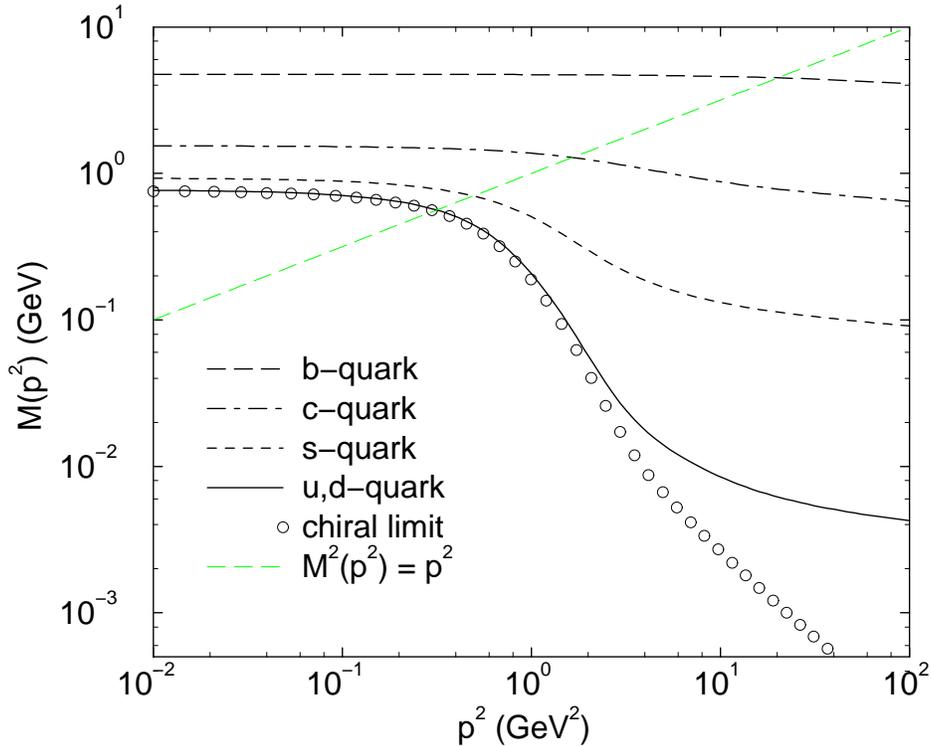,height=10cm}}
\parbox{40em}{\caption{Quark mass function obtained as a solution of
Eq.~(\protect\ref{gendse}) using the model of Eqs.~(\protect\ref{ouransatz}),
(\protect\ref{gk2}), and current-quark masses, fixed at $\zeta= 19\,$GeV:
$m_{u,d}^\zeta = 3.7\,$MeV, $m_s^\zeta = 82\,$MeV, $m_c^\zeta=0.58\,$GeV and
$m_b^\zeta=3.8\,$GeV.
The indicated solutions of $M^2(p^2)=p^2$ define the Euclidean
constituent-quark mass, $M^E_f$ in Eq.~(\protect\ref{euclidmass}), which
takes the values: $M^E_u=0.56\,$GeV, $M^E_s=0.70\,$GeV, $M^E_c= 1.3\,$GeV,
$M^E_b= 4.6\,$GeV.  (Adapted from
Refs.~\protect\cite{pieterrostock,mishaSVY}.)
\label{massfunction}}}
\end{figure}

In Fig.~\ref{massfunction} we present the renormalised dressed-quark mass
function, $M(p^2)$, obtained by solving Eq.~(\ref{dsemod}) using the model
and parameter values of Ref.~\cite{mr97},
Eqs.~(\ref{ouransatz}-\ref{params}), and also in the chiral limit and with
typical heavy-quark current-mass values.  In the presence of explicit chiral
symmetry breaking Eq.~(\ref{masanom}) describes the form of $M(p^2)$ for $p^2
> {\rm O}(1\,{\rm GeV}^2)$.  In the chiral limit, however, the ultraviolet
behaviour is given by
\begin{equation}
\label{Mchiral}
M(p^2) \stackrel{{\rm large}-p^2}{=}\,
\frac{2\pi^2\gamma_m}{3}\,\frac{\left(-\,\langle \bar q q \rangle^0\right)}
           {p^2
        \left(\sfrac{1}{2}\ln\left[p^2/\Lambda_{\rm QCD}^2\right]
        \right)^{1-\gamma_m}}\,,
\end{equation}
where $\langle \bar q q \rangle^0$ is the renormalisation-point-independent
vacuum quark condensate.  This behaviour too is characteristic of the QCD
renormalisation group~\cite{hdp76} and exhibits the power-law suppression
anticipated on page~\pageref{`without subtraction'}.  These results for the
large $p^2$ behaviour of the mass function are model independent; i.e., they
arise only because the DSE truncation is consistent with the QCD
renormalisation group at one loop.  (It has long been known that the
truncation defined by Eq.~(\ref{ouransatz}) yields results in agreement with
the QCD renormalisation group at one loop; e.g.,
Refs.~\cite{higashijimaportermelbourne,seattle}.)

The gauge invariant expression for the renormalisation-point-dependent vacuum
quark condensate was derived in Ref.~\cite{mrt98}:
\begin{equation}
\label{qbq0}
\,-\,\langle \bar q q \rangle_\zeta^0 := 
Z_4(\zeta^2,\Lambda^2)\, N_c {\rm tr}_{\rm D}\int^\Lambda_q\,
        S^{0}(q,\zeta)\,,
\end{equation}
where ${\rm tr}_D$ identifies a trace over Dirac indices only and the
superscript ``$0$'' indicates the quantity was calculated in the chiral
limit.  Substituting Eq.~(\ref{Mchiral}) into Eq.~(\ref{qbq0}), recalling
that $Z_4=Z_m$ in Landau gauge and using Eq.~(\ref{zmdef}), yields the
one-loop expression
\begin{equation}
\label{qbqzeta}
\langle \bar q q \rangle_\zeta^0 = \left(\sfrac{1}{2}\ln \zeta^2/\Lambda_{\rm
QCD}^2\right)^{\gamma_m} \, \langle \bar q q \rangle^0\,.
\end{equation}
Employing Eq.~(\ref{masanom}), this exemplifies the general result that
\begin{equation}
\label{mqbq}
m(\zeta)\,\langle \bar q q \rangle_\zeta^0 = 
\hat m \langle \bar q q \rangle^0\,;
\end{equation}
i.e., that this product is renormalisation point invariant and, importantly,
it shows that the behaviour expressed in Eq.~(\ref{Mchiral}) is exactly that
required for consistency with the gauge invariant expression for the quark
condensate.  A model, such as Ref.~\cite{diakonov}, in which the scalar
projection of the chiral limit dressed-quark propagator falls faster than
$1/p^4$, up to $\ln$-corrections, is only consistent with this quark
condensate vanishing, and it is this condensate that appears in the current
algebra expression for the pion mass~\cite{mrt98}.

Equation~(\ref{Mchiral}) provides a reliable means of calculating the quark
condensate because corrections are suppressed by powers of $\Lambda_{\rm
QCD}^2/\zeta^2$.  Analysing the asymptotic form of the numerical solution one
finds
\begin{equation}
\label{qbqM0}
-\langle \bar q q \rangle^0 = (0.227\,{\rm GeV})^3\,.
\end{equation}
Using Eq.~(\ref{qbqzeta}) one can define a one-loop evolved condensate
\begin{equation}
\label{qbq1}
\left.-\langle \bar q q \rangle_\zeta^0\right|_{\zeta=1\,{\rm GeV}} :=
-\left(\ln\left[1/\Lambda_{\rm QCD}\right]\right)^{\gamma_m} \, \langle \bar
q q \rangle^0 = (0.241\,{\rm GeV})^3\,.
\end{equation}
This can be directly compared with the value of the quark condensate employed
in contemporary phenomenological studies~\cite{derek}: $ (0.236\pm
0.008\,{\rm GeV})^3$.  The authors of Ref.~\cite{mr97} noted that increasing
$\omega \to 1.5\,\omega$ in ${\cal G}(k^2)$ increases the calculated value in
Eq.~(\ref{qbq1}) by $\sim 10$\%; i.e., the magnitude of the condensate is
correlated with the degree of infrared enhancement/strength in the effective
interaction.  That is unsurprising because it has long been known that there
is a critical coupling for DCSB; i.e., the kernel in the gap equation must
have an integrated strength that exceeds some critical
value~\cite{higashijimaportermelbourne}.  This is true in all fermion-based
studies of DCSB.

The renormalisation-point-invariant current-quark masses corresponding to the
$m_f(\zeta)$ in Fig.~\ref{massfunction} are obtained in the following
way: using Eq.~(\ref{qbq0}), direct calculation from the chiral limit
numerical solution gives
\begin{equation}
\langle \bar q q \rangle_{\zeta=19\,{\rm GeV}}^0 = - (0.275\,{\rm GeV})^3\,,
\end{equation}
and hence from the values of $m_f^\zeta \equiv m_f(\zeta)$ listed in
Fig.~\ref{massfunction} and Eqs.~(\ref{mqbq}), (\ref{qbqM0}), in MeV,
\begin{equation}
\begin{array}{cccc}
\hat m_{u,d}=6.60\,,\, &
\hat m_s = 147\,,\, &
\hat m_c = 1\,030\,,\, &
\hat m_b = 6\,760\,,
\end{array} 
\end{equation}
from which also follow one-loop evolved values in analogy with
Eq.~(\ref{qbq1}):
\begin{equation}
\label{oneloopmasses}
\begin{array}{cccc}
m_{u,d}^{1\,{\rm GeV}}= 5.5\,,\, &
m_s^{1\,{\rm GeV}} = 130\,,\, &
m_c^{1\,{\rm GeV}} = 860\,,\, &
m_b^{1\,{\rm GeV}} = 5\,700\,.
\end{array} 
\end{equation}

Figure~\ref{massfunction} highlights a number of qualitative aspects of the
quark mass function.  One is the difference in the ultraviolet between the
behaviour of $M(p^2)$ in the chiral limit and in the presence of explicit
chiral symmetry breaking.  In the infrared, however, the $u,d$-quark mass
function and the chiral limit solution are almost indistinguishable.  The
chiral limit solution is nonzero {\it only} because of the nonperturbative
DCSB mechanism whereas the $u,d$-quark mass function is purely perturbative
at $p^2>20\,$GeV$^2$.  Hence the evolution to coincidence of the chiral-limit
and $u,d$-quark mass functions makes clear the transition from the
perturbative to the nonperturbative domain.  It is on this nonperturbative
domain that $A(p^2,\zeta^2)$ differs significantly from one.  [This behaviour
and that of the light-quark mass function depicted in Fig.~\ref{massfunction}
have recently been confirmed in numerical simulations of
lattice-QCD~\cite{tonylatticequark}.]  A concomitant observation is that the
DCSB mechanism has a significant effect on the propagation characteristics of
$u,d,s$-quarks.  However, as evident in the figure, that is not the case for
the $b$-quark.  Its large current-quark mass almost entirely suppresses
momentum-dependent dressing, so that $M_b(p^2)$ is nearly constant on a
substantial domain.  This is true to a lesser extent for the $c$-quark.
 
To quantify the effect of the DCSB mechanism on massive quark propagation
characteristics the authors of Ref.~\cite{pieterrostock,mishaSVY} introduced
a single measure: ${\cal L}_f:=M^E_f/m_f^\zeta$, where $M^E_f$ is the
Euclidean constituent-quark mass, defined as the solution of
\begin{equation}
\label{euclidmass}
(M_f^E)^2 - M^2(p^2=(M_f^E)^2,\zeta^2) = 0\,.
\end{equation}
In this exemplifying model the Euclidean constituent-quark mass takes the
values listed in the figure, which have magnitudes and ratios consistent with
contemporary phenomenology; e.g., Refs.~\cite{pdg98,simon}, and 
\begin{equation}
\label{Mmratio}
\begin{array}{c|ccccc}
        f   &   u,d  &   s   &  c  &  b  \\ \hline
 {\cal L}_f &  150   &    10  &  2.2 &  1.2 
\end{array}\,.
\end{equation}
These values are representative and definitive: for light-quarks ${\cal
L}_{q=u,d,s} \sim 10$-$100$, while for heavy-quarks ${\cal L}_{Q=c,b} \sim
1$, and highlight the existence of a mass-scale characteristic of DCSB:
$M_\chi$.  The propagation characteristics of a flavour with $m_f^\zeta\leq
M_\chi$ are significantly altered by the DCSB mechanism, while for flavours
with $m_f^\zeta\gg M_\chi$ momentum-dependent dressing is almost irrelevant.
It is apparent and unsurprising that $M_\chi \sim 0.2\,$GeV$\,\sim
\Lambda_{\rm QCD}$.  This feature of the dressed-quark mass function provides
the foundation for a constituent-quark-like approximation in the treatment of
heavy-meson decays and transition form factors~\cite{mishaSVY}.

To recapitulate.  The quark DSE describes the phenomena of DCSB and the
concomitant dynamical generation of a momentum dependent quark mass function,
with the renormalisation-group-improved rainbow-truncation yielding
model-independent results for the momentum dependence of the mass function in
the ultraviolet.  For light quarks, defined by ${\cal L}_f\gg 1$, the
magnitude of the mass function in the infrared; i.e., for $p^2\lsim
1$-$2\,$GeV$^2$, is determined by the behaviour of the effective quark-quark
interaction on the same domain, and so is the value of the vacuum quark
condensate.  An infrared enhancement in this effective interaction is {\it
required} to describe observable light-meson phenomena.  We have only
provided a single illustration but it is supported by, e.g.,
Refs.~\cite{jain,fr96,fm96,dubravko,regAdelaide,pieterVM} and the observation
that with insufficient infrared integrated strength in the kernel of
Eq.~(\ref{gendse}) the quark condensate
vanishes~\cite{papa1,fredIR,axelIR,natale,fredIRnew}, which is a poor
starting point for light-hadron phenomenology.  The question now arises,
where does this strength come from?

\begin{figure}[t]
\centering{\
\epsfig{figure=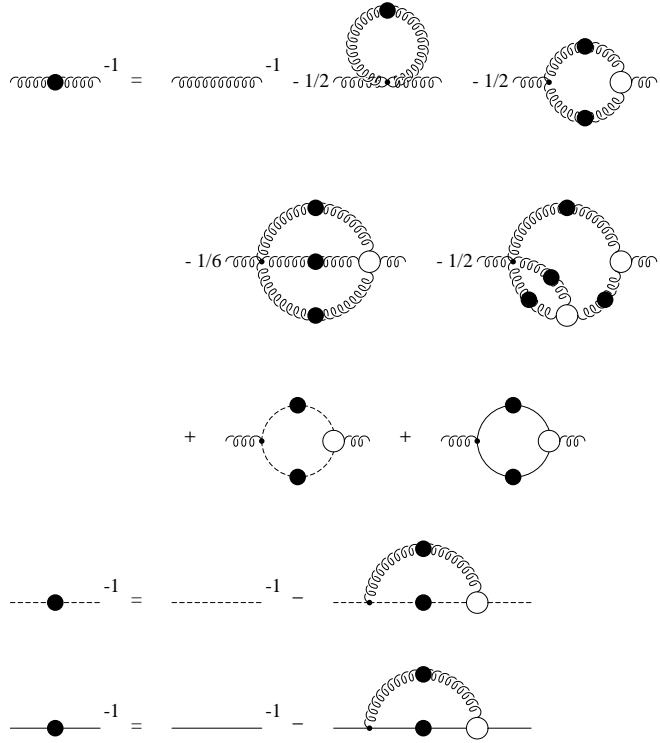,height=10cm}}
\parbox{40em}{\caption{\label{gluonDSE} From top to bottom, depictions of
the DSEs for the gluon (spring), ghost (dashed-line) and quark (solid-line)
$2$-point functions.  Following convention, a filled circle denotes a fully
dressed propagator and an open circle, a one-particle irreducible vertex;
e.g., the open circle in the first line represents the dressed-three-gluon
vertex.  The figure illustrates the interrelation between elements in the
tower of DSEs: the gluon propagator appears in the DSE for the quark and
ghost propagator; the ghost and quark propagator in the DSE for the gluon,
etc.  (Adapted from Ref.~\protect\cite{hauck}.)}}
\end{figure}
\addcontentsline{toc}{subsubsection}{\protect\numberline{ } {Gluon DSE}}
\subsubsect{Gluon DSE} 
Guidance here comes from studies of the DSE satisfied by the dressed-gluon
propagator, which is depicted in Fig.~\ref{gluonDSE}.  However, as we now
describe, these studies are inconclusive.

Early analyses~\cite{bbz} used the ghost-free axial gauge: $n\cdot A^a = 0$,
$n^2>1$, in which case the second equation in the figure is absent and two
independent scalar functions: $F_1$, $F_2$, are required to fully specify the
dressed-gluon propagator, cf. the covariant gauge expression in
Eq.~(\ref{gluoncovariant}), which requires only one function.  In the absence
of interactions: $F_1(k^2) = -1/k^2$, $F_2(k^2)\equiv 0$.  These studies
employed an {\it Ansatz} for the three-gluon vertex that doesn't possess a
particle-like singularity and neglected the coupling to the quark DSE.  They
also assumed $F_2\equiv 0$, even nonperturbatively, and ignored it in solving
the DSE.  The analysis then yielded
\begin{equation}
\label{enhanced}
F_1(k^2) \stackrel{k^2\to 0}{\sim} \frac{1}{k^4}\,;
\end{equation}
i.e., a marked infrared enhancement that can yield an area law for the Wilson
loop~\cite{west82} and hence confinement, and DCSB as described above {\it
without} fine-tuning.  This effect is driven by the gluon vacuum
polarisation, diagram three in the first line of Fig.~(\ref{gluonDSE}).  A
similar result was obtained in Ref.~\cite{atkinson83}.  However, a possible
flaw in these analyses was identified in Ref.~\cite{west83}, which argued
from properties of the spectral density in ghost-free gauges that $F_2$
cannot be zero but acts to cancel the enhancement in $F_1$.  [Preserving
$F_2$ yields a coupled system of equations for the gluon propagator that is
at least as complicated as that obtained in covariant gauges, which perhaps
outweighs the apparent benefit of eliminating ghost fields in the first
place.]

There have also been analyses of the gluon DSE using Landau gauge and those
of Refs.~\cite{mandelstam,bargadda,brownpennington,brown88b,brown89} are
unanimous in arriving at the covariant gauge analogue of
Eq.~(\ref{enhanced}), again driven by the gluon vacuum polarisation diagram.
In these studies {\it Ans\"atze} were used for the dressed-three-gluon
vertex, all of which were free of particle-like singularities.  However,
these studies too have weaknesses: based on an anticipated dominance of the
gluon-vacuum polarisation, truncations were implemented so that only the
third and fifth diagrams on the r.h.s. of the first equation in
Fig.~(\ref{gluonDSE}) were retained.  In covariant gauges there is {\it a
priori} no reason to neglect the ghost loop contribution, diagram six,
although perturbatively its contribution is estimated to be
small~\cite{brown89}.

Another class of Landau gauge studies are described in
Refs.~\cite{stingl12,stingl3}, which propose solving the DSEs via rational
polynomial {\it Ans\"atze} for the one-particle irreducible components of the
Schwinger functions appearing in Fig.~\ref{gluonDSE}; i.e., the self energies
and vertices.  This method attempts to preserve aspects of the organising
principle of perturbation theory in truncating the DSEs.  In concrete
calculations, for simplicity, only the first, third and sixth diagrams on the
r.h.s. of the first equation in Fig.~\ref{gluonDSE} survive, the last
[fermion] equation is neglected and the leading order solution of the ghost
equation has the appearance of the massless free propagator: $\sim 1/k^2$.
The analysis suggests that a consistent solution for the dressed-gluon
propagator is one that vanishes at $k^2=0$; i.e., in
Eq.~(\ref{gluoncovariant})
\begin{equation}
\label{vanishing}
d(k^2) \sim \frac{k^4}{k^4 + \gamma^4}\,.
\end{equation}
However, the associated polynomial Ansatz for the dressed-three-gluon vertex
exhibits particle-like singularities and this is characteristic of the
method.  The question of how this can be made consistent with the absence of
coloured bound states in the strong interaction spectrum is currently
unanswered.

Proponents of the result in Eq.~(\ref{vanishing}) claim support from
studies~\cite{zwanzigerold,zwanziger} of ``complete'' gauge fixing; i.e., in
the outcome of attempts to construct a Fadde$^\prime$ev-Popov-like
determinant that eliminates Gribov copies or ensures that the functional
integration domain for the gauge field is restricted to a subspace without
them.  Fixing a so-called ``minimal Landau gauge,'' which enforces a
constraint of integrating only over gauge field configurations inside the
Gribov horizon; i.e., on the simplest domain for which the Fadde\'ev-Popov
operator is invertible, the dressed-gluon $2$-point function is shown to
vanish at $k^2=0$.  However, the approach advocated in
Refs.~\cite{stingl12,stingl3} makes no use of the additional ghost-like
fields necessary to restrict the integration domain.

Thus far in this discussion of the gluon DSE we have reported nothing
qualitatively new and a more detailed review of the studies described can be
found in Ref.~\cite{cdragw}, Sec.~5.1.  What about contemporary studies?

The direct approach to solving the Landau gauge gluon DSE, pioneered in
Refs.~\cite{mandelstam,bargadda,brownpennington,brown88b,brown89}, has been
revived by two groups: ${\cal A}$, Refs.~\cite{hauck,hauckPRL,hauckAdelaide};
and ${\cal B}$, Refs.~\cite{bloch,bloch2,davidAdelaide,jacquesAdelaide}, with
the significant new feature that nonperturbative effects in the ghost sector
are admitted; i.e., a nonperturbative solution of the DSE for the ghost
propagator is sought in the form
\begin{equation}
G^{ab}(k) = - \delta^{ab}\,\frac{\varpi(k^2)}{k^2}\,\; [{\rm
without~interactions,~}\varpi(k^2) \equiv 1].
\end{equation}

These studies analyse a truncated gluon-ghost DSE system, retaining only the
third and sixth loop diagrams in the first equation of Fig.~(\ref{gluonDSE})
and also the second equation.  Superficially this is the same complex of
equations as studied in Refs.~\cite{stingl12,stingl3}.  However, the
procedure for solving it is different, arguably less systematic but also less
restrictive.  The difference between the groups is that ${\cal A}$ employ
{\it Ans\"atze} for the dressed-ghost-gluon and dressed-three-gluon vertices
constructed so as to satisfy the relevant Slavnov-Taylor identities while
${\cal B}$ simply use the bare, perturbative vertices.  Nevertheless they
agree in the conclusion that in this case the infrared behaviour of the gluon
DSE's solution is determined by the ghost loop alone: it overwhelms the gluon
vacuum polarisation contribution.  That is emphasised in
Ref.~\cite{davidAdelaide}, which eliminates every loop diagram in truncating
the first equation of Fig.~\ref{gluonDSE} {\it except} the ghost loop and
still recovers the behaviour of Ref.~\cite{jacquesAdelaide}.  That behaviour
is
\begin{equation}
\label{irAB}
\varpi(k^2) \sim \frac{1}{(k^2)^{\kappa}} ,\;
d(k^2)\sim (k^2)^{2\kappa}\,\;{\rm for~} k^2\lsim \Lambda^2_{\rm QCD}\,,\;
{\rm with~} 0.8\lsim \kappa \leq 1.
\end{equation}
Exact evaluation of the angular integrals that arise when solving the
integral equations gives the integer valued upper bound, $\kappa =
1$~\cite{bloch2}.  This corresponds to a dressed-gluon $2$-point function
that vanishes at $k^2=0$, although the suppression is very sudden with the
propagator not peaking until $k^2\approx\Lambda^2_{\rm QCD}$, where
\begin{equation}
(\left.d(k^2)/k^2)\right|_{k^2=\Lambda_{\rm QCD}^2} \sim 100/\Lambda_{\rm
QCD}^2\,; 
\end{equation}
i.e., it is very much enhanced over the free propagator.  [See, e.g.,
Ref.~\cite{hauck}, Fig.~12.]  $\kappa = 1$ also yields a dressed-ghost
propagator that exhibits a dipole enhancement analogous to that of
Eq.~(\ref{enhanced}).  A [renormalisation group invariant] strong running
coupling consistent with this truncations is:
\begin{equation}
\alpha(k^2) := \sfrac{1}{4\pi}\,g^2\,\varpi^2(k^2)\,d(k^2)
\end{equation}
and its value at $k^2=0$ is fixed by the numerical solutions:
\begin{equation}
\begin{array}{l|c|c}
                & {\cal A} & {\cal B} \\\hline
\alpha(k^2=0)   &   9.5       & \sim 4\;\mbox{or}\;12
\end{array}\,.
\end{equation}
[NB.  Group ${\cal A}$ approximates the angular integrals and uses vertex
{\it Ans\"atze}.  Group ${\cal B}$ uses bare vertices and in
Ref.~\cite{bloch} approximates the angular integrals to obtain $\alpha(0)\sim
12$, while in Ref.~\cite{bloch2} the integrals are evaluated exactly, which
yields $\alpha(0)=\sfrac{4}{3}\pi\approx 4.2$.]

The qualitative common feature is that the Grassmannian ghost loops act to
suppress the dressed-gluon propagator in the infrared.  That may also be said
of Refs.~\cite{stingl12,stingl3}.  [Indications that the quark loop, diagram
seven in Fig.~\ref{gluonDSE}, acts to oppose an enhancement of the type in
Eq.~(\ref{enhanced}) may here, with hindsight, be viewed as suggestive.]  One
aspect of ghost fields is that they enter because of gauge fixing via the
Fadde$^\prime$ev-Popov determinant.  Hence, while none of the groups
introduce the additional Fadde$^\prime$ev-Popov contributions advocated in
Refs.~\cite{zwanzigerold,zwanziger}, they nevertheless do admit ghost
contributions, and in their solution the number of ghost fields does not have
a qualitative impact.  Reference~\cite{zwanziger} also obtains a
dressed-propagator for the Fadde$^\prime$ev-Popov fields with a $k^2=0$
dipole singularity.  It contributes to the action via the term employed to
restrict the gauge field integration domain, in which capacity the dipole
singularity can plausibly drive an area law for Wilson loops.

Schwinger functions are the primary object of study in numerical simulations
of lattice-QCD and Refs.~\cite{latticegluon} report contemporary estimates of
the lattice Landau gauge dressed-gluon $2$-point function.  They are
consistent with a finite although not necessarily vanishing value of $d(k^2 =
0)$.  However, simulations of the dressed-ghost $2$-point function find no
evidence of a dipole singularity, with the ghost propagator behaving as if
$\varpi(k^2) = 1$ in the smallest momentum bins~\cite{schilling}.  [NB.
Since the quantitative results from groups ${\cal A}$ and ${\cal B}$ differ
and exhibit marked sensitivity to details of the numerical analysis, any
agreement between the DSE results for $\varpi(k^2)$ or $d(k^2)$ and the
lattice data on some subdomain can be regarded as fortuitous.]

The behaviour in Eqs.~(\ref{irAB}) also entails the presence of particle-like
singularities in extant {\it Ans\"atze} for the dressed-ghost-gluon,
dressed-three-gluon and dressed-quark-gluon vertices that are consistent with
the relevant Slavnov-Taylor identities.  [$\kappa = 1$ corresponds to an
ideal simple pole singularity.]  Hence while this behaviour may be consistent
with the confinement of elementary excitations, as currently elucidated it
also predicts the existence of coloured bound states in the strong
interaction spectrum.  Furthermore, while it does yield a strong running
coupling with $\alpha(k^2=0) \gsim 1$, that makes DCSB dependent on fine
tuning~\cite{fredIRnew}, and quantitative calculations based on the present
numerical solutions give a quark condensate only $\sim 5$\% of the value in
Eq.~(\ref{qbq1})~\cite{jacquesPrivate}.  Notwithstanding these remarks, the
studies of Refs.~\cite{hauck,hauckPRL,hauckAdelaide} and subsequently
Refs.~\cite{bloch,bloch2,davidAdelaide,jacquesAdelaide} are laudable.  They
have focused attention on a previously unsuspected qualitative sensitivity to
truncations in the gauge sector.

To recapitulate.  It is clear from
Refs.~\cite{papa1,fredIR,axelIR,natale,fredIRnew} that DCSB requires the
effective interaction in the quark DSE to be strongly enhanced at $k^2 \sim
\Lambda^2_{\rm QCD}$.  [Remember too that modern lattice
simulations~\cite{tonylatticequark} confirm the pattern of behaviour
exhibited by quark DSE solutions obtained with such an enhanced interaction.]
Studies of QCD's gauge sector indicate that gluon-gluon and/or gluon-ghost
dynamics can generate such an enhancement.  However, the qualitative nature
of the mechanism and its strength remains unclear: is it the gluon vacuum
polarisation or that of the ghost that is the driving force?  It is a
contemporary challenge to explore and understand this.

Finally, in discussing aspects of the gauge sector one might consider whether
instanton configurations play a role?  Instantons are solutions of the
classical equation of motion for the Euclidean gauge field.  As such they
form a set of measure zero in the gauge field integration space.
Nonetheless, they can form the basis for a semi-classical approximation to
the gauge field action and models based on this notion have been
phenomenologically successful~\cite{instantonreview}.  In this context we
note that the DSEs in Fig.~(\ref{gluonDSE}) are derived nonperturbatively and
their self-consistent solution includes the effects of all field
configurations.  Hence instanton-like configurations may contribute to the
form of the solution.  However, a successful description of observable
phenomena does not require that their contribution be quantified in a
particular truncation.  Nevertheless, one might estimate that a dilute liquid
of instantons, each with radius $\bar\rho \approx 1/(0.6\,$GeV), could
significantly effect the propagation characteristics of gluons only on the
domain of intermediate momenta: $k^2 \sim (0.6\,{\rm GeV})^2$.  Therefore
they cannot qualitatively affect the infrared aspects discussed in this
subsection.  They also make no contribution in the perturbative domain:
$k^2\gsim 1$--$2\,$GeV$^2$, where the perturbative matching inherent in the
DSEs is a strength that makes possible a unification of infrared and
ultraviolet phenomena, such as in the behaviour of bound state elastic and
transition form factors; e.g.,
Refs.~\cite{kevinpi0,mikepomeron,mrpion,klabucar,fizikaB,klabucar2}.

\addcontentsline{toc}{subsubsection}{\protect\numberline{ } {Confinement}}
\subsubsect{Confinement} 
Confinement is the failure to directly observe coloured excitations in a
detector: neither quarks nor gluons nor coloured composites.  The
contemporary hypothesis is stronger; i.e., coloured excitations cannot
propagate to a detector.  To ensure this it is sufficient that coloured
$n$-point functions violate the axiom of reflection positivity~\cite{gj81},
which is guaranteed if the Fourier transform of the momentum-space $n$-point
Schwinger function is not a positive-definite function of its arguments.
Reflection positivity is one of a set of five axioms that must be satisfied
if the given $n$-point function is to have a continuation to Minkowski space
and hence an association with a physical, observable state.  If an
Hamiltonian exists for the theory but a given $n$-point function violates
reflection positivity then the space of observable states, which is spanned
by the eigenstates of the Hamiltonian, does not contain anything
corresponding to the excitation(s) described by that Schwinger function.
[The violation of reflection positivity is not a necessary condition for
confinement~\cite{gastao}.  A text-book counterexample is massless
two-dimensional QED~\cite{schwingermodel} but in this case confinement of
electric charge and DCSB both arise as a peculiar consequence of the number
of dimensions.]

The free boson propagator does not violate reflection positivity:
\begin{equation}
\label{freeboson}
\Delta(x):= 
\int\frac{d^4 k}{(2\pi)^4}\,{\rm e}^{i k\cdot x}\,\frac{1}{k^2+m^2}
= \frac{1}{4\pi^2 x} \int_0^\infty d\ell \,J_1(\ell x)\,
\frac{\ell^2}{\ell^2 + m^2}
= \frac{m}{4\pi^2 x}\,K_1(m x)\,.
\end{equation}
Here $x:= (x\cdot x)^{1/2}>0$, $J_1$ is an oscillatory Bessel function of the
first kind and $K_1$ is the monotonically decreasing, strictly convex-up,
non-negative modified Bessel function of the second kind.  The same is true
of the free fermion propagator:
\begin{equation}
\label{freefermion}
S(x)= \int\frac{d^4 k}{(2\pi)^4}\,{\rm e}^{i k\cdot x}\,
\frac{m-i\gamma\cdot k }{k^2+m^2}
= (m-\gamma\cdot\partial ) \Delta(x) 
= \frac{m^2}{4\pi^2 x}
\left[ K_1(m x) +  \frac{\gamma\cdot x }{x} K_2(m x)\right]\,,
\end{equation}
which is also positive definite.  The spatially averaged Schwinger function
is a particularly insightful tool~\cite{fredIR,hollenberg}.  Consider the
fermion and let $T=x_4$ represent Euclidean ``time,'' then
\begin{equation}
\label{sigmasT}
\sigma_S(T):= \int d^3 x\, {\rm tr}_D S(\vec{x},T) =
\frac{1}{\pi}\int_0^{\infty}\,d\ell\,\frac{m}{\ell^2+m^2}\,\cos(\ell T) =
\sfrac{1}{2}\,{\rm e}^{-m T}.
\end{equation}
Hence the free fermion's mass can be easily obtained from the large $T$
behaviour of the spatial average:
\begin{equation}
\label{mTfree}
m \,T = - \lim_{T\to \infty} \ln \sigma_S(T)\,.
\end{equation}
[The boson analogy is obvious.]  This is just the approach used to determine
bound state masses in simulations of lattice-QCD.

For contrast, consider the dressed-gluon $2$-point function in
Eq.~(\ref{vanishing}):
\begin{equation}
\label{badDx}
D(x):= \int\frac{d^4 k}{(2\pi)^4}\,{\rm e}^{i k\cdot x}\, \frac{k^2}{k^4 +
\gamma^4} = \frac{1}{4\pi^2 x} \int_0^\infty d\ell \,J_1(\ell x)\,
\frac{\ell^4}{\ell^4 + \gamma^4}
= -\frac{\gamma}{4\pi^2 x} \, 
\left.\left(\frac{d}{dz}{\rm  ker}(z)\right)\right|_{z= \gamma x}\,,
\end{equation}
where ker$(z)$ is the oscillatory Thomson function.  $D(x)$ is not positive
definite and hence a dressed-gluon $2$-point function that vanishes at
$k^2=0$ violates the axiom of reflection positivity and is therefore not
observable; i.e., the excitation it describes is confined.  At asymptotically
large Euclidean distances
\begin{eqnarray}
D(x) & \stackrel{x\to\infty}{\propto} & \frac{\gamma^{1/2}}{x^{3/2}}\,
{\rm e}^{-\gamma x/\surd 2}\,
\left[\cos(\sfrac{1}{\sqrt{2}}\gamma x + \sfrac{\pi}{8}) 
+ \sin(\sfrac{1}{\sqrt{2}}\gamma x+ \sfrac{\pi}{8}) \right]\,.
\end{eqnarray}
Comparing this with Eq.~(\ref{freeboson}) one identifies a mass as the
coefficient in the exponential: $m_D =\gamma/\surd 2$.  [NB. At large $x$,
$K_1(x) \propto \exp(-x)/\surd x$.]  By an obvious analogy, the coefficient
in the oscillatory term is the {\it lifetime}~\cite{stingl12,stingl3}: $\tau
= 1/m_D$.  Both the mass and lifetime are tied to the dynamically generated
mass-scale $\gamma$, which, using
\begin{equation}
\frac{z}{z^2+\gamma^4} = \sfrac{1}{2}\, \frac{1}{z+i\gamma^2} +
\sfrac{1}{2}\,\frac{1}{z-i\gamma^2}\,,
\end{equation}
is just the displacement of the complex conjugate poles from the real-$k^2$
axis.  It is a general result that the Fourier transform of a real function
with complex conjugate poles is not positive definite.  Hence the existence
of such poles in a $n$-point Schwinger function is a sufficient condition for
the violation of reflection positivity and thus for confinement.  The
spatially averaged Schwinger function is also useful here.
\begin{equation}
\label{D(T)}
D(T) := \int d^3 x\, D(\vec{x},T) =
\frac{1}{\pi}\int_0^{\infty}\,d\ell\,\frac{\ell^2}{\ell^4+\gamma^4}\,
\cos(\ell T)
= \frac{1}{2\gamma}\,{\rm e}^{-\frac{1}{\surd 2} \gamma T}\,
\cos(\sfrac{1}{\surd 2}\gamma T + \sfrac{\pi}{4}) \,,
\end{equation}
and, generalising Eq.~(\ref{mTfree}), one can define a $T$-dependent mass:
\begin{equation}
\label{oscexamp}
m(T)\,T := -\ln D(T) = 
 \ln (2 \gamma) + \sfrac{1}{\surd 2} \gamma \,T 
- \ln\left[\cos(\sfrac{1}{\surd 2}\gamma T + \sfrac{\pi}{4})\right] \,.
\end{equation}
It exhibits periodic singularities whose frequency is proportional to the
dynamical mass-scale that is responsible for the violation of reflection
positivity.  If a dressed-fermion $2$-point function has complex conjugate
poles it too will be characterised by a $T$-dependent mass that exhibits such
behaviour.

This reflection positivity criterion has been employed to very good effect in
three dimensional QED~\cite{pieterQED3}.  First, some background.  QED$_3$ is
confining in the quenched truncation~\cite{mack}.  That is evident in the
classical potential
\begin{equation}
\label{Vrquench}
V(r):= \int_{-\infty}^{\infty}dx_3\,\int\frac{d^3 k}{(2\pi)^3}\, {\rm
e}^{i\vec{k}\cdot\vec{x} + ik_3 x_3}\,\frac{e^2}{k^2} =
\frac{e^2}{2\pi}\ln(e^2 r)\,,\;r^2=x_1^2+x_2^2\,,
\end{equation}
which describes the interaction between two static sources.  [NB. $e^2$ has
the dimensions of mass in QED$_3$.]  It is a logarithmically growing
potential, showing that the energy required to separate two charges is
infinite.  Furthermore, $V(r)$ is just a one-dimensional average of the
spatial gauge-boson $2$-point Schwinger function and it is not positive
definite, which indicates that the photon is also confined.

If now, however, the photon vacuum polarisation tensor is evaluated at order
$e^2$ using $N_f$ massless fermions then, using the notation of
Eq.~(\ref{gluoncovariant}), the photon propagator is characterised
by~\cite{appel} 
\begin{equation}
\label{Vrunquench0}
\frac{d(k^2)}{k^2} = \frac{1}{k^2+ \tilde\alpha k}\,,\;\mbox{from~}
\Pi(k^2) = \frac{\tilde\alpha}{k}\,,\;\tilde\alpha = N_f e^2 /8\,,
\end{equation}
and one finds~\cite{conrad}
\begin{equation}
V(r) = \frac{e^2}{4}
\left[\mbox{\bf H}_0(\tilde\alpha r) - N_0(\tilde\alpha r)\right]\,,
\end{equation}
where {\bf H}$_0(x)$ is a Struve function and $N_0(x)$ a Neumann function,
both of which are related to Bessel functions.  In this case $V(r)$ is
positive definite, with the limiting cases
\begin{equation}
V(r) \stackrel{r \approx 0}{\sim} - \ln(\tilde\alpha r)\,,\;\;
V(r) \stackrel{r \to \infty}{=} \frac{e^2}{2\pi} \,\frac{1}{\tilde\alpha r}\,,
\end{equation}
and confinement is lost in QED$_3$.  That is easy to understand: pairs of
massless fermions cost no energy to produce and can propagate to infinity so
they are very effective at screening the interaction.

With $d(k^2)=1/[1+\Pi(k^2)]$ and sensible, physical constraints on the form
of $\Pi(k^2)$, such as boundedness and vanishing in the ultraviolet, one can
show that~\cite{justin}
\begin{equation}
\label{Vrqed3}
V(r) \stackrel{r \to \infty}{=} \frac{e^2}{2\pi}\,
\frac{1}{1+\Pi(0)}\ln( e^2 r) + {\rm const.} + h(r)\,,
\end{equation}
where $h(r)$ falls-off at least as quickly as $1/r$.  Hence, the existence of
a confining potential in QED$_3$ just depends on the value of the vacuum
polarisation at the origin.  In the quenched truncation, $\Pi(0) = 0$ and the
theory is logarithmically confining.  With massless fermions,
$1/[1+\Pi(0)]=0$ and confinement is absent.  Finally, when the vacuum
polarisation is evaluated from a loop of massive fermions, whether that mass
is obtained dynamically via the gap equation or simply introduced as an
external parameter, one obtains $\Pi(0) < \infty$ and hence a confining
theory.

\begin{figure}[t]
\centering{\ \epsfig{figure=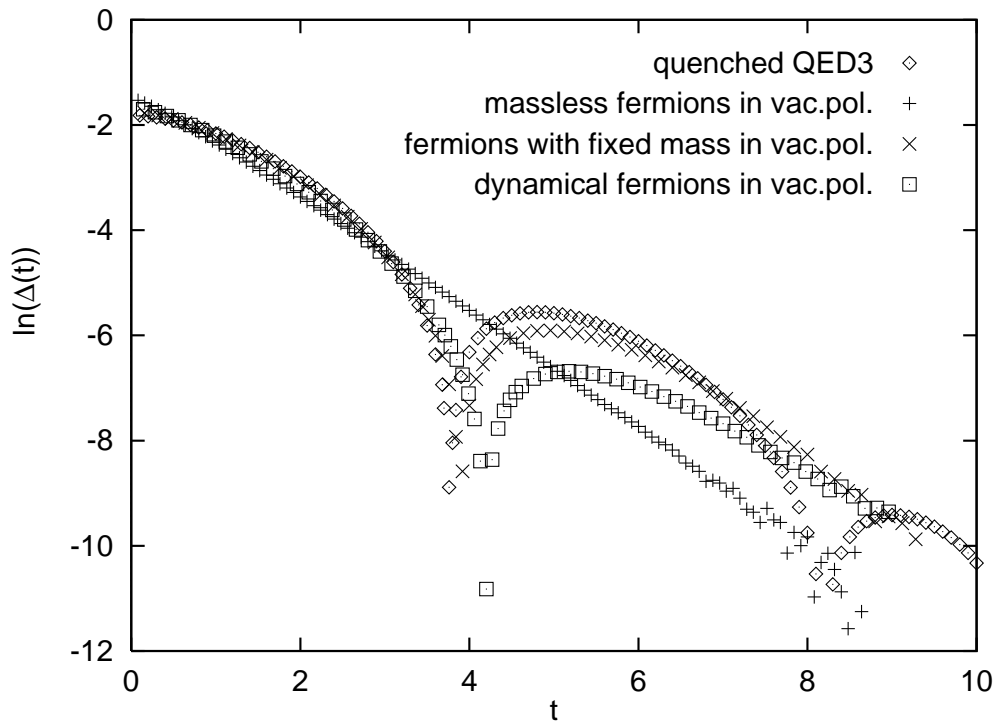,height=10cm}}
\parbox{40em}{\caption{\label{qed3fig} $\Delta(T):= -\sigma_S(T)$ from
Eq.~(\protect\ref{sigmasT}) for QED$_3$ with $2$ flavours of fermion.
$\Diamond$: confining, quenched theory; $+$, massless fermions used to
evaluate the photon vacuum polarisation tensor; $\times$, as before but with
fixed-mass fermions; $\Box$, again but fermions with a momentum-dependent
mass function.  (Adapted from Ref.~\protect\cite{pieterQED3}.)}}
\end{figure}

In Ref.~\cite{pieterQED3} the QED$_3$ gap equation is solved for all four
cases and the fermion propagator analysed.  The results are summarised by
Fig.~\ref{qed3fig}.  In the quenched theory, Eq.~(\ref{Vrquench}), the
dressed-fermion $2$-point function exhibits exactly those periodic
singularities that, via Eq.~(\ref{oscexamp}), are indicative of complex
conjugate poles.  Hence this feature of the $2$-point function, tied to the
violation of reflection positivity, is a clear signal of confinement in the
theory.  That is emphasised further by a comparison with the theory that is
unquenched via massless fermions in the vacuum polarisation,
Eq.~(\ref{Vrunquench0}).  As we have described, that theory is not confining
and in this case $\sigma_S(T)$ has the noninteracting, unconfined free
particle form in Eq.~(\ref{sigmasT}).  The difference could not be more
stark.  The remaining two cases exhibit the periodic singularities that
signal confinement, just as they should based on Eq.~(\ref{Vrqed3}).

At this point we note that any concern that the presence of complex conjugate
singularities in coloured $n$-point functions leads to a violation of
causality is misguided.  Microscopic causality only constrains the
commutativity of operators, and products thereof, that represent elements in
the space of observable particle states; i.e., the space spanned by
eigenstates of the Hamiltonian.  Since Schwinger functions that violate
reflection positivity do not have a continuation into that space there can be
no question of violating causality.  It is only required that ${\cal
S}$-matrix elements that describe colour-singlet to colour-singlet
transitions should satisfy the axioms, including reflection positivity.

The violation of reflection positivity by coloured $n$-point functions is a
sufficient condition for confinement.  However, it is not necessary, as the
example of planar, two-dimensional QCD shows~\cite{einhorn}.  There the
fermion two-point function exhibits particle-like singularities but the
colour singlet meson bound state amplitudes, obtained from a Bethe-Salpeter
equation, vanish at momenta coincident with the constituent-fermion mass
shell.  This excludes the pinch singularities that would otherwise lead to
bound state break-up and liberation of the constituents.  It is a realisation
of confinement via a failure of the cluster decomposition property
[CDP]~\cite{gj81,sw80}.  The CDP is a requirement that the difference between
the vacuum expectation value of a product of fields and all products of
vacuum expectation values of subsets of these fields must vanish faster than
any power.  [This is modified slightly in theories, like QED, with a
massless, asymptotic state: the photon.]  It can be understood as a statement
about charge screening and its failure means that, irrespective of the
separation between sources, the interaction between them is never negligible.
That is an appealing, intuitive representation of confinement.  Failure of
the CDP is an implicit basis for confinement in the bulk of QCD potential
models with; e.g., Refs.~\cite{ericold,franz} providing contemporary
illustrations.

Confinement is a more contentious issue than DCSB and its origin and
realisation less-well understood.  However, in this subsection we have
described a perspective that is common to many authors, and an interested
reader will find variations and more expansive discussions of various points
in; e.g.,
Refs.~\cite{zwanzigerold,zwanziger,gastao,brown89,stingl12,stingl3}, and
complementary perspectives in; e.g.,
Refs.~\cite{marshalldualreview,robertreview,tokilike,nora,matthias,eric,top}.
It is, of course, because confinement is poorly understood that its study and
modelling are important.  To presume otherwise is a misapprehension.

\subsect{Phenomenological Applications} 
\label{sectthree}
When Ref.~\cite{cdragw} was written the only application of DSEs to
observable phenomena consisted in the oft-repeated calculation of well-known
quantities, such as the pion mass and decay constant.
References~\cite{jain,pipi} represent a departure from that, and are
progenitors of the wide-ranging application of DSEs to observables newly
accessible at the current generation of experimental facilities.  Many of
these applications are reviewed in Refs.~\cite{revpeter,echaya,serdar} and
herein we only describe three recent, significant developments.

\addcontentsline{toc}{subsubsection}{\protect\numberline{ } {Light Mesons}}
\subsubsect{Light Mesons}
The model used to illustrate renormalisation and DCSB in Sec.~\ref{mr97sect}
has been applied to the calculation of vector meson masses and decay
constants~\cite{pieterVM}, and to elucidate the role of vector mesons in
connection with the electromagnetic pion form factor~\cite{mtpion}.  These
studies are important because, in concert with Ref.~\cite{mr97}, they
complete a DSE description of those light-mesons in the strong interaction
spectrum that are most often produced in reactions involving hadrons.  In so
doing they illustrate the efficacy of the renormalisation-group-improved
rainbow-ladder truncation for flavour non-singlet pseudoscalar and vector
mesons composed of light-quarks ($u$, $d$, $s$), and thereby that of the
systematic, Ward-Takahashi identity preserving truncation scheme introduced
in Ref.~\cite{truncscheme}.

The renormalised homogeneous Bethe-Salpeter equation for a bound state of a
dressed-quark and dressed-antiquark with total momentum $P$ is
\begin{equation}
\label{genbse}
[\Gamma_H(k;P)]_{tu} = \int^\Lambda_q \,
[\chi(q;P)]_{rs} K_{tu}^{rs}(q,k;P)\,,\;
\chi(q;P):= {\cal S}(q_+)\,\Gamma_H(q;P)\,{\cal S}(q_+)\,,
\end{equation}
with: $\Gamma_H(k;P)$ the Bethe-Salpeter amplitude [BSA], where $H$ specifies
the flavour structure of the meson; ${\cal S}(p):= {\rm
diag}[S_u(p),S_d(p),S_s(p)]$; $q_+ = q+\eta_P P$, $q_-= q - (1-\eta_P) P$;
and $r,\ldots, u$ represent colour-, Dirac- and flavour-matrix indices.
[$\eta_P\in[0,1]$ is the momentum partitioning parameter.  It appears in
Poincar\'e covariant treatments because, in general, the definition of the
relative momentum is arbitrary.  Physical observables, such as the mass, must
be independent of $\eta_P$ but that is only possible if the Bethe-Salpeter
amplitude depends on it.  $\eta_P=1/2$ for charge-conjugation eigenstates.]

In Eq.~(\ref{genbse}), $K^{rs}_{tu}(q,k;P)$ is the renormalised,
fully-amputated quark-antiquark scattering kernel, which, as we have seen,
also appears implicitly in Eq.~(\ref{gendse}) because it is the kernel of the
inhomogeneous DSE satisfied by $\Gamma_\nu(q;p)$.  $K^{rs}_{tu}(q,k;P)$ is a
$4$-point Schwinger function, obtained as the sum of a countable infinity of
skeleton diagrams.  It is two-particle-irreducible, with respect to the
quark-antiquark pair of lines and does not contain quark-antiquark to single
gauge-boson annihilation diagrams, such as would describe the leptonic decay
of a pseudoscalar meson.  [A connection between the fully-amputated
quark-antiquark scattering amplitude: $M = K + K ({\cal S}{\cal S}) K +
\ldots\,$, and the Wilson loop is discussed in Ref.~\cite{nora}.]  The
complexity of $K^{rs}_{tu}(q,k;P)$ is one reason why quantitative studies of
the quark DSE currently employ {\it Ans\"atze} for $D_{\mu\nu}(k)$ and
$\Gamma_\nu(k,p)$.  However, as illustrated by Ref.~\cite{mrt98}, the
complexity of $K^{rs}_{tu}(q,k;P)$ does not prevent one from analysing
aspects of QCD in a model independent manner and proving general results that
provide useful constraints on model studies.

Equation~(\ref{genbse}) is an eigenvalue problem and solutions exist only for
particular, separated values of $P^2$.  The eigenvector associated with each
eigenvalue: $\Gamma_H(k;P)$, the BSA, is a one-particle-irreducible,
fully-amputated quark-meson vertex.  In the flavour non-singlet pseudoscalar
channels the solutions having the lowest eigenvalues correspond to the $\pi$-
and $K$-mesons, while in the vector channels they correspond to the
$\omega$-, $\rho$- and $\phi$ mesons.

Following Ref.~\cite{truncscheme}, the renormalised inhomogeneous BSE for the
axial-vector vertex, consistent with the renormalisation--group-improved
quark DSE, Eqs.~(\ref{dsemod}) and (\ref{sigmod}), is
\begin{equation}
\Gamma_{5\mu}^l(k;P) = Z_2\,\sfrac{1}{2}\lambda^l\gamma_5\gamma_\mu -
\int^\Lambda_q {\cal G}((k-q)^2)\, D_{\alpha\beta}^{\rm free}(k-q)
\frac{\lambda^a}{2}\gamma_\alpha\, {\cal S}(q_+)\,\Gamma_{5\mu}^l(q;P)\,{\cal
S}(q_-)\, \frac{\lambda^a}{2}\gamma_\beta \,,
\end{equation}
where $\{\sfrac{1}{2}\lambda_F^l: l=1,\ldots,8 \}$ are the generators of
$SU(3)_{\rm flavour}$.  It is straightforward to verify that the axial-vector
Ward-Takahashi identity is satisfied; i.e., 
\begin{equation}
P_\mu\Gamma_{5\mu}^l(k;P) = {\cal S}^{-1}(k_+)\,\sfrac{1}{2}\lambda^li\gamma_5
+ \sfrac{1}{2}\lambda^li\gamma_5\,{\cal S}^{-1}(k_-)
-  M_\zeta\,i\Gamma_5^l(k;P) 
-i \Gamma_5^l(k;P) \, M_\zeta\,,
\end{equation}
where $M_\zeta= {\rm diag}[m_u(\zeta),m_d(\zeta),m_s(\zeta)]$ and the
renormalised pseudoscalar vertex satisfies its own inhomogeneous BSE:
\begin{equation}
\label{Gamma5T0}
\Gamma_5^l(k;P) = Z_4\,\sfrac{1}{2}\lambda^l\gamma_5 -
\int^\Lambda_q {\cal G}((k-q)^2)\, D_{\mu\nu}^{\rm free}(k-q)
\frac{\lambda^a}{2}\gamma_\mu\, {\cal S}(q_+)\,\Gamma_{5}^l(q;P)\,{\cal
S}(q_-)\, \frac{\lambda^a}{2}\gamma_\nu \,.
\end{equation}
[NB.  The product $M_\zeta\,\Gamma_5^l(k;P)$ is renormalisation point
independent.]

The pseudoscalar mesons appear as poles in both the axial-vector and
pseudoscalar vertices~\cite{mr97,mrt98} and equating pole residues yields the
homogeneous BSE
\begin{equation}
\label{bsemod}
\Gamma_H(k;P) + \int^\Lambda_q
{\cal G}((k-q)^2)\, D_{\mu\nu}^{\rm free}(k-q)
 \frac{\lambda^a}{2}\gamma_\mu\, {\cal S}(q_+)\,\Gamma_H(q;P)\,{\cal S}(q_-)\,
\frac{\lambda^a}{2}\gamma_\nu = 0\,.
\end{equation}
As is characteristic of homogeneous equations, the normalisation of the
solution is not fixed by this equation.  The canonical normalisation enforces
a requirement that the bound state contribution to the fully-amputated
quark-antiquark scattering amplitude: $M$, have unit residue.  In this
rainbow-ladder truncation that condition is expressed via
\begin{equation}
\label{pinorm}
2 P_\mu = \int^\Lambda_q {\rm tr} \left[ \bar\Gamma_H(q;-P) \, \frac{\partial
{\cal S}(q_+)}{\!\!\!\!\!\!\partial P_\mu}\, \Gamma_H(q;P)\, {\cal S}(q_-) +
\bar\Gamma_H(q;-P)\, {\cal S}(q_+)\, \Gamma_H(q;P)\, \frac{\partial {\cal
S}(q_-)}{\!\!\!\!\!\!\partial P_\mu}\right]\,,
\end{equation}
where 
\begin{equation}
\bar \Gamma_H(k,-P)^{\rm t} := C^{-1} \Gamma_H(-k,-P) C\,, 
\end{equation}
with $C=\gamma_2 \gamma_4$ the charge conjugation matrix:
\begin{equation}
\label{CCmtx}
C\gamma_\mu^{\rm t}C^\dagger = -\gamma_\mu\,;\;
[C,\gamma_5]=0\,, 
\end{equation}
and $X^{\rm t}$ denotes the matrix transpose of $X$.  The general form of a
pseudoscalar BSA is
\begin{equation}
\label{genpibsa}
\Gamma_H(k;P) = {\cal T}^H \gamma_5 \left[ i E_H(k;P) + \gamma\cdot P
F_H(k;P) \rule{0mm}{5mm} + \gamma\cdot k \,k \cdot P\, G_H(k;P) +
\sigma_{\mu\nu}\,k_\mu P_\nu \,H_H(k;P) \right]\,,
\end{equation}
where ${\cal T}^H$ is a matrix that describes the flavour content of the
meson; e.g., ${\cal T}^{\pi^+}= \sfrac{1}{2}(\lambda_F^1+i\lambda_F^2)$ and,
for bound states of constituents with equal current-quark masses, the scalar
functions $E$, $F$, $G$ and $H$ are even under $k\cdot P \to - k\cdot P$.
[NB.  Since the homogeneous BSE is an eigenvalue problem, $E_H(k;P)=
E_H(k^2,k\cdot P|P^2)$; i.e., $P^2$ is not a variable, instead it labels the
solution.  The same is true of each function.]

Equation~(\ref{bsemod}) also describes vector mesons, as can be shown by
considering the inhomogeneous equation for the renormalised vector vertex.
In general, twelve independent scalar functions are required to express the
Dirac structure of a vector vertex.  However, a vector meson bound state is
transverse:
\begin{equation}
P_\mu \Gamma_\mu^H(k^2,k\cdot P|P^2=-m_H^2)=0\,,
\end{equation}
where $m_H$ is the bound state's mass, and this constraint reduces to eight
the number of independent scalar functions.  One therefore has
\begin{equation}
\label{genvbsa}
\Gamma_\mu^H(k;P) = {\cal T}^H\,\sum_{l=1}^8\,O_\mu^l\,F_l(k;P)\,,
\end{equation}
with eight orthonormalised matrix covariants~\cite{pieterVM}
\begin{equation}
\label{defineOs}
\begin{array}{rcllcl}
O_\mu^1 &=& \gamma_\mu^{\rm T}\,, & \; O_\mu^2 &=& \sfrac{6}{\surd 5}
        \left(\hat k_\mu^{\rm T} \,\gamma^{\rm T}\cdot \hat k - \sfrac{1}{3}
        \gamma_\mu^{\rm T}\, \hat k^{\rm T}\cdot \hat k^{\rm T} \right)\,,\\
O_\mu^3 &=& 2 \hat k^{\rm T} \gamma\cdot \hat P\,, & \; O_\mu^4 &=& i\surd 2
\left(\gamma_\mu^{\rm T}\, \gamma\cdot \hat k^{\rm T}\, \gamma\cdot \hat P + 
\hat k_\mu^{\rm T}\, \gamma\cdot \hat P\right)\,,\\
O_\mu^5 &=& 2 \hat k_\mu^{\rm T}\,, & \; O_\mu^6 &=& \sfrac{i}{\surd 2}
\left(\gamma_\mu^{\rm T}\, \gamma^{\rm T}\cdot \hat k - \gamma^{\rm T}\,\cdot
\hat k\gamma_\mu^{\rm T}\right)\,,\\
O_\mu^7 + \sfrac{1}{\surd 2} O_\mu^8 &=& i\sqrt{\sfrac{ 3}{ 5}}\,
[1 + (\hat k \cdot \hat P)^2]
\left(\gamma_\mu^{\rm T}\,\gamma\cdot\hat P - \gamma\cdot\hat
P\,\gamma_\mu^{\rm T} \right) \,,& \;
O_\mu^8 &=& 2i\sqrt{\sfrac{6}{5}}\,\hat k_\mu^{\rm T}\,
        \gamma^{\rm T}\cdot \hat k \, \gamma\cdot \hat P\,,
\end{array}
\end{equation}
where: $ \gamma_\mu^{\rm T}:= \gamma_\mu + \gamma\cdot\hat P \, \hat P_\mu$,
$\hat P\cdot \hat P= -1$; and $\hat k\cdot \hat k = 1$, $\hat k_\mu^{\rm T}:=
\hat k_\mu + \hat k\cdot\hat P \, \hat P_\mu$.  With this decomposition the
magnitudes of the invariant functions, $F_l$, are a direct measure of the
relative importance of a given Dirac covariant in the BSA; e.g., one expects
$F_1$ to be the function with the greatest magnitude for $J^{PC}=1^{-\,-}$
bound states.

To calculate the meson masses, one first solves Eqs.~(\ref{dsemod}) and
(\ref{sigmod}) for the renormalised dressed-quark propagator.  This numerical
solution for $S(p)$ is used in the pseudoscalar BSE, Eq.~(\ref{bsemod}) under
the substitution of Eq.~(\ref{genpibsa}), which is a coupled set of four
homogeneous equations, one set for each meson; and the vector BSE,
Eq.~(\ref{bsemod}) with Eq.~(\ref{genvbsa}), a coupled set of eight equations
for each meson.  Solving the equations is a challenging numerical exercise,
requiring careful attention to detail, and two complementary methods were
both used in Refs.~\cite{mr97,pieterVM}.  While the numerical methods were
identical, the authors of Ref.~\cite{pieterVM} used a simplified version of
the effective interaction:
\begin{equation}
\label{gk2VM}
\frac{{\cal G}(k^2)}{k^2} = \frac{4\pi^2}{\omega^6} D k^2 {\rm
e}^{-k^2/\omega^2} + 4\pi\,\frac{ \gamma_m \pi} {\sfrac{1}{2} \ln\left[\tau +
\left(1 + k^2/\Lambda_{\rm QCD}^2\right)^2\right]} {\cal F}(k^2) \,,
\end{equation}
and varied the single parameter $D$ along with the current-quark masses:
$\hat m_u=\hat m_d$ and $\hat m_s$, in order to reproduce the observed values
of $m_{\pi}$, $m_K$ and $f_\pi$.  All other calculated results are {\it
predictions} in the sense that they are unconstrained.  (NB.  The Poincar\'e
invariant four-dimensional BSE is solved directly, eschewing the commonly
used artefice of a three-dimensional reduction, which introduces spurious
effects when imposing compatibility with Goldstone's theorem and also leads
to a misinterpretation of a model's parameters~\cite{alkoferbse}.)

Before reporting the results it is necessary to introduce the formulae for
the meson decay constants: $f_H$, which completely describe the strong
interaction contribution to a meson's weak or electromagnetic decay.
Following Ref.~\cite{mrt98}, the pseudoscalar meson decay constant is given
by
\begin{equation}
\label{caint}
\sfrac{1}{\surd 2}\,f_H P_\mu := \langle 0|\, \bar {\cal Q}({\cal T}^H)^{\rm
t} \gamma_\mu\gamma_5 {\cal Q} \,| H(P)\rangle = Z_2\,{\rm
tr}\int^\Lambda_k\, \left({\cal T}^H\right)^{\rm t} \gamma_5 \gamma_\mu
\,{\cal S}(k_+)\, \Gamma_H(k;P)\, {\cal S}(k_-)\,,
\end{equation}
where here ${\cal Q}={\rm column}(u,d,s)$.  The factor of $Z_2$ on the
r.h.s. ensures that $f_H$ is gauge invariant, and independent of the
renormalisation point and regularisation mass-scale; i.e., that $f_H$ is
truly an observable.  Equation~(\ref{caint}) is the pseudovector projection
of the unamputated Bethe-Salpeter wave function, $\chi(k;P)$, calculated at
the origin in {\it configuration space}.  As such, it is one field
theoretical generalisation of the ``wave function at the origin,'' which
describes the decay of bound states in quantum mechanics.  The analogous
expression for vector mesons is~\cite{mishaSVY}
\begin{eqnarray}
\nonumber
\sfrac{1}{\surd 2}\,f_H m_H \,\epsilon_\mu^\lambda(P) & := & \langle 0|\,
\bar {\cal Q}({\cal T}^H)^{\rm t} \gamma_\mu {\cal Q} \,| H(P)\rangle\\
& \Rightarrow & \sfrac{1}{\surd 2} f_H m_H = \sfrac{1}{3} Z_2\,{\rm
tr}\int^\Lambda_k\,\left({\cal T}^H\right)^{\rm t} \gamma_\mu \,{\cal
S}(k_+)\, \Gamma_\mu^H(k;P)\, {\cal S}(k_-)\,,
\end{eqnarray}
where $\epsilon_\mu^\lambda(P)$ is the vector meson's polarisation vector:
$P\cdot\epsilon^\lambda(P)=0$.

A best-fit is obtained~\cite{pieterVM} with
\begin{equation}
\label{DvalueVM}
D= (1.12\,{\rm GeV})^2\,,
\end{equation}
which is a 60\% increase over Eq.~(\ref{Dvalue}), as expected because the
single Gaussian term in Eq.~(\ref{gk2VM}) must here replace the sum of the
first two terms in Eq.~(\ref{gk2}); and renormalised current-quark masses
\begin{equation}
\label{paramsMT}
\begin{array}{cc}
m_{u,d}^{1\,{\rm GeV}} = 5.5\,{\rm MeV}\,,\; &
m_s^{1\,{\rm GeV}} = 124\,{\rm MeV}\,,
\end{array}
\end{equation}
which are little changed from the values used in Ref.~\cite{mr97},
Eq.~(\ref{oneloopmasses}).  [NB.  $\omega$ and $m_t$ are unchanged.  See the
discussion preceding Eq.~(\ref{Dvalue}).]  The results are presented in
Table~\ref{tableMesons}, and are characterised by a root-mean-square error
over predicted quantities of just $3.6\,$\%.  We emphasise that this is
obtained with a one-parameter model of the effective interaction.

\begin{table}[t]
\begin{center}
\parbox{40em}{\caption{Masses and decay constants [in GeV] of light vector
and flavour nonsinglet pseudoscalar mesons calculated using the
renormalisation-group-improved rainbow-ladder truncation.  The underlined
quantities were fitted.  The ``Obs.'' value of the vacuum quark condensate is
the global estimate of Ref.~\protect\cite{derek} and the masses are taken
from Ref.~\protect\cite{pdg98}, as are $f_\pi$, $f_K$.  The vector meson
decay constants are discussed in connection with
Eqs.~(\protect\ref{frho}-\protect\ref{fKast}).  For $\hat m_u=\hat m_d$, the
rainbow-ladder truncation gives $m_\omega=m_\rho$.  This degeneracy is lifted
by meson-loop self-energy contributions, such as
$\rho\to\pi\pi\to\rho$~\protect\cite{hollenberg,rhopipiKLM,rhopipiMAP}.  The
root-mean-square error over predicted quantities is just $3.6\,$\%.  (Adapted
from Ref.~\protect\cite{pieterVM}.)
\label{tableMesons}}}
\end{center}\vspace*{-1.0em}

\[
\begin{array}{l|ccccccccccc}
        & -(\langle \bar q q \rangle^0_{1\,{\rm GeV}})^{1/3} &
        m_\pi & m_K & m_\rho & m_{K^\ast} & m_\phi & f_\pi & f_K & f_\rho &
        f_{K^\ast} & f_\phi \\\hline
\rule{0mm}{1.2em}{\rm Obs.} & 0.236 & 0.139 & 0.496 & 0.770 & 0.892 & 1.020 &
        0.130 & 0.160 & 0.216 & 0.225 & 0.238\\
\rule{0mm}{1.2em}{\rm Calc.} & 0.242 & \underline{0.139} & \underline{0.496}
        & 0.747 & 0.956 & 1.088 & \underline{0.130} & 0.154 & 0.197 & 0.246 &
        0.255 \\\hline
\end{array}
\]
\end{table}

Experimental values of the pseudoscalar meson decay constants are obtained
directly via observation of their prominent $\beta$-decay mode.  However, for
the vector mesons this mode is not easily accessible and to proceed we note
that the rainbow-ladder truncation predicts ideal flavour mixing.  Using this
and isospin symmetry, one can relate the $f_\rho$ matrix element to that
describing $\rho\to e^+ e^-$ decay:
\begin{equation}
\label{frho}
\frac{m_{\rho}^2}{g_\rho}\,\epsilon^\lambda_\mu(P) 
:=\sfrac{1}{\surd 2} \, \langle 0| \bar {\cal Q} \,Q_e \,({\cal
T}^{\rho^0})^{\rm t}\gamma_\mu {\cal Q} | \rho^0_\lambda(P) \rangle
= \langle 0| \bar {\cal Q} ({\cal T}^{\rho^-})^{\rm t} \gamma_\mu {\cal Q} |
\rho^-_\lambda(p) \rangle
= \sfrac{1}{\surd 2} f_\rho m_\rho\,\epsilon^\lambda_\mu(P) \,,
\end{equation}
with $Q_e:= {\rm diag}[2/3,-1/3,-1/3]$; i.e., the quark's electromagnetic
charge matrix.  $\Gamma_{\rho^0\to e^+ e^-} = 6.77 \pm
0.32\,$keV~\cite{pdg98} $\Rightarrow$ $g_\rho = 5.03 \pm 0.12$ and hence
$f_\rho = 216 \pm 5\,$MeV.  For the $\phi$-meson
\begin{equation}
\frac{m_{\phi}^2}{g_\phi}\,\epsilon^\lambda_\mu(P) := \sfrac{1}{3}\,\langle
0| \bar s \gamma_\mu s | \phi_\lambda(P) \rangle :=
\sfrac{1}{3}\,f_\phi m_\phi\,\epsilon^\lambda_\mu(P)
\end{equation}
and hence $\Gamma_{\phi\to e^+ e^-} = 1.37 \pm 0.05\,$keV~\cite{pdg98}
$\Rightarrow$ $g_\phi= 12.9 \pm 0.2$ or $f_\phi = 238\pm 4\,$MeV.
$f_{K^\ast}$ follows from~\cite{pieterVM}
\begin{equation}
\label{fKast}
\Gamma_{\tau \to K^\ast\nu_\tau}/\Gamma_{\tau \to \rho\nu_\tau} = 0.051\;
\Rightarrow \; f_{K^\ast} = 1.042\,f_\rho\,.
\end{equation}

A number of other important observations are recorded in
Refs.~\cite{mr97,pieterVM}.  {\bf First}: The calculated values of observable
quantities are independent of the momentum partitioning parameter: $\eta_P$,
when all of the Dirac covariants, and their complete momentum dependence, are
retained; i.e., Poincar\'e invariance is manifest.  {\bf Second}: For
pseudoscalar mesons the leading $\gamma_5$-covariant is dominant but the
pseudovector components also play an important role; e.g., $f_K$ is $\lsim
30$\% smaller without them.  For the vector mesons, while $F_1$ is dominant,
$F_{2\ldots 5}$ are also important; e.g., $m_\rho$ is $\gsim 20$\% larger
without them.  {\bf Third}: Flavour nonsinglet pseudoscalar mesons obey a
mass formula~\cite{mr97,mrt98,miranskymunczek}, exact in QCD,
\begin{equation}
\label{massform}
f_H \,m_H^2 =  {\cal M}_H^\zeta \,r^\zeta_H\,,
\; \;{\cal M}_H^\zeta={\rm tr}_F
[ M_\zeta \{{\cal T}^H,({\cal T}^H)^{\rm t}\}]\,,
\end{equation}
where 
\begin{equation}
\label{rHres}
r^\zeta_H = 
-i\sqrt{2}\,Z_4\,{\rm tr}\int^\Lambda_k\, \left({\cal T}^H\right)^{\rm t}
\gamma_5 \,{\cal S}(k_+)\, \Gamma_H(k;P)\, {\cal S}(k_-)
:= - 2i\,\langle\bar q q \rangle_\zeta^H\, \frac{1}{f_H}
\end{equation}
is the gauge-invariant and cutoff-independent residue of the pion pole in the
pseudoscalar vertex.  As a residue, it is an analogue of $f_H$ and describes
the pseudoscalar projection of the unamputated Bethe-Salpeter wave function
calculated at the origin in configuration space.  For small current-quark
masses, Eq.~(\ref{massform}) yields the ``Gell-Mann--Oakes--Renner'' relation
as a corollary.  However, it is also valid for heavy-quarks and
predicts~\cite{mishaSVY,marisAdelaide} $m_H \propto \hat m_Q$ in the
heavy-meson domain, which is verified in the strong interaction spectrum.
{\bf Fourth}: The behaviour at large $k^2$ [ultraviolet relative momenta] is
model independent, determined as it is by the one-loop improved strong
running coupling.  {\bf Fifth}: For the pseudoscalar mesons, the asymptotic
behaviour of the subdominant pseudovector amplitudes [$F_H$, $G_H$] is
crucial for convergence of the integral describing $f_H$.  For the vector
mesons, the same is true of $F_{2\ldots 5}$.  This is also a
model-independent result because of the fourth point.

This subsection illustrates the reliability of the rainbow-ladder truncation
for light vector and flavour nonsinglet pseudoscalar mesons.  That is not an
accident but rather, as elucidated in Ref.~\cite{truncscheme}, it is the
result of cancellations between vertex corrections and crossed-box
contributions at each higher order in the quark-antiquark scattering kernel.
There are two other classes of light-meson: scalar and axial-vector.  A
separable model~\cite{conradsep} that expresses characteristics of the
rainbow-ladder truncation has been employed successfully in calculating the
masses and decay constants of $u$,$d$-quark axial-vector mesons~\cite{a1b1}.
Hence, while a more sophisticated study is still lacking, and is indeed
required, this suggests that the truncation can provide a good approximation
in this sector; i.e., that the conspiratorial cancellations are also
effective here.

In the scalar channel, however, the rainbow-ladder truncation is not certain
to provide a reliable approximation because the cancellations described above
do not occur~\cite{cdrqcII}.  This is entangled with the phenomenological
difficulties encountered in understanding the composition of scalar
resonances below
$1.4\,$GeV~\cite{pdg98,mikescalars1,mikescalars2,mikescalars3}.  For the
isoscalar-scalar meson the problem is exacerbated by the presence of timelike
gluon exchange contributions to the kernel, which are the analogue of those
diagrams expected to generate the $\eta$-$\eta^\prime$ mass splitting in BSE
studies~\cite{etareinhard}.  If the rainbow-ladder truncation is employed one
obtains ideal flavour mixing and degenerate isospin partners, and; e.g.,
\begin{equation}
\label{compile}
\begin{array}{lr|ccc}
                 & \mbox{Ref.} &(u/d)_{I=0,1} & u\bar s &  s\bar s \\\hline
{\rm calculated~mass}_{\rm\, in~GeV} 
                 & \protect\cite{jain} & 0.59        & 0.90 & 1.20\\
                 &  \protect\cite{pieterVM}
                                       & 0.67 & & \\
                 &  \protect\cite{conradsep}
                                       & 0.71        & 1.18  & 1.54  \\
                 &  \protect\cite{newsigmaT}
                                       & 0.59 & &\\\hline
{\rm averaged~mass}
                 &      & 0.64 \pm 0.06  & 1.04 \pm 0.20 & 1.37 \pm 0.24
\end{array}
\end{equation}
($\,0^{++}$ mesons containing at least one $s$-quark were not considered in
Refs.~\cite{pieterVM,newsigmaT}.)  Each model represented in this compilation
was constrained to accurately describe $\pi$- and $K$-meson observables, and
the standard deviation about their averaged vector meson masses is a factor
of $2$--$5$ smaller than that exhibited in the last row here.

In Eq.~(\ref{compile}) we do not compare directly with observed masses
because of the uncertainty in identifying the members of the scalar nonet.
We only note that: 1) the $u\bar s$ channel is least affected by those
corrections to the rainbow-ladder truncation that can alter the feature of
ideal flavour mixing and hence it may be appropriate to identify the $u\bar
s$ scalar with the $K_0^*(1430)$, in which case the mass is underestimated by
$\lsim 30\,$\%; and 2) an analysis of $\pi\pi$ data~\cite{mikescalars2}
identifies an isoscalar-scalar with
\begin{equation}
m_{0^{++}_{I=0}} \approx 0.46\,{\rm GeV}\,,\;\; \Gamma_{0^{++}_{I=0}\to
\pi\pi} \sim 0.22{\rm -}0.47\,{\rm GeV}\,.
\end{equation}
This supports ideal flavour mixing but, because $\Gamma_\sigma/m_\sigma$ is
large, suggests an accurate calculation of this state's mass will require a
Bethe-Salpeter kernel that explicitly includes couplings to the important
$\pi\pi$ mode.  In contrast, that coupling can be handled perturbatively in
the $\omega$-$\rho$ sector~\cite{hollenberg,rhopipiKLM,rhopipiMAP}.  If this
identification is correct then the mass estimate in Eq.~(\ref{compile}) is
$\lsim 40\,$\% too large.

The discussion here makes plain that the constituent-quark-like
rainbow-ladder scalar bound states are significantly altered by corrections
to the BSE's kernel.  That is good because it is consistent with observation:
understanding the scalar meson nonet is a complex problem.  (NB.  The
difficulties to be anticipated are illustrated; e.g., in
Ref.~\cite{carlshakin}.)  Using the DSEs this complexity is expressed in
unanswered questions.  For example, why do timelike gluon exchange
contributions to the kernel in the isoscalar-scalar channel not force a
deviation from ideal flavour mixing and, returning to the pseudoscalar
sector, what effect do they play in the $\eta$-$\eta^\prime$ mass splitting?

There are other contemporary questions.  For example, which improvements to
the rainbow-ladder kernel are necessary in order to study bound states
containing at least one heavy-quark?  The extent to which the cancellations
elucidated in Ref.~\cite{truncscheme} persist as the current-quark mass
evolves to values larger than $M_\chi$ is not known.  Even though the
rainbow-ladder truncation can provide an acceptable estimate of heavy-meson
masses; e.g., Ref.~\cite{jain}, improvements are necessary and potentials
derived from dual-QCD models; e.g., Refs.~\cite{marshalldualreview,tokilike},
have been applied more exhaustively~\cite{lewis}.  (Heavy-heavy-mesons are
also amenable to study via heavy-quark expansions~\cite{noraheavy}).  Other
composites admitted by QCD can also be considered.  The rainbow-ladder
truncation has recently been applied to ``exotic'' light-mesons, predicting a
$J^{PC}=1^{-+}$ meson with a mass of $\sim 1.4$ --
$1.5\,$GeV~\cite{bpprivate}.  [``exotic'' because such a $J^{PC}$ value is
unobtainable in the $q\bar q$ constituent quark model.]  However, there is a
dearth of DSE applications to the glueball spectrum, which has been explored
using other methods; e.g., Refs.~\cite{ericold,perry,dualQCDglueball}.  A
gluon-sector analogue of the rainbow-ladder truncation is an obvious starting
point and such studies must precede any exploration of hybrid quark-gluon
states~\cite{hybrids}.

That applications in all these areas are actively being pursued and
contemplated is an indication of a healthy discipline.

\addcontentsline{toc}{subsubsection}{\protect\numberline{ } {Electromagnetic
Pion Form Factor and Vector Dominance}} 
\subsubsect{Electromagnetic Pion Form Factor and Vector Dominance}
We have seen that the renormalisation group improved rainbow-ladder
truncation of the quark-antiquark scattering kernel: $K$, provides a good
understanding of colour singlet, mesonic spectral functions.  That is a key
success since describing these correlation functions is a core problem in
QCD~\cite{instantonreview}.  However, it is only a single step and one must
proceed from this foundation to the study of scattering observables, which
delve deeper into hadron structure.

The best such observables to study are elastic electromagnetic form factors,
because the probe is well understood, and the simplest ``target'' for a
theorist is the pion, as long as the theoretical framework accurately
describes DCSB.  The electromagnetic pion form factor is a much studied
observable but here, to be concrete, we focus on the application of DSEs to
this problem, and in that connection Ref.~\cite{grosspion} is a pilot.  As
our exemplar we choose Ref.~\cite{mtpion} because it is a direct application
of the effective interaction described in the previous subsection,
Eq.~(\ref{gk2VM}), and it is the most complete study to date.

In the isospin-symmetric limit the renormalised impulse approximation to the
pion's electromagnetic form factor is
\begin{eqnarray}
\label{pipiA}
\lefteqn{(p_1 + p_2)_\mu\,F_\pi(q^2):= \Lambda_\mu(p_1,p_2) }\\ & & \nonumber
= 2\, {\rm tr}\int_k^\Lambda \bar\Gamma_\pi(k;-p_2) {\cal S}(k_{++})\,
iQ_e\Gamma^\gamma_\mu(k_{++},k_{+-})\,{\cal S}(k_{+-})\,
\Gamma_\pi(k-q/2;p_1)\,{\cal S}(k_{--})\,,
\end{eqnarray}
$k_{\alpha\beta}:= k + \alpha p_1/2 + \beta q/2$ and $p_2:= p_1 + q$.  Here,
$\Gamma_\pi(k;P)$ is the pion BSA, which has the form in
Eq.~(\ref{genpibsa}), and ${\cal S}(k)={\rm diag}[S_{u=d}(k),S_{u=d}(k)]$.
No renormalisation constants appear explicitly in Eq.~(\ref{pipiA}) because
the renormalised dressed-quark-photon vertex: $\Gamma^\gamma_\mu$, satisfies
the vector Ward-Takahashi identity:
\begin{equation}
\label{vwti}
(p_1 - p_2)_\mu \, i\Gamma^\gamma_\mu(p_1,p_2) = 
S^{-1}(p_1) - S^{-1}(p_2)\,.
\end{equation}
Importantly, this also ensures current conservation:
\begin{equation}
(p_1-p_2)_\mu\,\Lambda_\mu(p_1,p_2)=0\,,
\end{equation}
and, using Eq.~(\ref{pinorm}), the correct normalisation of the form factor:
\begin{equation}
F(q^2=0)=1\,;
\end{equation}
i.e., combining the rainbow-ladder truncation of the quark-antiquark
scattering kernel with the impulse approximation yields a minimal, consistent
approximation~\cite{cdrpion}.  [NB.  As the $\pi^0$ is a charge conjugation
eigenstate, $F_{\pi^0}(q^2) \equiv 0$, $\forall q^2$.  The impulse
approximation yields this result.]

The only quantity in Eq.~(\ref{pipiA}) not already known is
$\Gamma^\gamma_\mu$.  It is the solution of an inhomogeneous BSE, which in
rainbow-ladder truncation is
\begin{equation}
\label{invectorbse}
\Gamma_{\mu}^\gamma(k;P) = Z_2\,\gamma_\mu - \int^\Lambda_q {\cal
G}((k-q)^2)\, D_{\alpha\beta}^{\rm free}(k-q) \frac{\lambda^a}{2}\gamma_\alpha\,
{\cal S}(q_+)\,\Gamma_{\mu}^\gamma(k;P) \,{\cal S}(q_-)\,
\frac{\lambda^a}{2}\gamma_\beta\,.
\end{equation}
It is straightforward to verify that Eq.~(\ref{vwti}) is satisfied, and owing
to this the general solution of Eq.~(\ref{invectorbse}) involves only eight
independent scalar functions and can be expressed as
\begin{equation}
\label{gvtxform}
\Gamma_{\mu}^\gamma(k;P) = \Gamma_\mu^{\rm BC}(k;P) +
\sum_{l=1}^8\,O_\mu^l\,F_l^\gamma(k;P)\,,
\end{equation}
where the matrices $O_\mu^l$ are given in Eq.~(\ref{defineOs})
and~\cite{bc80}
\begin{equation}
\label{bcvtx}
\Gamma_\mu^{\rm BC}(k;P)  =  i\Sigma_A(k_+^2,k_-^2)\,\gamma_\mu
+ (k_++k_-)_\mu\,\left[\sfrac{1}{2}i\gamma\cdot (k_++k_-) \,
\Delta_A(k_+^2,k_-^2) + \Delta_B(k_+^2,k_-^2)\right]\,;
\end{equation}
\begin{equation}
\Sigma_F(k_+^2,k_-^2)  =  \sfrac{1}{2}\,[F(k_+^2)+F(k_-^2)]\,,\;
\Delta_F(k_+^2,k_-^2) = \frac{F(k_+^2)-F(k_-^2)}{k_+^2-k_-^2}\,,
\end{equation}
with $F= A, B$; i.e., the scalar functions in the dressed-quark propagator.
$\Gamma_\mu^{\rm BC}(k;P)$ saturates the vector Ward-Takahashi identity, and
the remaining terms are transverse and hence do not contribute to the
r.h.s. of Eq.~(\ref{vwti}).

The importance of determining the dressed-quark-photon vertex from
Eq.~(\ref{invectorbse}) was recognised in Ref.~\cite{frankvertex}, where a
solution was obtained using the simple model of ${\cal G}(k^2)$ introduced in
Ref.~\cite{mn83}.  As is readily anticipated, the dressed-vertex exhibits
isolated simple poles at timelike values of $P^2$.  Each pole corresponds to
a $1^{-\,-}$ bound state, $P^2=-\!$ mass-squared, and its matrix-valued
residue is proportional to the bound state amplitude.  For $Q^2$ in the
neighbourhood of any one of these poles the behaviour of the pion form factor
is primarily determined by the manifestation of that bound state in the
$1^{-\,-}$ spectral density.  Vector meson dominance, in any of its forms, is
an {\it Ansatz} to be used for extrapolating outside of these neighbourhoods.
A direct solution of Eq.~(\ref{invectorbse}) obviates the need for such an
expedient and also any need to fabricate an interpretation of an off-shell
bound state.

\begin{figure}[t]
\centering{\ \epsfig{figure=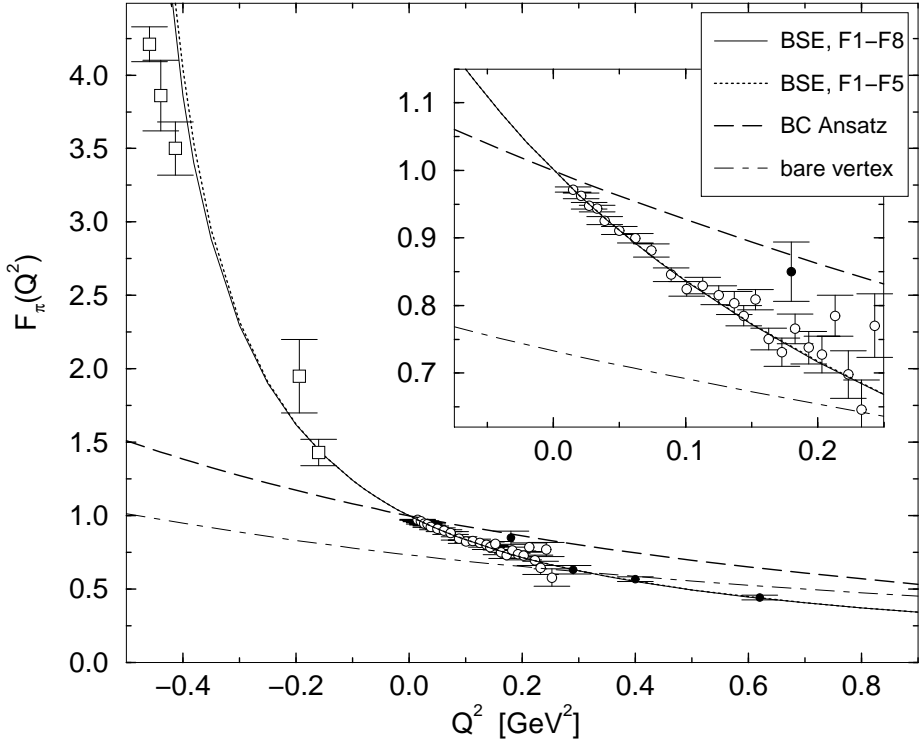,height=10cm}}\vspace*{0.1\baselineskip}

\parbox{40em}{\caption{Pion form factor calculated directly from the
interaction in Eq.~(\protect\ref{gk2VM}): solid line.  The data are: open
circles, Ref.~\protect\cite{pionffdat1}; squares,
Ref.~\protect\cite{pionffdat2}; filled circles,
Ref.~\protect\cite{pionffdat3}.  No parameters were varied to obtain the
result.  The other curves depict results from simplified calculations: dotted
line, retaining only the dominant vector covariants in
Eq.~(\protect\ref{gvtxform}), $F^\gamma_{1,\ldots,5}$; dashed line, retaining
only $\Gamma_\mu^{\rm BC}$ calculated with the numerical solution of the
quark DSE; dash-dot line, $\Gamma_{\mu}^\gamma(k;P) = \gamma_\mu$, which
violates the Ward-Takahashi identity.  (Adapted from
Ref.~\protect\cite{mtpion}.)\label{mtpionff}}}
\end{figure}

Using the interaction of Eq.~(\ref{gk2VM}), with its single parameter fixed
as discussed in connection with Eq.~(\ref{DvalueVM}), and solving numerically
for the renormalised dressed-quark propagator, pion BSA and
dressed-quark-photon vertex, Ref.~\cite{mtpion} obtains the pion form factor
depicted in Fig.~\ref{mtpionff} with
\begin{equation}
r_\pi^2 := -6 \left.\frac{d F_\pi(y)}{d y}\right|_{y=0}\!\Rightarrow
r_\pi = 0.68\,{\rm fm} \;\;{\rm cf.}\;\;r_\pi^{\rm Obs} = 0.663 \pm 0.006 \,.
\end{equation}
The complete calculation describes the data for both spacelike {\it and}
timelike momenta, plainly exhibiting the evolution to a simple pole
corresponding to the $\rho$-meson.  Here the $\rho$ is described by a simple
pole on the real-$P^2$ axis because the rainbow-ladder truncation excludes
the $\pi\pi$ contribution in the kernel.  However, that can be included
perturbatively~\cite{hollenberg,rhopipiKLM,rhopipiMAP} and estimates show it
yields no-more-than a $15\,$\% increase in $r_\pi$\cite{piloop}.  [NB.  As
illustrated by Refs.~\cite{wwbuck,conradkaon,mtkaon}, the extension to
$K$-meson form factors is straightforward but with the interesting new
feature that, unlike the neutral pion's elastic form factor,
$F_{K^0}(Q^2)\neq 0$, $\forall Q^2>0$, because the neutral kaons are not
charge-conjugation eigenstates.]

Proponents of vector meson dominance {\it Ans\"atze}~\cite{heath} have
historically claimed support in the accuracy of the estimate $r_\pi^2\approx
6/m_\rho^2$ and in this connection one can ask what alternative does the
direct calculation describe?  To address this Ref.~\cite{mtpion} observed
that, for $Q^2\in[-m_\rho^2,0.2\,{\rm GeV}^2]$, an {\it interpolation} of the
solution of Eq.~(\ref{invectorbse}) is provided by
\begin{equation}
\label{bcplus}
\Gamma_{\mu}^\gamma(k;Q) \approx \Gamma_\mu^{\rm BC}(k;Q) -
\frac{1}{g_\rho}\,\frac{Q^2}{Q^2+m_\rho^2}\,
\sum_{l=1}^5\,O_\mu^l\,F_l^\rho(k^2|Q^2=-m_\rho^2)\,,
\end{equation}
where 
\begin{equation}
F_l^\rho(k^2|Q^2=-m_\rho^2):= \frac{2}{\pi}\int_{-1}^{1}\,
dx\, \sqrt{1-x^2}\, F_l^\rho(k^2,ik\,m_\rho\, x|Q^2=-m_\rho^2)
\end{equation}
are the leading Chebyshev moments of the five dominant scalar functions in
the $\rho$-meson BSA and, following Ref.~\cite{mrt98}, $1/g_\rho$ is the
residue of the $\rho$-meson pole in the photon vacuum polarisation.
Substituting Eq.~(\ref{bcplus}) into Eq.~(\ref{pipiA}) yields
\begin{equation}
\label{Fpibcplus}
F_\pi(Q^2) \approx F_\pi^{\rm BC}(Q^2) -
\frac{g_{\rho\pi\pi}}{g_\rho}\,F_{\rho\pi\pi}(Q^2)\,\frac{Q^2}{Q^2+m_\rho^2}\,,
\end{equation}
where $g_{\rho\pi\pi}\,F_{\rho\pi\pi}(Q^2)$ is the impulse approximation to
the $\rho\,\pi\pi$ vertex, $F_{\rho\pi\pi}(Q^2=-m_\rho^2)=1$.  Implicit in
Eq.~(\ref{bcplus}) is a clear but not unique {\it definition} of an off-shell
$\rho$-$\bar q q$ correlation, which yields Eq.~(\ref{Fpibcplus}) as a
calculable approximation to the form factor wherein
\begin{equation}
\label{rpirho1}
(r_\pi^\rho)^2 := \frac{6}{m_\rho^2}\,
\frac{g_{\rho\pi\pi}}{g_\rho}\,F_{\rho\pi\pi}(0).  
\end{equation}
This is simply the vector meson dominance result {\it corrected} for
nonuniversality of the strong and electromagnetic $\rho$-meson couplings;
i.e., $g_\rho\neq g_{\rho\pi\pi}$, and, via $F_{\rho\pi\pi}(0)$, for the
nonpointlike nature of the off-shell $\rho$-$\bar q q$ correlation.
$F_{\rho\pi\pi}(Q^2)$ was not calculated in Ref.~\cite{mtpion} and therefore
here we make an estimate.  Typically $F_{\rho\pi\pi}(0) \approx
0.5$~\cite{rhopipiKLM} and using experimental values for the observables
[including $g_{\rho\pi\pi}=6.05$] we find
\begin{equation}
(r_\pi^\rho)^2 \sim  0.5\,r_\pi^2,
\end{equation}
which is a significant suppression with respect to the ``naive'' vector meson
dominance estimate.  

This illustration demonstrates that the pion charge radius is indeed
influenced by the $\rho$-pole's contribution to the $1^{-\,-}$ spectral
density.  However, that is unsurprising: the bound state poles {\it are} a
significant feature of the dressed-vertex.  More important is a realisation
that the separation in Eq.~(\ref{bcplus}) is completely arbitrary and hence
so is the fraction of the charge radius attributed to the ``off-shell
$\rho$-meson.''  One can shift any amount of strength between the two terms
and yet maintain an accurate interpolation of the dressed-vertex.  For
example,
\begin{equation}
\label{nugatory}
\frac{Q^2}{Q^2+m_\rho^2} = \frac{Q^2}{Q^2+m_\rho^2}\, {\rm
e}^{-\omega (1+Q^2/m_\rho^2)} + \frac{Q^2}{Q^2+m_\rho^2} \left[ 1 - {\rm
e}^{-\omega (1+Q^2/m_\rho^2)} \right]\,,
\end{equation}
is an apparently nugatory rearrangement. However, only the first term has a
pole at $Q^2+m_\rho^2=0$, and substituting Eq.~(\ref{nugatory}) into
Eq.~(\ref{bcplus}) and repeating the analysis one arrives at
\begin{equation}
\label{rhopi2}
(r_\pi^\rho)^2 = {\rm e}^{-\omega}\,\frac{6}{m_\rho^2}\,
\frac{g_{\rho\pi\pi}}{g_\rho}\,F_{\rho\pi\pi}(0)\, \stackrel{\omega =
1}{\sim}\, 0.2\,r_\pi^2\,.
\end{equation}
It is obvious now that $r_\pi^\rho$ can be made arbitrarily small while
preserving $r_\pi\approx r_\pi^{\rm Obs}$.  [NB.  Eq.~(\ref{rhopi2}) and the
procedure leading to it are no more contrived than Eq.~(\ref{rpirho1})
because Eq.~(\ref{nugatory}) can be interpreted as ``unfreezing'' the vector
meson BSA; i.e., of allowing the bound state correlation to be suppressed
off-shell: \mbox{$F_l^\rho(k^2|Q^2=-m_\rho^2) \to \exp(-\omega
[1+Q^2/m_\rho^2])\,F_l^\rho(k^2|Q^2=-m_\rho^2)$}.]

For large spacelike $Q^2$ the calculation of $F_\pi(Q^2)$ directly from
${\cal G}(k^2)$ is a computational challenge and for practical reasons the
study in Ref.~\cite{mtpion} was restricted to $Q^2\leq 1\,$GeV.  The
effective interaction doesn't appear explicitly in the impulse approximation,
Eq.~(\ref{pipiA}), only the dressed-quark propagator, pion Bethe-Salpeter
amplitude and dressed-quark-photon vertex.  Hence the numerical analysis is
markedly simplified if algebraic approximations to these Schwinger functions
are employed.

This was the approach adopted in Ref.~\cite{cdrpion}, which has since been
used in a wide range of applications; e.g.,
Refs.~\cite{mishaSVY,fizikaB,echaya,serdar,rhopipiMAP,fredFF,hecht}.  It is
efficacious because the vector and axial-vector Ward-Takahashi identities can
be used to motivate {\it Ans\"atze} for $\Gamma_\pi$ and $\Gamma_\mu^\gamma$
that are expressed solely in terms of $S$.  In the present context, algebraic
models for $S$ and $\Gamma_\pi$ have been developed~\cite{mrpion} that encode
the important qualitative aspects of the DSE and BSE solutions in
Ref.~\cite{mr97}.  These forms, along with $\Gamma_\mu^\gamma$ determined via
Eq.~(\ref{bcvtx}), make possible a calculation of $F_\pi(Q^2)$, $\forall
Q^2>0$, and an analytic analysis of the asymptotic behaviour.  One finds that
the pion's pseudoscalar covariant can alone provide a quantitative
description of $F_\pi(Q^2)$ for $Q^2\lsim 5\,$GeV$^2$.  However, beyond this
point the pseudovector covariants become important.  It is these terms in
Eq.~(\ref{genpibsa}) that ensure
\begin{equation}
\label{FpiUV}
Q^2\,F_\pi(Q^2) = {\rm const.},
\end{equation}
up to calculable $\ln (Q/\Lambda_{\rm QCD})^p$-corrections, and this
behaviour is unmistakable for $Q^2\gsim 10\,$GeV$^2$.  The anomalous
dimension: $p$, is determined by that of $F_\pi$ and $G_\pi$, which is a
model independent result.  [NB.  If the pseudovector covariants are
neglected, $Q^4\,F_\pi(Q^2) = {\rm const.}$.  Further, the dressed-quark
propagator obtained in Ref.~\cite{diakonov} necessarily yields results
inconsistent with Eq.~(\ref{FpiUV}).]  This application is an emphatic
demonstration of the DSEs' ability to provide a single and simultaneous
description of the infrared and ultraviolet aspects of observables.

\label{`anomalous processes'} That is also much in evidence in the study of
anomalous processes.  As first observed~\cite{jpopiejustin1justin2} in
connection with processes like $K^+ K^- \to \pi^+ \pi^0 \pi^-$, the
systematically truncated DSEs yield the anomalies of current algebra without
any model dependence; i.e., the anomalies remain a feature of the global
aspects of DCSB~\cite{witten}.  For the $\pi^0\to\gamma\gamma$ process this
feature was verified in Refs.~\cite{mrpion,cdrpion,bando} and for the
$\gamma\pi^\ast\to\pi\pi$ transition form factor, in Ref.~\cite{reinhardA}.
Consequently the DSEs provide a single framework wherein the value of such
transition form factors is fixed and model-independent at the soft-pion, zero
momentum transfer point, and their evolution is calculable on the entire
range of momentum transfer, reproducing the ultraviolet behaviour anticipated
from perturbative QCD.  These features have been elucidated and exploited in
Refs.~\cite{klabucar,fizikaB,klabucar2}.

\addcontentsline{toc}{subsubsection}{\protect\numberline{ } {Describing
Baryons}}  
\subsubsect{Describing Baryons}
Hitherto we have described the application of DSEs to the study of mesonic
observables, which requires and illustrates a contemporary understanding of
the two-body problem in quantum field theory.  Baryons, as a three-body
problem, pose a greater challenge and historically they have been described
using constituent-quark-like models, which remain an important contemporary
tool; e.g., Refs.~\cite{simon,serdarthomasfritzsimula}.  We identify a
beginning of progress with a direct assault on the baryon problem in a
realisation~\cite{regdq,hugodq} that field theoretical models of the strong
interaction admit the construction of a meson-diquark auxiliary-field
effective action and thereby a description of baryons as loosely-bound
quark-diquark composites.

A head-on DSE approach begins with a relativistic Fadde$^\prime$ev equation,
which can be derived~\cite{regfe} by exploiting the fact that single gluon
exchange between two quarks is attractive in the colour-antitriplet channel,
whether or not the gluon is dressed.  Indeed, using a rainbow-ladder
truncation of the quark-quark scattering kernel, one obtains nonpointlike,
colour-antitriplet diquark [quark-quark] bound states in the strong
interaction spectrum~\cite{justindq}.  However, as demonstrated in
Ref.~\cite{truncscheme}, this is a flaw of the rainbow-ladder truncation:
higher order terms in the kernel ensure that the quark-quark scattering
matrix does not exhibit the simple poles that correspond to asymptotic
states.  Nevertheless, studies with improved kernels~\cite{jacquesdq} do
support a physical interpretation of the spurious rainbow-ladder diquark
masses.  Denoting the mass by $m_{qq}$, then $\ell_{qq}:= 1/m_{qq}$
represents the range over which a true diquark correlation in this channel
can persist {\it inside} a baryon.  In this sense they are
``pseudo-particle'' masses and can be used to estimate which two-body
correlations should be retained in solving the Fadde$^\prime$ev equation.
(NB.  Gluon mediated interactions are repulsive in the colour-sextet
quark-quark channel, just as they are in the colour-octet meson
channel~\cite{justindq}.)

Reference~\cite{conradsep} tabulates calculated values of these
pseudo-particle masses, from which we extract:
\begin{equation}
\label{diquarkmasses}
\begin{array}{l|cccccccc}
(qq)_{J^P}           & (ud)_{0^+} & (us)_{0^+}  & (uu)_{1^+}& (us)_{1^+} 
        & (ss)_{1^+} & (uu)_{1^-} & (us)_{1^-} & (ss)_{1^-}\\\hline
 m_{qq}\,({\rm GeV}) & 0.74       & 0.88        & 0.95      & 1.05 
        & 1.13       & 1.47 & 1.53 & 1.64
\end{array}
\end{equation}
The mass ordering is characteristic and model-independent, and indicates that
an accurate study of the nucleon should retain the scalar and pseudovector
correlations: $(ud)_{0^+}$, $(uu)_{1^+}$, because for these $m_{qq} \lsim
m_N$.  This expectation is verified in calculations, where one
finds~\cite{gcm98,ishii,raAB} that including the $(uu)_{1^+}$ correlation
yields a nucleon whose mass is a welcome $\sim 33\,$\% less-than that of a
quark$+$scalar-diquark-only nucleon.  We note that
$m_{(ud)_{0^+}}/m_{(uu)_{1^+}} = 0.78$~cf.~$0.76 = m_N/m_\Delta$ and hence
one might anticipate that the presence of diquark correlations in baryons is
likely to provide a straightforward explanation of the $N$-$\Delta$
mass-splitting.  Lattice estimates, where available~\cite{latticediquark},
agree with these results.

\begin{figure}[t]
\centering{\ \epsfig{figure=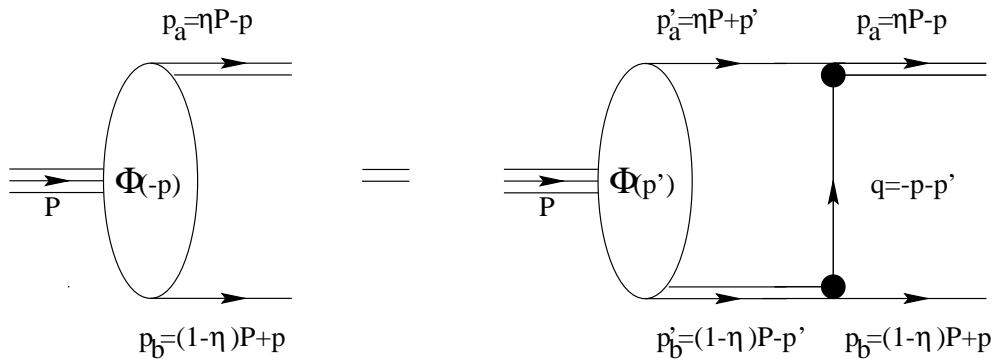,height=4.8cm}}

\parbox{40em}{\caption{A quark-diquark Fadde$^\prime$ev equation for a baryon
with total momentum $P$: single line, quark propagator; double line, diquark
propagator; $\Phi(p)$, quark-diquark Bethe-Salpeter amplitude.  [For
nonpointlike diquarks, the filled circle represents a diquark Bethe-Salpeter
amplitude.]  Here binding is effected by a participant quark leaving the
diquark and joining with the dormant quark to form another diquark; i.e., via
iterated quark exchange.  (Adapted from Ref.~\protect\cite{reinhard}.)
\label{raFE}}}
\end{figure}

Reference~\cite{reinhard} is an extensive study of the octet and decuplet
baryon spectrum based on a quark-diquark Fadde$^\prime$ev equation.  It
represents the nucleon as a composite of a quark and {\it pointlike} diquark,
which are bound together by a repeated exchange of roles between the dormant
and diquark-participant quarks, as depicted in Fig.~\ref{raFE}, and employs
the confined-particle representation of the dressed-quark and -diquark
propagators advocated in Ref.~\cite{conradsep}.  Seven parameters appear in
the model and, unfortunately, variations are permitted in some of them that
allow the calculated results to override intuitive expectations.  For
example, the $N$-$\Delta$ mass-splitting is fitted by adjusting the relative
strength of the quark-diquark couplings in the scalar and pseudovector
channels whilst simultaneously enforcing $m_{(ud)_{0^+}}=m_{(uu)_{1^+}}$.
Bethe-Salpeter equation studies show that this is erroneous; i.e., these
masses cannot be equal, and they and the couplings are not independent.
Hence this means of generating the $N$-$\Delta$ mass-splitting is unlikely to
be completely correct.  Nevertheless, the importance of the study is a
demonstration that an accurate description of the spectrum is possible, and
indeed no calculated mass is more-than $1$\% away from its experimental
value.  Improvements are now possible and they will build on the lessons
Ref.~\cite{reinhard} provides; e.g, replacing undetermined parameters with
values obtained in precursor calculations.  This is an important frontier.

Even before such thorough relativistic bound state calculations, the notion
that diquark correlations in baryons could be significant found support in
the analysis of scattering observables~\cite{anselmino}.  As remarked
earlier, these observables delve deep into hadron structure and so are a
perennial focus of theory and experiment.  With a growing understanding of
the nature of diquark correlations, detailed nucleon models are now being
applied in the nonperturbative evaluation of nucleon structure functions;
e.g., Ref.~\cite{piller}, and nucleon form
factors~\cite{bentzFad,keiner,jacquesnucleon,pichowskydiquark,myriad}.
As an exemplar, we review a calculation of the nucleon's scalar form factor,
which also yields the nucleon $\sigma$-term~\cite{myriad}.  This is an
important application because it illustrates the only method that allows an
unambiguous off-shell extrapolation in the estimation of meson-nucleon form
factors and thereby provides an analogue for the discussion of vector meson
dominance, Eqs.~(\ref{bcplus}-\ref{rhopi2}).

Underlying Ref.~\cite{myriad} is the observation that nucleon propagation is
described by a $6$-point Schwinger function
\begin{equation}
G_{\alpha\alpha^\prime}^{\tau\tau^\prime}(R-R^\prime)
:= \langle \Psi_\alpha^\tau(R) 
\bar\Psi_{\alpha^\prime}^{\tau^\prime}(R^\prime)\rangle\,,
\end{equation}
where 
\begin{eqnarray}
\label{PsiGrass}
\Psi_\alpha^\tau(R) & := & \int\prod_{i=1}^3 d^4x_i\,
\psi(x_i-R;\alpha_i,\tau_i;\alpha,\tau)
\varepsilon_{abc} \,q_a(x_1;\alpha_1,\tau_1)\,
q_b(x_2;\alpha_2,\tau_2)\, q_c(x_3;\alpha_3,\tau_3),\\
\bar\Psi_\alpha^\tau(R) & := & \int\prod_{i=1}^3 d^4x_i\,
\psi(x_i-R;\alpha_i,\tau_i;\alpha,\tau)^\ast
\varepsilon_{abc} \,\bar q_a(x_1;\alpha_1,\tau_1)\,
\bar q_b(x_2;\alpha_2,\tau_2)\, \bar q_c(x_3;\alpha_3,\tau_3),
\end{eqnarray}
with: $q_a(x_1;\alpha_1,\tau_1)$, etc., quark Grassmann variables;
$\alpha_i$, $\alpha$ quark and nucleon Dirac subscripts; $\tau_i$, $\tau$ the
isospin analogues; and $\varepsilon_{abc}$ ensuring colour neutrality.  In
these expressions $\psi(x_i-R;\alpha_i,\tau_i;\alpha,\tau)$ describes the
distribution of quarks in the nucleon and; e.g., it can represent a nucleon
Fadde$^\prime$ev amplitude.  The electromagnetic interaction of this nucleon
is described by the current
\begin{equation}
\label{emcurrentN}
J_\mu(R^\prime-R_0,R_0-R) = -\langle \bar\Psi(R^\prime)\,\bar q(R_0)
iQ_e\gamma_\mu q(R_0)\, \Psi(R)\rangle\,.
\end{equation}

To proceed, Ref.~\cite{myriad} writes
\begin{equation}
\psi(x_i-R;\alpha_i,\tau_i;\alpha,\tau) = \int\prod_{i=1}^3\,\frac{d^4
p_i}{(2\pi)^4} \psi(p_i;\alpha_i,\tau_i;\alpha,\tau)\,
\exp\left[-i\sum_{i=1}^3 p_i\cdot (x_i-R)\right]\,
\end{equation}
and employs a product {\it Ansatz} for the nucleon amplitude
\begin{equation}
\label{Psi}
\psi(p_i;\alpha_i,\tau_i;\alpha,\tau)= 
\delta^{\tau \tau_3}\,\delta_{\alpha \alpha_3}\,\psi(p_1+p_2,p_3)\,
\Delta(p_1+p_2)\,
\Gamma_{\alpha_1 \alpha_2}^{\tau_1 \tau_2}(p_1,p_2) \,,
\end{equation}
where $\psi(\ell_1,\ell_2)$ is a Bethe-Salpeter-like amplitude characterising
the relative-momentum dependence of the correlation between diquark and quark
[$\sim\Phi$ in Fig.~\ref{raFE}], $\Delta(K)$ describes the pseudo-particle
propagation characteristics of the diquark, and
\begin{eqnarray}
\label{gdq}
\Gamma_{\alpha_1 \alpha_2}^{\tau_1 \tau_2}(p_1,p_2) & = &
(C i\gamma_5)_{\alpha_1 \alpha_2}\, (i\tau^2)^{\tau_1\tau_2}\,
\Gamma (p_1,p_2)
\end{eqnarray}
represents the momentum-dependence, and spin and isospin character of the
diquark correlation; i.e., it corresponds to a Bethe-Salpeter-like amplitude
for what here is a nonpointlike diquark.  Complete antisymmetrisation is not
explicit in this product {\it Ansatz} but that is effected in $\Psi$ via the
contraction with the Grassmann elements, Eq.~(\ref{PsiGrass}).

\begin{figure}[t]
\begin{center}
\parbox{35em}{
\epsfig{figure=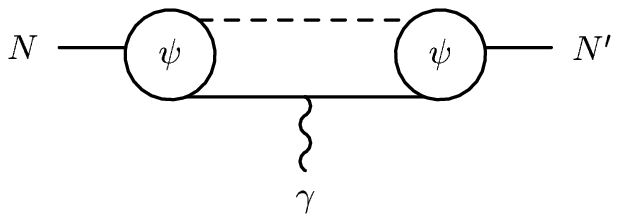,height=2.0cm}\vspace*{-3.8\baselineskip}

\hspace*{\fill}\epsfig{figure=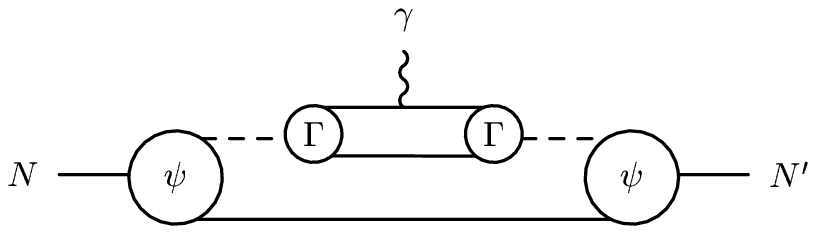,height=2.0cm}

\epsfig{figure=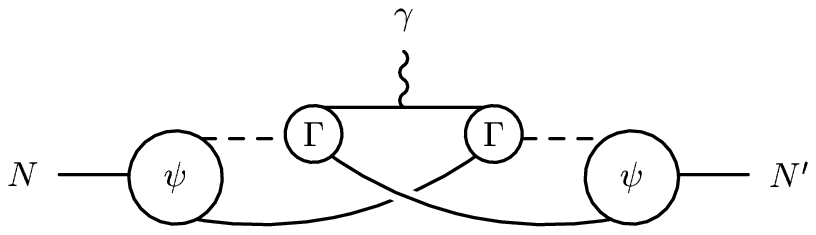,height=2.0cm}\vspace*{0.6\baselineskip}

\hspace*{\fill}\epsfig{figure=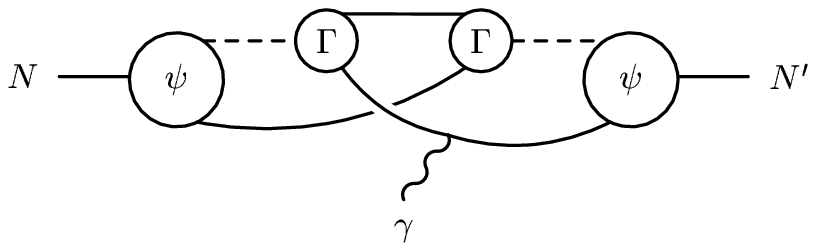,height=2.0cm}\vspace*{-1.2\baselineskip}

\epsfig{figure=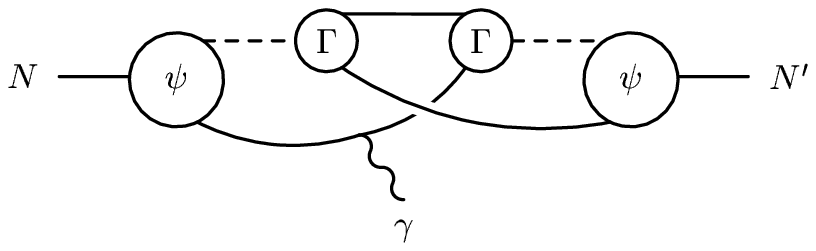,height=2.0cm}}
\parbox{40em}{\caption{Using the product {\it Ansatz} of
Eq.~(\protect\ref{Psi}), the impulse approximation to the electromagnetic
current, Eq.~(\protect\ref{emcurrentN}), requires the calculation of these
five contributions.  $\psi$: $\psi(\ell_1,\ell_2)$ in (\protect\ref{Psi});
$\Gamma$: Bethe-Salpeter-like diquark amplitude in (\protect\ref{gdq});
dotted line: $\Delta(K)$, diquark propagator in (\protect\ref{dprop}); solid
line: $S(q)$, quark propagator in (\protect\ref{qprop}).  The lowest three
diagrams, which describe the interchange between the dormant quark and the
diquark participants, effect the antisymmetrisation of the nucleon's
Fadde$^\prime$ev amplitude.  Current conservation follows because the
photon-quark vertex is dressed, given in (\protect\ref{bcvtx}).
\label{jacquesPic}}}
\end{center}
\end{figure}

The impulse approximation to the electromagnetic current in this model can
now be obtained directly from Eq.~(\ref{emcurrentN}).  This $8$-point
Schwinger function expresses a product of eight Grassmann variables and, via
an analogue of Wick's theorem, that can be reduced to a sum of products of
four $2$-point dressed-quark Schwinger functions and the dressed-quark-photon
vertex, with momenta and indices correlated via Eq.~(\ref{Psi}).  Using the
$1 \leftrightarrow 2$ particle exchange symmetry exhibited explicitly by
Eq.~(\ref{Psi}) one arrives at the result depicted in Fig.~\ref{jacquesPic}.

The product {\it Ansatz} is completely specified once explicit forms for the
functions are given and Ref.~\cite{myriad} employs
\begin{eqnarray}
\label{littlepsi}
\psi(\ell_1,\ell_2) & = & \frac{1}{{\cal N}_\Psi}\,{\cal
F}(\ell^2/\omega_\psi^2)\,,\;\ell := \sfrac{1}{3}\,\ell_1 -
\sfrac{2}{3}\,\ell_2\,,\\ 
\Gamma(q_1,q_2) & = & 
 \frac{1}{{\cal N}_\Gamma}\,
        {\cal F}(q^2/\omega_\Gamma^2)\,,\;q:=
\sfrac{1}{2}\,q_1-\sfrac{1}{2}\,q_2\,, \\
\label{dprop}
\Delta(K)  & = & \frac{1}{m_\Delta^2}\,{\cal F}(K^2/\omega_\Gamma^2)\,,\\
{\cal F}(y) & = & \frac{1- {\rm e}^{-y}}{y}\,,
\end{eqnarray}
whose parameters were fixed by fitting the proton's charge form factor on
$Q^2\in[0,3]\,$GeV$^2$~\cite{jacquesnucleon}:
\begin{equation}
\label{paramsNFF}
\begin{array}{l|ccc}
               & \;\omega_\psi & \omega_\Gamma & m_\Delta \\\hline
\mbox{in GeV}  &  \;0.20  & 1.4   & 0.63
\end{array}\,,\;\;\;\;
\begin{array}{l|ccc}
               & \;1/\omega_\psi & 1/\omega_\Gamma & 1/m_\Delta \\\hline
\mbox{in fm}  &  0.99   & 0.14 & 0.31
\end{array}\,.
\end{equation}
The two tables demonstrate the internal consistency of the model.
$d_\Gamma:= 1/\omega_\Gamma$ is a measure of the mean separation between the
quarks constituting the scalar diquark and $d_\psi:= 1/\omega_\psi$ is the
analogue for the quark-diquark separation.  $d_\Gamma<d_\psi$ is necessary if
the quark-quark clustering interpretation is to be valid.
$\ell_{(ud)_{0^+}}= 1/m_\Delta$ is a measure of the range over which the
diquark persists and that must be significantly less than the nucleon's
diameter.  [${\cal N}_\Psi$ and ${\cal N}_\Gamma$ are the nucleon and $(ud)$
diquark normalisation constants, which are defined and calculated analogously
to Eq.~(\ref{pinorm}), and ensure composite electric charges of $1$ for the
proton and $1/3$ for the diquark.]

As is plain in Fig.~\ref{jacquesPic}, the calculation also involves the
dressed-quark propagator and, based on the success of applications such as
Refs.~\cite{mishaSVY,fizikaB,echaya,serdar,rhopipiMAP,fredFF,hecht}, the
following algebraic parametrisation is employed in
Refs.~\cite{jacquesnucleon,myriad}:
\begin{eqnarray}
\label{qprop}
S(p)  & = &  -i\gamma\cdot p\, \sigma_V(p^2) + \sigma_S(p^2)\,,\\
\label{ssm}
\bar\sigma_S(x)  & = &  2\,\bar m \,{\cal F}(2 (x+\bar m^2))
+ {\cal F}(b_1 x) \,{\cal F}(b_3 x) \,
\left[b_0 + b_2 {\cal F}(\epsilon x)\right]\,,\\
\label{svm}
\bar\sigma_V(x) & = & \frac{1}{x+\bar m^2}\,
\left[ 1 - {\cal F}(2 (x+\bar m^2))\right]\,,
\end{eqnarray}
$x=p^2/\lambda^2$, $\bar m$ = $m/\lambda$, $\bar\sigma_S(x) =
\lambda\,\sigma_S(p^2)$ and $\bar\sigma_V(x) = \lambda^2\,\sigma_V(p^2)$.
The mass-scale, $\lambda=0.566\,$GeV, and parameter values
\begin{equation}
\label{tableA} 
\begin{array}{ccccc}
   \bar m& b_0 & b_1 & b_2 & b_3 \\\hline
   0.00897 & 0.131 & 2.90 & 0.603 & 0.185 
\end{array}\;,
\end{equation}
were fixed in a least-squares fit to light-meson
observables\cite{conradkaon}.  [$\epsilon=10^{-4}$ in (\ref{ssm}) acts only
to decouple the large- and intermediate-$p^2$ domains.]  This simple form
captures the essential features of direct solutions of Eq.~(\ref{gendse}).
It represents the dressed-quark propagator as an entire function, which is
inspired by the algebraic solutions in Refs.~\cite{entire,entireCJB} and
ensures confinement via the means described in Sec.~\ref{mr97sect}; and
exhibits DCSB with
\begin{equation}
-\langle \bar qq \rangle_{1\,{\rm GeV}^2}  = 
\lambda^3\,\frac{3}{4\pi^2}\,
\frac{b_0}{b_1\,b_3}\,\ln\frac{1}{\Lambda_{\rm QCD}^2}
= (0.221\,{\rm GeV})^3\,,
\end{equation}
which is calculated directly from Eqs.~(\ref{zmdef}), (\ref{qbq0}) after noting
that Eqs.~(\ref{ssm}), (\ref{svm}) yield the chiral limit quark mass function of
Eq.~(\ref{Mchiral}) with $\gamma_m = 1$.  This is a general feature; i.e.,
the parametrisation exhibits asymptotic freedom at large-$p^2$ omitting only
the additional $\ln( p^2/\Lambda_{\rm QCD}^2)$-suppression, which is a useful
but not necessary simplification.  As we see here, this omission introduces
model artefacts that are easily identified and accounted for.

Reference~\cite{jacquesnucleon} obtains a good description of the proton's
charge and magnetic form factors, and also the neutron's magnetic form
factor.  The nonpointlike nature of the scalar diquark correlation makes
possible a magnetic moment ratio of $|\mu_n/\mu_p| = 0.55$, cf. $0.68$
experimentally.  This ratio is always less-than $ 0.5$ when the correlation
is pointlike.  The neutron's charge form factor is poorly described.  The
charge-radius-squared is negative, consistent with the data, but is $60$\%
too large.  That defect results primarily from neglecting the contribution of
the axial-vector diquark.

The nucleon's scalar form factor is 
\begin{equation}\label{sigma}
\sigma(q^2)\, \bar u(P') u(P):= \langle P'|\, m (\bar u u + \bar d
d)\,|P\rangle \,,
\end{equation}
where the nucleon spinors satisfy: 
\begin{equation}
\label{spinors}
\gamma\cdot P \, u(P) = i M u(P)\,,\;
\bar u(P)\,\gamma\cdot P = i M \bar u(P)\,,
\end{equation}
with the nucleon mass $M=0.94\,$GeV, $q=(P^\prime-P)$ and $R= (P^\prime +
P)$. The $\pi N$ $\sigma$-term is just $\sigma(q^2=0)$, which is the
in-nucleon expectation value of the explicit chiral symmetry breaking term in
the QCD Lagrangian.  The general form of a fermion-scalar vertex is
\begin{equation}
\Lambda\!\!\!_{\mbox{\large\boldmath $1$}}(q,P) = f_1 + i \gamma\cdot q \,f_2 + i
\gamma\cdot R \,f_3 + i \sigma_{\mu\nu} R_\mu q_\nu \,f_4 \,,\;
f_i=f_i(q^2,R^2)
\end{equation}
since $q\cdot R=0$ for elastic processes.  However, using
Eqs.~(\ref{spinors}) the scalar current simplifies:
\begin{eqnarray}
J_{\mbox{\large\boldmath $1$}}(P',P) 
& :=& \bar u(P') \Lambda_{\mbox{\large\boldmath $1$}}(q,P) u(P) 
 =  s(q^2) \,\bar u(P') \,u(P) \,,\\
\label{sigmaadd}
s(q^2) & = & f_1 - 2 M f_3 + q^2 f_4 \,.
\end{eqnarray}

The impulse approximation to $J_{\mbox{\large\boldmath $1$}}(P',P)$ is also
given by the five diagrams in Fig.~\ref{jacquesPic} but with the
dressed-quark-photon vertex replaced by the dressed-quark-scalar vertex,
which is the solution of an inhomogeneous BSE analogous to
Eq.~(\ref{invectorbse}).  In Ref.~\cite{myriad}, to hasten an exemplifying
result, the scalar vertex equation was solved using the
Goldstone-theorem-preserving separable model of Ref.~\cite{conradsep}.  In
that model the BSE assumes the form
\begin{equation}\label{IBS}
\Gamma_{\mbox{\large\boldmath $1$}}(k;Q) = \mbox{\large\boldmath $1$} -
\sfrac{4}{3} \int \frac{d^4 q}{(2\pi)^4} \Delta(k-q) \gamma_\mu S(q_+)
\Gamma_{\mbox{\large\boldmath $1$}}(q;Q) S(q_-) \gamma_\mu \,,
\end{equation}
with the interaction
\begin{equation}\label{SepAn}
\Delta(k-q) = G(k^2)G(q^2) + k\cdot q \, F(k^2) F(q^2)\,,
\end{equation}
where $F(k^2)$, $G(k^2)$ are regularised forms of the $A$, $B$ obtained from
Eqs.~(\ref{ssm}), (\ref{svm}).  [NB. The regularisation ensures convergence of
necessary integrals in the separable model.]  The solution of Eq.~(\ref{IBS})
is
\begin{equation}\label{Ga1Sol}
\Gamma_{\mbox{\large\boldmath $1$}}(k;Q) = \mbox{\large\boldmath $1$}  +
t_1(Q^2)\, G(k^2)
+ \, i\,t_2(Q^2) \,F(k^2)\,\frac{k\cdot Q \,\gamma\cdot Q}{Q^2}
+ \,i\,t_3(Q^2) F(k^2)\,  \gamma\cdot k \,.
\end{equation}

The $\sigma$-term is only sensitive to the vertex at $Q^2=0$, where the
solution reduces to
\begin{equation}\label{Ga1Sol0}
\Gamma_{\mbox{\large\boldmath $1$}}(k;Q)|_{Q^2=0} = \mbox{\large\boldmath
$1$}  + t_1(0)\, G(k^2) + t_3(0)\, F(k^2)\, i \gamma\cdot k \,,
\end{equation}
with $t_1(0)=0.242\,$GeV, $t_3(0)=-0.0140\,$GeV.  Calculation shows that at
$k^2=0$ the $t_1$-term is 6-times larger than the bare term; i.e., it is
dominant in the infrared.  That is to be expected because it represents the
effect of the nonperturbative DCSB mechanism in the solution.  This and the
other $t_i$-terms vanish as $k^2\to \infty$, which is just a manifestation of
asymptotic freedom.

What makes the inhomogeneous scalar BSE important here is that it has a
solution for all $Q^2$ and that solution exhibits a pole at the
$\sigma$-meson mass; i.e., in the neighbourhood of $(-Q^2)= m_\sigma^2=
(0.715\,{\rm GeV})^2$
\begin{equation}
\label{scalarvtx}
\Gamma_{\mbox{\large\boldmath $1$}}(k;Q) = {\it regular}\; + 
\frac{n_\sigma m_\sigma^2}{Q^2+m_\sigma^2}\,\Gamma_\sigma(k;Q)\,,
\end{equation}
where {\it regular} indicates terms that are regular in this neighbourhood
and $\Gamma_\sigma(k;Q)$ is the canonically normalised $\sigma$-meson
Bethe-Salpeter amplitude, whose form is exactly that of
$(\Gamma_{\mbox{\large\boldmath $1$}}(k;Q) - \mbox{\large\boldmath $1$})$ in
Eq.~(\ref{Ga1Sol}).  This is qualitatively identical to the behaviour of the
solution of Eq.~(\ref{invectorbse}) discussed above in connection with
$F_\pi(Q^2)$ and elucidated in Refs.~\cite{mtpion,frankvertex}.  The simple
pole appears in each of the functions $t_i(Q^2)$ and a pole fit yields, with
$m$ the current-quark mass,
\begin{equation}
m \,n_\sigma = 3.3\,{\rm MeV}\,.
\end{equation}
$n_\sigma m_\sigma^2$ is the analogue of the residue of the $\pi$-pole in the
pseudoscalar vertex, Eq.~(\ref{rHres}): $-\sqrt{2}\,\langle \bar q
q\rangle_\pi/f_\pi$, and its flow under the renormalisation group is
identical.  $m\, n_\sigma$ is renormalisation point independent and its value
can be compared with
\begin{equation}
\frac{-m \langle \bar q
q\rangle_\pi}{\sfrac{1}{\sqrt{2}}f_\pi}\,\frac{1}{m_\sigma^2} = 
3.6\,{\rm MeV};
\end{equation}
i.e., the magnitude of $n_\sigma$ is typical of effects driven by dynamical
chiral symmetry breaking.  The $\sigma$-meson-$\bar q q$ coupling is defined
on-shell 
\begin{equation}
g_{\sigma \bar q q}:= \left.\Gamma_\sigma(0;Q)\right|_{Q^2=-m_\sigma^2} =
12.6\,,
\end{equation}
and its magnitude can be placed in context via a comparison with $g_{\pi \bar
q q}=11.8$, which is defined analogously.

The expectation value in Eq.~(\ref{sigma}) is obtained with 
\begin{equation}
\label{sigmavtx}
\Gamma_{m}(k;Q) = m \,\Gamma_{\mbox{\large\boldmath $1$}}(k;Q)
\end{equation}
as the probe vertex in Fig.~\ref{jacquesPic}, and using the solution for
$\Gamma_{\mbox{\large\boldmath $1$}}$ described here the calculated
$\sigma$-term is~\cite{myriad}
\begin{equation}
\label{sigmavalue}
\sigma/M_N = 0.015\,.
\end{equation}
No parameters were varied to obtain this result, which may be compared with a
recent lattice computation~\cite{latticesigma}: $\sigma/M_N = 0.019\pm 0.05$,
calculated using an extrapolation in the current-quark mass.  Alternative
extrapolation methods can lead to larger values; e.g., $\sigma/M_N =
0.047$--$0.059$~\cite{dereksigma}, which are also suggested by some
phenomenological analyses~\cite{vw91,carlsigma}.  [The $\sigma$-term is not
directly accessible via experiment but a value is theoretically inferred by
extrapolating $\pi N$ scattering data using dispersion
relations~\cite{knecht}: $\sigma/M_N = 0.047$--$0.076$.]  Simple estimates
indicate that the result in Eq.~(\ref{sigmavalue}) increases with decreasing
$m_\sigma$, and a reduction in $m_\sigma$ is a likely consequence of using an
improved kernel in the inhomogeneous scalar BSE.  Thus Eq.~(\ref{sigmavalue})
is an excellent first estimate.

\begin{figure}[t]
\centering{\
\epsfig{figure=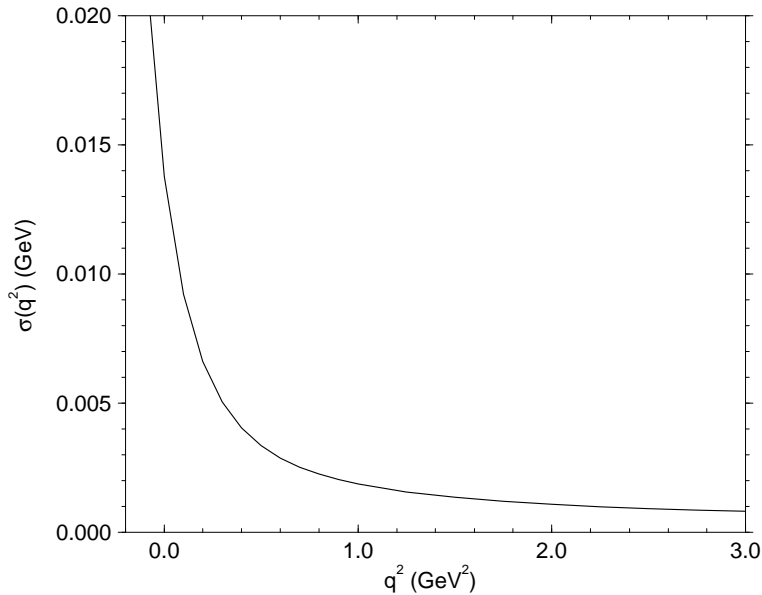,height=8cm}} \vspace*{\baselineskip} 
\parbox{40em}{\caption{$\sigma(q^2)$ as calculated in
Ref.~\protect\cite{myriad} using the solution of the inhomogeneous scalar
BSE, Eq.~(\protect\ref{IBS}).  The rapid increase with decreasing $q^2$ is
associated with the evolution to the $\sigma$-meson pole.  A similar feature
was also encountered in the calculation of the pion's electromagnetic form
factor, see Fig.~\protect\ref{mtpionff} on page~\protect\pageref{mtpionff}
and the associated discussion.  On this scale, $\sigma(q^2)$ calculated
without the $t_{2,3}(Q^2)$ contributions is indistinguishable from the full
calculation.  (Adapted from Ref.~\protect\cite{myriad}.)\label{sigFF}}}
\end{figure}

The nucleon's scalar form factor is depicted in Fig.~\ref{sigFF}, where the
evolution to the $\sigma$-meson pole is evident.  Fitting ($t = -q^2$)
\begin{equation}
\label{atpole}
\sigma(t) = g_{\sigma NN}\,\frac{m \,n_\sigma}{1-t/m_\sigma^2}\,,\;
t\in [0.1,0.5]\,{\rm GeV}^2\,,
\end{equation}
which isolates the residue associated with $\Gamma_m(k;Q)$, one obtains the
on-shell coupling: $g_{\sigma NN} = 27.3$.  A direct calculation using the
solution of the homogeneous Bethe-Salpeter equation yields $g_{\sigma NN} =
27.7$, in agreement within Monte-Carlo errors.  Equation~(\ref{atpole}) alone
overestimates the magnitude of the calculated $\sigma(t)$ everywhere {\it
except} in the neighbourhood of the pole.

As the lowest-mass pole-solution of a rainbow-ladder BSE, Eq.~(\ref{IBS}),
the $\sigma$-meson described here is not obviously related to the scalar
meson introduced in phenomenological nucleon-nucleon
potentials~\cite{machleidt,vincent} and meson exchange models~\cite{harry} to
mock-up two-pion exchange.  However, a coupling relevant to such models can
be estimated by introducing $g_\sigma(t)$:
\begin{equation}
\label{onshellg}
\sigma(t) =: g_{\sigma}(t)\,\frac{m\,n_\sigma}{1-t/m_\sigma^2}\,,
\end{equation}
where a fit to the calculated $\sigma(t)$ in Fig.~\ref{sigFF} yields
\begin{equation}
g_{\sigma}(t) = 1.61 + 
2.61\,\frac{1}{(1-t/\Lambda_\sigma^2)^{10}}\,,\;\;
\Lambda_\sigma=1.56\,{\rm GeV}\,,
\end{equation}
with the large exponent merely reflecting the rapid evolution from bound
state to {\it continuum dominance} of the vertex in the spacelike region.
$g_\sigma(t)$ describes the $t$-dependent nucleon to
scalar-quark-antiquark-correlation coupling strength and at the
mock-$\sigma$-mass: $m_\sigma^{2\pi}=0.5\,$GeV,
\begin{equation}
g_{\sigma}:= g_{\sigma}((m_\sigma^{2\pi})^2) = 9.3\,,
\end{equation}
which may be compared with a phenomenologically inferred value: $g_\sigma =
10$~\cite{Friman}.  This gives a microscopic interpretation of the
mock-$\sigma$.  [It is important to note that $g_{\sigma}(4\,m_\pi^2) = 5.2$
and hence this comparison is meaningful on a relevant phenomenological
domain.]  Further, $g_\sigma(q^2\to\infty)=1.61$ so that $\sigma(q^2)$ is
well approximated by a single monopole for $q^2>1\,$GeV$^2$.  However, the
residue is very different from the on-shell value.  The scalar radius of the
nucleon is obtained from
\begin{equation}
\label{radsig}
\langle r_{\sigma NN}^2 \rangle := -\frac{6}{\sigma}\, 
\left.\frac{d \sigma(q^2)}{dq^2}\right|_{q^2=0} = (0.89\,\mbox{fm})^2\,,
\end{equation}
which may be compared with an inferred value~\cite{Friman}: $(1.2\,{\rm fm})^2$.

Reference~\cite{myriad} calculates a range of non-electromagnetic nucleon
form factors and provides a uniformly good description, yielding
meson-nucleon form factors that are ``soft'' and couplings that generally
agree well with those employed in meson-exchange models.  Where there are
discrepancies with these models or with experiment, a plausible cause and
means for its amelioration was readily identified.  The study demonstrates
that it is realistic to hope for useful constraints on meson-exchange models
from well-moulded models of hadron structure.  The analysis of the nucleon
$\sigma$-term is particularly important because it illustrates the only
method that allows an unambiguous off-shell extrapolation in the estimation
of meson-nucleon form factors: one must employ solutions of the {\it
inhomogeneous} BSEs to describe correlations in a channel when the total
momentum is not in the neighbourhood of a bound state pole.  Improvements to
the simple quark$+$scalar-diquark nucleon model are being developed and their
application to a wide range of hadronic phenomena is being explored.  These
studies hold the promise of providing a much-needed standard model of the
nucleon.

\sect{Nonzero Density and Temperature} 
Identifying the onset and properties of a quark gluon plasma [QGP] is the
primary objective of studies at nonzero temperature and density. Such a
plasma is expected to be characterised by light quarks and gluons propagating
freely over distances $\sim 10$-times larger than the proton.  The
exploration of QCD in this domain requires a knowledge of equilibrium
statistical field theory, for which Sec.~6 of Ref.~\cite{echaya} is a primer
and Refs.~\cite{ESFTtexts} provide a graduate-level introduction.  Here the
application of DSEs provides our primary medium and we will describe the
necessary and related concepts.

Section~\ref{sectionDSEs} reviews the application of DSEs at zero density and
temperature.  It shows that the framework is efficacious on this domain and
well constrained.  Both are important because little is known about the
domain of high density and/or temperature, and much of what we do know comes
from the extrapolation of models.  Since the DSEs provide a nonperturbative
framework that admits the simultaneous study of DCSB and confinement they are
well suited to explore the phase transition that yields the QGP.
Furthermore, as they also accurately describe bound states, they can be used
to explore the response of hadron properties to extremes of density and
temperature; i.e., to elucidate signals of QGP formation.  These features
also mean that the DSEs can address all the phenomena accessible in
lattice-QCD simulations and therefore they can be used to confirm the results
of such studies.  Importantly too the lattice simulations can be used to
constrain the model-dependent elements of DSE studies, making possible a
reliable DSE extrapolation into the domain of nonzero chemical potential QCD,
for example, which is inaccessible in contemporary lattice-QCD.

\subsect{In-medium Essentials}
\label{secInMedium}
\addcontentsline{toc}{subsubsection}{\protect\numberline{ } {Equation of State}}
\subsubsect{Equation of State}
The $T\neq 0$ analogue of the partition function in Eq.~(\ref{partitionZ}) is
\begin{equation}
\label{partitionZT}
{\cal Z}_{(\mu,T)}[\bar\eta,\eta,J]= \int d\mu_\beta(\bar
q,q,A,\bar\omega,\omega)\,
\exp\int_0^\beta\! d\tau\int\! d^3x 
\left[ \bar q \,\eta + \bar \eta \,q + J_\mu^a \,A_\mu^a\right],
\end{equation}
with $x_4\to \tau$, $\beta:= 1/T$ and the measure
\begin{eqnarray}
\label{measureZT}
\lefteqn{d\mu_\beta(\bar q,q,A,\bar\omega,\omega) := }\\
&& \nonumber \prod_{\vec{x}\tau\in[0,\beta]} \prod_{\phi} {\cal D} \bar
q_\phi(\vec{x},\tau) {\cal D} q_\phi(\vec{x},\tau) \prod_{a} {\cal D} \bar
\omega^a(\vec{x},\tau) {\cal D} \omega^a(\vec{x},\tau) \prod_{\mu} {\cal
D}A_\mu^a(\vec{x},\tau)
\exp(-S_\beta[\bar q,q,A]-S_\beta^g[\bar\omega,\omega,A])\,,
\end{eqnarray}
where $S_\beta[\bar q,q,A]$, $S_\beta^g[\bar \omega,\omega,A]$ are the
actions of Eqs.~(\ref{ActionQCD}), (\ref{ghostaction}) but for the obvious
bounding of the $\tau$-integral.  The frame specified by separating the
spacetime integral into a product of a compact integral over Euclidean time
and the volume integral is not prescribed by the theory.  This separation
breaks the theory's original $O(4)$ symmetry to $O(3)$ and is effected by
introducing an arbitrary, normalised, spacelike vector, $u_\mu$, with $u\cdot
p=0$ defining the $T=0$-hyperplane.  Here we have employed a conventional
choice for the heat bath vector: $u=(\vec{0},1)$.  [NB.  The zero temperature
theory is recovered by taking the limit $T\to 0$; i.e., $\beta \to \infty$,
in Eqs.~(\ref{partitionZT}), (\ref{measureZT}).]  The chemical potential is
introduced as a Lagrange multiplier
\begin{equation}
S_\beta[\bar q,q,A] \to S_{\mu,\beta}[\bar q,q,A]:= S_\beta[\bar q,q,A]
- \int_0^\beta\! d\tau\int\! d^3x \,\mu \,\bar q \gamma_4 q\,.
\end{equation}
With $\mu\neq 0$, $S_{\mu,\beta}[\bar q,q,A]$ is not Hermitian and hence
$d\mu_{\mu,\beta}$ is not a probability measure.  That makes numerical
simulations of lattice-regularised $\mu\neq 0$ QCD very difficult; in fact,
there is currently no satisfactory algorithm~\cite{barbour}.  The functional
integral in Eq.~(\ref{partitionZT}) is evaluated on the space of Grassmann
fields for which $q(\vec{x},0)=-q(\vec{x},\beta)$; i.e., over fields
satisfying antiperiodic boundary conditions, and over gauge fields satisfying
periodic boundary conditions.  These boundary conditions are an essential
aspect of the difference between fermion and boson statistics.

The partition function is particularly important in equilibrium statistical
field theory.  As usual, it yields the DSEs in a straightforward manner,
however, more than that, all the thermodynamic functions are obtained from
the partition function.  The thermodynamic potential and pressure densities
are
\begin{equation}
\label{EOS}
-\omega(\mu,T) = p(\mu,T) = \frac{1}{\beta V} \ln {\cal Z}_{(\mu,T)}\,,
\end{equation}
where $\beta V$ is the four-volume normalising factor.  The stable phase of
the system is that in which the pressure is maximal or equivalently the
thermodynamic potential energy is minimised.  The expression for the pressure
is the Equation of State [EOS] and from that one immediately obtains the
baryon number and entropy densities:
\begin{equation}
\label{densities}
\rho(\mu,T) = \left.\frac{\partial}{\partial \mu}\, p(\mu,T)\right|_{T\,{\rm
fixed}}\!,\;\;\; 
s(\mu,T)= \left.\frac{\partial}{\partial T}\, p(\mu,T)\right|_{\mu\,{\rm
fixed}}
\end{equation}
and also the energy density:
\begin{equation}
\epsilon(\mu,T) = -p(\mu,T) + \mu\,\rho(\mu,T) + T\,s(\mu,T)\,.
\end{equation}
The calculation of these quantities in QCD is a contemporary focus; e.g.,
Refs.~\cite{edwinechaya,bastiscm,pressures}.

A simple estimate of the pressure due to dressed-quarks is obtained via the
``steepest descent'' approximation, which yields
\begin{equation}
\label{pSigma}
p_{\Sigma}(\mu,T) = \frac{1}{\beta V}\left\{ {\rm TrLn}\left[\beta
S^{-1}\right] - \sfrac{1}{2}{\rm Tr}\left[\Sigma\,S\right]\right\},
\end{equation}
where $S$ is the solution of the gap equation with $\Sigma$ the associated
self energy, and ``Tr'' and ``Ln'' are extensions of ``tr'' and ``$\ln$'' to
matrix-valued functions.  Equation~(\ref{pSigma}) is just the auxiliary field
effective action~\cite{haymaker}, which yields the free fermion pressure in
the absence of interactions; i.e., when $\Sigma \equiv 0$.  At this level of
truncation the total pressure receives an additive contribution from
dressed-gluons:
\begin{equation}
p_{\Delta}(\mu,T) = -\frac{1}{\beta V} \,\sfrac{1}{2} {\rm TrLn}\left[
\beta^2 D_{\mu\nu}^{-1}\right] \,,
\end{equation}
where $D_{\mu\nu}$ is the $T\neq 0$ dressed-gluon $2$-point function, and
this yields the free-gluon pressure in the absence of interactions.

\addcontentsline{toc}{subsubsection}{\protect\numberline{ } {Propagators}}
\subsubsect{Propagators}
As just described, the introduction of temperature to Euclidean QCD breaks
the original $O(4)$ symmetry to $O(3)$.  In this case the most general form
of the dressed-quark propagator is~\cite{kalash}
\begin{equation}
\label{reallygeneral}
S(p,u)^{-1} = 
i\gamma\cdot p \,a(p^2,[u\cdot p]^2) + b(p^2,[u\cdot p]^2) + i \gamma\cdot u
\,u\cdot p\,c(p^2,[u\cdot p]^2) + i \sigma_{\mu\nu} p_\mu u_\nu \,
d(p^2,[u\cdot p]^2)\,,
\end{equation}
where $u_\mu=(\vec{0},1)$, and the surviving $O(3)$ invariance of
Eq.~(\ref{reallygeneral}) is apparent upon inspection.  Even at $T\neq 0$,
the chiral-limit dressed-quark propagator should satisfy
\begin{equation}
V(\alpha)\,S(p,u)^{-1}\,V(\alpha) = S(p,u)^{-1}\,,\;\;
V(\alpha) = \exp\left(i\gamma_5\,\sfrac{1}{2}\lambda_F^l \alpha^l\right)\,,
\end{equation}
if chiral symmetry is realised in the Wigner-Weyl mode.  Hence $b \equiv 0
\equiv d$ in this case.  Furthermore, since the term involving $d$ does not
appear in the free fermion action of thermal field theory, there are no
perturbative divergences in this function's evaluation and $d$ is therefore
power-law suppressed in the ultraviolet.  [cf.~The discussion of the chiral
limit fermion mass function on page~\pageref{`without subtraction'}.]  These
features, along with the vector exchange nature of the interaction, underly
the fact that $d(p^2,u\cdot p^2)$ only plays a minor role in the solution of
the QCD gap equation.  Hereafter we neglect it and assume the following as
the most general form of the dressed-quark $2$-point function:
\begin{eqnarray}
\label{generalenough}
S(\vec{p},\tilde\omega_k) & = & 
\frac{1}{i\vec{\gamma}\cdot\vec{p}\,A(\vec{p}\,^2,\tilde\omega_k^2) +
B(\vec{p}\,^2,\tilde\omega_k^2) + i \gamma_4\,\tilde\omega_k
C(\vec{p}\,^2,\tilde\omega_k^2)} \,,\\
& = & -i\vec{\gamma}\cdot\vec{p}\,\sigma_A(\vec{p}\,^2,\tilde\omega_k^2) +
\sigma_B(\vec{p}\,^2,\tilde\omega_k^2)- i
\gamma_4\,\tilde\omega_k\sigma_C(\vec{p}\,^2,\tilde\omega_k^2)\,,
\end{eqnarray}
where $\tilde\omega_k = \omega_k + i\mu$, with $\mu$ the chemical potential,
and $\omega_k= (2 k + 1)\,\pi T$, $k\in Z\!\!\!\!Z$, are the fermion
Matsubara frequencies, which ensure $q(\vec{x},0)=-q(\vec{x},\beta)$.  The
scalar functions: ${\cal F}= A$, $B$, $C$, are complex and satisfy
\begin{equation}
\label{FFstar}
{\cal F}(\vec{p}\,^2,\tilde\omega_k^2)^\ast = 
{\cal F}(\vec{p}\,^2,\tilde\omega_{-k-1}^2)\,.
\end{equation}
[NB.  The functions $A$, $C$ cannot be neglected.  For $T=0$, $A=C$ and
$A-1\not \equiv 0$; e.g., Refs.~\cite{mr97,tonylatticequark}, and the
momentum dependence of $A$ can conspire with that of $B$ to ensure
confinement of dressed-quarks~\cite{entireCJB}.  Furthermore, as will become
clear, the $(\vec{p}\,^2,\tilde\omega_k^2)$-dependence of $A$, $C$ has a
marked effect on the behaviour of bulk thermodynamic quantities.]

The dressed-quark $2$-point function satisfies a gap equation
\begin{equation}
S(p_{\omega_k})^{-1} = Z_2^A \,i\vec{\gamma}\cdot \vec{p} + Z_2 \,
(i\gamma_4\,\tilde\omega_k + m_{\rm bare})\, + \Sigma^\prime(p_{\omega_k} )\,,
\end{equation} 
which can be derived in the usual way from the partition function,
Eq.~(\ref{partitionZT}).  Here we have introduced a shorthand notation:
$p_{\omega_k}= (\vec{p},\tilde\omega_k)$, and the regularised self energy is
\begin{eqnarray}
\label{sigmap}
\Sigma^\prime(p_{\omega_k}) & = & i\vec{\gamma}\cdot
  \vec{p}\,\Sigma_A^\prime(p_{\omega_k} ) +
  i\gamma_4\,\omega_k\,\Sigma_C^\prime(p_{\omega_k} ) +
  \Sigma_B^\prime(p_{\omega_k})\,, \\
\label{regself}
\Sigma_{\cal F}^\prime(p_{\omega_k}) & = & \sfrac{1}{4}{\rm tr}\,{\cal P}_{\cal
F}\int_{l,q}^{\bar\Lambda}\,
\sfrac{4}{3}\,g^2\,D_{\mu\nu}(\vec{p}-\vec{q},\omega_k-\omega_l) \,
\gamma_\mu S(q_{\omega_l})\Gamma_\nu(q_{\omega_l};p_{\omega_k})\,,
\end{eqnarray}
where the renormalisation factors are ${\cal P}_A:=
-(Z_1^A/|\vec{p}|^2)i\vec{\gamma}\cdot \vec{p}$,
${\cal P}_B:= Z_1$,
${\cal P}_C:= -(Z_1/\omega_k)i\gamma_4$, 
and $\int_{l,q}^{\bar\Lambda}:=\, T
\,\sum_{l=-\infty}^\infty\,\int^{\bar\Lambda}d^3q/(2\pi)^3$, with
$\int^{\bar\Lambda}$ representing a translationally invariant regularisation
of the three-dimensional integral and $\bar\Lambda$ is the regularisation
mass-scale.  Since introducing $\mu$, $T$ does not generate qualitatively new
divergences, the regularisation and renormalisation proceeds just as in the
absence of the medium so that the renormalised self energies are
\begin{equation}
\label{renself}
\begin{array}{rcl}
{\cal F}(p_{\omega_k};\zeta) & = & 
\xi_{\cal F} + \Sigma_{\cal F}^\prime(p_{\omega_k};{\bar\Lambda})
    - \Sigma_{\cal F}^\prime(\zeta^-_{\omega_0};{\bar\Lambda})\,,
\end{array}
\end{equation}
with $\zeta$ the renormalisation point, $(\zeta^-_{\omega_0})^2 := \zeta^2 -
\omega_0^2$, $\xi_A = 1 = \xi_C$, and $\xi_B=m_R(\zeta)$.  [cf.\
Eqs.~(\ref{dsemod}-\ref{arenbren}).]

The regularised self energy is expressed in terms of the renormalised
dressed-gluon $2$-point function and dressed-quark-gluon $3$-point function.
At $T\neq 0$ in Landau gauge the complete expression of the former requires
two $O(3)$-scalar functions [cf.\ Eq.~(\ref{gluoncovariant})]
\begin{equation}
\label{gluonpropT}
g^2 D_{\mu\nu}(p_{\Omega_k}) = 
P_{\mu\nu}^L(p_{\Omega_k} ) \Delta_F(p_{\Omega_k} ) + 
P_{\mu\nu}^T(p_{\Omega_k}) \Delta_G(p_{\Omega_k}  ) \,,
\end{equation}
where $p_{\Omega_k}:= (\vec{p},\Omega_k)$, $\Omega_k = 2 k \pi T$ is the
Matsubara frequency for bosons, and
\begin{eqnarray}
P_{\mu\nu}^T(p_{\Omega_k}) & := &\left\{
\begin{array}{ll}
0, &  \mu\;{\rm and/or} \;\nu = 4,\\
\displaystyle
\delta_{ij} - \frac{p_i p_j}{p^2}, &  \mu,\nu=i,j\,=1,2,3\;,
\end{array}\right.\\
P_{\mu\nu}^L(p_{\Omega_k}) & = & 
\delta_{\mu\nu}- \frac{p_\mu p_\nu}{p^2} - P_{\mu\nu}^T(p_{\Omega_k}) \,.
\end{eqnarray}
In the absence of interactions:
$\Delta_F(p_{\Omega_k}^2)=\Delta_G(p_{\Omega_k}) = 1/p_{\Omega_k}^2$.  The
QCD analogue of a Debye [or electric screening] mass appears as a
$T$-dependent contribution to $\Delta_F$, the gluon's longitudinal
polarisation function.  The complete expression of the renormalised
dressed-quark-gluon $3$-point function requires at least fifty-four scalar
functions [cf.\ twelve for $T=0$] but, since the theory is renormalisable,
ultraviolet divergences are only encountered in evaluating two of them.  It
follows from this that sensible, character-preserving truncations are
possible.

\addcontentsline{toc}{subsubsection}{\protect\numberline{ } {Order
Parameters}}
\subsubsect{Order Parameters}
In order to demarcate the QGP phase it is necessary to identify order
parameters; i.e., find operators $X_i$ whose expectation values are nonzero
in the normal phase: $\langle X_i \rangle \neq 0$, but vanish in the QGP:
$\langle X_i \rangle \equiv 0$.  This is often a difficult task.  Having
identified such operators then, if $\langle X_i \rangle$ is discontinuous at
the transition point we will describe the transition as first order;
otherwise we classify it as second order.

In Sec.~\ref{mr97sect} we described the phenomenon of DCSB and its connection
with the purely nonperturbative generation of a scalar term in the
dressed-quark self energy when the current-quark mass is zero.  One
characteristic of this effect is the large magnitude, Eq.~(\ref{qbqM0}), of
the vacuum quark condensate that appears in Eqs.~(\ref{Mchiral}),
(\ref{qbq0}), which has many observable consequences; e.g., it is the primary
cause of what is naively an unexpectedly large $\pi$-$\rho$ mass splitting.
DCSB is a defining feature of the normal phase of QCD.  If, with increasing
$\mu$ and/or $T$, the condensate vanishes at some particular value of these
parameters then a new phase of QCD has been reached.  It is a phase in which
chiral symmetry is realised explicitly and, as weakly interacting systems
cannot support condensate formation, this is plainly a good defining property
of a QGP.  The discussion of the chiral limit,
Eqs.~(\ref{chiralA}-\ref{qbq1}), also makes plain that it is not possible to
have a nonzero quark condensate unless the dressed-quark mass function,
$M(p_{\omega_k})$, is nonzero too.  Therefore the magnitude of this function
at any single point is also an order parameter appropriate for characterising
the chiral symmetry restoring aspect of the QGP transition.  In fact, in DSE
studies it is by far the most direct, since it appears in all the integrands
that describe other quantities.

We also discussed an arguably more significant aspect of cold, sparse QCD,
namely confinement.  It is easy to build a model that exhibits DCSB without
confinement; e.g., Refs.~\cite{vw91,njltwo,sk92,bb95}.  However, building a
covariant model with confinement and an uncomplicated realisation of all
aspects of DCSB is a challenging task.  As emphasised by the study of QED$_3$
in Ref.~\cite{pieterQED3}, that challenge is met with a modicum of success
using the DSE truncation scheme explored in Ref.~\cite{truncscheme} and the
realisation of confinement via a violation of reflection positivity by
coloured $n$-point functions.  This also exposes a simple deconfinement order
parameter appropriate to both light- and heavy-quarks, whose application is
illustrated in Fig.~\ref{qed3fig}.  To make it plain we return to $D(T)$
defined in Eq.~(\ref{D(T)}).  When calculated from a Schwinger function with
complex conjugate poles, $D(T)$ has at least one zero.  Denote its position
by $T_0^{z_1}$ and define
\begin{equation}
\label{conforder}
\kappa_0:= 1/T_0^{z_1}\,.
\end{equation}
If, with increasing $\mu$ and/or $T$, one observes $\kappa_0$ approaching
zero then the [first] zero in $D(T)$ is moving to $T=\infty$.  This evolution
of $\kappa_0$ corresponds to the limiting case when there is no zero at all,
which is just the free, unconfined particle described by $\sigma_S(T)$ in
Eqs.~(\ref{sigmasT}).  Thus $\kappa_0$ is a good order parameter for
deconfinement~\cite{axelT}: it is nonzero in the confined phase, and remains
nonzero until the Schwinger function evolves into one that respects
reflection positivity and corresponds to an asymptotic state.  It is clear
that deconfinement is also a good defining property of a QGP.

\addcontentsline{toc}{subsubsection}{\protect\numberline{ } {Critical Behaviour}}
\subsubsect{Critical Behaviour}
In second order transitions the length scale associated with correlations in
the system diverges as the order parameter vanishes and a range of critical
exponents can be defined that characterise the behaviour of macroscopic
properties at the transition point.  In the context of QCD this can be
elucidated via the free energy, which here we write as $f(t,h)$, where
$t:=T/T_c-1$ is the reduced temperature, with $T_c$ a putative critical
temperature, and $h:= \beta m_R$ is the source of explicit chiral symmetry
breaking measured in units of the temperature.  [We omit the chemical
potential for now so that $f=\epsilon= (T^2/V) (\partial \ln {\cal
Z}_{(T)}/\partial T)$.]  $h$ plays a role analogous to that of an external
magnetic field applied to a ferromagnet.  Since correlation lengths diverge
it follows that for $t,h\to 0$ the free energy is a generalised homogeneous
function; i.e.,
\begin{equation}
f(t,h) = \frac{1}{b}\,f(t \,b^{y_t},h \,b^{y_h}) \,.
\end{equation}
As a consequence the ``magnetisation'' behaves as follows: 
\begin{equation}
M(t,h) := \left.\frac{\partial\, f(t,h)} {\!\!\!\!\!\!\partial
h}\right|_{t\;{\rm fixed}}\!\!,\;\; \; 
M(t,h) = b^{y_h-1}\,M(t \,b^{y_t},h \,b^{y_h})\,.
\end{equation}
[NB.  Using the partition function one finds easily that $M(t,h)$ is just the
vacuum quark condensate.  In these formulae it can be replaced by any
equivalent order parameter.]  The scaling parameter: $b$, is arbitrary and
along the trajectory $|t| b^{y_t}= 1$ one has
\begin{equation}
\label{Mtzero}
M(t,h)  = |t|^{(1-y_h)/y_t}\,M({\rm sgn}(t), h \,|t|^{-y_h/y_t})\,; \;\;
{\rm i.e.,}\;\;
M(t,0)  \propto |t|^\beta\,,\; \beta:= \frac{1-y_h}{y_t}\,.
\end{equation}
Alternatively, along the trajectory $h b^{y_h}= 1$
\begin{equation}
\label{Mhzero}
M(t,h)  = h^{(1-y_h)/y_h}\,M(t \,h^{-y_t/y_h},1)\,; \;\;
{\rm i.e.,}\;\;
M(0,h) \propto h^{1/\delta}\,,\; \delta:= \frac{y_h}{1-y_h}\,.
\end{equation}

Equations~(\ref{Mtzero}), (\ref{Mhzero}) quantify the behaviour to be
expected of an order parameter at a second order transition.  They also
introduce two critical exponents and, using the renormalisation group,
scaling laws can be derived that relate all other such exponents to $\beta$
and $\delta$~\cite{cpbook}.  It is widely conjectured that the values of
these exponents are fully determined by the dimension of space and the nature
of the order parameter.  This is the notion of {\it universality}$\,$; i.e.,
that the critical exponents are {\it independent} of a theory's microscopic
details and hence all theories can be grouped into a much smaller number of
universality classes according to the values of their critical exponents.  If
this is the case, the behaviour of a complicated theory near criticality is
completely determined by the behaviour of a simpler theory in the same
universality class.  In mean-field theories
\begin{equation}
\beta^{\rm MF}= \sfrac{1}{2}\,,\;\;  \delta^{\rm MF} = 3.0\,.
\end{equation}
The success of the nonlinear $\sigma$-model in describing long-wavelength
pion dynamics underlies a conjecture~\cite{pisarski} that chiral symmetry
restoration at $T\neq 0$ in 2-flavour QCD is a second order transition with
the theory lying in the universality class characterised by the
3-dimensional, $N=4$ Heisenberg magnet [$O(4)$ model] whose critical
exponents are~\cite{neqfour}:
\begin{equation}
\beta^H= 0.38 \pm 0.01\,,\;\;
\delta^H = 4.82 \pm 0.05\,.
\end{equation}

As an alternative to Eqs.~(\ref{Mtzero}), (\ref{Mhzero}), the critical exponents
can be determined by studying the pseudocritical behaviour of the chiral and
thermal susceptibilities:
\begin{equation}
\label{defchi}
\chi_h(t,h) := \left.\frac{\partial\, M(t,h)} {\!\!\!\!\!\!\partial
h}\right|_{t\;{\rm fixed}}\!,\;\;\;
\chi_t(t,h) := \left.\frac{\partial\, M(t,h)} {\!\!\!\!\!\!\partial
t}\right|_{h\;{\rm fixed}}\!,
\end{equation}
which are smooth functions of $t$ with maxima at the pseudocritical points:
$t_{\rm pc}^h$ and $t_{\rm pc}^t$, where
$ t_{\rm pc}^h \propto\, h^{1/(\beta \delta)}\, \propto\, t_{\rm pc}^t $.
Since $\beta\delta >0$, the pseudocritical points approach the critical
point: $t=0$, as $h\to 0^+$ and 
\begin{equation}
\label{chislopes}
\chi_h^{\rm pc}  :=  \chi_h(t_{\rm pc}^h,h) \propto h^{-z_h}\,,\;
        z_h:= 1 - \frac{1}{\delta} \,,\;\;
\chi_t^{\rm pc}  :=  \chi_t(t_{\rm pc}^t,h) 
        \propto h^{-z_t}\,,
        \;z_t:= \frac{1}{\beta\delta}\,(1-\beta)\,.
\end{equation}
[The appendix of Ref.~\cite{arne1} provides a demonstration.]  Mean field
behaviour corresponds to
\begin{equation}
z_h^{\rm MF} = \sfrac{2}{3}\,,\;\;
z_t^{\rm MF} = \sfrac{1}{3}\,.
\end{equation}
Analysing the susceptibilities can provide more accurate results when
numerical noise is anticipated, which is why the method has been used in
lattice-QCD analyses~\cite{edwinechaya}.

\subsect{Quark Gluon Plasma Phase}
\label{DAmodelsection}
\addcontentsline{toc}{subsubsection}{\protect\numberline{ } {Temperature}}
\subsubsect{Temperature} 
We are now in a position to describe the application of DSEs to demarcating
the QGP and exploring its properties.  For the moment we persist with $\mu=0$
and explore the temperature induced transition because that is the domain on
which the most complete studies exist [in all approaches].

The properties of the class of rainbow-ladder models are elucidated in
Ref.~\cite{arne2}.  This class is defined by Eq.~(\ref{rainbow}) with the
mutually consistent $T\neq 0$ constraints: $Z_1 = Z_2$, $Z_1^A = Z_2^A$.
[NB.  The rainbow truncation is the leading term in a $1/N_c$-expansion of
$\Gamma_\nu(q_{\omega_l};p_{\omega_k})$.]  The form of
$D_{\mu\nu}(p_{\Omega_k})$ has no bearing on whether or not a model is in the
class.  To be concrete, Ref.~\cite{arne2} considered models defined by
\begin{eqnarray}
\label{uvpropf}
\Delta_F(p_{\Omega_k}) & = & {\cal D}(p_{\Omega_k};m_g)\,,\;
\Delta_G(p_{\Omega_k})  =  {\cal D}(p_{\Omega_k};0)\,,\\
\label{delta}
 {\cal D}(p_{\Omega_k};m_g) & := & 
        2\pi^2 D\,\sfrac{2\pi}{T}\delta_{0\,k} \,\delta^3(\vec{p}) 
        + {\cal D}_{\rm M}(p_{\Omega_k};m_g)\,,
\end{eqnarray}
with three choices of ${\cal D}_{\rm M}(p_{\Omega_k};m_g)$.  The first term
in ${\cal D}(p_{\Omega_k};m_g)$ is a simple $T\neq 0$ generalisation of the
distribution in Eq.~(\ref{gk2}).  It represents the long-range piece of the
effective interaction and the ${\cal D}_{\rm M}(p_{\Omega_k};m_g)$ are chosen
so as not to pollute that.  (NB. $\Delta_G$ does not contain a simple
$T$-dependent mass.  Such ``magnetic'' screening is more complicated than
electric screening because of infrared divergences in QCD~\cite{ESFTtexts}.)
The three choices are: \\
\hspace*{1em} A) ${\cal D}_{\rm M:=A}(p_{\Omega_k};m_g) \equiv 0$, with
$D:=\eta^2/2$ and $\eta = 1.06$ fixed~\cite{mn83} by fitting $\pi$- and
$\rho$-meson masses at $T=0$.  $m=12\,$MeV yields $m_\pi=140\,$MeV.  ${\cal
D}_{\rm A}$ yields an ultraviolet finite model and hence the renormalisation
point and cutoff can be removed simultaneously.  Nevertheless it exhibits
many features in common with more sophisticated {\it
Ans\"atze}~\cite{bastiscm}.\\
\hspace*{1em} B) The model of Ref.~\cite{axelT}, obtained with $D:= (8/9) \,
m_t^2$ and
\begin{equation}
\label{modelfr}
{\cal D}_{\rm M:=B}(p_{\Omega_k};m_g) = \sfrac{16}{9}\pi^2\,
\frac{1-{\rm e}^{- s_{\Omega_k} /(4m_t^2)}}
        {s_{\Omega_k} }\,,
\end{equation}
$s_{\Omega_k}:= p_{\Omega_k}^2+ m_g^2$, where, following
Ref.~\cite{reinhardCoulomb}, $m_g^2= 4\pi^2 \gamma_m (N_c/3+N_f/6) = (8/3)\,
\pi^2 T^2$ [$N_f=3$].  The mass-scale $m_t=0.69\,{\rm GeV}=1/0.29\,{\rm fm}$
marks the boundary between the perturbative and nonperturbative domains, and
was also fixed~\cite{fr96} by fitting $\pi$- and $\rho$-meson properties at
$T=0$.  At $\zeta=9.5\,$GeV, $m_R=1.1\,$MeV yields $m_\pi=140\,$MeV.  This
model adds a Coulomb-like short-range interaction to that of
Ref.~\cite{bastiscm}, thereby marginally improving its ultraviolet
behaviour.\\
\hspace*{1em} C) A minimal $T\neq 0$ extension of the effective interaction
in Eq.~(\ref{gk2}) with the only modification being $k^2 \to s_{\Omega_k}$,
$m_g^2 = (16/5)\, \pi^2 T^2$, in the last two terms.  [Recall that this
model's parameters were fitted with $N_f=4$].  As described above, ${\cal
D}_{\rm C}$ ensures that the model preserves the one-loop renormalisation
group behaviour of QCD.  To further explore model-dependence, two parameter
sets were employed in Ref.~\cite{arne2}, which differ only in the value of
$\omega$: $_1\omega= 0.6\,m_t$, $_2\omega= 1.2\,m_t$.  They are equivalent,
yielding the same values for pion observables: with $_1\omega$, a
renormalisation point invariant current-quark mass of $\hat m = 6.6\,$MeV
yields $m_\pi=140\,$MeV; while with $_2\omega$, $\hat m = 5.7\,$MeV effects
that.

Two chiral order parameters were employed in analysing chiral symmetry
restoration
\begin{equation}
\label{chiforms}
{\cal X}:= B(\vec{p}=0,\omega_0), \; 
{\cal X}_C:= \frac{B(\vec{p}=0,\omega_0)}{C(\vec{p}=0,\omega_0)}.
\end{equation}
They should be equivalent and the onset of that equivalence determines the
$h$-domain on which Eqs.~(\ref{chislopes}) are valid.  These being {\it bona
fide} order parameters is useful and particularly important because it means
that the lowest Matsubara frequency completely determines the character of
the chiral phase transition, as conjectured in Ref.~\cite{jackson}.
Furthermore, it was numerically verified in Ref.~\cite{arne2} that in the
chiral limit and for $t \sim 0$: $f_\pi \propto \langle\bar q q\rangle
\propto {\cal X}(t,0)$; i.e., these quantities are all {\it bona fide} order
parameters.

\begin{table}[t]
\begin{center}
\parbox{40em}{\caption{Critical temperature for chiral symmetry restoration
and critical exponents characterising the second-order transition in the four
exemplary models.  Averaging over the independent models yields $T_c = 154
\pm 24\,$MeV.  (Adapted from Ref.~\protect\cite{arne2}.)
\label{critthings} }}
\[
\begin{array}{lcccc}
{\rm Model} & {\rm A} & {\rm B} & {\rm C}_{_1\omega} & {\rm C}_{_2\omega}
            \\\hline
 T_c (MeV)  &  169 & 174 & 120 & 152  \\
 z_h        &  0.666 & 0.67  \pm  0.01  & 0.667  \pm   0.001 
                        &0.669  \pm  0.005 \\
 z_t        &  0.335 & 0.33  \pm  0.02  & 0.333  \pm   0.001 & 0.33  \pm  0.01 \\\hline  
\end{array}
\]
\end{center}
\end{table}

The results for the critical temperature and exponents are presented in
Table~\ref{critthings}, where the quoted critical temperatures are easily
determined using either Eqs.~(\ref{Mtzero}), (\ref{Mhzero}) or $t_{\rm pc}^h
\propto\, h^{1/(\beta \delta)}\, \propto\, t_{\rm pc}^t$.  The models are all
constrained by observables in two-flavour QCD and the critical temperature
agrees with the current estimate from simulations of lattice-QCD with two
dynamical light-quark flavours~\cite{edwinechaya}.  (Quenched lattice-QCD
exhibits a first order [deconfining] transition at a much higher critical
temperature: $T_c^{\rm quench} =
270(5)\,$MeV~\cite{edwinechaya,edwinTrento}.)  Studies using simple versions
of the Nambu--Jona-Lasinio model typically yield higher values: $T_c\gsim
200\,$MeV~\cite{sk92}.  However, that is likely a model artefact.
Modifications that make possible a relaxation of the overly restrictive
connection between the model's mass-scale and the dressed-quark mass; such as
$1/N_c$-[or meson-loop-]corrections or replacing the contact interaction by a
separable form, can easily effect a reduction of the critical temperature by
$\lsim 25$\%; e.g., Refs.~\cite{bastiO2,bastiO1,bruno,broniowski}.  This sort
of reduction brings it in-line with the estimate in Table~\ref{critthings}.

\begin{figure}[t]
\centering{\ \epsfig{figure=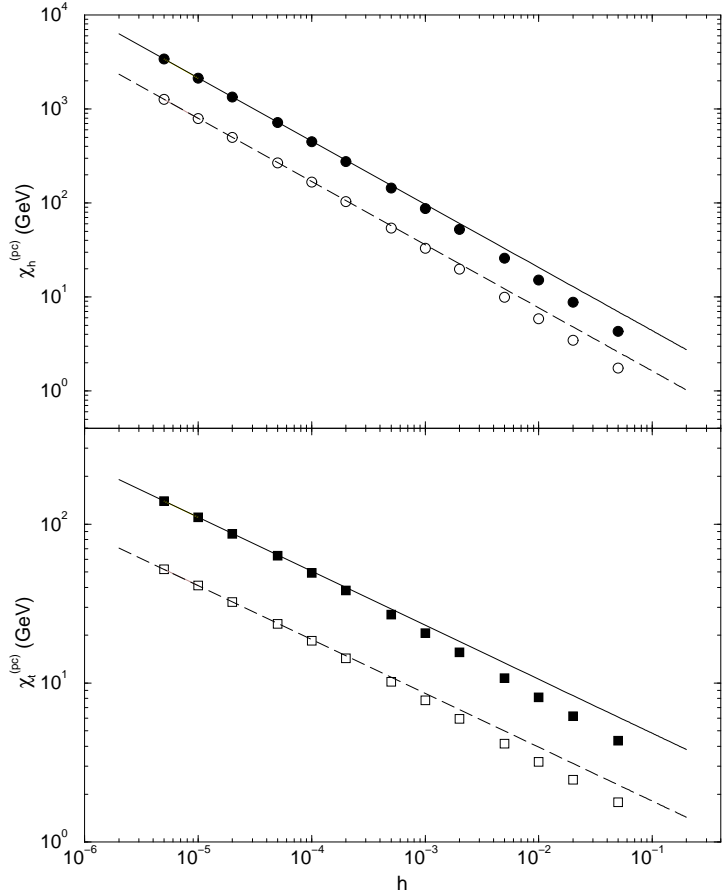,height=12.0cm}}
\parbox{40em}{\caption{$h$-dependence of the maxima of the chiral (circles)
and thermal (squares) susceptibilities calculated with the effective
interaction of Eq.~(\protect\ref{delta}) using ${\cal D}_C$ with $\omega =
1.2\, m_t$.  The filled symbols are obtained from ${\cal X}$ and the open
from ${\cal X}_C$, Eq.~(\protect\ref{chiforms}).  The slope of the straight
lines is given in Table~\protect\ref{critthings} and they are drawn through
the two smallest $h$-values, which is explained in the discussion of
Eq.~(\protect\ref{zi}).  (Adapted from Ref.~\protect\cite{arne2}.)
\label{figchi}}}
\end{figure}

\begin{figure}[t]
\centering{\ \epsfig{figure=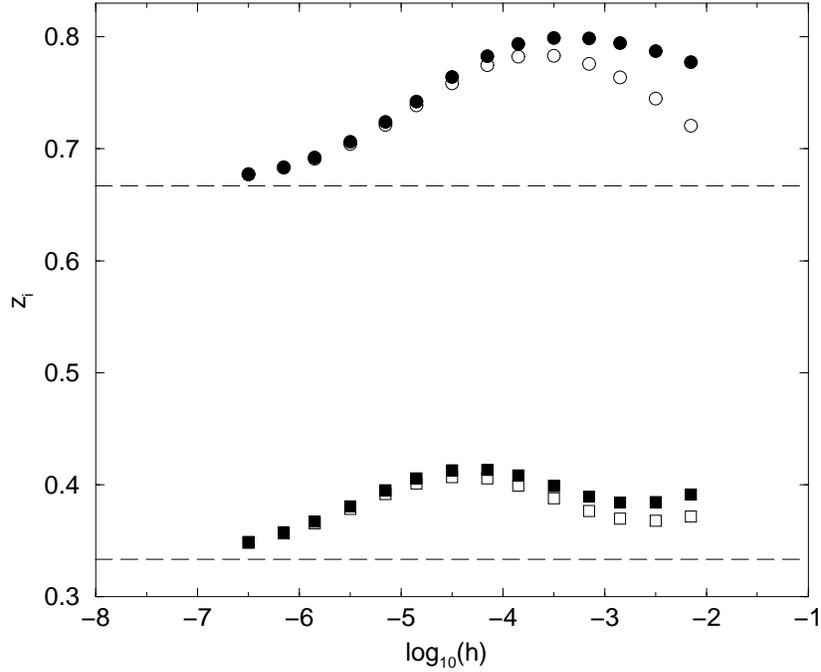,height=9.0cm}}
\parbox{40em}{\caption{$z_i^h$ (circles) and $z_i^t$ (squares) from
Eq.~(\protect\ref{zi}) calculated with the effective interaction of
Eq.~(\protect\ref{delta}) using ${\cal D}_B$ in Eq.~(\protect\ref{modelfr}):
filled symbols - ${\cal X}$, open symbols - ${\cal X}_C$.  The dashed lines
are the mean field values: $z_h=2/3$, $z_t=1/3$.  (Adapted from
Ref.~\protect\cite{arne2}.)
\label{findiff}}}
\end{figure}

The behaviour of the pseudocritical maxima in the chiral and thermal
susceptibilities is illustrated in Fig.~\ref{figchi}.  It is plain that the
results exhibit curvature and hence the scaling relations of
Eq.~(\ref{chislopes}) are not satisfied on the entire domain.  The domain on
which they are valid may be established by defining a ``local'' critical
exponent:
\begin{equation}
\label{zi}
z_i:= \,-\,\frac{ \ln \chi^{\rm pc}_i - \ln \chi^{\rm pc}_{i+1}}
        {\ln h_i - \ln h_{i+1}}\,,
\end{equation}
where $(h_i,\chi^{\rm pc}_i)$ and $(h_{i+1},\chi^{\rm pc}_{i+1})$ are
adjacent data pairs.  $h$ lies in the scaling region when $z_i$ is
independent of the order parameter.  Applying this to the ${\cal D}_{\rm B}$
model yields the results in Fig.~\ref{findiff}, which indicate clearly that
the scaling relations are not valid until
\begin{equation}
\label{mfr}
\log_{10} (h/h_u)< -7\,, 
\end{equation}
where $h_u:= m_R/T_c$ is defined with the current-quark mass that gives
$m_\pi =140\,$MeV in this model.  Applying the same method to the
renormalisation-group improved models, ${\cal D}_{\rm C}$, shows that in
these cases the scaling relations are only valid for
\begin{eqnarray}
\label{mmrammrb}
_1\omega \! : \,\log_{10} (h/h_u)< -5\,, &\; & _2\omega \! : \,\log_{10}
(h/h_u)< -6\,;
\end{eqnarray}
i.e., for current-quark masses six orders of magnitude smaller than those of
real $u$,$d$-quarks.  

While these are model studies the result is also likely to be true in QCD;
i.e., while the critical temperature is relatively easy to determine, very
small current-quark masses may be necessary to accurately calculate the
critical exponents from the chiral and thermal susceptibilities.  If that is
the case, the calculation of these exponents via numerical simulations of
lattice-QCD will not be feasible.  The lattice-to-lattice variation of the
critical exponents described in Refs.~\cite{edwinechaya,edwinTrento}, with
even a first order transition being possible, could be a signal of this.

It is clear from Table~\ref{critthings} that each of the models considered in
Ref.~\cite{arne2} is mean field in nature.  As the long-range part of the
effective interaction is identical in each case and the correlation length
diverges as $t\to 0$, that may not be surprising.  The models are
distinguished by the feature that they describe the long-wavelength dynamics
of QCD very well, with mesons represented as nonpointlike quark-antiquark
composites.  Coulomb gauge models~\cite{reinhardCoulomb} also describe mesons
as composites and exhibit mean field critical exponents.  The long-range part
of the interaction there corresponds directly to the regularised Fourier
amplitude of a linearly rising potential and therefore it is in no simple way
equivalent to the distribution in Eq.~(\ref{gk2}).  Separable models
with~\cite{bastiO1}: ${\cal D}(p_{\omega_k}-q_{\omega_l};m_g)\propto
g(|\vec{p}|)\,g(|\vec{q}|)$, where $g(|\vec{p}|)$ is a non-increasing
function of its argument, can also provide a good description of
long-wavelength pion dynamics.  They too describe mesons as composites and
exhibit an explicit fermion substructure, and they also have mean field
critical exponents, as do models with
$g=g(\omega_k^2+\vec{p}\,^2)$~\cite{pctTrento}.  [NB.  The
Nambu--Jona-Lasinio model can be viewed as a separable model obtained with
$g(|\vec{p}|)= \theta(1-|\vec{p}|/ _3\!\Lambda)$, where $_3\!\Lambda$ is a
cutoff parameter.  Also, in the rainbow-truncation DSE model of
Ref.~\cite{japan1}, which employs an infrared-finite effective interaction,
the critical exponent obtained from ${\cal X}$ is $1/2$.  However, a
different exponent is reported to be obtained from $\langle \bar q q\rangle$.
We expect that is erroneous, as these must be equivalent order parameters.]

All these examples are members of the class of rainbow truncation models.
Therefore the results indicate that chiral symmetry restoration in this class
is a mean field transition.  This is an essential consequence of the fermion
substructure in the rainbow gap equation: non-trivial critical exponents
require the appearance of an infrared divergence in this equation.  However,
in rainbow truncation the self energy is infrared-regulated by the zeroth
fermion Matsubara frequency: $\omega_0 = \pi T\gg 0$ in the vicinity of the
transition.  For the same reason chiral symmetry restoration at finite-$T$ in
QCD will be a mean field transition {\it unless} $1/N_c$-corrections to the
vertex are large for $T\simeq T_c$.  There are, however, examples in which
that is certainly true~\cite{stephanov1}.  Infrared divergences might be
anticipated if bosonic modes dominate the gap equation near $T_c$ because the
zeroth boson Matsubara frequency vanishes.  It is therefore important to note
that pseudoscalar meson-like fluctuations are nonperturbative
$1/N_c$-corrections to the dressed-vertex.  They do not contribute to the
dressed-gluon $2$-point function.  Therefore a true determination of the
critical exponents must await the accurate incorporation of these
$1/N_c$-corrections.

We have seen that chiral symmetry restoration in $2$-flavour QCD occurs at
$T\approx 150\,$MeV via a second order transition whose critical exponents
are not yet reliably determined.  What of deconfinement?  In QCD with two
dynamical flavours the Wilson-line/Polyakov-loop is not strictly a good order
parameter for deconfinement because of flux tube breaking mediated by
quark-antiquark pair formation.  However, in simulations the Polyakov loop's
susceptibility develops a peak at the same temperature as the chiral
susceptibility and this has been interpreted as evidence of coincident
deconfinement and chiral symmetry restoration~\cite{edwinTrento}.

\begin{figure}[t]
\centering{\ \epsfig{figure=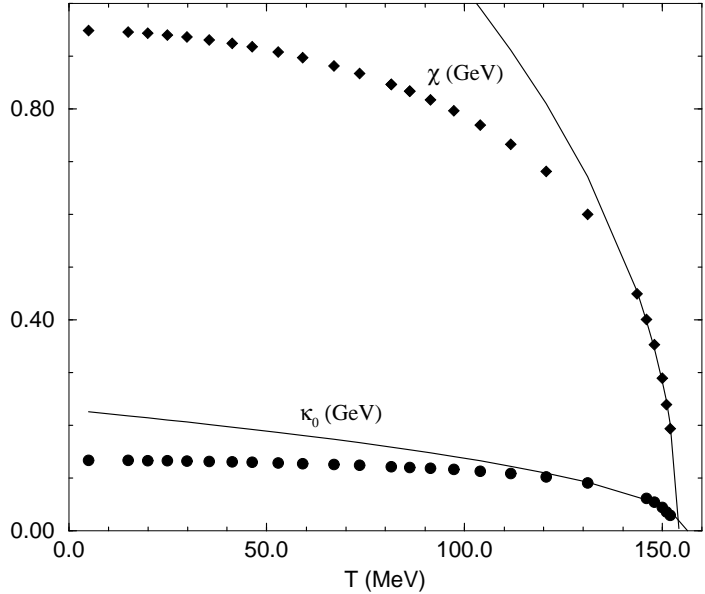,height=8.0cm}}
\parbox{40em}{\caption{Chiral and confinement order parameters calculated in
Ref.~\protect\cite{axelT} using ${\cal D}_B$ of Eq.~(\protect\ref{modelfr})
but with an {\it erroneous} value of $m_g^2 = 8\pi^2 T^2$, which is why the
coincident deconfinement and chiral symmetry restoring transitions here occur
at $T_c\approx 155\,$MeV, cf. Table~\protect\ref{critthings}.  The solid
lines are fits: ${\cal X} \propto (-t)^{1/2} \propto \kappa_0$.
\label{confPic}}}
\end{figure}

$\kappa_0$ in Eq.~(\ref{conforder}) provides an order parameter that can be
used to explore deconfinement via the gap equation.  It was exploited in
Ref.~\cite{axelT} where the deconfinement transition in model ``B)'' was
studied.  In this case the Schwinger function analysed is
\begin{equation}
\sigma_S(r) := \sfrac{2}{\pi}\int_0^\infty\,dp\,p\,\sin(p r)\,
\sigma_B(p^2,\omega_0^2)\,,
\end{equation}
where $\sigma_B$ is calculated in the chiral limit.  Figure~\ref{confPic}
depicts the behaviour of $\kappa_0(T)$ obtained from $\sigma_S(r)$ and
clearly the deconfinement order parameter is nonzero in the normal phase.
$\kappa_0(T)$ vanishes at the same temperature as does ${\cal X}$.  Hence, in
this model, the deconfinement and chiral symmetry restoring transitions are
coincident [to better-than $1$\%] and analysis that parallels
Ref.~\cite{arne2} shows that deconfinement is also a mean field transition.
(This improves over the critical exponents estimated in Ref.~\cite{axelT}.)
Whether the transitions are always coincident is unknown, however, that is
not unlikely given the role the scalar piece of the dressed-quark self energy
plays in determining both transitions.

The rainbow-ladder class of models fails to exhibit any flavour dependence in
the transition to a QGP, a flaw it shares; e.g., with random matrix
models~\cite{jackson}.  It is anticipated that with three light quark
flavours the chiral symmetry restoring transition is first
order~\cite{pisarski} but becomes second order if the mass of the
``$s$-quark'' is raised above a presently-undetermined critical value.  This
can be expected because for an infinitely heavy $s$-quark one recovers the
two-flavour theory with its second order transition.  (A summary of lattice
results is presented in Ref.~\cite{edwinTrento}.)  A coupling between
flavours is necessary to achieve such a pattern of symmetry breaking and, as
seen above, that appears only when $1/N_c$-corrections to the vertex are
incorporated.

\addcontentsline{toc}{subsubsection}{\protect\numberline{ } {Chemical
Potential}}
\subsubsect{Chemical Potential}
The introduction of a chemical potential moves us to a domain in which the
only available nonperturbative information comes from models.  For $\mu\neq
0$ the dressed-quark self energies in general acquire an imaginary part
driven by the magnitude of $\mu$; e.g., Refs.~\cite{bastiscm,japan1,gregp}.
This effect is not observed in explorations of the Nambu--Jona-Lasinio model
or indeed in any model in which the interaction is energy-independent; i.e.,
instantaneous.  That follows from Eq.~(\ref{FFstar}): an instantaneous
interaction cannot yield an energy-dependent solution and hence $M=M^\ast
\Rightarrow M\in \rule{1pt}{0.7em}{\rm R}$.  It is an unrealistic limitation
since the interaction in the normal [confined] phase of QCD is not
instantaneous.

An extensive exploration of the phase structure {\it and} thermodynamics of
two-flavour QCD with $\mu\neq 0\neq T$ is presented Ref.~\cite{bastiscm},
which employs the ${\cal D}_{\rm A}$ model.  ${\cal D}_{\rm A}$ is an
infrared-dominant model that does not represent well the behaviour of
$D_{\mu\nu}(\vec{p},\Omega_k)$ away from $|\vec{p}|^2+ \Omega_k^2 \approx 0$
and that introduces model-dependent artefacts.  However, there is significant
merit in its algebraic simplicity and, since the artefacts are easily
identified, the model remains useful as a means of elucidating many of the
qualitative features of more sophisticated Ans\"atze.  Therefore we use it
here as an exemplar.

In this model the gap equation is 
\begin{equation}
\label{mndse}
S(p_{\omega_k})^{-1} = S_0(p_{\omega_k})^{-1} + \sfrac{1}{4}\eta^2\gamma_\nu
        S(p_{\omega_k} ) \gamma_\nu
\end{equation}
and the simplicity is now apparent: the model allows the reduction of an
integral equation to an algebraic equation, which is always a useful step
when the goal is to develop an intuitive understanding of complicated
phenomena.  Substituting Eq.~(\ref{generalenough}) yields
\begin{eqnarray}
\label{beqnfour}
\eta^2 m^2 & = & B^4 + m B^3 + \left(4 p_{\omega_k}^2 - \eta^2 - m^2\right)
        B^2 -m\,\left( 2\,{{\eta }^2} + {m^2} + 4\,p_{\omega_k}^2 \right)B
        \,, \\ A(p_{\omega_k}) & = & C(p_{\omega_k}) = \frac{2
        B(p_{\omega_k})}{m +B(p_{\omega_k})}\,,
\end{eqnarray}
and it is now apparent that neglecting $A$, $C$, as in
Refs.~\cite{japan1,japan2}, is a poor approximation in the presence of DCSB;
i.e., on the domain where $B(p_{\omega_k})\gg m$, which must at least
introduce quantitative errors.

In the chiral limit: $m=0$, Eq.~(\ref{beqnfour}) reduces to a quadratic
equation for $B(p_{\omega_k})$, which has two qualitatively distinct
solutions.  The ``Nambu-Goldstone'' solution, for which
\begin{eqnarray}
\label{ngsoln}
B(p_{\omega_k}) & = &\left\{
\begin{array}{lcl}
\sqrt{\eta^2 - 4 p_{\omega_k}^2}\,, & &
\rule{1pt}{0.7em}{\rm R}(p_{\omega_k}^2) < \sfrac{\eta^2}{4}\\ 0\,, & & {\rm
 otherwise}
\end{array}\right.\\
C(p_{\omega_k}) & = &\left\{
\begin{array}{lcl}
2\,, & & \rule{1pt}{0.7em}{\rm R}(p_{\omega_k}^2)<\sfrac{\eta^2}{4}\\
\sfrac{1}{2}\left( 1 + \sqrt{1 + 2 \eta^2/p_{\omega_k}^2}\right)
\,,& & {\rm otherwise}\,,
\end{array}\right.
\end{eqnarray}
describes a phase in which: 1) chiral symmetry is dynamically broken, because
one has a nonzero quark mass-function, $B(p_{\omega_k})$, in the absence of a
current-quark mass; and 2) the dressed-quarks are confined, because the
dressed-quark $2$-point Schwinger function violates reflection positivity.
The alternative ``Wigner'' solution, for which
\begin{equation}
\hat B(p_{\omega_k})  \equiv  0\,,\;\;
\label{wsolnC}
\hat C(p_{\omega_k}) = \sfrac{1}{2}\left( 1 + \sqrt{1 +
2\eta^2/p_{\omega_k}^2 }\right)\,,
\end{equation}
characterises a phase in which chiral symmetry is not broken and the
dressed-quarks are not confined.

\begin{figure}[t]
\centering{\ \epsfig{figure=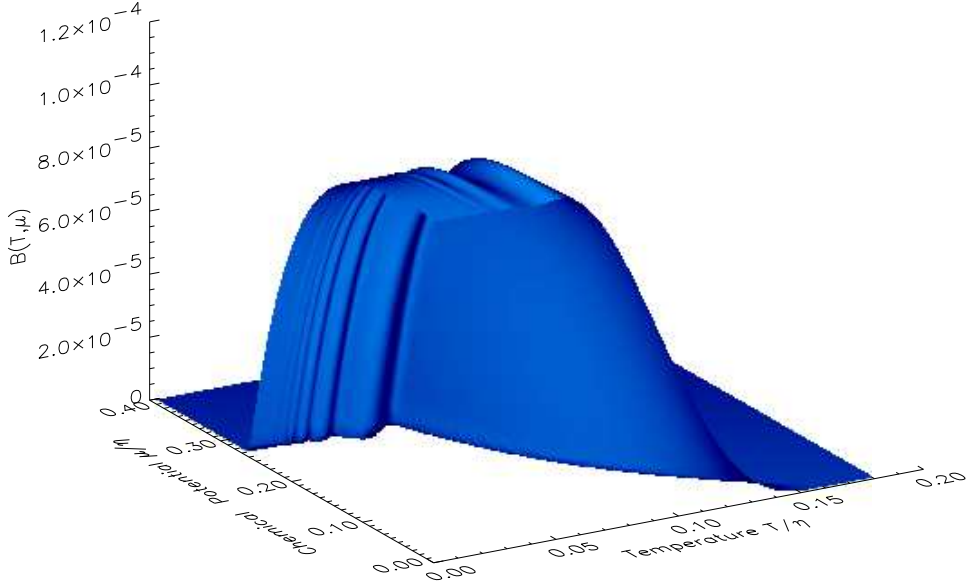,height=9.0cm}}
\parbox{40em}{\caption{${\cal B}(\mu,T)$ in Eq.~(\protect\ref{BmuTMN}):
${\cal B}(\mu,T)>0$ marks the domain: ${\cal D}$, of confinement and DCSB,
with the line ${\cal B}(\mu,T)=0$ demarcating the phase boundary.  At
$\mu=0$, the transition is second order with mean field critical exponents
and $T_c=169\,$MeV, using $\eta=1.06\,$GeV.  For all nonzero $\mu$ the
transition is first order with a tricritical point at $(\mu=0,T_c)$.  At
$T=0$, $\mu_c= 300\,$MeV.  The value is $25$\% larger; i.e.,
$\mu_c=375\,$MeV, in the model defined by ${\cal D}_{\rm B}$, which improves
the ultraviolet behaviour of the effective interaction~\protect\cite{gregp},
and in a two-flavour, ultrarelativistic Fermi gas at this chemical potential
the baryon number density is $2.9\,\rho_0$, $\rho_0=0.16\,$fm$^{-3}$.
(Adapted from Ref.~\protect\cite{bastiscm}.)
\label{bagpresMN}}}
\end{figure}

To explore the phase transition one can consider the relative stability of
the confined and deconfined phases, which is measured by their difference in
pressure.  Using Eq.~(\ref{pSigma}) alone, because the gluon contributions
cancel, that difference is
\begin{eqnarray}
{\cal B}(\mu,T) & := & p_{\Sigma_{\rm NG}}(\mu,T) - p_{\Sigma_{\rm
W}}(\mu,T)\,,\\
& = &4 N_c \int_{l,q}^{\bar\Lambda}
\left\{ 
\ln\left[ \frac{|\vec{p}|^2 A^2 + \tilde\omega_k^2 C^2 + B^2}
{|\vec{p}|^2 \hat A^2 + \tilde\omega_k^2 \hat C^2}\right]
+  |\vec{p}|^2 \left(\sigma_{A} - \hat\sigma_{A}\right)
+  \tilde\omega_k^2  \left(\sigma_{C} - \hat\sigma_{C}\right)\right\}\,,
\end{eqnarray}
which in the particular case considered here yields
\begin{equation}
\label{BmuTMN}
{\cal B}(\mu,T)  =  \eta^4\,2 N_c N_f \frac{\bar T}{\pi^2}\sum_{l=0}^{l_{\rm
        max}} 
        \int_0^{\bar\Lambda_l}\,dy\,y^2\,
        \left\{\Real\left(2 \bar p_l^2 \right) 
                - \Real\left(\frac{1}{C(\bar p_l)}\right)
- \ln\left| \bar p_l^2 C(\bar p_l)^2\right|        \right\}\,,
\end{equation}
with: $\bar T=T/\eta$, $\bar \mu=\mu/\eta$; $\bar\omega^2_{l_{\rm max}}\leq
\sfrac{1}{4}+\bar\mu^2$, $\bar\Lambda_l^2 = \bar\omega^2_{l_{\rm
max}}-\bar\omega_l^2$, $\bar p_l = (\vec{y},\bar\omega_l+i\bar\mu)$.  ${\cal
B}(\mu,T)$ defines a $(\mu,T)$-dependent ``bag constant''~\cite{cr85} and in
this model 
\begin{equation}
\label{bagvalue}
{\cal B}(0,0)= (0.102\,\eta)^4 = (0.109\,{\rm GeV})^4\,,
\end{equation}
which can be compared with the value $\sim (0.145\,{\rm GeV})^4$ commonly
used in bag-like models of hadrons.  The positive value indicates that the
confining phase of the model with DCSB is stable at $\mu=0=T$.  The phase
boundary in the $(\mu,T)$-plane is defined by the line ${\cal B}(\mu,T)=0$ ,
which is evident in Fig.~\ref{bagpresMN}.  In this model the transition is
first order {\it except} at $\mu=0$, which is clear from the nonzero
derivative at the phase boundary, and the deconfinement and chiral symmetry
restoring transitions are coincident, which is a consequence of the character
of the solutions of the gap equation, Eqs.~(\ref{ngsoln}--\ref{wsolnC}).  The
$T=0$ transitions are also coincident in the model obtained with ${\cal
D}_{\rm B}$, Eq.~(\ref{modelfr})~\cite{gregp}.  The chiral order parameters
increase with increasing $\mu$, as they do in all models that preserve the
momentum dependence of the dressed-quark self energies; e.g,
Refs.~\cite{japan1,gregp,japan2}, typically reaching a value $\lsim 20$\%
larger than that at $\mu=0=T$.  This is an anticipated result of confinement:
as long as $\mu<\mu_c$, each additional quark must be paired with an
antiquark thereby increasing the density of condensate pairs.  It cannot be
observed in models that omit the momentum dependence of the dressed-quark
self energy, such as those with an instantaneous interaction.

The criterion of maximal pressure was also employed in Ref.~\cite{japan2},
wherein the model effective interaction saturates in the infrared and
exhibits one-loop improved ultraviolet behaviour analogous to that in
Eq.~(\ref{gk2}).  The calculated $\mu=0$ critical temperature agrees with the
estimate in Table~\ref{critthings} and the $T=0$ critical chemical potential:
$\mu_c=422\,$MeV, is just $13$\% larger than that estimated in
Ref.~\cite{gregp}.  A similar interaction was employed in Ref.~\cite{japan1},
wherein $(\mu=0,T_c)$ agrees with Table~\ref{critthings}.  However, the
estimate of $(\mu_c,T=0)$ is $60$\% larger than in Refs.~\cite{gregp,japan2}
and is likely too high.  The usual range is $\mu_c\in[300,400]\,$MeV; e.g.,
see also Refs.~\cite{blaschkebiel,diakonov2}.

As clear in Fig.~\ref{bagpresMN}, the chiral phase boundary traces out a
curve in the $(\mu,T)$-plane.  In Ref.~\cite{japan2}, beginning at
$(\mu_c,T=0)$, this curve describes a line of first order critical points
that ends with a tricritical point at $(\mu_{\rm tc}\approx 0.5\,
\mu_c,T_{\rm tc} \approx 0.8 \,T_c)$.  For $T>T_{\rm tc}$ the transition is
second order.
In comparison with the ${\cal D}_{\rm A}$ model, this is a shift of the
tricritical point away from the $\mu=0$ axis.  We expect such a relocation to
be a feature of all models with a better treatment of the ultraviolet
behaviour of the effective interaction; e.g., there are preliminary
indications that in the separable model of Ref.~\cite{pctTrento} a
tricritical point is located at $(\mu_{\rm tc}\approx 0.4\, \mu_c,T_{\rm tc}
\approx 0.9 \,T_c)$.
The existence of a tricritical point has observable
consequences~\cite{stephanov2}.  However, if it occurs at a value of $\mu$ as
large as that estimated in these models it may be difficult to detect using
the relativistic heavy ion collider [RHIC], which will focus on the domain of
low baryon number density.

With ${\cal B}(\mu,T)>0$, then, in the exploration of hadronic observables
using the rainbow-ladder truncation, $p_{\Sigma_{\rm NG}}(\mu,T)$ describes
the free energy of the ``vacuum'' or ``ground state,'' relative to which all
excitations are measured $\forall\,(\mu,T)\in {\cal D}$; i.e., in the domain
of confinement and DCSB.  Therefore $p_{\Sigma_{\rm NG}}$ is not directly
observable and does not contribute to the thermodynamic pressure.
Nevertheless, it does evolve with changes in $(\mu,T)$, as evident in
Fig.~\ref{bagpresMN}, and this evolution simply reflects the
$(\mu,T)$-dependence of the dressed-quark self energies, upon which it
depends.  [NB.  If it did not evolve there could never be a phase
transition.]  The changes in the dressed-quark self energy represent the
$(\mu,T)$-dependent rearrangement of the vacuum and are indirectly
observable; e.g., in the behaviour of hadron masses and decay constants.
Hence, on the domain ${\cal D}$, they are evident in changes of the only true
contributions to the pressure; i.e., those of the colour singlet hadronic
correlations.  At the phase boundary, denoted by $\partial {\cal D}$:
$\left. p_{\Sigma_{\rm NG}}(\mu,T)\right|_{\partial {\cal D}} =
\left. p_{\Sigma_{\rm W}}(\mu,T)\right|_{\partial {\cal D}}$.  This is the
last vacuum reference point: hereafter dressed-quarks and -gluons are the
active degrees of freedom.  

\begin{figure}[t]
\centering{\ \epsfig{figure=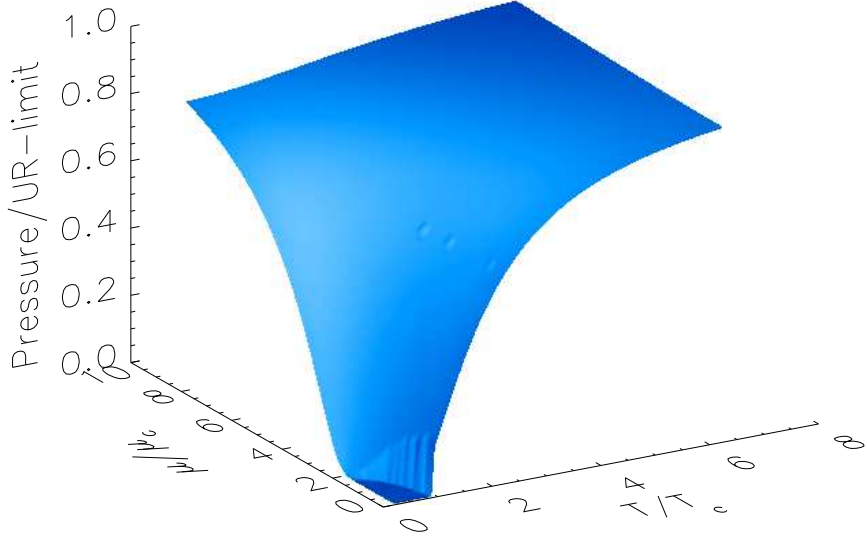,height=8.0cm}}
\parbox{40em}{\caption{$P_q(\mu,T)$ in Eq.~(\protect\ref{qpres}) calculated
using the ${\cal D}_{\rm A}$ and normalised by the pressure of an
ultrarelativistic Fermi gas, Eq.~(\protect\ref{sbpres}).  This ratio
approaches one slowly; e.g., at $(\mu=0,T\sim 2\,T_c)$ or $(\mu\sim
3\,\mu_c,T=0)$, $P_q(\mu,T)$ is only one-half of the free particle pressure.
The retardation is a consequence of the persistence into the deconfined
domain of nonperturbative effects in the dressed-quark's vector self energy.
Note the ``mirroring'' of the $T$-dependence in the evolution with $\mu$,
which we anticipate is a general feature.  (Adapted from
Ref.~\protect\cite{bastiscm}.)
\label{pressurePic}}}\vspace*{-3ex}
\end{figure}

The dressed-quarks' contribution to the pressure is
\begin{equation}
\label{qpres}
P_q(\mu,T) = \theta({\cal D})\left\{ p_{\Sigma_{\rm W}}(\mu,T) -
        \left.p_{\Sigma_{\rm W}}(\mu,T)\right|_{\partial {\cal D}}\right\},
\end{equation}
where $\theta( {\cal D})$ is a step function, equal to one for $(T,\mu)\in
{\cal D}$.  This pressure can be calculated numerically and is depicted in
Fig.~\ref{pressurePic} for model ${\cal D}_{\rm A}$.  As evident in
Eq.~(\ref{wsolnC}), nonperturbative, momentum-dependent modifications of the
dressed-quark's vector self energy persist into the deconfined domain and
this feature retards the approach of the pressure to the ultrarelativistic
limit:
\begin{equation}
\label{sbpres}
p_{\rm UR}(\mu,T)= 
        N_c N_f \frac{1}{12\pi^2}\left(
         \mu^4 + 2 \pi^2  \mu^2   T^2 + \sfrac{7}{15}\pi^4
                          T^4\right)\,.
\end{equation}
A qualitatively similar result is observed in numerical simulations of $T\neq
0$ lattice-QCD~\cite{edwinechaya}.  Such a correspondence is impossible in
models where the dressed-quark vector self energy is omitted.  

The pressure in Eq.~(\ref{qpres}) is incomplete.  It is zero in the confined
domain where, as remarked above, the only contribution is that of colour
singlet hadronic correlations.  Those contributions can be estimated; e.g.,
using the hadronisation techniques of Ref.~\cite{cahillalone}, but that has
not yet been done.  In the deconfined domain, one must add the contribution
of the dressed-gluons.  However, hitherto the model studies have employed
``frozen'' gluons; i.e., the structure of the effective interaction does not
evolve with $(\mu,T)$.  $T\neq 0$ lattice-QCD simulations suggest that to be
a good approximation, at least until very near the phase
boundary~\cite{edwinechaya}, and Ref.~\cite{reinhardCoulomb} indicates that
changes in the interaction in this vicinity have little effect on the
properties of the transitions.  That is good because improving on the frozen
approximation is difficult, requiring information about the effective
interaction in a new domain.  A first step is to allow the mass-scale
characterising the infrared behaviour of the interaction to drop suddenly and
significantly near the phase boundary, thereby mocking-up the $T$-dependence
of the string tension.  Such a study is presently lacking.

Not withstanding these weaknesses, the pressure depicted in
Fig.~\ref{pressurePic} still provides a rudimentary model for the
dressed-quark pressure that exhibits significant new qualitative features;
e.g., the persistence into the deconfined domain of nonperturbative effects
in the vector self energy yields an EOS much softer than that of a bag model.
In this capacity it has been used to explore the structure and stability of
nonstrange quark stars~\cite{quarkstar1}.
References~\cite{bastiscm,gregp,solitongcm} indicate that a transition to
quark matter should occur at approximately three-times nuclear matter
density, and the comparison between results obtained with the EOS depicted in
Fig.~\ref{pressurePic} and a bag model EOS shows; e.g., that the prediction
for the maximum stable radius of a pure quark star is model-independent:
$8$--$10\,$km, also in agreement with estimates based on a colour dielectric
model~\cite{drago}.  The maximum mass, however, is very sensitive to the EOS.
The quark core described by Eq~(\ref{qpres}) must be surrounded by an
hadronic crust and that will also affect the properties of the star.  The
problem rapidly becomes involved, and determining the composition of dense
astrophysical objects and those signals that point to the realisation of
quark matter in their core is an important focus of contemporary research;
e.g., Refs.~\cite{pctTrento,drago,weber,davidqqstar}.

\addcontentsline{toc}{subsubsection}{\protect\numberline{ } {Superfluidity in
Quark Matter}}
\subsubsect{Superfluidity in Quark Matter}
In Sec.~\ref{sectthree} we described aspects of diquark correlations and the
manner in which they help to understand baryon properties.  Another aspect
follows from the meson-diquark auxiliary-field effective
action~\cite{regdq,hugodq}: the steepest descent approximation to the vacuum
pressure suggests the possibility of diquark condensation; i.e., quark-quark
Cooper pairing, and that was first explored using a simple version of the
Nambu--Jona-Lasinio model~\cite{kahana}.  A nonzero chemical potential
promotes Cooper pairing in fermion systems and the possibility that it is
exhibited in quark matter was considered using the rainbow-ladder truncation
of the gap equation~\cite{bl84}.  Interest in that possibility has been
renewed~\cite{krishna}.  A quark-quark Cooper pair is a composite boson with
both electric and colour charge, and hence superfluidity in quark matter
entails superconductivity and colour superconductivity.  However, the last
feature makes it difficult to identify an order parameter that can
characterise a transition to the superfluid phase: the Cooper pair is gauge
dependent and an order parameter is ideally describable by a gauge-invariant
operator.

As we saw above, at $\mu=0=T$ QCD exhibits a nonzero quark-antiquark
condensate but it is undermined by increasing $\mu$ and $T$, and there is a
domain of the $(\mu,T)$-plane, evident in Fig.~\ref{bagpresMN}, for which
$\langle\bar q q\rangle=0$.  Increasing $T$ also opposes Cooper pairing.
However, since increasing $\mu$ promotes it, there may be a
(large-$\mu$,low-$T$)-subdomain in which quark matter exists in a superfluid
phase.  While that domain may not be accessible at RHIC, such conditions
might exist in the core of dense astrophysical objects~\cite{bl84}, which
could undergo a transition to superfluid quark matter as they cool.
Unambiguous signals of such a superfluid phase are actively being sought;
e.g, Refs.~\cite{pctTrento,davidqqstar}.

We take Ref.~\cite{jacquesdq} as our exemplar, in which it was observed that a
direct means of determining whether a SU$_c(N$) gauge theory supports $0^+$
diquark condensation is to study the gap equation satisfied by
\begin{equation}
\label{sinv}
\BD(p,\mu) := 
{\cal S}(p,\mu)^{-1} =\left(
\begin{array}{cc}
D(p,\mu) & \Delta^i(p,\mu)\, \gamma_5  \lambda_{\wedge}^i \\
 -\Delta^i(p,-\mu)\, \gamma_5  \lambda_{\wedge}^i
        & C D(-p,\mu)^{\rm t} C^\dagger
\end{array}\right).
\end{equation}
Here $T=0$, for illustrative simplicity and because temperature can only act
to destabilise a Cooper pair, and, with $\omega_{[\mu]}= p_4+i\mu$,
\begin{equation}
\label{DCA}
D(p,\mu) = i \vec{\gamma}\cdot \vec{p}\, A(\vec{p}\,^2,\omega_{[\mu]}^2) +
B(\vec{p}\,^2,\omega_{[\mu]}^2 ) + i \gamma_4 \,\omega_{[\mu]}
\,C(\vec{p}\,^2,\omega_{[\mu]}^2 )\,;
\end{equation}
i.e., the inverse of the dressed-quark $2$-point function,
Eq.~(\ref{generalenough}).  In Eq.~(\ref{sinv}), $\{\lambda_{\wedge}^i$,
$i=1\ldots n^\wedge_c$, $n^\wedge_c= N_c (N_c-1)/2\}$ are the antisymmetric
generators of SU$_c(N_c)$ and $C$ is the charge conjugation matrix,
Eq.~(\ref{CCmtx}).  Using such a gap equation to study superfluidity makes
unnecessary a truncated bosonisation, which in all but the simplest models is
a procedure difficult to improve systematically.

In addition to the usual colour, Dirac and isospin indices carried by the
elements of $\BD(p,\mu)$, the explicit matrix structure in Eq.~(\ref{sinv})
exhibits a quark bispinor index and is made with reference to
\begin{equation}
\label{QQCD}
Q(x) :=  \left(\begin{array}{c}
                        q(x)\\
                        \underline q(x):= \tau^2_f\, C\, \bar q^{\rm t}
                         \end{array} \right)\,,\;\;
\bar Q(x)  := \left(\begin{array}{cc}
                        \bar q(x)\;\;
                        \bar {\underline q}(x):= q^{\rm t} \,C\,\tau^2_f
\end{array} \right),
\end{equation}
where $\{\tau_f^i: i=1,2,3\}$ are Pauli matrices, Eq.~(\ref{PauliMs}), that
act on the isospin index.  (Only two-flavour theories are considered in this
exemplar.  Additional possibilities open in three-flavour
theories~\cite{wilczekgossip}.)

As we have seen, a nonzero quark condensate: $\langle \bar q q \rangle \neq
0$, is represented in the solution of the gap equation by
$B(\vec{p}\,^2,\omega_{[\mu]}^2 )\not\equiv 0$.  The new but analogous
feature here is that diquark condensation is characterised by
$\Delta^i(p,\mu)\not\equiv 0$, for at least one $i$.  That is clear if one
considers the quark piece of the QCD Lagrangian density: $L[\bar q,q]$.  It
is a scalar and hence $L[\bar q,q]^{\rm t}= L[\bar q,q]$.  Therefore $L[\bar
q,q] \propto L[\bar q,q] + L[\bar q,q]^{\rm t}$, and that sum, and hence the
action, can be re-expressed in terms of a diagonal matrix using the bispinor
fields in Eq.~(\ref{QQCD}): $\bar Q \, {\rm diag}[D,C D^{\rm t} C^\dagger]
Q$.  It is plain now that a dynamically-generated lower-left element in
$\BD(p,\mu)$, the inverse of the dressed-bispinor propagator, corresponds to
a $\bar {\underline q} q$ [diquark] correlation.

The bispinor gap equation can be written in the form
\begin{equation}
\label{bispinDSE}
\BD(p,\mu)  =  \BD_0(p,\mu) 
+ \left( 
\begin{array}{cc}
\Sigma_{11}(p,\mu) & \Sigma_{12}(p,\mu)\\
\gamma_4\,\Sigma_{12}(-p,\mu)\,\gamma_4 & C \Sigma_{11}(-p,\mu)^{\rm t}
C^\dagger  
\end{array} \right),
\end{equation}
where in the absence of a diquark source term
\begin{equation}
\BD_0(p,\mu) = (i\gamma\cdot p + m )\tau_Q^0 - \mu\,\tau_Q^3\,,
\end{equation}
with $m$ the current-quark mass, and the additional Pauli matrices:
$\{\tau_Q^\alpha,\alpha = 0,1,2,3\}$, act on the bispinor indices.  The
structure of $\Sigma_{ij}(p,\mu)$ specifies the theory and, in practice, also
the approximation or truncation of it.

Two colour QCD [QC$_2$D] provides an important and instructive example.  In
this case 
$ \Delta^i \lambda_\wedge^i =\Delta \tau^2_c $
in Eq.~(\ref{sinv}), with $\sfrac{1}{2}\vec{\tau_c}$ the generators of
$SU_c(2)$, and it is useful to employ a modified bispinor
\begin{eqnarray}
Q_2(x) &:= &\left(\begin{array}{c}
                         q(x)\\
\underline q_2:=\tau^2_c\,\underline q(x)  \end{array} \right),\;
%
\end{eqnarray}
with $\bar Q_2$ the obvious analogue of $\bar Q$ in Eq.~(\ref{QQCD}).  Now
the Lagrangian's fermion-gauge-boson interaction term is simply
\begin{equation}
\bar Q_2(x) \,\sfrac{i}{2} g \gamma_\mu \tau_c^k\tau_Q^0
\,Q_2(x)\,A_\mu^k(x) 
\end{equation}
because SU$_c(2)$ is pseudoreal; i.e.,  
$ \tau^2_c\left(-\vec{\tau}_c\right)^{\rm t}\tau^2_c = \vec{\tau}_c \,,\;$ 
and the fundamental and conjugate representations are equivalent.  (That the
interaction term takes this form is easily seen using $L[\bar q,q]^{\rm t}=
L[\bar q,q]$.)

Using the pseudoreality of SU$_c(2)$ it can be shown that, for $\mu=0$ and in
the chiral limit, the general solution of the bispinor gap equation
is~\cite{jacquesdq}
\begin{equation}
\label{sinvgen2}
\BD(p) = i\gamma\cdot p\, A(p^2) + {\cal V}(-\mbox{\boldmath$\pi$})\, {\cal
M}(p^2)\,,\;\;
{\cal V}(\mbox{\boldmath$\pi$}) = 
\exp\left\{i \gamma_5\, \sum_{\ell=1}^5\,T^\ell\, \pi^\ell\right\}
= {\cal V}(-\mbox{\boldmath$\pi$})^{-1} \,,
\end{equation}
where $\pi^{\ell=1,\ldots,5}$ are arbitrary constants and $\{T^{1,2,3}=
\tau_Q^3 \otimes \vec{\tau_f},\, T^4= \tau_Q^1\otimes \tau_f^0,\, T^5=
\tau_Q^2\otimes\tau_f^0 \}$, $\{T^i,T^j\}=2 \delta^{ij}$, so that
\begin{equation} 
\label{Sp}
{\cal S}(p) = \frac{-i\gamma\cdot p A(p^2) + {\cal V}(\mbox{\boldmath$\pi$})
{\cal M}(p^2)} {p^2 A^2(p^2) + {\cal M}^2(p^2)}\,
:= \,-i\gamma\cdot p \,\sigma_V(p^2) + {\cal
V}(\mbox{\boldmath$\pi$})\,\sigma_S(p^2)\, .
\end{equation}
[$\mbox{\boldmath$\pi$}=(0,0,0,0,-\sfrac{1}{4}\pi)$ produces an inverse
bispinor propagator with the simple form in Eq.~(\ref{sinv}).]  That the gap
equation is satisfied for any constants $\pi^\ell$ signals a vacuum
degeneracy: if the interaction supports a mass gap, then that gap describes a
five-parameter continuum of degenerate condensates:
\begin{equation}
\label{cndst}
\langle \bar Q_2 
{\cal V}(\mbox{\boldmath$\pi$}) Q_2\rangle \neq 0\,,
\end{equation}
and there are 5 associated Goldstone bosons: 3 pions, a diquark and an
anti-diquark.  In this construction, Eq.~(\ref{sinvgen2}), one has a simple
elucidation of a necessary consequence of the Pauli-G\"ursey symmetry of
QC$_2$D.  For $m\neq 0$, the gap equation requires
${\rm tr}_{FQ} \left[ T^i {\cal V} \right] = 0\,,$
so that in this case only $\langle \bar Q_2 Q_2 \rangle \neq 0$ and the
Goldstone bosons are now massive but remain degenerate.

For $\mu\neq 0$ the general solution of the gap equation has the form 
\begin{equation}
\label{BDmu} 
\BD(p,\mu)= \left(
\begin{array}{cc}
D(p,\mu) & \gamma_5\,\Delta(p,\mu)  \\
- \gamma_5\Delta^\ast(p,\mu) & C D(-p,\mu) C^\dagger
\end{array}
\right)\,,
\end{equation}
and in the {\it absence} of a diquark condensate; i.e.,
for $\Delta \equiv 0$,
\begin{equation}
[U_B(\alpha), \BD(p,\mu)]=0\,, \; 
U_B(\alpha):= {\rm e}^{i \alpha \tau_Q^3 \otimes \tau_f^0}\,,
\end{equation}
which is a manifestation of baryon number conservation in QC$_2$D.  In this
case the explicit form of the gap equation is complicated but its features
and those of its solutions are easily illustrated using the model ${\cal
D}_{\rm A}$ introduced above; e.g., it becomes clear that a chemical
potential promotes fermion-pair condensation, at the expense of
fermion-antifermion pairs, and $\Delta \in \Real$, $\forall \mu$.  The
rainbow-truncation gap equation was solved using this model in
Ref.~\cite{jacquesdq}, and the relative stability of the quark- and
diquark-condensed phases measured via the pressure difference
\begin{equation}
\label{deltaP}
\delta p(\mu) := p_{\Sigma[B=0,\Delta]}(\mu) - p_{\Sigma[B,\Delta=0]}(\mu)\,,
\end{equation}
where the pressure here is obtained as an obvious generalisation of
Eq.~(\ref{pSigma}).  $\delta p(\mu)$ can be expressed in terms of
$\mu$-dependent bag constants
\begin{equation}
{\cal B}_B(\mu) := p_{\Sigma[B,\Delta=0]}(\mu) -
p_{\Sigma[B=0,\Delta=0]}(\mu)\,,\;\;
{\cal B}_\Delta(\mu) := p_{\Sigma[B=0,\Delta]}(\mu) -
p_{\Sigma[B=0,\Delta=0]}(\mu)\,,
\end{equation}
which measure the stability of a quark- or diquark-condensed vacuum relative
to that with chiral symmetry realised in the Wigner-Weyl mode.  [NB.
Improving on rainbow-ladder truncation may yield quantitative changes of
$\lsim 20$\% in the exemplary results that follow, however, the pseudoreality
of QC$_2$D and the equal dimension of the colour and bispinor spaces, which
underly the theory's Pauli-G\"ursey symmetry, ensure that the entire
discussion remains qualitatively unchanged.]

The $(m,\mu)=0$ degeneracy of the quark and diquark condensates,
Eq.~(\ref{cndst}), is manifest in
\begin{equation}
{\cal B}_B(0) = {\cal B}_\Delta(0) = (0.13 \,m_{J=1})^4\,,
\end{equation}
where $m_{J=1}$ is the $(m,\mu)=0$ mass of the model's vector meson,
calculated in rainbow-ladder truncation.  Increasing $\mu$ at $m=0$ and
excluding diquark condensation, chiral symmetry is restored at
\begin{equation}
\mu_{2c}^{B,\Delta=0} = 0.40 \,m_{J=1}\,,
\end{equation}
where ${\cal B}_B(\mu) = 0$.  However,
$ \forall \mu > 0$: $\delta p(\mu) >0$ and ${\cal B}_\Delta(\mu)>0\,, $
with 
\begin{equation}
{\cal B}_\Delta(\mu_{2c}^{B,\Delta=0}) = (0.20\, m_{J=1})^4 > {\cal
B}_\Delta(0). 
\end{equation}
Therefore the vacuum is unstable with respect to diquark condensation for all
$\mu> 0$, and this is always dynamically preferred over quark condensation.

$(B=0, \Delta\neq 0)$ in Eq.~(\ref{BDmu}) corresponds to
$\mbox{\boldmath$\pi$}=(0,0,0,0,\sfrac{1}{2}\pi)$ in Eq.~(\ref{cndst}); i.e.,
$ \langle \bar Q_2 i\gamma_5\tau_Q^2 Q_2\rangle \neq 0 $.
The usual chiral [$SU_A(2)$] transformations are realised via
\begin{equation}
\BD(p,\mu) \to V(\vec{\pi})\, \BD(p,\mu) \,V(\vec{\pi}) \,,
\;\; V(\vec{\pi}):= {\rm e}^{i\gamma_5 \vec{\pi}\cdot\vec{T}}\,,\;
\vec{\pi}=(\pi^1,\pi^2,\pi^3)\,,
\end{equation}
and therefore, since the anticommutator $\{\vec{T},T^{4,5}\}=0$, a diquark
condensate does not break chiral symmetry.  However, $(B=0, \Delta\neq 0)$
does yield a dressed-bispinor propagator that violates reflection positivity
and hence the model exhibits confinement to arbitrarily large densities.
[NB.  Reference~\cite{jacquesdq} employs a frozen gluon approximation.  In a
more realistic analysis, the $\mu$-dependence of $\eta$, the mass-scale
characterising the model, would be significant for $\mu \sim
\mu_{2c}^{B,\Delta=0}$, and $\eta \to 0$ as $\mu\to \infty$, which would
ensure deconfinement at large-$\mu$.]  Finally, although the $\mu\neq 0$
theory is invariant under
\begin{equation}
 Q_2 \to U_B(\alpha) \,Q_2\,,\; \bar Q_2 \to \bar Q_2\,U_B(-\alpha) \,, 
\end{equation}
which is associated with baryon number conservation, the diquark condensate
breaks this symmetry:
\begin{equation}
\langle \bar Q_2 i\gamma_5\tau_Q^2 Q_2\rangle \to \cos(2 \alpha)\, \langle
\bar Q_2 i\gamma_5\tau_Q^2 Q_2\rangle - \sin(2 \alpha)\,\langle \bar Q_2
i\gamma_5\tau_Q^1 Q_2\rangle\,.
\end{equation}
Hence, for $(m=0,\mu\neq 0)$, one Goldstone mode remains.  (These symmetry
breaking patterns and the concomitant numbers of Goldstone modes in QC$_2$D
are also described in Ref.~\cite{stephanov3}.)

For $m\neq 0$ and small values of $\mu$, the gap equation only admits a
solution with $\Delta \equiv 0$; i.e., diquark condensation is blocked and
this is because the current-quark mass is a source of quark condensation.
However, with increasing $\mu$ a diquark condensate is generated and the
${\cal D}_{\rm A}$ model exhibits the following minimum chemical potentials
for diquark condensation:
\begin{equation}
m= 0.013\,m_{J=1}  \Rightarrow  \mu^{\Delta \neq 0} =
0.051\,m_{J=1}\,,\;\;
m= 0.13\,m_{J=1}  \Rightarrow  \mu^{\Delta \neq 0} = 0.092\,m_{J=1}\,.
\end{equation}
This retardation of diquark condensation by a nonzero current-quark mass can
also be seen in Ref.~\cite{bergesbiel}.

The exploration of superfluidity in true QCD encounters two differences: the
dimension of the colour space is greater than that of the bispinor space and
the fundamental and conjugate representations of the gauge group are not
equivalent.  The latter is of obvious importance because it entails that the
quark-quark and quark-antiquark scattering matrices are qualitatively
different. As we saw above in connection with Eq.~(\ref{diquarkmasses}),
these differences ensure that colour singlet meson bound states exist but
[necessarily coloured] diquark bound states do not.

$n_c^\wedge=3$ in QCD and hence in canvassing superfluidity it is necessary
to choose a direction for the condensate in colour space; e.g.,
$\Delta^i\lambda^i_\wedge = \Delta\, \lambda^2$ in Eq.~(\ref{sinv}), so that
\begin{equation}
\label{sinvQCD}
\BD(p,\mu) = \left(
\begin{array}{c|c}
D_\|(p,\mu) P_\| + D_\perp(p,\mu) P_\perp 
        & \Delta(p,\mu) \gamma_5 \lambda^2 \\ \hline
- \Delta(p,-\mu) \gamma_5 \lambda^2 \mbox{\rule{0mm}{1.0em}}
        & C D_\|(-p,\mu)C^\dagger P_\| + C D_\perp(p,\mu)C^\dagger P_\perp 
\end{array}\right)\,,
\end{equation}
where $P_\|=(\lambda^2)^2$, $P_\perp + P_\| = {\rm diag}({1,1,1})$, and
$D_\|$, $D_\perp$ are defined via obvious generalisations of
Eqs.~(\ref{sinv}), (\ref{DCA}).  (NB.  It is this selection of a direction in
colour space that opens the possibility for colour-flavour locked diquark
condensation in a theory with three effectively-massless quarks; i.e.,
current-quark masses $\ll \mu$~\cite{krishna1}.)  In Eq.~(\ref{sinvQCD}) the
evident, demarcated block structure makes explicit the bispinor index.  Here
each block is a $3\times 3$ colour matrix and the subscripts: $\|$, $\perp$,
indicate whether or not the subspace is accessible via $\lambda_2$.  The
bispinors associated with this representation are given in Eqs.~(\ref{QQCD})
and in this case the Lagrangian's quark-gluon interaction term is
\begin{eqnarray}
\label{bispingamma}
\bar Q(x) i g\Gamma_\mu^a Q(x) A_\mu^a(x)\,,\;\; \Gamma_\mu^a & = &
\left(\begin{array}{c|c} \sfrac{1}{2}\gamma_\mu \lambda^a & 0
\\[0.2\parindent]\hline 0 \mbox{\rule{0mm}{1.0em}} &
-\sfrac{1}{2}\gamma_\mu (\lambda^a)^{\rm t}\end{array}\right)\,.
\end{eqnarray}

It is straightforward to derive the gap equation at arbitrary order in the
truncation scheme of Ref.~\cite{truncscheme} and it is important to note that
because
\begin{equation}
D_\|(p,\mu) P_\| + D_\perp(p,\mu) P_\perp = \lambda^0 \left\{\sfrac{2}{3}
D_\|(p,\mu) + \sfrac{1}{3} D_\perp(p,\mu) \right\} +
\sfrac{1}{\sqrt{3}}\lambda^8 \left\{ D_\|(p,\mu) - D_\perp(p,\mu)\right\}
\end{equation}
the interaction: $\Gamma_\mu^a {\cal S}(p,\mu)\Gamma_\nu^a$, necessarily
couples the $\|$- and $\perp$-components.  That interplay is discarded in
models that ignore the vector self energy of quarks, which, as we have
repeatedly seen, is a qualitatively important feature of
QCD~\cite{entireCJB,bastiscm,kisslinger,thoma}.

\begin{figure}[t]
\centering{\ \epsfig{figure=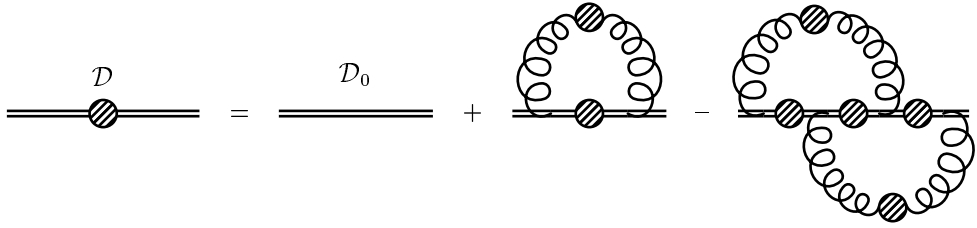,height=3.0cm}}\vspace*{0.5\baselineskip}

\parbox{40em}{\caption{Illustration of the dressed-ladder vertex-corrected
gap equation, which is the next-to-leading-order in the truncation scheme of
Ref.~\protect\cite{truncscheme}.  [The ladder truncation is obtained by
dropping the last diagram.]  Each bispinor quark-gluon vertex is bare, given
by Eq.~(\protect\ref{bispingamma}), but the shaded circles mark quark and
gluon $2$-point functions that are dressed.  The corresponding truncation in
the relevant Bethe-Salpeter equations ensures the absence of diquark bound
states in the strong interaction spectrum.  (Adapted from
Ref.~\protect\cite{jacquesdq}.)
\label{dressed}}}
\end{figure}

Reference~\cite{jacquesdq} explored the possibility of diquark condensation
in QCD via the gap equation using the model defined by ${\cal D}_{\rm A}$ in
both the rainbow and a vertex-corrected truncation.  The latter is
illustrated in Fig.~\ref{dressed}.  For $\mu=0$ the rainbow-ladder truncation
yields
\begin{equation}
\label{RLresults}
\begin{array}{ccc}
m_\omega^2 = m_\rho^2  =  \sfrac{1}{2}\,\eta^2\,, &
\langle \bar q q \rangle^0  = (0.11\,\eta)^3\,,&
{\cal B}_B(\mu=0)  =  (0.10\,\eta)^4\,,
\end{array}
\end{equation}
and momentum-dependent vector self energies, Eq.~(\ref{ngsoln}), which lead
to an interaction between the $\|$- and $\perp$-components of $\BD$ that
blocks diquark condensation~\cite{bastitrento}.  This is in spite of the fact
that
$ \lambda^a \lambda^2 (-\lambda^a)^{\rm t} = \sfrac{1}{2}\lambda^a
\lambda^a, $
which entails that the rainbow-truncation quark-quark scattering kernel is
purely attractive and strong enough to produce diquark bound
states~\cite{justindq}.  [Remember that in the colour singlet meson channel
the rainbow-ladder truncation gives the colour coupling $\lambda^a
\lambda^a$.]  For $\mu\neq 0$ and in the {\it absence} of diquark
condensation, we saw in connection with Fig.~(\ref{bagpresMN}) that the model
exhibits coincident, first order chiral symmetry restoring and deconfining
transitions at
\begin{equation}
\mu_{c,\,{\rm rainbow}}^{B,\Delta=0} = 0.28 \,\eta = 0.3\,{\rm GeV}\,.
\end{equation}

For $\mu\neq 0$, however, the rainbow-truncation gap equation admits a
solution with $\Delta(p,\mu)\not\equiv 0$ and $B(p,\mu)\equiv 0$.  $\delta
p(\mu)$ in Eq.~(\ref{deltaP}) again determines whether the stable ground
state is the quark-condensed or superfluid phase.  With increasing $\mu$,
${\cal B}_B(\mu)$ decreases, very slowly at first, and ${\cal
B}_{\Delta}(\mu)$ increases rapidly from zero.  That evolution continues
until
\begin{equation}
\mu_{c,\,{\rm rainbow}}^{B=0,\Delta} = 0.25 \,\eta = 0.89\,\mu_{c,\,{\rm
rainbow}}^{B,\Delta=0}\,,
\end{equation}
where ${\cal B}_\Delta(\mu)$ becomes greater-than ${\cal B}_B(\mu)$.  This
signals a first order transition to the superfluid ground state and at the
boundary
\begin{equation}
\label{qqqbq}
\langle \bar Q i\gamma_5\tau_Q^2\lambda^2 Q\rangle_{\mu=\mu_{c,\,{\rm
rainbow}}^{B=0,\Delta}} 
= (0.65)^3\,\langle \bar Q Q\rangle_{\mu=0}\,.
\end{equation}
The chemical potential at which the switch to the superfluid ground state
occurs is consistent with other estimates made using models comparable to the
rainbow-truncation
class~\cite{blaschkebiel,diakonov2,bergesbiel,schaeferbiel}, as is the
magnitude of the gap at this
point~\cite{blaschkebiel,diakonov2,krishna1,shuryakbiel}.

A question that now arises is: how sensitive is this phenomenon to the nature
of the quark-quark interaction?  As we discussed in connection with
Eq.~(\ref{diquarkmasses}), the inhomogeneous ladder BSE exhibits
particle-like singularities in the $0^+$ diquark channels and such states do
not exist in the strong interaction spectrum.  Does diquark condensation
persist when a truncation of the gap equation is employed that does not
correspond to a BSE whose solutions exhibit diquark bound states?  The vertex
corrected gap equation depicted in Fig.~\ref{dressed} is just such a
truncation and it was also studied in Ref.~\cite{jacquesdq}.

In this case there is a $\Delta \not \equiv 0 $ solution even for $\mu=0$,
and using ${\cal D}_{\rm A}$
\begin{equation}
\begin{array}{ccc}
m_\rho^2  =  (1.1)\, \,m_\rho^{2\;{\rm ladder}}, &
\; \langle \bar Q Q\rangle  =  (1.0)^3\,\langle \bar Q Q\rangle^{\rm
rainbow} ,&
\; {\cal B}_B  =  (1.1)^4\, {\cal B}_B^{\,{\rm rainbow}}\,,
\end{array}
\end{equation}
where the rainbow-ladder results are given in Eqs.~(\ref{RLresults}), and
\begin{equation}
\langle \bar Q i\gamma_5\tau_Q^2\lambda^2 Q\rangle = (0.48)^3\,\langle \bar Q
 Q\rangle\,,\;\; {\cal B}_\Delta = (0.42)^4\,{\cal B}_B \,.
\end{equation}
The last result shows, unsurprisingly, that the quark-condensed phase is
favoured at $\mu=0$.  {\it Precluding} diquark condensation, the solution of
the vertex-corrected gap equation exhibits coincident, first order chiral
symmetry restoring and deconfinement transitions at
\begin{equation}
\mu_c^{B,\Delta=0} = 0.77\,\mu_{c,\,{\rm rainbow}}^{B,\Delta=0}\,.
\end{equation}

Admitting diquark condensation, however, the $\mu$-dependence of the bag
constants again shows there is a transition to the superfluid phase, here at
\begin{equation}
\mu_{c}^{B=0,\Delta} =  0.63\, \mu_{c}^{B,\Delta=0}\,,\;{\rm with}\;
\langle \bar Q i\gamma_5\tau_Q^2\lambda^2 Q \rangle_{\mu =
0.63\,\mu_{c}^{B,\Delta=0}} = (0.51)^3\,\langle \bar Q Q\rangle_{\mu=0}\,.
\end{equation}
Thus the material step of eliminating diquark bound states leads only to
small quantitative changes in the quantities characterising the still extant
superfluid phase.

Solving the inhomogeneous BSE for the $0^+$ diquark vertex in the
quark-condensed phase provides additional insight~\cite{jacquesdq}.  At
$\mu=0$ and zero total momentum: $P=0$, the additional [confining]
contributions to the quark-quark scattering kernel generate an enhancement in
the magnitude of the scalar functions in the Bethe-Salpeter amplitude.
However, as $P^2$ evolves into the timelike region, the contributions become
repulsive and block the formation of a diquark bound state.  Conversely,
increasing $\mu$ at any given timelike-$P^2$ yields an enhancement in the
magnitude of the scalar functions, and as $\mu\to \mu_{c}^{B,\Delta=0}$ that
enhancement becomes large, which suggests the onset of an instability in the
quark-condensed vacuum.  This ``robustness'' of scalar diquark condensation
is consistent with the observations in Ref.~\cite{jackson1}.  However, the
studies described herein do not obviate the question of whether the diquark
condensed phase is stable with-respect-to dinucleon
condensation~\cite{birse}, which requires further attention.  [NB.  As
remarked above, the inclusion of temperature undermines a putative diquark
condensate and existing studies~\cite{blaschkebiel,bergesbiel} suggest that
it will disappear for $T \gsim 60$--$100\,$MeV.  However, such temperatures
are high relative to that anticipated inside dense astrophysical objects,
which may therefore provide an environment for detecting quark matter
superfluidity.]

Finally, this discussion illustrates that, in some respects; such as the
transition point and magnitude of the gap, the phase diagram of QC$_2$D is
quantitatively similar to that of QCD.  That is a useful observation because
the simplest superfluid order parameter is gauge invariant in QC$_2$D, and
the fermion determinant is real and positive, which makes tractable the
exploration of superfluidity in QC$_2$D using numerical simulations of the
lattice theory~\cite{hands}.  Hence, the results of those studies may provide
an additional, reliable guide to the nature of quark matter superfluidity.

\sect{Density, Temperature and Hadrons}
\subsect{Masses and Widths}
\addcontentsline{toc}{subsubsection}{\protect\numberline{ } {Temperature}}
\subsubsect{Temperature}
Hitherto we have canvassed the bulk thermodynamic properties of QCD at
nonzero $(\mu,T)$.  However, the terrestrial formation of a QGP will be
signalled by changes in the observable properties of those colour singlet
mesons that reach detectors.  In this connection, a primary feature of the
QGP is chiral symmetry restoration and, since the properties of the pion
(mass, decay constant, other vertex residues, etc.) are tied to the dynamical
breaking of chiral symmetry, an elucidation of the $T$-dependence of these
properties is important; particularly since a prodigious number of pions is
produced in heavy ion collisions.  Also important is understanding the
$T$-dependence of the properties of the scalar analogues and chiral partners
of the pion in the strong interaction spectrum.  For example, should the mass
of a putative light isoscalar-scalar
meson~\cite{mikescalars1,mikescalars2,mikescalars3,jacquesscalar} fall below
$2\,m_\pi$, the strong decay into a two pion final state can no longer
provide its dominant decay mode.  In this case electroweak processes will be
the only open decay channels below $T_c$ and the state will appear as a
narrow resonance~\cite{stephanov2}.  Analogous statements are true of
isovector-scalar mesons.

The effective interaction denoted above as ${\cal D}_{\rm C}$ in
Eq.~(\ref{delta}), with $\omega=1.2\,m_t$, has been employed~\cite{newsigmaT}
in an exploration of the $T$-dependence of scalar and pseudoscalar meson
properties.  As we saw in Sec.~\ref{sectthree}, mesons appear as simple poles
in $3$-point vertices {\it and} these vertices alone provide information
about the persistence of correlations away from the bound state pole, which
can be useful in studying the $T$-evolution of a system with deconfinement.
For $T\neq 0$ and two flavours, the ladder-truncation of the inhomogeneous
BSE for the isovector $0^{-+}$ vertex is
\begin{equation}
\label{psvtx}
\Gamma_{\rm ps}^i(p_{\omega_k};P_0;\zeta)  =  Z_4
\,\sfrac{1}{2}\tau^i\gamma_5
- \int_{l,q}^{\bar\Lambda} \,\sfrac{4}{3}\, g^2
D_{\mu\nu}(p_{\omega_k}-q_{\omega_l})
\gamma_\mu S(q_{\omega_l}^+)\,\Gamma_{\rm ps}^i(q_{\omega_l};P_0;\zeta)\,
S(q_{\omega_l}^-)\,\gamma_\nu\,,
\end{equation}
where $q_{\omega_l}^\pm= q_{\omega_l}\pm P_0/2$, and with $P_0=(\vec{P},0)$
this is the equation for the zeroth Matsubara mode.  [This is an extension of
Eq.~(\ref{Gamma5T0}).  Remember,
$m_R(\zeta)\,\Gamma_{\rm ps}^i(p_{\omega_k};P_{\Omega_n};\zeta)$
is renormalisation point independent.]

The solution of Eq.~(\ref{psvtx}) has the form [hereafter the
$\zeta$-dependence is often implicit]
\begin{eqnarray}
\label{psvtxform}
\lefteqn{\Gamma_{\rm ps}^i(p_{\omega_k};\vec{P}) = }\\
&& \nonumber
\sfrac{1}{2}\tau^i\gamma_5\,\left[i E_{\rm ps} (p_{\omega_k};\vec{P})
+ \vec{\gamma}\cdot\vec{P} \,F_{\rm ps} (p_{\omega_k};\vec{P})
+\vec{\gamma}\cdot\vec{p}\,\vec{p}\cdot\vec{P}\,
        G^{\|}_{\rm ps} (p_{\omega_k};\vec{P})
+ \gamma_4\omega_k\,\,\vec{p}\cdot\vec{P}\,
G^\perp_{\rm ps} (p_{\omega_k};\vec{P})
\right]\,,
\end{eqnarray}
where the $T\neq 0$ analogues of the $\sigma_{\mu\nu}$-like contributions in
Eq.~(\ref{genpibsa}) are omitted because they play a negligible role at
$T=0$~\cite{mr97}.  The breaking of $O(4)$ symmetry is responsible for making
this $T\neq 0$ amplitude more complicated than its zero-temperature
counterpart.  For the higher Matsubara frequencies the form is still more
complicated, with three additional terms.  The scalar functions in
Eq.~(\ref{psvtxform}) exhibit a simple pole at $\vec{P}^2+m_\pi^2 = 0$ so
that
\begin{equation}
\label{IHVpspole}
\Gamma_{\rm ps}^i(p_{\omega_k};\vec{P}) = 
\frac{r_{\pi}(\zeta)}{\vec{P}^2 + m_\pi^2}\,
\Gamma_\pi^i(p_{\omega_k};\vec{P}) + {\rm regular,}
\end{equation}
where again {\it regular} means terms regular at {\it this} pole and
$\Gamma_\pi^i(p_{\omega_k};\vec{P})$ is the bound state pion Bethe-Salpeter
amplitude, canonically normalised via the obvious $T\neq 0$ extension of
Eq.~(\ref{pinorm}).

The pole position in Eq.~(\ref{IHVpspole}) determines the spatial
``screening-mass'' of the pion's zeroth Matsubara mode and its inverse
describes the persistence length of that mode at equilibrium in the heat
bath.  There is a screening mass and amplitude for each mode, and each mode's
amplitude is canonically normalised.  The full $T\neq 0$ bound state
propagator can be calculated via any polarisation tensor that receives a
contribution from the bound state but only once all the screening masses have
been determined.  (For the pion: the pseudoscalar or pseudovector
polarisations will serve~\cite{mrt98}.)  The propagator so obtained is
defined only on a discrete set of points along what might be called the
imaginary-energy axis and the ``pole-mass;'' i.e., the mass that yields the
bound state energy pole for $\vec{p}\sim 0$, is obtained only after an
analytic continuation of the propagator onto the real-energy axis.  [The loss
of $O(4)$ invariance for $T\neq 0$ means that, in general, the pole mass and
screening masses are unequal.]  That continuation is not unique but an
unambiguous result is obtained by requiring that it yield a function that is
bounded at complex-infinity and analytic off the real axis~\cite{polemasses}.
From this description it is nonetheless clear that the screening masses
completely specify the properties of $T\neq 0$ bound states.  Furthermore,
the often used calculational expedient of replacing the meson Matsubara
frequencies by a continuous variable: $\Omega_n \to -i \nu$, and the
identification of the energy scale thus obtained with a pole mass, is seen to
be merely an artefice.  However, since this prescription yields the correct
result for free particle propagators, it {\it might} provide an illustrative
guide.

The residue in Eq.~(\ref{IHVpspole}) is 
\begin{equation}
\label{rpi}
\delta^{ij}\, ir_{\pi} = \,Z_4\,{\rm tr}\int_{l,q}^{\bar\Lambda}\,
\sfrac{1}{2}\tau^i\,\gamma_5 \chi_\pi^j(q_{\omega_l};\vec{P})\,,
\end{equation}
where $\chi_\pi(q_{\omega_l};\vec{P}):= S(q_{\omega_l}^+)
\Gamma_\pi(q_{\omega_l};\vec{P}) S(q_{\omega_l}^-)$ is the unamputated
Bethe-Salpeter wave function.  [Ref.~\cite{newsigmaT} employs the $f_\pi =
92\,$MeV normalisation, cf. Eq.~(\ref{rHres})] Substituting
Eq.~(\ref{IHVpspole}) into Eq.~(\ref{psvtx}) and equating pole residues
yields the homogeneous pion BSE, which provides the simplest way to obtain
the bound state amplitude.  As already noted, the pion also appears as a pole
in the axial-vector vertex and there the residue is the leptonic decay
constant
\begin{equation}
\label{fpi}
\delta^{ij}\, \vec{P}\,f_\pi  = 
Z_2^A\,{\rm tr}\int_{l,q}^{\bar\Lambda}\,
\sfrac{1}{2}\tau^i\,\gamma_5 \vec{\gamma}\,
\chi_\pi^j(q_{\omega_l};\vec{P})\,.
\end{equation}
At $T=0$ in this ${\cal D}_{\rm C}$ model the calculated chiral limit values
are [remember: $f_\pi=92\,$MeV normalisation]
\begin{equation}
\begin{array}{ccc}
f_\pi^0 = 0.088\,{\rm GeV}\,, &
\; -\langle \bar q q \rangle^0_{1\,{\rm GeV}^2} = (0.235\,{\rm GeV})^3\,, 
& 
\; r_\pi^0(1\, {\rm GeV}^2)=(0.457\,{\rm GeV})^2\,.
\end{array}
\end{equation}

The analogue of Eq.~(\ref{psvtx}) for the $0^{++}$ vertex is 
\begin{equation}
\label{scavtx}
\Gamma_{\rm s}^\alpha(p_{\omega_k};P_0;\zeta)  =  Z_4
\,\sfrac{1}{2}\tau^\alpha\mbox{\large\boldmath $1$}
- \int_{l,q}^{\bar\Lambda} \,\sfrac{4}{3}\, g^2
D_{\mu\nu}(p_{\omega_k}-q_{\omega_l})
\gamma_\mu S(q_{\omega_l}^+)\,\Gamma_{\rm
s}^i(q_{\omega_l};P_{\Omega_n};\zeta)\, S(q_{\omega_l}^-)\,\gamma_\nu\,,
\end{equation}
where $\alpha= 0,1,2,3$.  However, as discussed in connection with
Eq.~(\ref{compile}), the combination of rainbow and ladder truncations is not
certain to provide a reliable approximation in the scalar sector.
Nevertheless, in the absence of an improved, phenomenologically efficacious
kernel, Ref.~\cite{newsigmaT} employed Eq.~(\ref{scavtx}) in the expectation
that it would provide some qualitatively reliable insight, an approach
justified {\it a posteriori}.

The solution of Eq.~(\ref{scavtx}) has the form 
\begin{eqnarray}
\label{scvtxform}
\lefteqn{\Gamma_{\rm s}^i(p_{\omega_k};\vec{P}) = }\\
&& \nonumber
\sfrac{1}{2}\tau^\alpha\,\mbox{\large\boldmath $1$}\,
\left[ E_{\rm s} (p_{\omega_k};\vec{P})
+ i \vec{\gamma}\cdot\vec{p}\,G^{\|}_{\rm s} (p_{\omega_k};\vec{P})
+ i \gamma_4\omega_k\,G^\perp_{\rm s} (p_{\omega_k};\vec{P})
+ i \vec{\gamma}\cdot\vec{P}\,\vec{p}\cdot\vec{P}\,
F_{\rm s} (p_{\omega_k};\vec{P})
\right]\,,
\end{eqnarray}
where here the requirement that the neutral mesons be charge conjugation
eigenstates shifts the $\vec{p}\cdot\vec{P}$ term cf. the $0^{-\,+}$
amplitude in Eq.~(\ref{psvtxform}).  As already observed in connection with
Eq.~(\ref{scalarvtx}), the scalar functions in Eq.~(\ref{scvtxform}) exhibit
a simple pole at $\vec{P}^2+m_{\sigma}^2 = 0$ with residue
\begin{equation}
\label{rsc}
\delta^{\alpha\beta}\, r_{\sigma} = Z_4\,{\rm tr}\int_{l,q}^{\bar\Lambda}\,
\sfrac{1}{2}\tau^\alpha\,\chi_{\sigma}^\beta(q_{\omega_l};\vec{P})\,.
\end{equation}
However, since a $V-A$ current cannot connect a $0^{++}$ state to the vacuum,
the scalar meson does not appear as a pole in the vector vertex; i.e,
\begin{equation}
 \delta^{\alpha\beta}\, \vec{P}\,f_{\sigma} = Z_2^A\,{\rm
tr}\int_{l,q}^{\bar\Lambda}\, \sfrac{1}{2}\tau^\alpha\, \vec{\gamma}\,
\chi_\sigma^\beta(q_{\omega_l};\vec{P})\equiv 0\,.
\end{equation}
The homogeneous equation for the scalar bound state amplitude is obtained, as
usual, from Eq.~(\ref{scavtx}) by equating pole residues.

\begin{figure}[t] 
\centering{\
\epsfig{figure=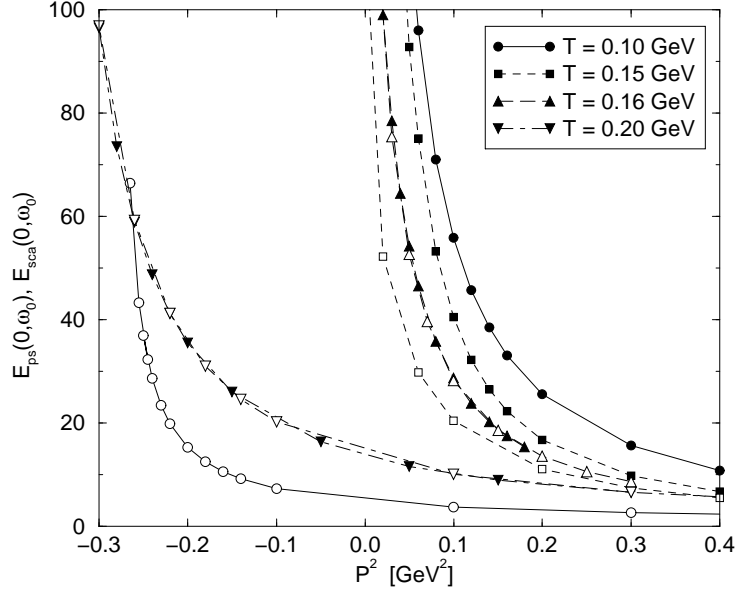,height=8.0cm}}\vspace*{0.5\baselineskip} 

\parbox{40em}{\caption{Leading Dirac amplitude for the pseudoscalar (shaded
symbols) and scalar (open symbols) vertices [$E$ in
Eqs.~(\protect\ref{psvtxform}), (\protect\ref{scvtxform})] evaluated at
\mbox{$(\vec{p}=0,\omega_0)$} and plotted as function of $\vec{P}^2$ in the
chiral limit; i.e., $E(\vec{p}=0,\omega_0;\vec{P}^2)$, for a range of
temperature values.  The bound state poles are evident in each case.
(Adapted from Ref.~\protect\cite{newsigmaT}.)
\label{ihbse}}}
\end{figure}

The $T$-dependence of the pole positions in the solution of the inhomogeneous
BSEs is illustrated in Fig.~\ref{ihbse}, from which it is evident that: 1) at
the critical temperature, $T_c=152\,$MeV in Table~\ref{critthings}, one has
degenerate, massless pseudoscalar and scalar bound states; and 2) the bound
states persist above $T_c$, becoming increasingly massive with increasing
$T$.  These features are also observed in numerical simulations of
lattice-QCD~\cite{edwinechaya}.  The bound state amplitudes are obtained from
the homogeneous BSEs.  Above $T_c$, all but the leading Dirac amplitudes:
$E_\pi$, $E_\sigma$, vanish and the surviving amplitudes are pointwise
identical.  These results indicate that the chiral partners are {\it locally}
identical above $T_c$, they do not just have the same mass.

The results are easily understood algebraically.  The BSE is a set of coupled
homogeneous equations for the Dirac amplitudes.  Below $T_c$ each of the
equations for the subleading Dirac amplitudes has an ``inhomogeneity'' whose
magnitude is determined by $B_0$, the scalar piece of the quark self energy,
which is dynamically generated in the chiral limit.  $B_0$ vanishes above
$T_c$ eliminating the inhomogeneity and allowing a trivial, identically zero
solution for each of these amplitudes.  Additionally, with $B_0\equiv 0$ the
kernels in the equations for the dominant pseudoscalar and scalar amplitudes
are identical, and hence so are the solutions.  It follows from these
observations that the Goldberger-Treiman-like relation~\cite{mrt98}
\begin{equation}
\label{gtlrel}
f_\pi^0 E_\pi(p_{\omega_k};0) = B_0(p_{\omega_k})
\end{equation}
is satisfied for all $T$ only because both $f_\pi^0$ and $B_0(p_{\omega_k})$
are equivalent order parameters for chiral symmetry restoration.  

\begin{figure}[t]
\epsfig{figure=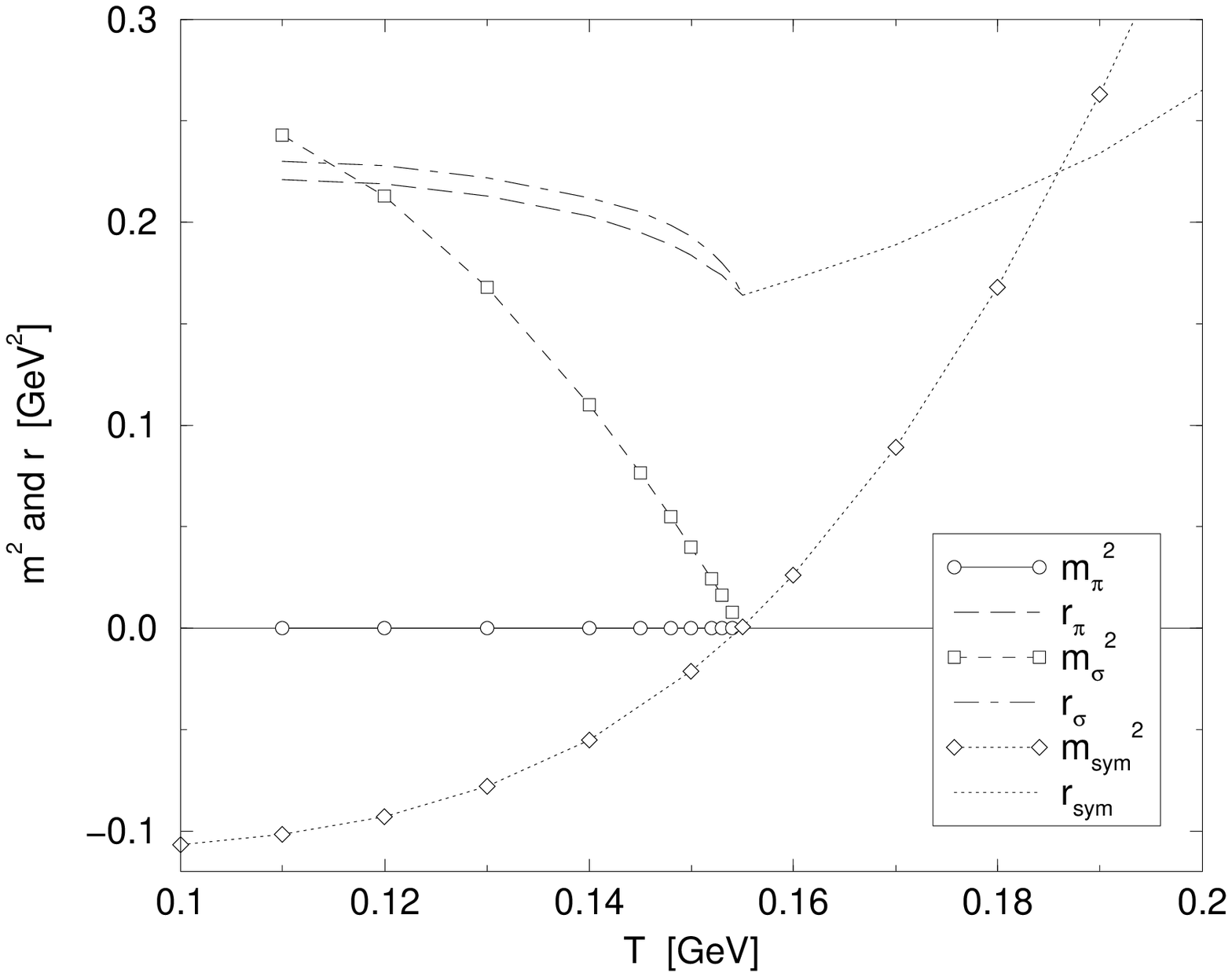,height=17.5em}\hspace*{\fill}

\vspace*{-18.8em}

\hspace*{\fill}\epsfig{figure=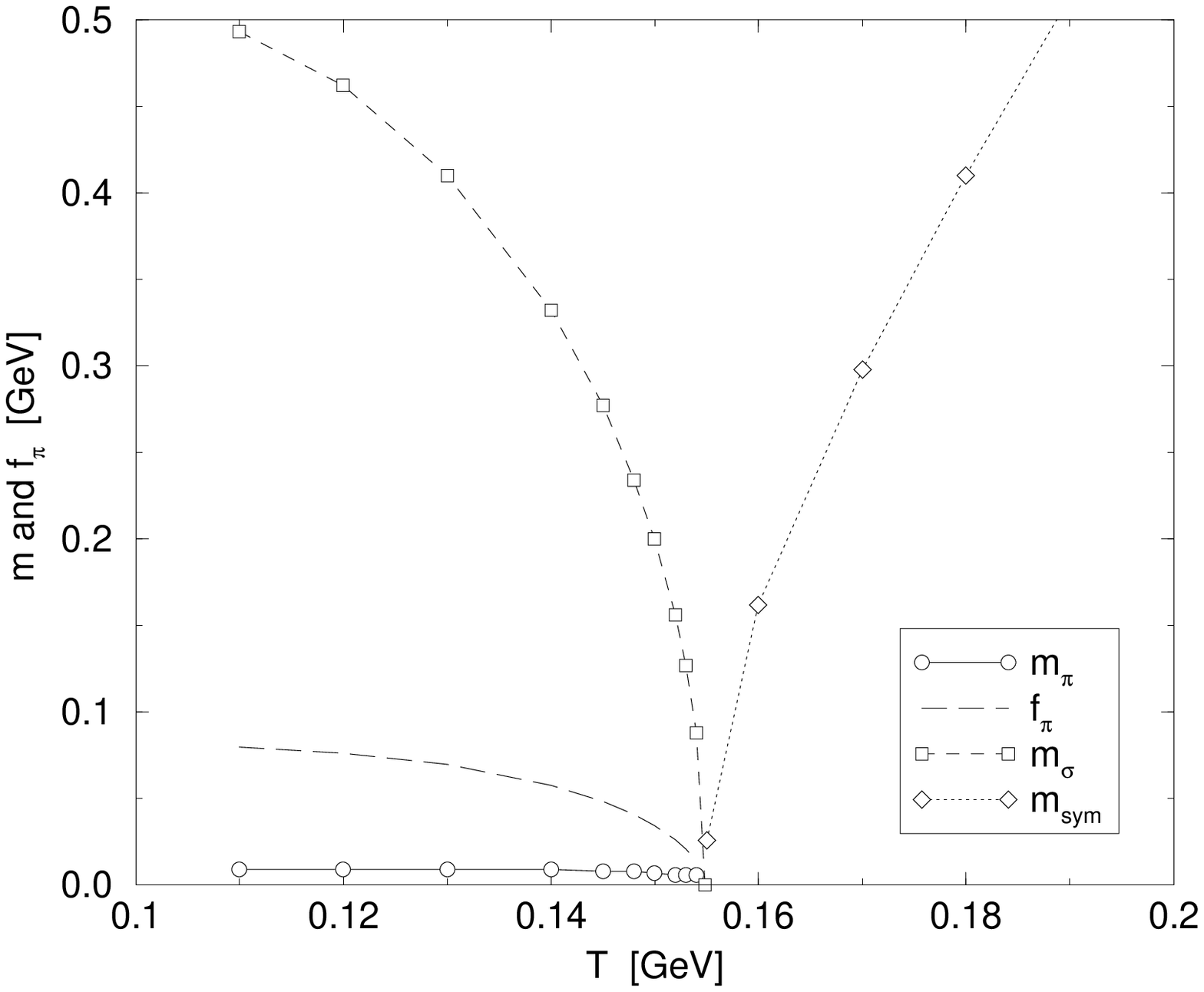,height=17.5em} \vspace*{-1em}

\begin{center}
\parbox{40em}{\caption{\label{obs0}$\hat m = 0$ results.  {\it Left Panel}:
$T$-dependence of the meson masses-squared and pole-residue matrix elements,
Eqs.~(\protect\ref{rpi}), (\protect\ref{rsc}), along with the mass-squared:
$m_{\rm sym}^2$, and residue: $r_{\rm sym}$, calculated in the chirally
symmetric $B_0\equiv 0$ phase.  For $T\leq T_c$, $m_{\rm sym}^2<0$, which is
a signal of the instability of the chirally symmetric phase at low $T$.  For
$T>T_c$, $m_\pi^2=m_\sigma^2=m_{\rm sym}^2$.  {\it Right Panel}:
$T$-dependence of the masses and pion decay constant,
Eq.~(\protect\ref{fpi}).  $m_\pi=0$ within numerical error.  (Adapted
from Ref.~\protect\cite{newsigmaT}).}}\vspace*{-1em}
\end{center}
\end{figure}

Using the calculated bound state amplitudes and dressed-quark propagators,
the $T$-dependence of the matrix elements in Eqs.~(\ref{rpi}), (\ref{fpi}),
(\ref{rsc}) follows.  It is depicted in Fig.~\ref{obs0} for the chiral limit.
Below $T_c$ the scalar meson residue in the scalar vertex, $r_\sigma$ in
Eq.~(\ref{rsc}), is a little larger than the residue of the pseudoscalar
meson in the pseudoscalar vertex, $r_\pi$ in Eq.~(\ref{rpi}).  However, they
are nonzero and equal above $T_c$, which is an algebraic consequence of $B_0
\equiv 0$ and the vanishing of the subleading Dirac amplitudes.
As a {\it bona fide} order parameter for chiral symmetry restoration
\begin{equation}
f_\pi^0 \propto \sqrt{-t}\,,\; t= (T/T_c - 1) \lsim 0\,,
\end{equation}
as anticipated from the discussion accompanying Table~\ref{critthings}.  The
same result was obtained in Ref.~\cite{reinhardCoulomb}, where it was
observed too that
\begin{equation}
\frac{1}{ (r_\pi^0)_{\rm em}} \propto f_\pi^0\,;
\end{equation}
i.e., the pion charge radius diverges at $T_c$ in the chiral limit.  This is
plausible but Ref.~\cite{newsigmaT} did not attempt its verification.
Reference~\cite{newsigmaT} did confirm the behaviour of the ratio of pole
residues observed in Ref.~\cite{arne2}:
\begin{equation}
\frac{r_\pi^0(\zeta)}{f_\pi^0} \propto \frac{1}{\sqrt{-t}}\,,\; t \lsim 0\,.
\end{equation}
The concomitant results: $f_\pi^0 = 0$ and $r_\pi^0 \neq 0$ for $t>0$,
demonstrate that the pion disappears as a pole in the axial-vector
vertex~\cite{axelT} but persists as a pole in the pseudoscalar vertex.
Evident also in Fig.~\ref{obs0} is that
\begin{equation}
m_\sigma^0 \propto \sqrt{-t}\,,\;\; t \lsim 0:
\end{equation}
$m_\sigma^2$ follows a linear trajectory for $t\lsim 0$.  Such behaviour in
the isoscalar-scalar channel might be anticipated because this channel has
vacuum quantum numbers and hence the bound state is a strong interaction
analogue of the electroweak Higgs boson~\cite{mikescalars3}.  

\begin{figure}[t]
\centering{\
\epsfig{figure=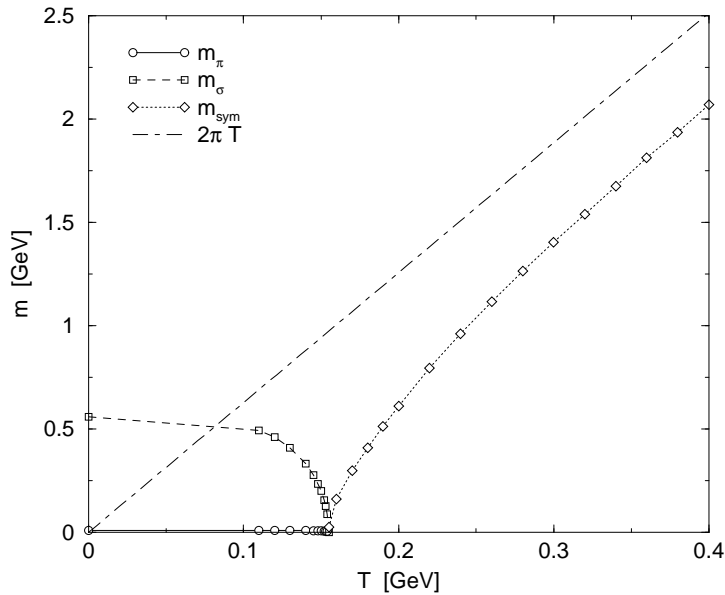,height=8.0cm}}

\parbox{40em}{\caption{\label{screen}$T$-dependence of the meson masses for
large $T$ with $\hat m=0$, see Fig.~\protect\ref{obs0}.  $m_\pi = m_\sigma$
for $T>T_c$ and $m/(2\pi T) \to 1^-$.  This behaviour persists with $\hat m
\neq 0$, as illustrated in Refs.~\protect\cite{pctTrento,peterNew}, and
observed in lattice-QCD simulations~\protect\cite{edwinechaya}.)  (Adapted
from Ref.~\protect\cite{newsigmaT}.)}}
\end{figure}

$m_{\rm sym}^2$ in Fig.~\ref{obs0} is the mass obtained when the chirally
symmetric solution of the quark DSE is used in the BSE.  [$B_0 \equiv 0$ is
always a solution in the chiral limit.]  For $t>0$, $m_{\rm sym}^2(T)$ is the
unique meson mass-squared trajectory.  However, for $t<0$, $m_{\rm sym}^2<0$;
i.e., the solution of the BSE in the Wigner-Weyl phase exhibits a tachyonic
solution.  [cf. The Nambu-Goldstone phase masses: $m_\sigma^2 > m_\pi^2 =
0$.]  By analogy with the $\sigma$-model, this tachyonic mass indicates the
instability of the Wigner-Weyl phase below $T_c$ and translates into the
statement that the pressure is not maximal in this phase.
Figure~\ref{screen} depicts the evolution of this [common] meson mass at
large $T$.  As expected in a gas of weakly interacting quarks and gluons
\begin{equation}
\label{screeneq}
\frac{m_{0^\pm{\rm meson}}}{2\, \omega_0} \to 1^-\,,
\end{equation}
where $\omega_0 = \pi T$ is a quark's zeroth Matsubara frequency and
``screening mass.''  Similar behaviour is observed for the $\rho$-meson mass
in Ref.~\cite{peterNew} and can be demonstrated algebraically using the
${\cal D}_{\rm A}$ model for the effective interaction~\cite{rhomuT}.

\begin{figure}[t]
\epsfig{figure=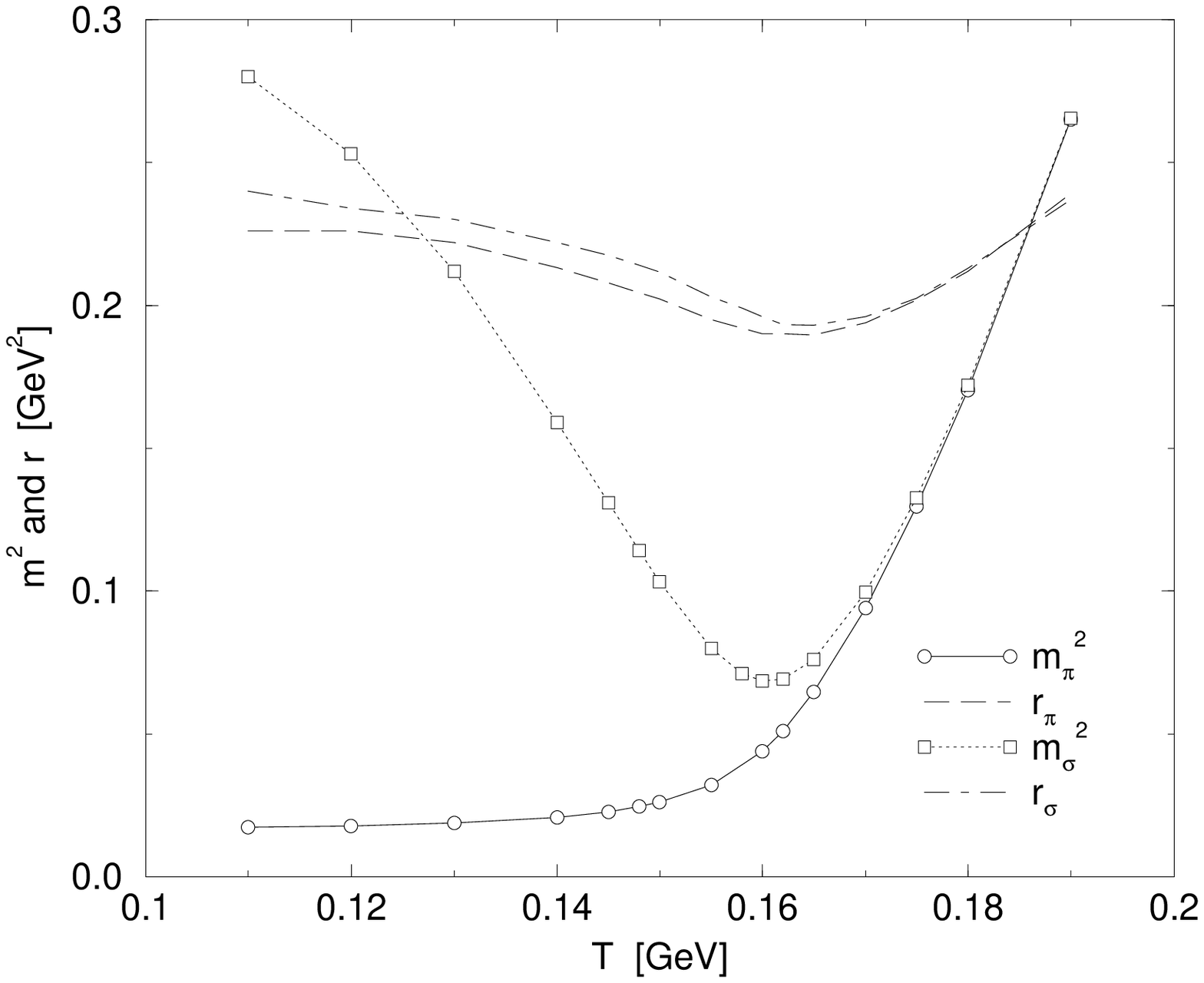,height=17.5em}\hspace*{\fill}

\vspace*{-18.7em}

\hspace*{\fill}\epsfig{figure=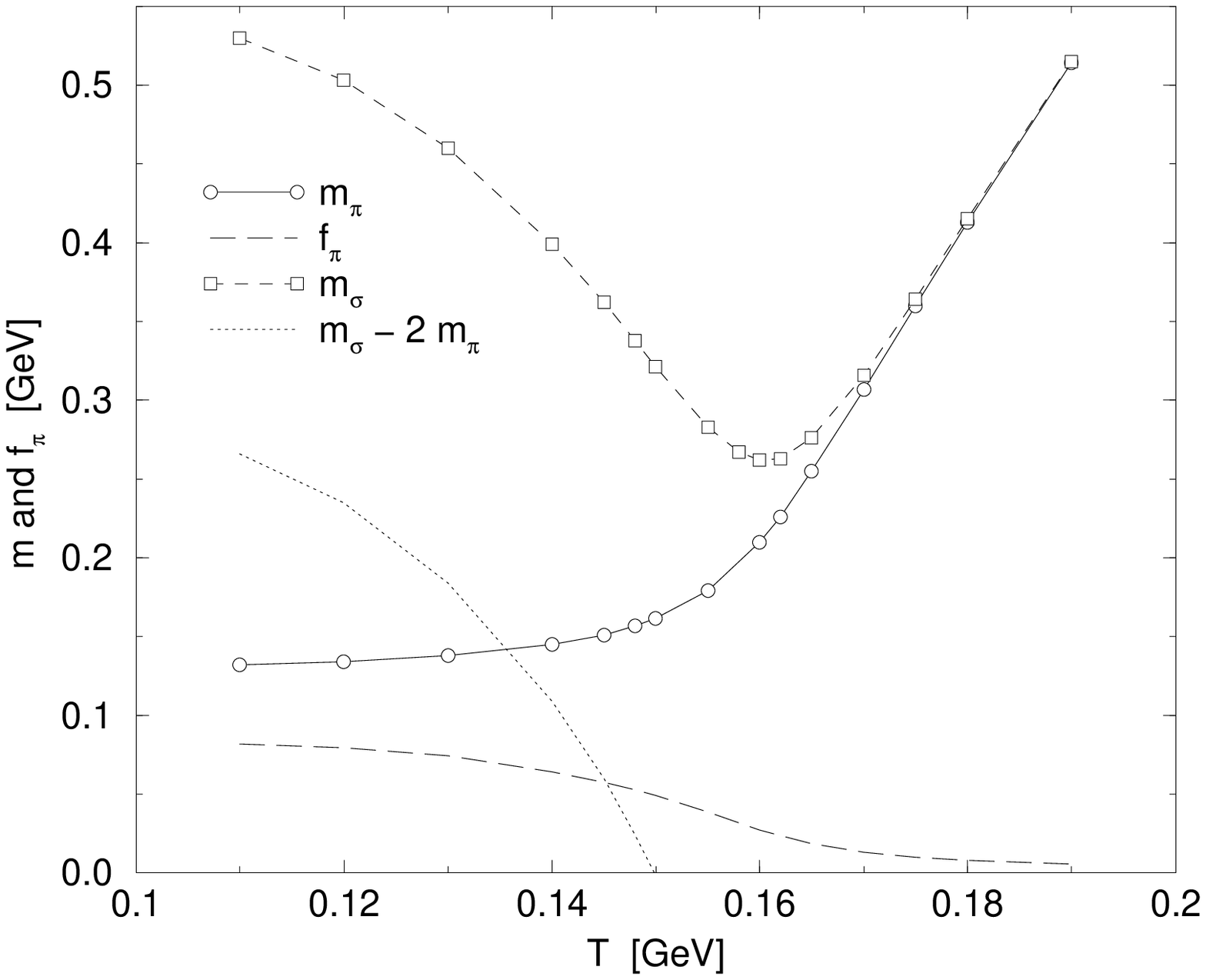,height=17.5em}\vspace*{-1em}

\begin{center}
\parbox{40em}{\caption{\label{obsm} $T$-dependence of the meson masses and
pole-residue matrix elements,
Eqs.~(\protect\ref{rpi},\protect\ref{fpi},\protect\ref{rsc}), for $\hat m
\neq 0$.  That the transition has become a crossover is evident in the
behaviour of $f_\pi$.  The meson masses become indistinguishable at $T\sim
1.2 \,T_c$, a little before the local equivalence is manifest, which is
unsurprising given that the mass is an integrated quantity.  The small
difference between $r_\sigma$ and $r_\pi$ below $T_c$ is again evident and
they assume a common value at the same temperature as the masses.  (Adapted
from Ref.~\protect\cite{newsigmaT}.)}}\vspace*{-1em}
\end{center}
\end{figure}

The results described hitherto were all calculated in the chiral limit.  The
extension to $\hat m \neq 0$ is straightforward although calculations with
the renormalisation group improved rainbow-ladder truncation become more time
consuming.  That is why simpler models, such as employed in
Refs.~\cite{pctTrento,peterNew}, can be useful.  For $\hat m \neq 0$, chiral
symmetry restoration with increasing $T$ is exhibited as a crossover rather
than a phase transition.  The solutions of the inhomogeneous BSEs again
exhibit a pole for all $T$, with the bound state amplitudes obtained from the
associated homogeneous equations.  In this case the scalar and pseudoscalar
bound states are locally identical for $T \gsim \sfrac{4}{3} T_c$.
Figure~\ref{obsm} is the $\hat m\neq 0$ analogue of Fig.~\ref{obs0}.  An
important result is that the axial-vector Ward-Takahashi identity is
satisfied, both above and below the chiral transition temperature, which was
demonstrated in Ref.~\cite{newsigmaT} via Eq.~(\ref{massform}): the two sides
remain equal $\forall\, T$.  The Gell-Mann--Oakes-Renner formula, however,
which involves $r_\pi^0(\zeta)$, fails for $t > - 0.1$, as observed too in
the separable model study of Ref.~\cite{yura2}.  

\begin{figure}[t]
\centering{\
\epsfig{figure=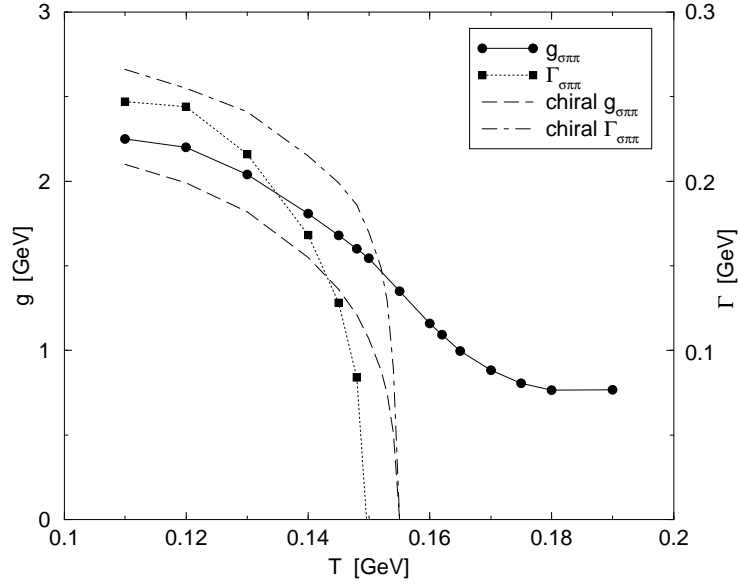,height=8.0cm}}

\parbox{40em}{\caption{\label{widtha}$T$-dependence of the
iso\-scalar-scalar-$\pi\pi$ coupling and width, both in the chiral limit and
for realistic light current-quark masses. The phase space factor
$(1-(2m_\pi/m_\sigma)^2)^{1/2}$ is $\theta(T_c-T)$ in the chiral limit but
nontrivial for $\hat m \neq 0$, vanishing at $T\approx 0.98 \,T_c$; i.e.,
this decay channel closes at a temperature just $2$\% less-than $T_c$.
(Adapted from Ref.~\protect\cite{newsigmaT}.)}}
\end{figure}

As one can anticipate from Sec.~\ref{sectthree}, the calculated $\sigma$ and
$\pi$ bound state amplitudes and dressed-quark propagator also make possible
a study of two-body decays.  For example, the impulse approximation to the
isoscalar-scalar-$\pi\pi$ coupling is described by the matrix element
\begin{eqnarray}
\label{sigpipi}
\lefteqn{g_{\sigma\pi\pi} := \langle \pi(\vec{p_1})\pi(\vec{p_2}) |
\sigma(\vec{p})\rangle} \\ && \nonumber = 2 N_c{\rm
tr}_D\int_{l,q}^{\bar\Lambda}\,
\Gamma_\sigma(k_{\omega_l};\vec{p})\,S_u(k_{++}) \,
i\Gamma_\pi(k_{0+};-\vec{p_1})\,
S_u(k_{+-})\,i\Gamma_\pi(k_{-0};-\vec{p_2})\,S_u(k_{--})\,,
\end{eqnarray}
$k_{\alpha\beta}= k_{\omega_l}+(\alpha/2)\vec{p_1}+(\beta/2)\vec{p_2}$, in
terms of which the width is
\begin{equation}
\label{sigwidth}
\Gamma_{\sigma\to(\pi\pi)}  = 
\sfrac{3}{2}\,g_{\sigma\pi\pi}^2\,\frac{\sqrt{1-4m_\pi^2/m_\sigma^2}}{16\,\pi\,
m_{\sigma}}\,.
\end{equation}
The coupling and width obtained from Eq.~(\ref{sigpipi}) are depicted in
Fig.~\ref{widtha}, which indicates that both vanish at $T_c$ in the chiral
limit.  Again this can be traced to $B_0 \to 0$.  For $\hat m \neq 0$, the
coupling reflects the crossover but that is not observable because the width
vanishes just below $T_c$ where the isoscalar-scalar meson mass falls below
$2 m_\pi$ and the phase space factor vanishes.  [See the right panel of
Fig.~\ref{obsm}.]  The evolution $m_\sigma \to 2 m_\pi$ may, however, be
observable via an enhancement in the $\pi\pi\to\gamma\gamma$
cross-section~\cite{volkov}.

Additionally, the particular properties of the $\pi^0\to\gamma\gamma$ decay,
which is mediated by the ``triangle anomaly,'' [see the paragraph after
Eq.~(\ref{FpiUV}) on page~\pageref{`anomalous processes'}] make interesting
the behaviour of this process at $T\neq 0$.  At $T=0$, the anomalous
contribution to the divergence of the axial-vector vertex is saturated by the
pseudoscalar piece, $E_\pi$, of the pion Bethe-Salpeter
amplitude~\cite{mrpion}
\begin{equation}
\label{pi0gg}
\hat T_{\mu\nu}(k_1,k_2) = {\rm tr}\int_{l,q}^{\bar\Lambda}\,
S(q_1)\,\gamma_5 \tau^3 i E_\pi(\hat q;-P)\,
S(q_2)\,iQ_e\Gamma_\mu(q_2,q_{12})\, S(q_{12})\,i
Q_e\Gamma_\nu(q_{12},q_1)\,,
\end{equation}
where here, just to be specific, $k_1=(\vec{k}_1,0)$, $k_2=(\vec{k}_2,0)$,
$P=k_1+k_2$, $q_1= q_{\omega_l}-k_1$, $q_2= q_{\omega_l}+k_2$, $\hat
q=\sfrac{1}{2}(q_1+q_2)$, $q_{12}= q_{\omega_l} - k_1 + k_2$.
Equation~(\ref{pi0gg}) involves the dressed-quark-photon vertex:
$\Gamma_\mu$, which also appeared in the calculation of $F_\pi(Q^2)$,
described in Sec.~\ref{sectthree}.  As we saw, quantitatively reliable
numerical solutions of the $T=0$ vector vertex equation are now
available~\cite{mtpion}.  However, this anomalous coupling is insensitive to
details and an accurate result requires only that the dressed vertex satisfy
the vector Ward-Takahashi identity.  For $T=0$ with real photons,
Eq.~(\ref{pi0gg}) is expressed in terms of one scalar function:
\begin{equation}
\label{tmunu}
\hat T_{\mu\nu}(k_1,k_2) = \frac{\alpha_{\rm em}}{\pi}\,
\epsilon_{\mu\nu\rho\sigma}\,k_{1\rho}\,k_{2\sigma}\,{\cal T}(0)\,.
\end{equation}
Now, as long as $\Gamma_\mu$ satisfies the Ward-Takahashi identity,
Eq.~(\ref{vwti}), one finds algebraically in the chiral limit
\begin{equation}
 f_\pi^0 \,{\cal T}(0) := g_{\pi^0\gamma\gamma}=1/2\,,
\end{equation}
completely independent of model details~\cite{mrpion,cdrpion,bando}, as
required since at $T=0$ the anomalies are a feature of the global aspects of
DCSB~\cite{witten}.  This value reproduces the experimental width.
[Remember, the normalisation in this subsection yields $f_\pi=92\,$MeV.]

The $T\neq 0$ calculation requires only a valid extension of
Eq.~(\ref{bcvtx}) and one such is
\begin{eqnarray}
i \vec{\Gamma}(q_{\omega_{l_1}},q_{\omega_{l_2}}) & = & 
 \Sigma_A(q_{\omega_{l_1}}^2,q_{\omega_{l_2}}^2)\,i\vec{\gamma} + 
(\vec{q}_1+\vec{q}_2)\,
[\sfrac{1}{2}\,i G(q_{\omega_{l_1}},q_{\omega_{l_2}})
        + \Delta_B(q_{\omega_{l_1}}^2,q_{\omega_{l_2}}^2)],\\
i\Gamma_4(q_{\omega_{l_1}},q_{\omega_{l_2}})& = &  
\Sigma_C(q_{\omega_{l_1}}^2,q_{\omega_{l_2}}^2)\,i\gamma_4 +
(\omega_{l_1}+\omega_{l_2})\, [\sfrac{1}{2}\,i
G(q_{\omega_{l_1}},q_{\omega_{l_2}}) +
\Delta_B(q_{\omega_{l_1}}^2,q_{\omega_{l_2}}^2)],\\
G(q_{\omega_{l_1}},q_{\omega_{l_2}}) & = &  \vec{\gamma}\cdot(\vec{q}_1 +
\vec{q}_2) \Delta_A(q_{\omega_{l_1}}^2,q_{\omega_{l_2}}^2)  + \gamma_4
(\omega_{l_1} + \omega_{l_2})
\Delta_C(q_{\omega_{l_1}}^2,q_{\omega_{l_2}}^2).
\end{eqnarray}
It is a particular case of the {\it Ansatz} in Ref.~\cite{cabo} and satisfies
the $T\neq 0$ vector Ward-Takahashi identity
\begin{equation}
(q_{\omega_{l_1}}-q_{\omega_{l_2}})_\mu\,
i\Gamma_\mu(q_{\omega_{l_1}},q_{\omega_{l_2}})
= S^{-1}(q_{\omega_{l_1}}) - S^{-1}(q_{\omega_{l_2}}).
\end{equation}
For $T\neq 0$ the tensor structure of Eq.~(\ref{tmunu}) survives to the
extent that, with $k_1$, $k_2$ as defined, it ensures one of the photons is
longitudinal (a plasmon) and the other transverse, with 
\begin{equation} 
\hat T_{i4}(k_1,k_2) = \frac{\alpha_{\rm em}}{\pi}\,
(\vec{k_1}\times\vec{k_2})_i\,{\cal T}(0)\,.
\label{anomT}
\end{equation}

\begin{figure}[t]
\centering{\ \epsfig{figure=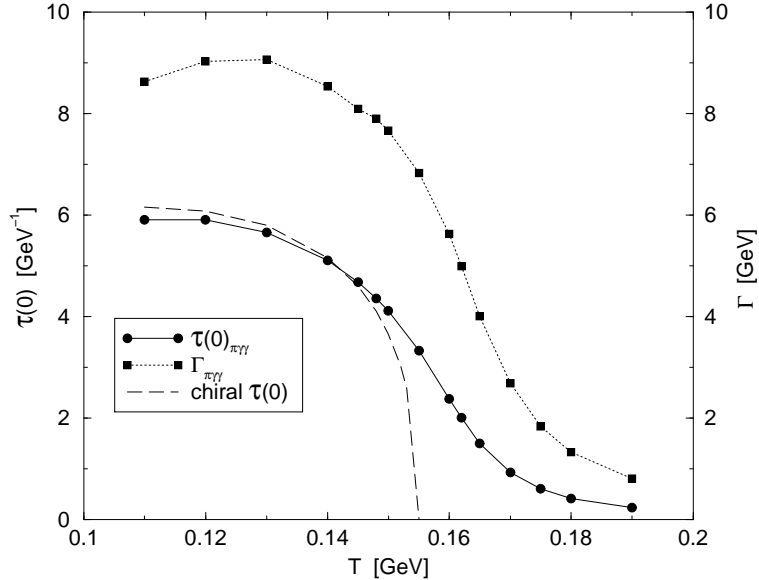,height=8.0cm}}

\parbox{40em}{\caption{\label{widthb} $T$-dependence of the coupling ${\cal
T}(0)$ in Eq.~(\protect\ref{anomT}) and the $\pi^0\to \gamma\gamma$ width:
$\Gamma_{\pi\to\gamma\gamma}= \alpha_{\rm em}^2 m_\pi^3 |{\cal T}(0)|^2/(16
\pi^3)$, which is identically zero in the chiral limit because $m_\pi=0$.
(Adapted from Ref.~\protect\cite{newsigmaT}.)}}
\end{figure}

The $T$-dependence of the anomalous coupling follows from that of ${\cal
T}(0)$, and it and the $T$-dependence of the width are depicted in
Fig.~\ref{widthb}.  In the chiral limit the interesting quantity is: ${\cal
T}(0)=g^0_{\pi^0\gamma\gamma}/f_\pi^0$, and obvious in the figure is that it
vanishes at $T_c$.  (It vanishes with a mean field critical
exponent~\cite{newsigmaT}.)  Thus, in the chiral limit, the coupling to the
dominant decay channel closes for both charged {\it and} neutral pions.
These features were anticipated in Ref.~\cite{pisarski1}.  Further, the
calculated ${\cal T}(0)$ is monotonically decreasing with $T$, supporting the
perturbative O($T^2/f_\pi^2)$ analysis in Ref.~\cite{pisarski2}.  For $\hat m
\neq 0$ both the coupling: $g_{\pi^0\gamma\gamma}/f_\pi$, and the width
exhibit the crossover, with a slight enhancement in the width as $T\to T_c$
due to the increase in $m_\pi$.  This is similar to the results of
Ref.~\cite{yura}, although the $T$-dependence depicted here is much weaker
because the pion mass approaches twice the $T \neq 0$ free-quark
screening-mass from below, {\it never} reaching it, Eq.~(\ref{screeneq});
i.e., the continuum threshold is not crossed.

The $T$-dependence of meson properties illustrated here is robust: it agrees
with the results obtained in lattice simulations when there is an overlap,
and with the results of other models.  The {\it local} equivalence exhibited
by isovector chiral partners above $T_c$ might be expected as a general
feature of the bound state spectrum in the Wigner-Weyl phase.  However, an
explicit demonstration is numerically challenging; e.g., in the $\rho$-$a_1$
complex the bound state amplitudes have eight independent functions even at
$T=0$, compared with the four in the pseudoscalar and scalar amplitudes at
$T\neq 0$.

Adding a third light quark introduces one qualitatively new aspect: the
$\eta$-$\eta^\prime$ system and the restoration of $U_A(1)$ symmetry, which
can affect the order of the chiral transition~\cite{pisarski}.  As already
observed, it is necessary to move beyond the rainbow-ladder truncation before
that can be addressed using the DSEs.  The question has been explored in
lattice simulations but the results are not currently conclusive: the mass
splittings used to characterise the symmetry breaking might become smaller
near $T_c$~\cite{UA1yes} but strong, topological arguments can nevertheless
be made in favour of the non-restoration of $U_A(1)$ symmetry~\cite{UA1no}.
Much remains to be done and improved models can be useful.

\addcontentsline{toc}{subsubsection}{\protect\numberline{ } {Chemical
Potential}}
\subsubsect{Chemical Potential}
The relation between chemical potential and baryon number density can only be
determined after the EOS is known; i.e., via Eq.~(\ref{densities}).  As
described in connection with Eq.~(\ref{qpres}) on page~\pageref{qpres}, the
dressed-quarks and -gluons contribute nothing to the EOS in the confined
domain, even though they dominate it in the QGP, and the only true
contributions to the pressure in the confined domain are those of colour
singlet bound states.  This physical requirement is overlooked in many model
explorations of the density dependence of meson properties.  For example, in
applications of the Nambu--Jona-Lasinio model the EOS for a free fermion gas
is used to {\it define} a baryon number density.  While this is the EOS for
that non-confining model it is not a good model for QCD's EOS.

The application of DSEs in calculating the $\mu$-dependence of hadron
properties is rudimentary.  However, even that is significant given the
problem is currently inaccessible in lattice simulations.  The model obtained
with ${\cal D}_{\rm A}$ in Eq.~(\ref{delta}) again provides a useful,
algebraic exemplar.  As described after Eq.~(\ref{bagvalue}), the QGP
transition is first order at $(\mu,T=0)$ and the chiral order parameters
increase with increasing $\mu$ when the dressed-quark self energy is momentum
dependent.  Mechanically, the latter is an obvious consequence of analyticity
in the neighbourhood of the real axis: any function, $O(4)$ invariant at
$\mu=0=T$, has the expansion
\begin{equation}
\label{mechanics}
f(\vec{p}^{\,2},\tilde\omega_k^2) \stackrel{T\sim 0\sim \mu}{=} 
f(\vec{p}^{\,2},0) + \tilde\omega_k^2\,
\left.\frac{\partial f(\vec{p}^{\,2},y)}{\partial y}\right|_{y =
\tilde\omega_k^2=0} + \ldots\,,
\end{equation}
and since $\Real [\tilde\omega_k^2 ] = \omega_k^2 - \mu^2$ then, if $\Real[
f(\vec{p}^{\,2},\tilde\omega^2)]$ decreases with $T^2$, the derivative is
negative and $\Real[ f(\vec{p}^{\,2},\tilde\omega^2)]$ must increase with
$\mu^2$.  [Only the real-part is important because the imaginary-part of
physical quantities vanishes after summing over the Matsubara frequencies.
The derivative is zero in models with an instantaneous interaction.]

\begin{figure}[t]
\centering{\ \epsfig{figure=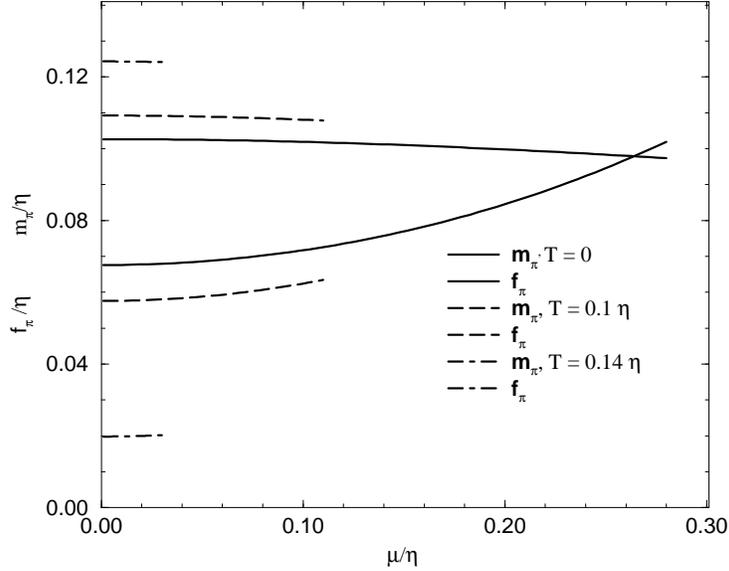,height=9.0cm}}\vspace*{-1em}

\parbox{40em}{\caption{\label{fpimpi} The pion's weak decay constant,
Eq.~(\protect\ref{npialg}), as a function of $\mu$ for a range of values of
$T$: $f_\pi$ increases with $\mu$ and decreases with $T$. [$f_\pi(\mu_c,T=0)=
1.51 \, f_\pi(0,0)$.  In Ref.~\protect\cite{gregp}, which employs model
${\cal D}_{\rm B}$, $f_\pi(\mu_c,0)= 1.25 \, f_\pi(0,0)$.] The pion's mass is
plotted too.  It falls slowly and uniformly with $\mu$ [$m_\pi(\mu_c,0)= 0.95
\, m_\pi(0,0)$] but increases with $T$.  (Adapted from
Ref.~\protect\cite{rhomuT}.)}}
\end{figure}

Equation~(\ref{mechanics}) is exemplified in the behaviour of the pion's
leptonic decay constant, which using the algebraic ${\cal D}_{\rm A}$ model
is simply expressed:
\begin{eqnarray}
\label{npialg}
f_\pi^2 & = & \eta^2 \frac{16 N_c }{\pi^2} \bar T\,\sum_{l=0}^{l_{\rm max}}\,
\frac{\bar\Lambda_l^3}{3} \left( 1 + 4 \,\bar\mu^2 - 4 \,\bar\omega_l^2 -
\sfrac{8}{5}\,\bar\Lambda_l^2 \right)\,,
\end{eqnarray}
where the notation is specified in connection with Eq.~(\ref{BmuTMN}) on
page~\pageref{BmuTMN}.  Its behaviour is depicted in Fig.~\ref{fpimpi}, as is
that of the pion's mass.  $m_\pi$ is almost {\it insensitive} to changes in
$\mu$ and only increases slowly with $T$, until $T$ is very near the critical
temperature, as already seen in Fig.~\ref{obsm}, which was calculated with
the renormalisation-group-improved ${\cal D}_{\rm C}$ model.  The
insensitivity to $\mu$ mirrors that to $T$ and is the result of compensating
changes in $r_\pi(\zeta)$ and $f_\pi$; i.e., a consequence of the
axial-vector Ward-Takahashi identity.  All these features are also evident in
Refs.~\cite{japan1,japan2}.

A first step in exploring the properties of the $\rho$-meson is solving the
the vector meson BSE.  That too takes a particularly simple form in the
${\cal D}_{\rm A}$ model~\cite{rhomuT}:
\begin{equation}
\label{bse}
\Gamma_M(p_{\omega_k};\vec{P})= - \frac{\eta^2}{4}\, 
\Real\left\{\gamma_\mu\, S(p_{\omega_k} +\sfrac{1}{2} \vec{P})\,
\Gamma_M(p_{\omega_k} ;\check P_\ell)\, S(p_{\omega_k} -\sfrac{1}{2}
\vec{P})\,\gamma_\mu\right\}\,.
\end{equation}
There are two solutions: one longitudinal and one transverse with-respect-to
$\vec{P}\,$:
\begin{equation}
\Gamma_\rho = \left\{
\begin{array}{l}
\gamma_4 \,\theta_{\rho+} \\
\left(
\vec{\gamma} - \sfrac{1}{|\vec{P}|^2}\,\vec{P} \vec{\gamma}\cdot\vec{P}\right)\,
        \theta_{\rho-}
\end{array}
\right.\,,
\end{equation}
where $\theta_{\rho+}$ labels the longitudinal and $\theta_{\rho-}$ the
transverse solution.  Substituting, one finds that for $\hat m =0$
\begin{equation}
\label{mminus}
m_{\rho-}^2 = \sfrac{1}{2}\,\eta^2,\;\mbox{{\it independent} of $\mu$ and $T$.}
\end{equation}
Even for nonzero current-quark mass, $m_{\rho-}$ changes by less than 1\% as
$\mu$ and $T$ are increased from zero toward their critical values.  This
insensitivity is just what one would expect for the transverse mode:
remember, there is no constant mass shift in the transverse polarisation
tensor for a gauge-boson.

\begin{figure}[t]
\centering{\ \epsfig{figure=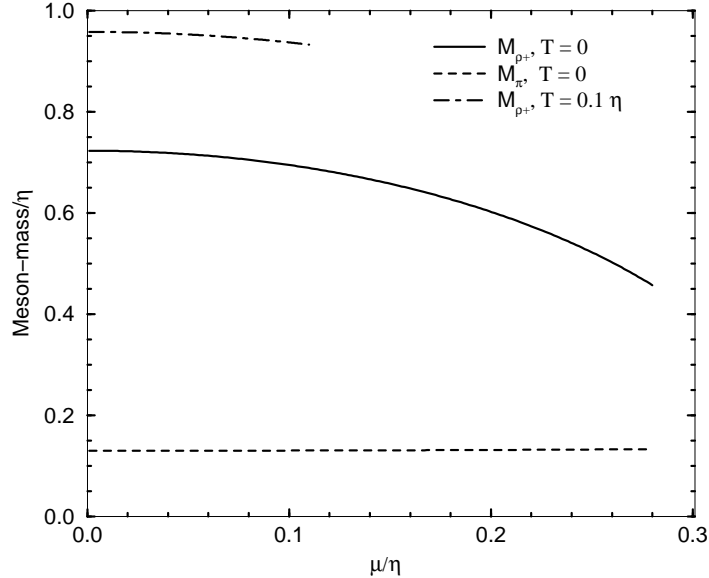,height=9.0cm}}\vspace*{-1em}

\parbox{40em}{\caption{\label{pirhomass} $M_{\rho+}$ and $m_\pi$ as a
function of $\bar\mu$ for $\bar T = 0, 0.1$.  On the scale of this figure,
$m_\pi$ is insensitive to this variation of $T$, which does not take approach
near to $(\mu=0,T_c)$.  The current-quark mass is $m= 0.011\,\eta$, which for
$\eta=1.06\,$GeV yields $M_{\rho+}= 770\,$MeV and $m_\pi=140\,$MeV at
$T=0=\mu$.  (Adapted from Ref.~\protect\cite{rhomuT}.)}}
\end{figure}

For the longitudinal component one obtains in the chiral limit:
\begin{equation}
\label{mplus}
m_{\rho+}^2 = \sfrac{1}{2} \eta^2 + 4 (\pi^2 T^2 - \mu^2 )\,,
\end{equation}
where the anticorrelation between the response of $m_{\rho+}$ to $T$ and its
response to $\mu$ is plain.  From Eq.~(\ref{mplus}) and continuity at the
second-order phase boundary it follows that
\begin{equation}
\frac{m_{\rho+}^2}{(2\pi T)^2} \,\stackrel{T\to\infty}{\to} \,1^+\,,
\end{equation}
which is analogous to Eq.~(\ref{screeneq}).  Here, however, the limit is
approached from above because $m_{\rho^+}\neq 0$ in the chiral limit and {\it
increases} with $T$.  (NB.  It is only because this model exhibits
confinement that such a result is possible.  Studying the $\rho$-meson in
non-confining dressed-quark-based models requires that some means be employed
to suppress or eliminate the $\rho \to \bar q q$ threshold; e.g.,
Ref.~\cite{wambachNc}.  However, that artefice is merely an indigent
expression of confinement.  For the transverse component of the $\rho$,
$m_{\rho^-}^2/(2\omega_0)^2\to 1^-$ because of Eq.~(\ref{mminus}); e.g.,
Ref.~\cite{peterNew}.)  The $(\mu,T)$-dependence of the $\rho$-meson mass is
depicted in Fig.~\ref{pirhomass} and; e.g., at $T=0$, $M_{\rho+}(\mu_c)
\approx 0.65\, M_{\rho+}(\mu=0)$.  As observed in the introduction, though,
the connection between $\mu$-dependence and baryon-density-dependence cannot
be determined until the EOS is calculated.  Without it one can only observe
that, in a two-flavour free-quark gas, the $T=0$ critical chemical potential
corresponds to $3\rho_0$, see Fig.~\ref{bagpresMN} on
page~\pageref{bagpresMN}.  Therefore, at $1$--$2\,\rho_0$ a mass reduction
less-than this should be anticipated, plausibly no more than
$25$\%~\cite{rhomuT}.

The $T$-dependence described above was also observed in the confining model
of Refs.~\cite{pctTrento,peterNew}, wherein too it was found that, because of
the $T$-dependence of $m_\pi$, $m_{\rho^-}$, the dominant $2\pi$-decay mode
of the $\rho_-$ meson mode is phase-space blocked for $T/T_c>1.2$.  [cf.\ The
$2\pi$ mode of the isoscalar-scalar discussed in connection with
Eq.~(\ref{sigwidth}).]  The $T$-dependence of the $\rho$-meson's dilepton
decay width was also considered: it is suppressed by a factor of $0.9$ in the
vicinity of $(\mu=0,T_c)$.  However, its $\mu$-dependence is yet to be
explored.

The anticorrelation, anticipated in Eq.~(\ref{mechanics}), between the $\mu$
and $T$ dependence of mass-dimension-two observables; such as ${\cal X}$,
$f_\pi$, $m_{\rho^+}$, etc., is apparent.  It entails that, in these cases at
least, there is a trajectory in the $(\mu,T)$-plane along which the
observables are constant~\protect\cite{bastihirschegg}.  It also means that
observables calculated using the rainbow-ladder truncation do not exhibit a
$\mu$-scaling law of the type conjectured for baryon-number-density in
Ref.~\cite{brown}.

We anticipate that the mass of the $a_1$-meson [the chiral partner of the
$\rho$] will decrease with increasing $T$ so that it can evolve to meet the
increasing $\rho$-mass, with $m_{\rho^+} = m_{a_1^+}$ at the phase boundary.
However, since model ${\cal D}_{\rm A}$ is defective in not supporting an
axial-vector bound state, that remains to be verified.  (It fails to support
a scalar bound state too~\cite{mn83}.)  As remarked in Sec.~\ref{sectthree}
and Ref.~\cite{a1b1}, much remains to be learnt about axial-vector mesons, in
which connection confinement is an important element.  Determining the
$\mu$-dependence of $m_{a_1}$ is particularly interesting given the $\mu$
cf.\ $T$ anticorrelation exemplified above.

In the studies described here, and also in lattice simulations, $\mu$ is an
intensive thermodynamic parameter whose presence modifies the propagation
characteristics of dressed, confined particles, and this modification is
transmitted to the observable hadrons they comprise.  It is clear from the
existence of a critical quark chemical potential, below which asymptotic
quark states cannot be produced, that in the confined domain there is no
simple proportionality between the quark chemical potential and the chemical
potential associated with colour singlet baryons.  In approaches based on
elementary hadronic degrees of freedom; e.g., those reviewed in
Ref.~\cite{wambach}, this consideration is bypassed.  Colour singlet baryon
density is introduced directly via the expedient of in-medium elementary
meson and nucleon propagators, which are then employed in calculating the
myriad nuclear-matter many-body loop integrals that contribute to observable
processes.  The approach has a long history and yields a useful
phenomenology.  However, it ignores the dressed-quark level effects described
above and also questions, such as, just what is represented by an off-shell
hadron propagator?  [cf.\ The discussions associated with
Eqs.~(\ref{bcplus})--(\ref{rhopi2}), page~\pageref{bcplus}, and
(\ref{scalarvtx})--(\ref{radsig}), page~\pageref{scalarvtx}.]  We judge that
in understanding QCD at nonzero baryon density it is important to uncover the
nature of the relationship between these approaches.

\subsect{Hadronic Signatures of the Quark-Gluon Plasma}
A QGP existed approximately one microsecond after the big bang, and primary
goals of current generation experiments at CERN and Brookhaven are the
terrestrial recreation of the plasma and an elucidation of its properties.  A
number of signals for QGP formation at high temperature and low baryon number
density have been suggested and here we briefly review three of them.  They
and others are discussed more extensively in Refs.~\cite{signatures,eskola}.

$J/\Psi$~{\it Suppression.}\hspace*{1em}
Reference~\cite{MS86} proposed ``\ldots that $J/\Psi$ suppression in nuclear
collisions should provide an unambiguous signature of quark-gluon plasma
formation.''  The reasoning is simple.  A $c \bar c$-pair produced in a hard
parton-parton interaction will evolve into a $J/\Psi$-meson if the colour
interaction is sufficiently strong to effect binding.  That will always be
the case {\it unless} the $c \bar c$-pair is produced in a heat-bath of
deconfined, colour-carrying excitations that [Debye-]screen the $c \bar
c$-attraction; i.e., unless the $c \bar c$-pair is produced in a QGP.  QGP
formation occurs only for temperatures greater than some critical value.
Hence the $J/\Psi$ production cross-section should evolve smoothly with
controllable experimental parameters; such as, projectile and target mass
numbers, and impact parameter, until a QGP is produced, when a dramatic
suppression should follow.

\begin{figure}[t]
\vspace*{4em}

\centering{\ \hspace*{-4em}\epsfig{figure=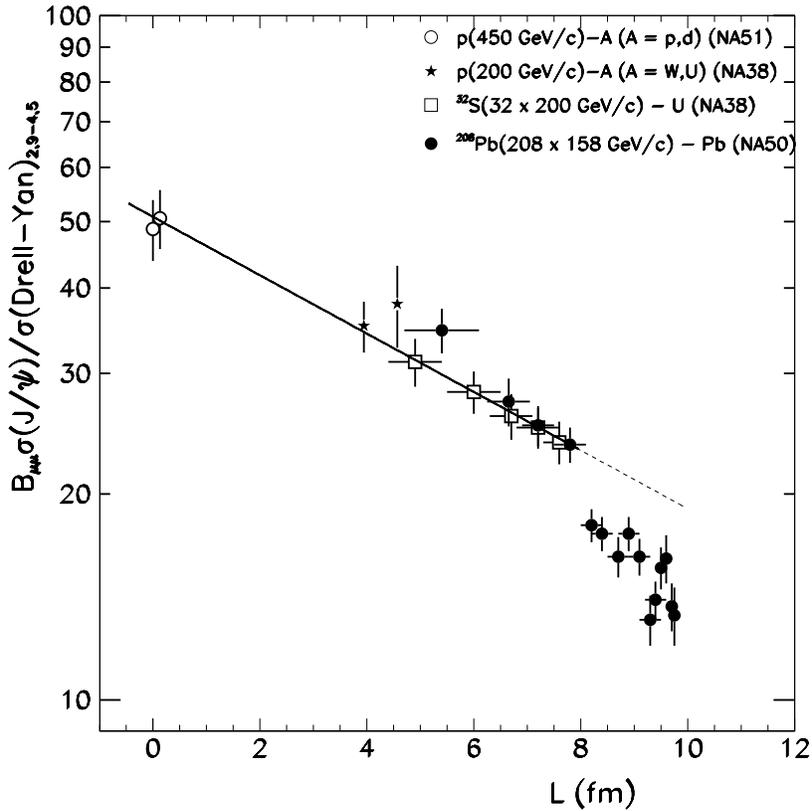,height=14.0cm}}
\vspace*{-12em} 

\parbox{40em}{\caption{\label{jpsifig} Ratios of $J/\Psi$-to-Drell-Yan
cross-sections as a function of the average nuclear path: $L$, traversed by
the $c \bar c$-pair, which is a model-dependent but calculated quantity.  The
straight line is the scaling relation of Eq.~(\protect\ref{ABscaling}).
(Adapted from Ref.~\protect\cite{JPsiTrento}.)}}
\end{figure}

Following this suggestion, the $J/\Psi$ production cross-section has been
systematically explored in relativistic heavy ion collisions at CERN using
the Super Proton Synchrotron [SpS].  As illustrated in Fig.~\ref{jpsifig},
all results are described by a simple scaling law:
\begin{equation}
\label{ABscaling}
(A_{\rm projectile}\times B_{\rm target})^\alpha\,, \;  \alpha=0.92 \pm 0.01\,,
\end{equation}
where $A$, $B$ are the mass numbers, {\it except} those for Pb-Pb collisions.
The scaling relation is easily understood as a consequence of nuclear
absorption~\cite{Gerschel} and is termed ``normal'' $J/\Psi$-suppression.
While the Pb-Pb data agree with this normal pattern for peripheral
collisions; i.e., $L\leq 8\,$fm, that is not the case for the most central
collisions, which correspond to an energy density range of
$2$--$3\,$GeV$/$fm$^3$.  Hitherto, using standard in-medium hadronic tools,
this ``anomalous'' suppression is inexplicable.  However, an explanation can
be found in the transition to a QGP; e.g., Refs.~\cite{Nardi,BlaTandy}, and
this signal has recently been claimed~\cite{heinz} as ``\ldots evidence for
the creation of a new state of matter in Pb-Pb collisions at the CERN SPS.''

\begin{figure}[t]
\vspace*{-4em}
\centering{\ \epsfig{figure=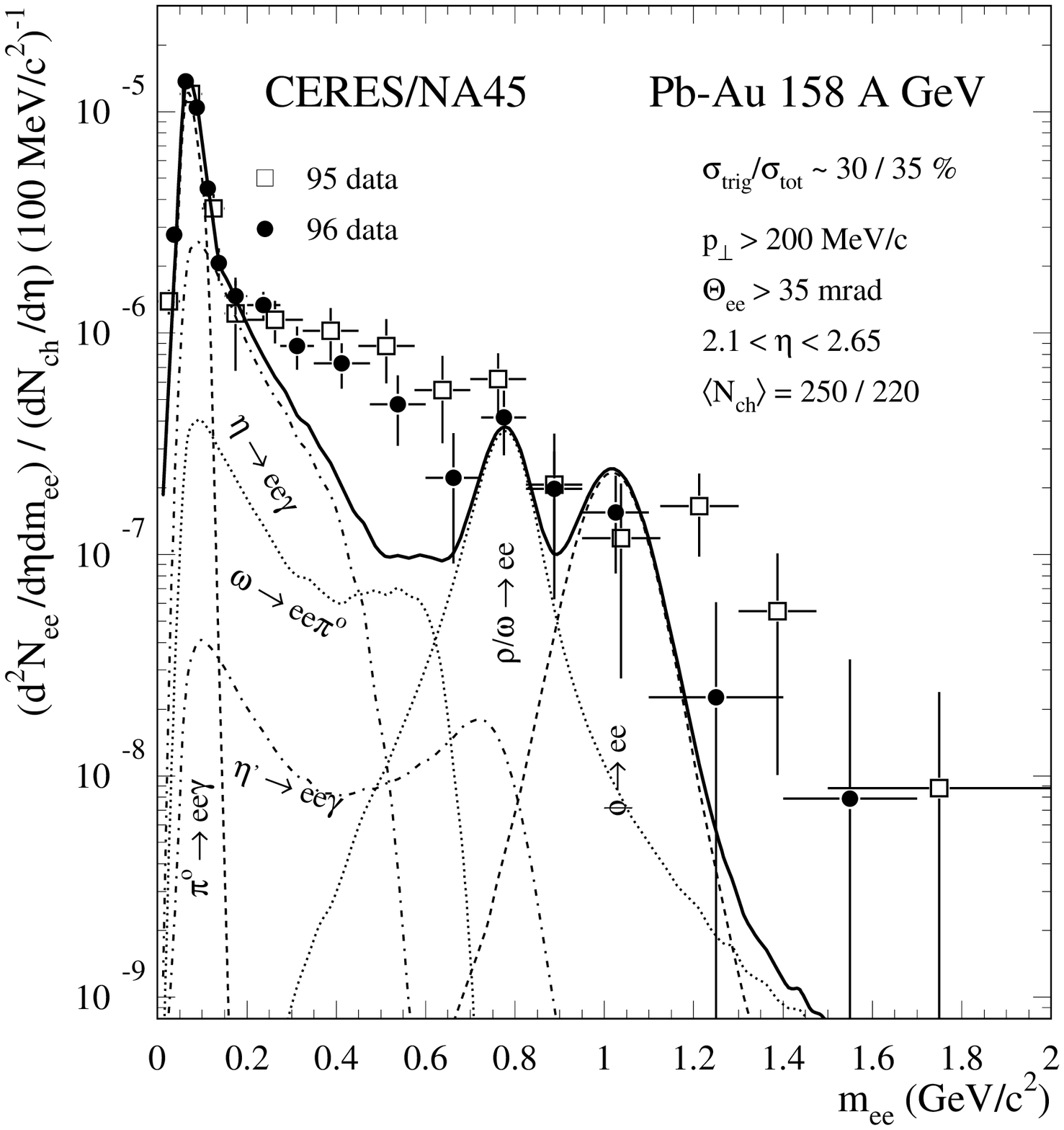,height=11.25cm}}

\parbox{40em}{\caption{\label{lowmassfig} Recent Pb-beam data collected by
the CERES group.  Two independent data sets: 1995 and 1996, are compared to
the dilepton yield expected in proton-nucleus collisions.  The excess is
characterised by an enhancement factor: 3.3, 1996; 2.5, 1995.  It is an
interesting effect but not a direct signal of QGP formation.  (Adapted
from Ref.~\protect\cite{drees}.)}}
\end{figure}

{\it Low-mass Dilepton Enhancement.}\hspace*{1em}
Leptons produced in a relativistic heavy ion collision escape the interaction
region without attenuation by strong interactions, which means they are a
probe of phenomena extant in the early phase of the collision.  Lepton pair
production data collected in relativistic S-Au and Pb-Au collisions at the
CERN SpS exhibits an excess with-respect-to proton-nucleus data in the
``low-mass'' region: $0.25\,{\rm GeV}\lsim M_{e^+ e^-}\lsim 0.70\,{\rm GeV}$,
which is illustrated in Fig.~\ref{lowmassfig}.  

Since the clear excess appears above the $\pi\pi$ threshold and below the
position of the $\rho$-meson peak in proton-nucleus collisions, the search
for an explanation has focused on exploring the in-medium properties of the
$\rho$-meson.  Adequate explanations are found in: 1) collisional broadening;
i.e., the $\rho$ has a shorter lifetime in an hadron-rich medium, and an
increase in the $\rho$-meson's width due to in-medium-modified hadron-loop
contributions to the $\rho$-meson's self-energy~\cite{wambach,cassing}; and
2) a simple reduction in the $\rho$-meson mass~\cite{cassing,brown2}.  The
phenomenological models require large values of temperature [$T\sim
0.15\,$GeV] and baryon density [$\rho \sim 1$--$2\,\rho_0$] to describe the
data.  On this domain the $\rho$-meson screening masses, calculated in
Refs.~\cite{peterNew,rhomuT} and described in connection with
Fig.~\ref{pirhomass}, are {\it increased} by $\lsim 25$\% and, while
$\Gamma_{\rho \to e^+ e^-}$ is roughly unchanged, $\Gamma_{\rho \to \pi\pi}$
is much reduced~\cite{peterNew}.  These constraints and effects are ignored
in contemporary analyses, even though such features can affect photon
production rates~\cite{georgefai}.

{\it Strangeness Enhancement.}\hspace*{1em}
The lifetime of a terrestrial QGP can only be of the order: $\tau \sim
5$--$10\,$fm, which is much too short for weak interactions to be important.
Hence strangeness, once produced, can only disappear through $s$-$\bar s$
annihilation.  Such events are unlikely unless there is a super-abundance of
strangeness.  Therefore strangeness carrying reaction products in the debris
are a good probe of the conditions created by a relativistic heavy ion
collision.

\begin{figure}[t]
\centering{\ \epsfig{figure=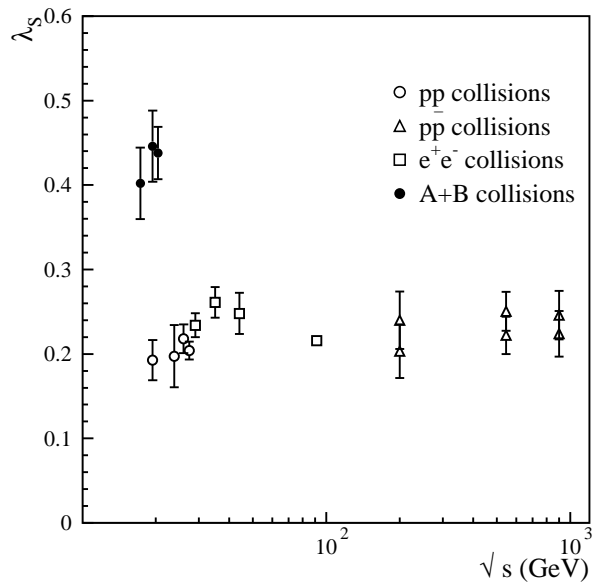,height=10.6cm}}\vspace*{-7em}

\parbox{40em}{\caption{\label{globalS} By the measure in
Eq.~(\protect\ref{strange}), global strangeness production is enhanced by a
factor of $\sim 2$ in A-B [S-S, S-Ag, Pb-Pb] collisions cf. $e^+$-$e^-$,
$p$-$p$, $p$-$\bar p$.  The data are from Refs.~\protect\cite{strangedat}.
(Adapted from Ref.~\protect\cite{BECATTINI}).}}
\end{figure}

The amount of strangeness produced in collisions can be quantified via the
ratio~\cite{BECATTINI}
\begin{equation}
\label{strange}
\lambda_s^{AA} \equiv \frac{2\langle s+\bar s\rangle}{\langle u+\bar
u\rangle + \langle d+\bar d\rangle}\,,
\end{equation}
where $\langle s +\bar s\rangle$, etc., are the mean multiplicities of newly
produced valence quark-antiquark pairs at primary hadron level before
resonance decays.  Experimentally, as illustrated in Fig.~\ref{globalS},
\begin{equation}
\lambda_s^{AA} \simeq 2 \lambda_s^{pp}\,.
\end{equation}
In addition to this global enhancement, specific enhancements in the yields
of $K$, $\bar K$, $\Lambda$, $\bar \Lambda$, etc., have been observed in CERN
experiments~\cite{strangedat2}.  Detailed analyses indicate that quark
degrees-of-freedom are necessary to describe the strangeness enhancement,
with the suggestion that it may be indicative of QGP formation~\cite{eskola}.

\sect{Toward a Kinetic Description}
\label{SECkinetic}
Hitherto we have described features of cold, sparse QCD: the nature of DCSB,
confinement and bound state properties, and then the effect that the
intensive thermodynamic parameters chemical potential and temperature have on
these phenomena.  In the latter we applied the methods of equilibrium
statistical field theory and elucidated properties of the QGP phase.  The
terrestrial creation of this QGP, however, is expected to be effected via
relativistic heavy ion collisions, which initially yield a quantum system far
from equilibrium.  This system must then evolve to form the plasma, and the
study of that evolution and the signals that characterise the process are an
important contemporary aspect of QGP research.

\subsect{Preliminaries}
\label{prelim}
The energy density in an ideal gas of eight gluons and two flavours of
massless quarks is
\begin{equation}
\epsilon_{g+u+d} = (2\times 8)\, \frac{\pi^2 T^4}{30} + (2\times 3\times
2)\,\frac{7\pi^2 T^4}{120} = \frac{37 \pi^2 T^4}{30}\,.
\end{equation}
As we saw in connection with Table~\ref{critthings}, the critical temperature
for QGP formation is $T_c\simeq 0.15\,$GeV and at this temperature:
$\epsilon_{g+u+d}= 0.8\,$GeV/fm$^3$.  Construction of RHIC at Brookhaven
National Laboratory is complete and it will soon provide counter-circulating,
colliding $100\,$A$\,$GeV $\,^{197}\!$Au beams to generate a total
centre-of-mass energy of $40\,$TeV, which corresponds to an initial energy
density: $\epsilon \sim 10$--$100\,$GeV/fm$^3$.  The Large Hadron Collider
[LHC] project at CERN is scheduled for completion in 2005.  Plans are for it
to provide $^{208}$Pb-$^{208}$Pb collisions with $\sqrt{s} \gsim 2\,000\,$TeV
and a consequent initial energy density $\epsilon \gsim 1\,000\,$GeV/fm$^3$.
With these energy scales RHIC and LHC should certainly provide the conditions
necessary for QGP formation.

Control over the conditions produced in a relativistic heavy ion collision
can only be exerted via two experimental parameters: the beam/target
properties and, to some extent, the impact parameter.  They can be used
together to analyse the debris collected in the detectors.  It is in the
behaviour of this debris, summed over many events, that signals of the
evolution and formation of a QGP must be found.  Predicting just what the
signals are requires an understanding of these processes, including and
perhaps especially their the non-equilibrium aspects.

\begin{figure}[t]
\vspace*{-1.5em}

\centering{\ \psfig{figure=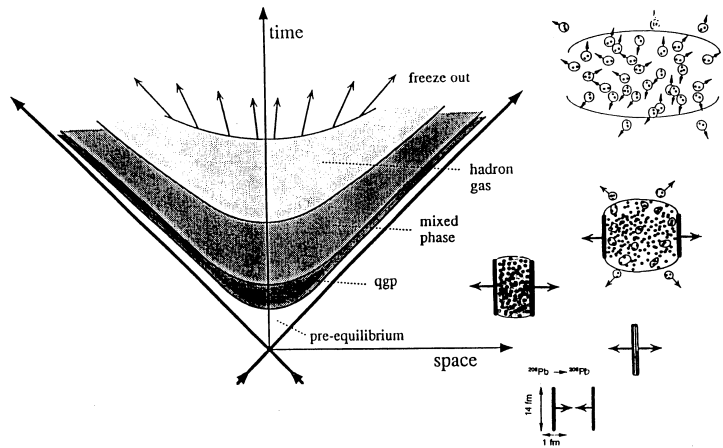,height=10cm,angle=-90}}

\parbox{40em}{\caption{\label{fighic} Spacetime evolution of a relativistic
heavy ion collision.  Two [Lorentz contracted] disc-like nuclei collide and
pass through each other, transforming a large amount of their kinetic energy
into the potential energy of a highly excited spacetime volume.  This energy
is dissipated via particle production, which is a non-equilibrium process.
The system evolves to equilibrium, expanding and passing through the QGP
phase, hadronisation and freeze-out, where the observable hadronic debris is
created.  (Adapted from Ref.~\protect\cite{bastithesis}.)}}
\end{figure}

Figure~\ref{fighic} illustrates the spacetime evolution of a relativistic
heavy ion collision.  All stages subsequent to the formation of a thermalised
QGP are adequately addressed using hydrodynamical models~\cite{hydro}.
However, in this approach the plasma's initial conditions [energy density,
temperature, etc.] and EOS must be specified.  These initial conditions can
only be reliably determined by following the complete evolution of the system
produced in the collisions; i.e., by understanding the pre-equilibrium stage.
Furthermore, the very existence of a pre-equilibrium phase can lead to
signals of QGP formation; e.g., plasma oscillations, a disoriented chiral
condensate~\cite{DCC}, and out-of-equilibrium photon and dilepton
emission~\cite{lerran}.

The QGP is a hot, equilibrated, many-parton agglomeration, and in recent
years two main approaches have been employed in describing how such a system
might be produced in a relativistic heavy ion collision.  In the perturbative
parton picture~\cite{Gribovpartons}, the colliding nuclei are visualised as
pre-formed clouds of quarks and gluons.  The collision proceeds via rapid,
multiple, short-range parton-parton interactions, which generate entropy and
transverse energy in a cascade-like process.  In the string
picture~\cite{anders83}, after passing through one another, the colliding
nuclei are imagined to stretch a high energy-density flux tube between them,
which decays via a nonperturbative particle-antiparticle production process.
Each approach has its merits and limitations, and their simultaneous pursuit
provides complimentary results.  The analysis of data proceeds via one of the
many Monte-Carlo event generators that have been developed for both
pictures~\cite{partonMC,Geiger95,stringMC}.

{\it Flux Tube Model and Schwinger Mechanism.}\hspace*{1em}
One intuitively appealing, semi-classical picture of confinement is provided
by the notion that it is effected via the formation of a small-diameter
colour flux tube between colour sources; and there is evidence in lattice
simulations for the appearance of such flux tubes between
heavy-quark-antiquark pairs; e.g., Ref.~\cite{rwhay}.  This is a motivation
for the string-like models just introduced, which have been used to study
particle production in $e^-$-$e^+$, $p$-$p$ and $p$-nucleus
collisions~\cite{anders83,nuss}.  A flux tube yields a linearly-rising,
confining quark-antiquark potential: $V_{q \bar q}(r) = \sigma\, r$, where
the string tension: $\sigma$, can be estimated in static-quark lattice
simulations.  An overstretched flux tube can be viewed as a strong background
field.  As such it destabilises the QCD vacuum, which is corrected by
particle-antiparticle production via a process akin to the Schwinger
mechanism in QED~\cite{SCH,greiner}.  We use this particle production
mechanism as the primary medium for our discourse.

\begin{figure}[t]
\centering{\ \epsfig{figure=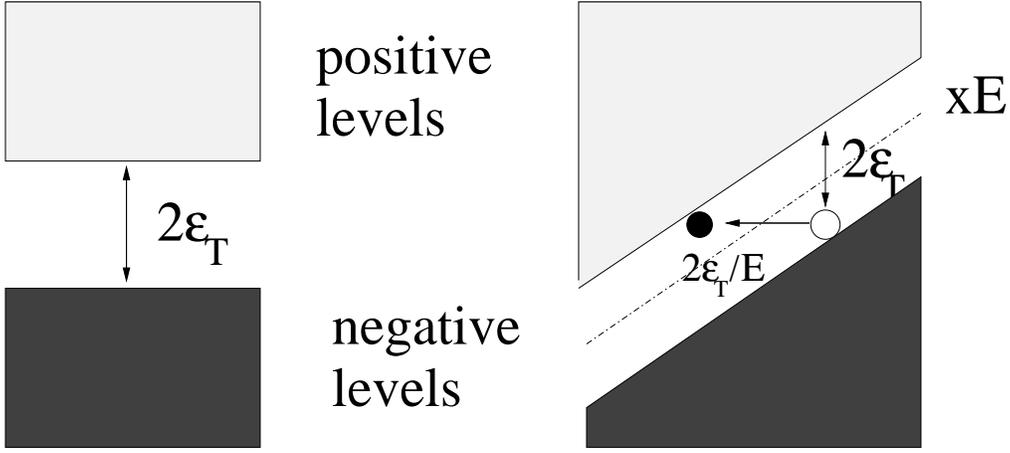,height=6cm}}

\parbox{40em}{\caption{\label{diracsea} {\it Left Panel}: For $E=0$ the
vacuum is characterised by a completely filled negative-energy Dirac sea and
an unoccupied positive-energy continuum, separated by a gap: $2\,
\varepsilon_{\rm T}= 2\,(m^2 + p_\perp^2)^{1/2}$.  {\it Right Panel}:
Introducing a constant external field: $\vec{E} = \hat{e}_x E$, which is
produced by a potential: $A^0= -E \vec{x}$, tilts the energy levels.  In this
case a particle in the negative-energy sea will tunnel through the gap with a
probability $\sim \exp(-\pi\varepsilon_T^2/eE)$.  Succeeding, it will be
accelerated by the field in the $-x$-direction, while the hole it leaves
behind will be accelerated in the opposite direction.  The energy-level
distortion is increased with increasing $E$ and hence so is the tunnelling
probability.  [NB.  $\tau_{\rm tu}\sim\varepsilon_T/eE$ is a measure of the
time between tunnelling events.  $eE$ can be related to the flux-tube
string-tension.]
}}
\end{figure}

The fermion production rate for a constant, homogeneous electric field $E$ in
a flux tube is~\cite{nuss}
\begin{equation}
\label{WKB}
S(p_\perp)=\frac{dN}{dtdVd^2p_\perp}=
|eE|\ln\bigg[1+\exp\bigg(-\frac{2\pi(m^2+p_\perp^2)}{|eE|}\bigg)\bigg],
\end{equation}
where $m$ is the mass and $e$ the charge of the produced particles, and it is
plain from this formula that the rate increases with increasing $E$ and is
suppressed for large $m$ and/or $p_\perp$.  Particle-antiparticle production
via this mechanism is analogous to a tunnelling process in quantum mechanics,
as illustrated in Fig.~\ref{diracsea}.  [NB.  The Schwinger
formula~\cite{SCH} follows immediately by integrating Eq.~(\ref{WKB}) over
the transverse momentum: $\int d^2p_\perp$.]

What happens once the particles are produced?  Naturally, they are
accelerated in the tube by the field.  This produces currents, which generate
an electric field that works to screen the flux-tube [background] field.  In
the absence of other effects, the net field vanishes and particle production
stops.  However, the currents persist, now generating a field and renewed
particle production that opposes their own existence.  That continues until
the net current vanishes.  But at this point there is again a strong electric
field~\ldots~and the process repeats itself.  This is the ``back-reaction''
phenomenon and the obvious consequence is time-dependent fields and currents;
i.e., plasma oscillations characteristic of the system.  In recent years its
affects have been studied in both boson and fermion production~\cite{Back}
and we exemplify the process in Sec.~\ref{backsection}.

There are a number of shortcomings in extant applications of the flux-tube
particle-production picture to QGP formation: 1) The non-Abelian nature of
the chromoelectric field is often ignored because the QCD analogue of the
Schwinger mechanism is poorly understood.  Instead the flux tube is
represented via a classical electromagnetic potential; 2) On the whole,
finite volume effects are neglected, with the electric field assumed to be
homogeneous in space.  A more realistic description would account for the
geometry of the interaction region.  Some steps have been taken in this
direction, with a consideration of effects produced by a cylindrical
boundary~\cite{juda} and those in a finite-length flux tube with a confining
transverse potential~\cite{pavelwong}.  Going further, the geometrical
representation of a flux tube can be replaced, allowing the flux tube profile
itself to be a dynamical quantity, whose presence and stability is affected
by the charged particle currents~\cite{ben}; 3) The time-evolution of the
system is described using either mean-field theory, which retains quantum
effects but makes problematic an exploration of the effects of collisions, or
a transport equation~\cite{bialas}, which neglects quantum effects.  Below we
will describe a partial reconciliation of these approaches; and 4) Little
attention has been paid to particular non-Abelian features in transport
equations~\cite{Elze89}.  A study~\cite{bahl2} of Wong's
equation~\cite{qcdwong} is one step in the direction of explicitly including
colour algebra effects.

Quantum field theory can be applied directly to out-of-equilibrium plasma
phenomena; for example, Refs.~\cite{Back,cooperboyaknollgies}.  However,
kinetic equations provide an appealing alternative because of their intuitive
character.  This approach begins with a transport equation
\begin{equation}
\label{one}
p^\mu\,\frac{\partial f}{\partial q^\mu}- Q\, p^\mu F_{\mu\nu}\,
\frac{\partial f}{\partial p_\nu}=S(p,q)+C(p,q)\,,
\end{equation}
where: $f$ is the single particle distribution function, which gives the
ensemble fraction of particles in a given phase-space cell [$p$ is
four-momentum and $q$ is a spacetime four-vector]; $F_{\mu\nu}$ is the field
strength tensor; and $Q$ is the particles' charge.  The source term:
$S(p,q)$, describes the production of particle-antiparticle pairs and
$C(p,q)$ is the collision term, whose origin is intuitively obvious but which
is difficult to approximate accurately.  

A Boltzmann equation of this form has been widely applied to particle
production using a source term of the type in Eq.~(\ref{WKB}) both with and
without a collision term, and in the following we exemplify this.  However,
to provide an introductory overview, Refs.~\cite{bialas} employ a classical,
Abelian electric field in the source term but completely ignore the effect of
collisions, $C\equiv 0$ in Eq.~(\ref{one}).  The effect of collisions,
represented via a ``relaxation time approximation,'' [described in connection
with Eq.~(\ref{collterm})] has been considered in an hydrodynamical
approximation to Eq.~(\ref{one}), Refs~\cite{KM,GKM}, with the influence of
back-reactions neglected in the former but explored in the latter.  A
comparison between the transport equation and mean-field theory approaches
has also been made~\cite{Back} and the results are remarkably similar.  That,
however, is problematic since; e.g., the application of the Schwinger source
term in the presence of a rapidly changing electric field is {\it a priori}
unjustified and, although quantum field theory with its manifest microscopic
time reversal invariance must underly the behaviour all quantum systems,
experience confirms that systems far from equilibrium exhibit macroscopically
irreversible behaviour that is amenable to treatment using [inherently
time-irreversible] kinetic theory.  The nature of the connection has recently
been established~\cite{Rau,gsi,kme}.

At this point we re-emphasise that flux-tube models describe the
nonperturbative production of soft partons.  The production of hard and
semi-hard partons is described by perturbative QCD and that mechanism is
explored in Refs.~\cite{minijet}.  Simultaneously incorporating both types of
particle production is challenging but Ref.~\cite{nayak} is a step in that
direction.  Therein hard and semihard partons are produced via ``minijet
gluons'' and provide the initial conditions necessary to solve the transport
equations, and the subsequent evolution of the plasma is described by a
classical but non-Abelian transport equation.  Collisions are accounted for
using a relaxation time approximation but quantum effects in the source term
are neglected and, as will become apparent, they can be important in strong
fields.

\subsect{Quantum Vlasov Equation}
The derivation of a {\it quantum} Vlasov equation in Refs.~\cite{Rau,gsi,kme}
provides a connection between the quantum field theoretical and transport
equation approaches to particle production and plasma evolution, and shows
that the particle source term is intrinsically non-local in time; i.e., {\it
non-Markovian}.  Therefore calculating the plasma's properties at any given
instant requires a complete knowledge of the history of the formation
process.  

The derivation begins with the Dirac [Klein-Gordon] equation for fermions
[bosons] in an external, time-dependent, spatially homogeneous vector
potential: $A_\mu$, in Coulomb gauge: $A_0=0$, taken to define the $z$-axis:
${\vec A} = (0,0,A(t))$.  The corresponding electric field is
\begin{equation}
E(t) = -{\dot A}(t)=-\frac{dA(t)}{dt}\,,
\end{equation}
also along the $z$-axis.  The vacuum instability created by this external
field is corrected via particle-antiparticle production, Fig.~\ref{diracsea},
which is a time-dependent process.  The transition from the in-state to the
instantaneous, quasi-particle state at time $t$ is achieved by a
time-dependent Bogoliubov transformation, which effects the diagonalisation
of the system's Hamiltonian.  By this means one obtains a kinetic equation
for the single particle distribution function
\begin{equation}
f(\vec{P},t)= \langle 0 | a^\dagger_{\vec{P}}(t)\,a_{\vec{P}}(t)|0\rangle\,,
\end{equation}
which is defined as the vacuum expectation value, in the time-dependent
basis, of creation and annihilation operators: $a^\dagger_{\vec{P}}(t)$,
$a_{\vec{P}}(t)$, for particle states at time $t$ with three-momentum
$\vec{P}$.  That equation is
\begin{equation}
\label{10}
\frac{df_\pm(\wvp,t)}{dt}=\frac{\partial f_\pm(\wvp,t) }{\partial
t}+eE(t)\frac{\partial f_\pm(\wvp,t)}{\partial P_\parallel(t)}=
\frac{1}{2}{\cal W}_\pm(t)\int_{-\infty}^t dt'{\cal W}_\pm(t')
[1 \pm 2 f_\pm(\wvp,t')]\cos[x(t',t)]\,,
\end{equation}
where the lower [upper] sign in Eq.~(\ref{10}) corresponds to fermion [boson]
pair creation.  The momentum is defined as $\wvp=(p_1,p_2,P_\parallel(t))$,
with the longitudinal [kinetic] momentum $P_\parallel(t)=p_\parallel-eA(t)$,
$p_\parallel=p_3$.  [NB.  $eE(t) = {\dot P_\parallel}(t)$, the particle
velocity attained via acceleration by the field $E(t)$.]  For
fermions~\cite{Rau,gsi} and bosons~\cite{gsi,kme} the transition amplitudes
are
\begin{equation}
\label{12}
{\cal W}_-(t)=\frac{eE(t)\varepsilon_\perp}{\omega^2(t)}\,,\;\;
{\cal W}_+(t)=\frac{eE(t)P_\parallel(t)}{\omega^2(t)}\,,
\end{equation}
where the transverse energy
$\varepsilon_\perp=\sqrt{m^2+\vec{p}_\perp^{\,2}}$, $\vp_\perp = (p_1,p_2)$,
and $\omega(t)=\sqrt{\varepsilon_\perp^2+P_\parallel^2(t)}$ is the total
energy.  In Eq.~(\ref{10}),
\begin{equation}
\label{30} 
x(t^\prime,t) = 2[\Theta(t)-\Theta(t^\prime)]\,,\;\;
\Theta(t) = \int^t_{-\infty}dt^\prime \omega(t^\prime)\,,
\end{equation}
is the dynamical phase difference.  Equation~(\ref{10}) is the precise
analogue of directly solving QED with an external field in mean-field
approximation, as done; e.g., in Ref.~\cite{Back}.  The physical content is
therefore equivalent and, in particular, the fundamental quantum character is
preserved.  The appeal of this kinetic equation, however, is that it
simplifies: the identification of reliable approximations; widespread
applications; and numerical studies.

This quantum Vlasov equation has three qualitatively important features: 1)
The source is non-Markovian for two reasons -- (i) the source term on the
r.h.s. requires complete knowledge of the distribution function's evolution
from $t_{-\infty}\rightarrow t$; and (ii) the integrand is a non-local
function of time, which is apparent in the coherent phase oscillation term
$\cos x(t^\prime,t)$ and reflects the quantum nature of the source term; 2)
Particles are produced with a momentum distribution cf.\ the Schwinger source
term, which produces particles with zero longitudinal momentum; and 3) the
production rate is affected by the particles' statistics, as evident in
Eqs.~(\ref{12}) and also in the sign appearing in the statistical factor $[1
\pm 2f_\pm]$, which leads to different phase space occupation~\cite{basti}.

These features can have a material impact on the solution of the kinetic
equation, and their importance depends on the field strength and the
time-scales characterising the production process.  The first time-scale is
set by the Compton wavelength of the produced particles:
\begin{equation}
\label{tau}
\tau_{\rm qu} \sim \frac{1}{\varepsilon_\perp}\,.
\end{equation}
Events taking place over times less-than this expose the negative-energy
elements in particle wave-packets, a core quantum field theoretical feature.
This is the time-scale of the rapid oscillations generated by the dynamical
phase difference in $\cos x(t^\prime,t)$.  The high frequencies involved mean
that the main contributions arise when $t\sim t^\prime$, and therefore a
local approximation can be justified for weak field plasmas~\cite{kme}.  We
identified a second time-scale in Fig.~\ref{diracsea}; i.e., the time taken
for a particle to tunnel through the barrier.  It can also be motivated by
considering the transition amplitudes in Eqs.~(\ref{12}), which for a
constant electric field can be written in the form; e.g.,
\begin{equation}
{\cal W}_-(t)=\frac{\varepsilon_\perp/eE}
{(t-p_\parallel/eE)^2+(\varepsilon_\perp/eE)^2}\,,
\end{equation}
which is a Lorentzian characterised by the time-scale [width]
\begin{equation}
\label{tcl}
\tau_{\rm tu} \sim \frac{\varepsilon_\perp}{eE}\,.
\end{equation}
This is also the time required to accelerate a charged particle to the speed
of light in an electric field.  

The times scales in Eqs.~(\ref{tau}) and (\ref{tcl}) are comparable when $eE
\sim \varepsilon_\perp^2$.  For weak fields, $\tau_{\rm qu} \ll \tau_{\rm
tu}$ and, in the integrand on the r.h.s. of Eq.~(\ref{10}), the $\cos
x(t^\prime,t)$ oscillations occur on a time-scale so-much shorter than
variations in the other elements that a stationary phase approximation is
valid.  That approximation yields a local source term and this important
limit of Eq.~(\ref{10}) was discussed in Ref.~\cite{kme}.  

$\tau_{\rm tu} \gsim \tau_{\rm qu}$ for strong fields and hence the
particle's Compton wavelength extends across the gap.  In this case wave-like
[quantum] interference effects become important in the behaviour of $f(p,q)$,
which changes at a rate comparable with $\cos x(t^\prime,t)$, so that a
stationary phase approximation is not valid.  Can this strong field scenario
be relevant to contemporary relativistic heavy ion collisions?  It is easy to
make an estimate.  The field strength in a flux tube is commensurate with the
QCD string tension; i.e., $eE \sim \sigma \approx 0.4\,$GeV$^2 \sim (2.5\,
\Lambda_{\rm QCD})^2$, and $\varepsilon_\perp \sim 2$--$3\,\Lambda_{\rm QCD}$
is achievable.  Therefore the answer is yes: effects arising from the
non-Markovian structure of the source term are likely to be exhibited in
contemporary relativistic heavy ion collisions.

\begin{figure}[t]
\centering{\ \epsfig{figure=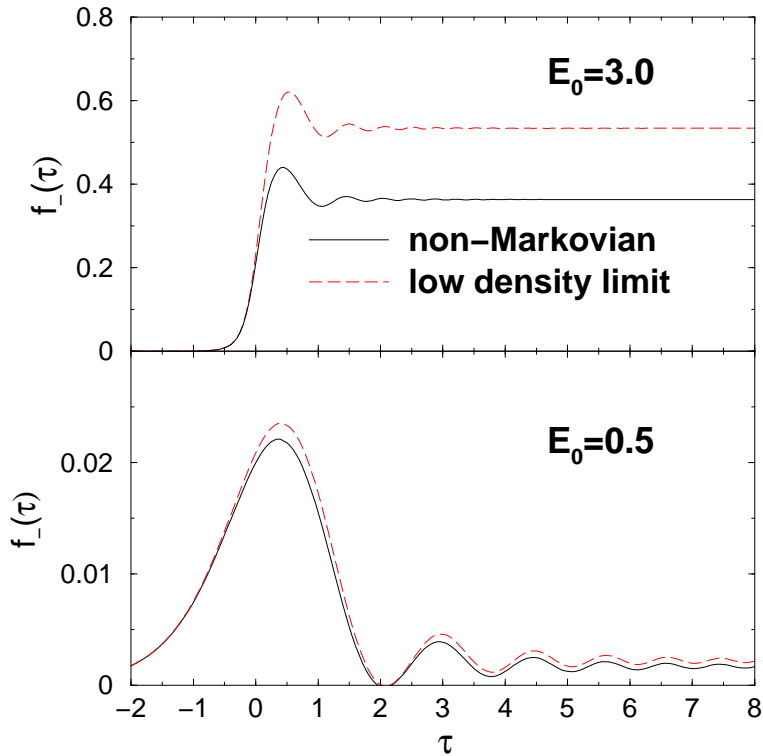,height=10cm,angle=-90}}\vspace*{1em}

\parbox{40em}{\caption{\label{fermcomp} Time-evolution in a constant electric
field of the complete single fermion distribution function cf.\ that obtained
in the low-density limit, Eq.~(\protect\ref{lowdensity}), with
$p_\|=0=p_\perp$ and initial condition $f(t\rightarrow -\infty)=0$.  {\it
Upper Panel}: Strong field, $eE_0/m^2=3.0$; {\it Lower Panel}: Weak field,
$eE_0/m^2=0.5$.
(Energy unit: $m\approx \Lambda_{\rm QCD}$.  Adapted from
Ref.~\protect\cite{PRD}.)}}
\end{figure}

This was illustrated in Ref.~\cite{PRD}, wherein a comparison was made
between the complete solution and that obtained in the low density limit;
i.e., retaining $\cos x(t^\prime,t)$ in the source term but making the
replacement $[1 \pm 2 f_\pm] \to 1$, which is the assumption
\begin{equation}
\label{lowdensity}
f_\pm \simeq 0\,,\;{\rm  almost~everywhere,}
\end{equation}
so that the source term becomes
\begin{equation}
\label{lds}
S^0_\pm(\vp,t) = \sfrac{1}{2} {\cal W}_\pm(t)\int_{-\infty}^tdt'{\cal
W}_\pm(t')\cos x(t^\prime,t)\,.
\end{equation} 
Using Eq.~(\ref{lds}), the solution of Eq.~(\ref{10}) is
\begin{equation}
f^0_\pm(\vp,t)=\int_{-\infty}^{\,t}\,dt^\prime S_\pm^0(\vp,t^\prime)\,,
\end{equation}
and is depicted for fermions in Fig.~\ref{fermcomp}. 
Particle production begins at $t=0$ because, with $p_\|=0$, the transition
amplitudes in Eq.~(\ref{12}) are maximal at this point [the vector potential
is zero], and the distribution function rapidly approaches a Schwinger-like
asymptotic value:
\begin{equation}
f^{\rm full}(t\rightarrow \infty) = \exp(-\pi\varepsilon_\perp^2/eE)
=
\begin{array}{ccc} e^{-\pi/3} & \approx & 0.35\; ({\rm strong})\\ 
e^{-2 \pi} & \approx & 0.0019\; ({\rm weak})
\end{array}\,.
\end{equation}
The qualitative features of the results depicted in Fig.~\ref{fermcomp} are
easily understood.  A fermion, once produced with a certain momentum, retains
it because there are no further interactions [collisions are ignored].  The
number density is greater for strong fields because they produce more
particles and this plainly means that the low-density limit will smoothly
become invalid with increasing field strength.  The low-density limit
overestimates the fermion distribution function in strong fields because it
eliminates Pauli blocking.  The opposite effect is seen for
bosons~\cite{PRD}.  For very strong fields: $f_- \to 1$, again because of
Pauli blocking, but there is no such bound on the boson distribution
function.  Finally, since $\sigma/\varepsilon_\perp^2 > 3$ is achievable in
contemporary relativistic heavy ion collisions, the low-density limit is
quantitatively unreliable.

\subsect{Back-reactions}
\label{backsection}
The illustration above assumed a constant electric field.  Allowing the more
realistic case of a time-dependent field introduces another time-scale,
namely that characterising the response of the field to the system's
evolution.  This brings us to the phenomenon of back-reactions, which have
been explored in connection with models in cosmology and recently much in
connection with QGP evolution.  In both cases the particles produced by the
strong background field modify that field: in cosmology it is the
time-dependent gravitational field, which couples via the masses, and in a
QGP, it is the chromoelectric field affected by the partons' colour charge.

The effect of feedback is incorporated by solving Maxwell's equation, which
for a spatially homogeneous but time dependent electric field is just
\begin{equation}
\label{max}
\dot E(t)= - j(t)\,.
\end{equation}
The total field is a sum of two terms: an external field, $E_{ex}(t)$,
excited by an external current, $j_{ex}(t)$, such as might represent a heavy
ion collision [that is a model input]; and an internal field, $E_{in}(t)$,
generated by the internal current, $j_{in}(t)$, which characterises the
behaviour of the particles produced.  The internal current has two
components~\cite{KM}: continued spontaneous production of charged particle
pairs creates a polarisation current, $j_{pol}(t)$, that depends on the
particle production rate, $S(\vp,t)$; and the motion of the existing
particles in the plasma generates a conduction current, $j_{cond}(t)$, that
depends on their momentum distribution, $f(\vp,t)$.  The internal field is
therefore obtained from
\begin{equation}\label{1.14}
-{\dot E}_{in}(t)= j_{in}(t)= j_{cond}(t)+j_{pol}(t)\,.
\end{equation}

In mean field approximation the currents can be obtained directly from the
constraint of local energy-density conservation: $\dot \epsilon = 0$, where
\begin{equation}
\label{totalenergy}
\epsilon(t) = \sfrac{1}{2} E^2(t) 
+ 2 \int\!\sfrac{d^3p}{(2\pi)^3}\,\omega(\vec{p},t)\,f(\vec{p},t)\,,
\end{equation}
and the factor of $2$ accounts for antiparticles.  For bosons the constraint
yields
\begin{equation}
\label{eEdiv}
{\dot E}(t) = -2e\int\!\sfrac{d^3p}{(2\pi)^3}\,\frac{p_\parallel-eA(t)}
{\omega(\vp,t)}\bigg[f(\vp,t)+\frac{\omega(\vp,t)}{{\dot
\omega}(\vp,t)}\frac{df(\vp,t)}{dt}\bigg]\,,
\end{equation}
and one easily identifies the currents
\begin{equation}
\label{jcond}
j_{cond}(t)=2e\int\!\sfrac{d^3p}{(2\pi)^3}\,
\frac{p_\parallel-eA(t)}{\omega(\vp,t)} f(\vp,t)\,,\;\;
j_{pol}(t)=\frac{2}{E(t)}\int\!\sfrac{d^3p}{(2\pi)^3}\,\omega(\vp,t)S(\vp,t)\,.
\end{equation}
Equation~(\ref{1.14}) now yields the internal field.  This construction has
been used extensively to study back-reactions; e.g.,
Refs.~\cite{Back,blochVE}.

The expression for the polarisation current exhibits the usual short-range
divergence associated with charge renormalisation, however, its
regularisation and renormalisation is straightforward~\cite{Back,blochVE}.
That accomplished, Maxwell's equation assumes the form
\begin{eqnarray}
\label{Edot}
\lefteqn{-{\ddot A}^\pm(t) = {\dot E}^\pm(t)= - j^{ex}(t)}\\
\nonumber && -g_\pm e
\int\!\sfrac{d^3P}{(2\pi)^3}\,\frac{P_\parallel(t)}{\omega(\wvp,t)}
\left[f_\pm(\wvp,t) + \frac{1}{2}\left\{ 2\frac{S(\wvp,t)}{{\cal
W}_\pm(\wvp,t)}-\frac{e\, \dot E^\pm(t)\,P_\parallel(t) }
{4\,\omega^4(\vec{P},t)} \right\}
\bigg(\frac{\epsilon_\perp}{P_\parallel(t)}\bigg)^{g_\pm-1}\right]\,,
\end{eqnarray}
where $g_-=2$, $g_+=1$, and all fields and charges are understood to be fully
renormalised.

\begin{figure}[t]
\centering{\ \epsfig{figure=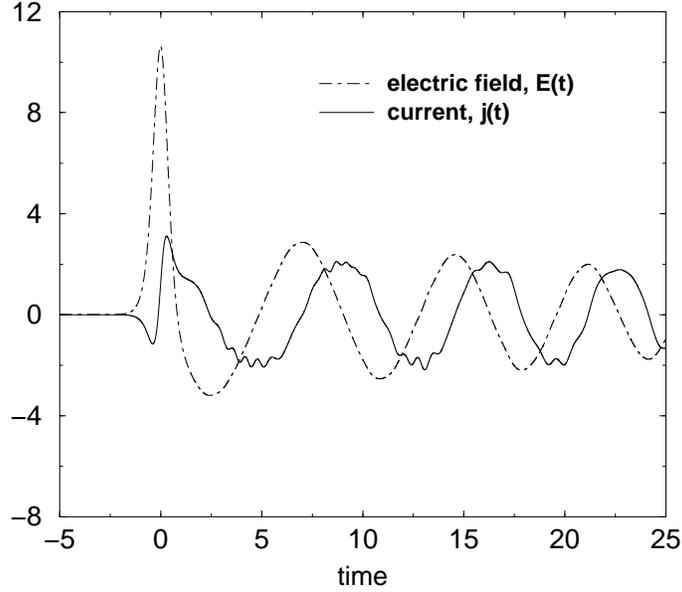,height=9cm,angle=-90}}
\parbox{40em}{\caption{\label{figplasma} Time evolution for bosons of the
electric field and current for an impulse external field,
Eq.~(\protect\ref{impulse}), with $A_0=10.0$, $b=0.5$ and the coupling
$e^2=5$.  The plasma oscillations driven by the back-reactions are evident.
A complete explanation of the qualitative features evident here is given
after Eq.~(\protect\ref{impulse}).  (Energy unit: $m\approx\Lambda_{\rm
QCD}$.  Adapted from Ref.~\protect\cite{blochVE}).}}
\end{figure}

The effect of back-reactions can now be illustrated by solving this equation
in concert with Eq.~(\ref{10}).  [Collisions are still neglected: $C\equiv 0$
in Eq.~(\ref{one}).]  A relativistic heavy ion collision can be mimicked by
an impulse profile for the external field~\cite{blochVE}:
\begin{equation}
\label{impulse}
E_{ex}(t)=\,-\,\frac{A_0}{b}\, {\rm sech}^{2}(t/b)\,,
\end{equation}
which ``switches-on'' at $t\sim -2 b$ and off at $t\sim 2 b$, with a maximum
magnitude of $E_{\rm max}=A_0/b$ at $t=0$.  Once this field has vanished only
the induced internal field remains to create particles and affect their
motion.  The calculated field and current profiles are depicted in
Fig.~\ref{figplasma}, and the qualitative features are easily understood.
The impulse electric field is evident at $t\simeq 0$.  It produces particles
and accelerates them, and their motion generates an internal current that
opposes the impulse field.  Shortly after the ``collision'' the current
reaches a short-lived plateau, when the total field vanishes and particle
production halts temporarily.  At about this time the external,
collision-mimicking field ``turns-off.''  Nevertheless, in its absence, the
total field grows in magnitude but now acts in the opposite direction,
decelerating the existing particles, causing new particles to be produced and
accelerating them in the new direction.  The effect of that is clear, after a
time the total current must vanish.  A pattern is now established and, in the
absence of other influences such as collisions or radiation, it repeats
itself, reaching a steady state once the wash from the collision-mimicking
impulse configuration disappears completely.  The oscillations characteristic
of a plasma with field-current feedback have now set-in.  The oscillation
period is the new time scale:
\begin{equation}
\tau_{\rm pl},\,{\rm the~plasma~period.}
\end{equation}
It decreases with increasing field strength.  

\begin{figure}[t]
\centering{\ \epsfig{figure=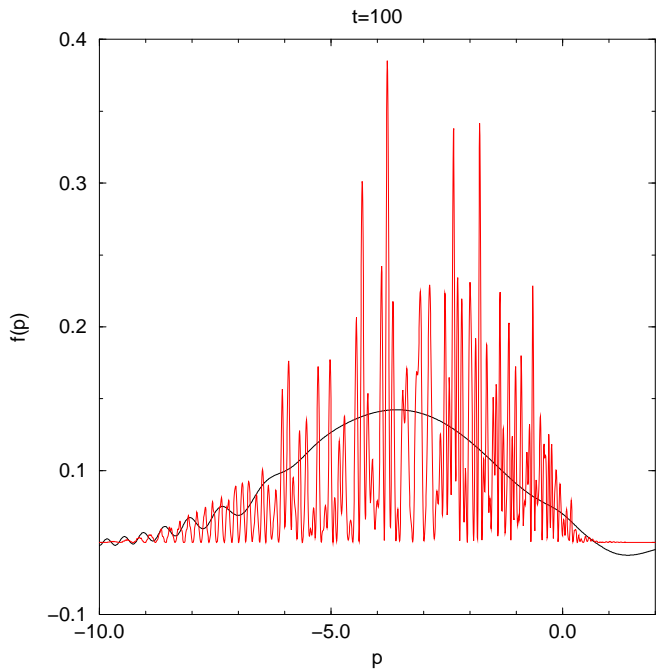,height=9cm}}
\parbox{40em}{\caption{\label{yuvalspic} $f(p,t={\rm fixed})$.  The rapid
oscillations are a quantum coherence effect, a manifestation of interference
between effects on the $\tau_{\rm qu}$ and $\tau_{\rm tu}$ time-scales.
Employing a local [or stationary phase or Markov] approximation to the source
term, these effects are averaged-out. (Energy unit: $m\approx\Lambda_{\rm
QCD}$.  Adapted from Ref.\protect\cite{kme}.)}}
\end{figure}
One more feature of the results is the high frequency oscillations evident at
the current's peaks and troughs.  They are {\bf not} a numerical artefact and
become more pronounced with increasing values of $eE/m^2$; i.e., when
$\tau_{\rm qu} \simeq \tau_{\rm tu}$.  This makes plain that they are a
non-Markovian feature and result from interference between effects on the
tunnelling and quantum time-scales.  Of course, as illustrated via an
analogous feature in Fig.~\ref{yuvalspic}, they disappear if a local
approximation to the source term is used~\cite{kme} because the stationary
phase approximation suppresses such interference effects.

One additional observation is important here.  Our example employed an
electric field whose magnitude, $\propto A_0$, is large and hence the
tunneling time, $\tau_{tu}$, is small, being inversely proportional to $eE$.
The period of the plasma oscillations, $\tau_{pl}$, also decreases with
increasing $A_0$ but nevertheless, as clear in Fig.~\ref{figplasma},
$\tau_{tu} \ll \tau_{pl}$.  Thus, in contrast to the effect it has on the
production process~\cite{PRD}, the temporal nonlocality of the non-Markovian
source term is unimportant to the collective plasma
oscillation~\cite{blochVE}.  This is the reason why kinetic equations with a
simple source term of the form in Eq.~({\ref{WKB}) are successful in
describing plasma oscillations~\cite{Back}.  Whether or not these
oscillations are observable in relativistic heavy ion collisions depends on
the effect of dissipative processes, which we now discuss.

\subsect{Collisions and Evolution of the Quark-Gluon Plasma}
Thus far we have illustrated the phenomena of quantum particle production in
strong fields and field-current feedback.  The stationary state is
unrealistic because the dissipative processes: ``collisions,'' that lead to
thermalisation have been neglected.  [Collisions can also lead to particle
production and effect hadronisation but we defer that discussion.]  

In the presence of collisions the kinetic equation takes the form
\begin{equation}
\label{coll}
\frac{df_\pm(\vp,t)}{dt}=S_\pm(\vp,t) + C_\pm(\vp,t)\,.
\end{equation}
A simple and widely-used model~\cite{GKM,flor,bhal,eis,memory} for the
collision term is
\begin{equation}
\label{collterm}
C_\pm(\vp,t) =\frac{f_\pm^{eq}(\vp,t) - f_\pm(\vp,t)}{\tau_{\rm r}}\,,
\end{equation}
where $\tau_{\rm r}$ is the ``relaxation time''  and
$f^{eq}$ is the thermal equilibrium distribution function
\begin{equation}
\label{feq} 
f_\pm^{eq}(\vp,t) = \frac{1}{\exp[\omega(\vp,t)/T(t)]\mp 1}\,, 
\end{equation}
with $T(t)$ a time-dependent temperature profile, discussed on
page~\pageref{`temperatureprofile'}.  The relaxation time is a fourth
time-scale and it is plain that plasma oscillations can only be observable if
$\tau_{\rm pl}\ll \tau_{\rm r}$; i.e., if collisions take place much-less
frequently than oscillations.

In many of the exploratory calculations hitherto undertaken, $\tau_{\rm r}$
is a constant.  That might be argued to be inadequate because the collision
time is supposed to characterise a system's thermalisation, a process which
is interdependent with time-evolving quantities such as density and
temperature.  A more realistic representation might therefore employ a
time-dependent $\tau_{\rm r}$, which is calculated self-consistently as the
plasma evolves.  Reference~\cite{nayak2}, employing collisions in a
gluon-minijet plasma, is a step in that direction.  However, from another
perspective, any sophisticated collision term should be derived from, and
justified by, an underlying microscopic theory (see; e.g.,
Refs.~\cite{Elze89,Rau,Boterman}.); and, furthermore, a relaxation time
approximation of any sort can only be valid under conditions of
quasi-equilibrium, which cannot be justified {\it a priori} in the presence
of strong fields.  A truly realistic collision term will introduce
non-Markovian effects in addition to those already present in the source
term, and exemplary studies exist in connection with: relativistic heavy ion
collisions~\cite{Cgreiner}; collective effects in nuclear
matter~\cite{schuck}; nuclear fragmentation~\cite{Colonna}; and the damping
rates of giant dipole resonances~\cite{fuhrman}.  These studies make clear
that even a binary collision approximation, as characteristic of a
Boltzmann-Uehling-Uhlenbeck kinetic equation, can be inadequate under extreme
conditions; e.g., in the presence of strong fields and/or when the particle
density is high.  These are precisely the conditions relevant to QGP
formation.  The patent complexity justifies the use in exploratory,
illustrative studies of the simple $\tau_{\rm r}=\,$constant relaxation time
approximation.  Improvements are a contemporary challenge.

\begin{figure}[t]
\centering{\ \epsfig{figure=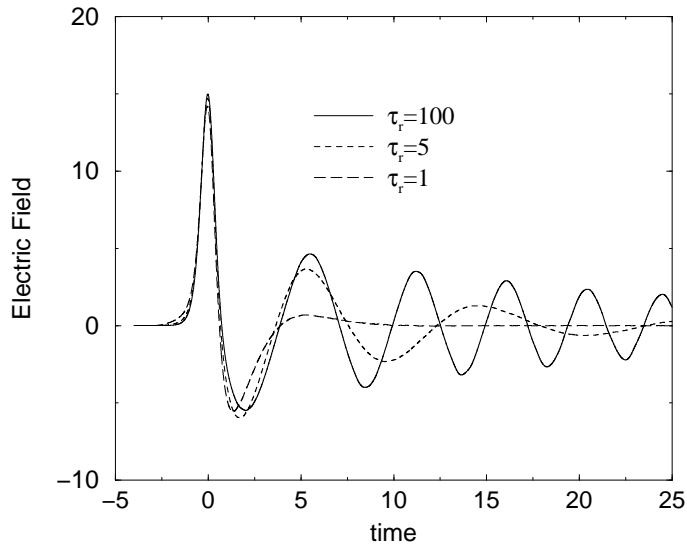,height=9cm,angle=-90}}
\parbox{40em}{\caption{\label{fig12} Time evolution for bosons of the
electric field obtained using different relaxation times in the collision
term of Eq.~(\protect\ref{collterm}), and with the impulse external field,
Eq.~(\protect\ref{impulse}), where $A_0=7.0$, $b=0.5$ and the coupling
$e^2=4$.  (Energy unit: $m\approx\Lambda_{\rm QCD}$.  Adapted from
Ref.~\protect\cite{blochVE}.)}}
\end{figure}

As intuition suggests, collisions effect a damping of the distribution
functions' time-dependent structure.  That is illustrated in
Fig.~\ref{fig12}, with this particular calculation obtained using an {\it
Ansatz} for the temperature profile~\cite{blochVE}
\begin{equation}
\label{TAnsatz}
T(t) = T_{eq} + (T_m - T_{eq})\, {\rm e}^{-t^2/t_0^2}\,,
\end{equation}
where $T_{eq}=\Lambda_{\rm QCD}$, $T_m=2\,T_{eq}$, $t_0^2=10/\Lambda_{\rm
QCD}$.  [The definition and meaning of $T(t)$ is discussed below.]  As
evident in the figure, for large relaxation times the plasma oscillations are
unaffected, as might be anticipated because this is the collisionless limit.
However, for $\tau_{\rm r}\sim \tau_{\rm pl}$, the collision term has a
significant effect, with both the amplitude and frequency of the plasma
oscillations being damped.  Finally, there is a value of $\tau_{\rm r}$ below
which oscillations are not possible, just as in the case of an overdamped
harmonic oscillator, and the system evolves quickly and directly to thermal
equilibrium.  The time taken by the plasma to thermalise depends on the ratio
$\tau_{\rm pl}/\tau_{\rm r}$, being longer for larger values.

Why use the temperature profile in Eq.~(\ref{TAnsatz}) and, indeed, what is
temperature in a system far from equilibrium?\label{`temperatureprofile'} The
notion of temperature is introduced via an assumption of quasi-equilibrium at
each time $t$, which is only truly valid if the fields are not too strong;
i.e., only as long as $\tau_{\rm tu}\gg\tau_{\rm pl}$.  This temporally local
temperature can be calculated self-consistently with the evolution of the
distribution functions, an approach which represents a slight improvement
over employing {\it Ans\"atze}, such as Eq.~(\ref{TAnsatz}), and, in fact,
can provide an {\it a posteriori} justification for that expedient.  The
calculation of $T(t)$ can, however, proceed in a number of ways so that the
resulting profile is not unique but the differences are only small and
quantitative.

To illustrate the definition of a temperature profile we follow
Ref.~\cite{memory}.  The total energy density in the evolving plasma is given
in Eq.~(\ref{totalenergy}), where the second term is the particle
contribution: $\epsilon_f(t)$, and the particle number density is
\begin{equation}
\label{number}
n(t) = 2\int\!\sfrac{d^3p}{(2\pi)^3}\,f_-(\vp,t)\,.
\end{equation}
An intuitive definition of the local temperature is to require that, at each
$t$, the average energy-per-particle in the evolving plasma is that of a a
quasi-equilibrium gas; i.e.,
\begin{equation}
\label{eneneq}
\frac{\epsilon_f(t)}{n(t)} = \frac{\epsilon^{eq}(t)}{n^{eq}(t)}\,,
\end{equation}
where
\begin{equation}
\label{neq}
\epsilon^{eq}(t) = 2\,\int\!\sfrac{d^3p}{(2\pi)^3}\,
\omega(\vp,t)f_-^{eq}(\vp,t)\,, \;\;
n^{eq}(t) = 2\,\int\!\sfrac{d^3p}{(2\pi)^3}\, f_-^{eq}(\vp,t)\,.
\end{equation}
With $f_-^{eq}$ given in Eq.~(\ref{feq}), Eq.~(\ref{eneneq}) is an implicit
equation for $T(t)$, which must be solved simultaneously with
Eqs.~(\ref{Edot}) and (\ref{coll})~\cite{memory}.  

\begin{figure}[t]
\centering{\ \epsfig{figure=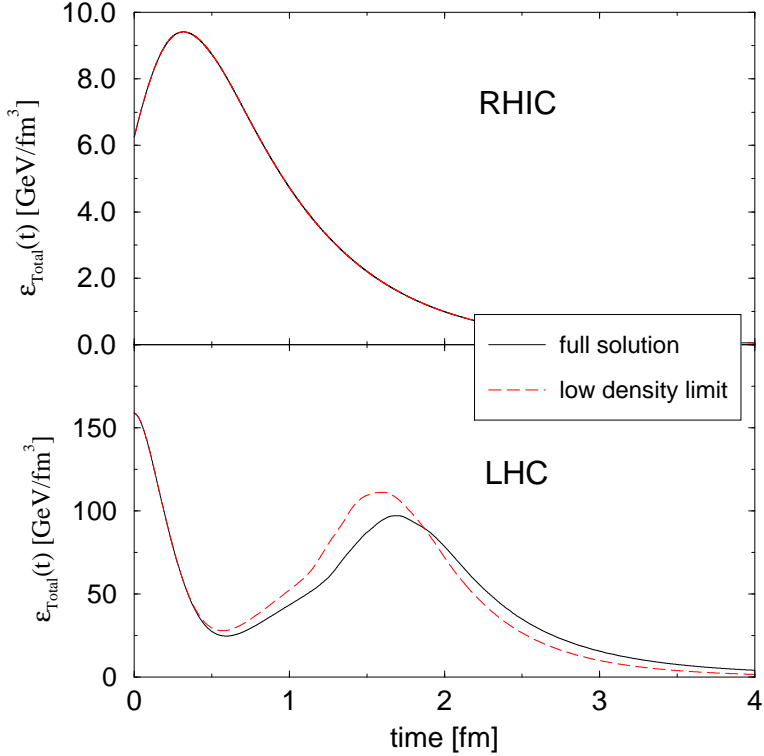,height=10cm,angle=-90}}\vspace*{1em}

\parbox{40em}{\caption{\label{fig1} Time evolution of the fermion's energy
density for RHIC (upper panel) and LHC (lower panel) conditions.  For RHIC
conditions, the result obtained using the low density approximation,
Eq.~(\protect\ref{lds}), is indistinguishable from the complete solution.
That is not the case for LHC-like conditions.  (Adapted from
Ref.~\protect\cite{memory}.)}}
\end{figure}

In Ref.~\cite{memory}, this system of equations was solved with
$m=\Lambda_{\rm QCD}$, $e=2$, and $\tau_{\rm r}= 1/\Lambda_{\rm QCD} \approx
1\,$fm, and using two exemplary impulse field configurations:
Eq.~(\ref{impulse}), with $b=0.5/\Lambda_{\rm QCD} \approx 0.5\,$fm and
either $A_0^{RHIC} = 4\,\Lambda_{\rm QCD}$ or $A_0^{LHC} = 20\,\Lambda_{\rm
QCD}$.  As evident in Fig.~\ref{fig1}, $A_0^{RHIC}$ yields a calculated
initial energy-density characteristic of RHIC conditions [see
Sec.~\ref{prelim}] while $A_0^{LHC}$ gives a very much greater value
characteristic of the initial energy-density expected at LHC.  For RHIC
conditions the energy-density rises rapidly but, after reaching a maximum,
decays monotonically.  The low density approximation is valid here.  For LHC
conditions, with an initial energy-density twenty-times larger, the situation
is different: the solution obtained in the low density limit is only a
qualitative guide to the plasma's evolution and plasma oscillations are
evident on {\it observable} time-scales.  These oscillations retard the
equilibration of the plasma, roughly doubling the time taken cf. RHIC-like
conditions.  This application makes evident an expected terrestrial hierarchy
of time-scales:
\begin{equation}
\tau_{\rm qu}\sim\tau_{\rm tu}<\tau_{\rm pl}\sim\tau_{\rm r}\,.
\end{equation}

In a very intense field the produced particles travel almost at the speed of
light and hence the plasma oscillation period can be estimated via the
ultrarelativistic formula~\cite{Kluger:1993}:
\begin{equation}
\sfrac{1}{2}\tau^{UR}_{\rm pl} \approx \frac{\sqrt{n_{\rm max}\, m+ E_{\rm
max}^2}}{n_{\rm max}\,e}\approx 2\,,
\end{equation}
using the input and LHC-like results of Ref.~\cite{memory}, which yield
$n_{\rm max}\simeq 12/\Lambda_{\rm QCD}^3$ from Eq.~(\ref{number}).  That
this is a reliable guide for LHC-like conditions is evident in
Fig.~\ref{fig1}, remembering that $\epsilon(t) \sim E^2(t)$ so that the peaks
exhibited in the lower panel are separated by $\sfrac{1}{2}\tau_{\rm pl}$.

\begin{figure}[t]
\centering{\ \epsfig{figure=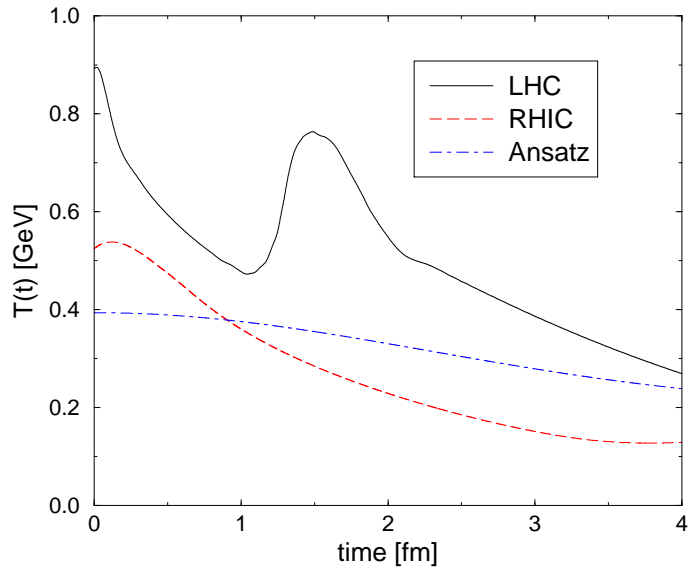,height=9cm,angle=-90}}
\parbox{40em}{\caption{\label{temperaturepic} Time evolution of the
temporally local quasi-equilibrium temperature for RHIC and LHC initial
conditions.  For comparison we also plot the {\it Ansatz} in
Eq.~(\protect\ref{TAnsatz}), which is a fair guide for RHIC-like conditions.
(Adapted from Ref.~\protect\cite{memory}.)}}
\end{figure}
The calculated temperature profile is depicted in Fig.~\ref{temperaturepic}.
Under RHIC-like conditions, an initial temperature of $T(t=0)\sim 0.5\,$GeV
is reached and the temperature decreases monotonically.  The LHC-like source
conditions yield an initial temperature twice as large: $T^{\rm LHC}(t=0)
\sim 0.9\,$GeV, and the temperature fluctuates in tune with the energy
density.

As described above, the exemplary impulse models in Ref.~\cite{memory} set
initial energy density scales appropriate for RHIC and LHC.  An improvement
over such {\it Ans\"atze} is to calculate the initial conditions, as done;
e.g., in Refs.~\cite{nayak,nayak2} for the case of a gluon minijet plasma.
Even with this improvement, however, the results obtained for observable
quantities; such as, $\epsilon(t)$, $n(t)$, the temperature and equilibration
time, are semi-quantitatively identical and all features are qualitatively
unchanged.

It is a uniform prediction that plasma oscillations are present under
LHC-like conditions and they may manifest themselves in the dilepton
spectrum.  The dilepton production rate is (see; e.g., Ref.~\cite{wongHIC})
\begin{equation}
\label{dilepton}
\frac{d\,N_{l^+l^-}}{dt\,d^3x}\sim \int d^3p_1d^3p_2\,
f(E_1)\,f(E_2)\,\sigma(M)\,v_{12}\,,
\end{equation}
where $f$ is the distribution function of the charged elements in the plasma,
$\sigma(M)$ is the cross-section for particle-antiparticle annihilation into
a dilepton pair with invariant mass $M$ and $v_{12}$ is the standard
phase-space factor [a ``relative particle-antiparticle velocity''].  The
distribution function responsible for the features in Fig.~\ref{fig1} itself
exhibits structure and, according to Eq.~(\ref{dilepton}), that will be
transmitted to the dilepton spectrum.  The oscillations characterising the
pre-equilibrium phase of the plasma could then be exhibited as repeated
dilepton bursts.  However, this signal will only be detectable if it is
strong enough relative to other process, such as Drell-Yan.  That
quantitative question is currently unanswered.

\subsect{Hadronisation}
The penultimate stage of QGP evolution is hadronisation; i.e., the
re-establishment of the conditions that prevail at zero density and
temperature.  That process in not instantaneous and hence here too a
dynamical approach is necessary, one that provides the means of describing
phenomena such as bound state formation, and the onset of DCSB and
confinement.  Consequently the development of a transport theory in which
such effects are directly accessible is an important current focus.

Relativistic transport theory and hydrodynamics have long been useful tools
in the study of non-equilibrium states of matter at nonzero density and
temperature.  They have been employed extensively in relativistic nuclear
physics, both in the intermediate energy region~\cite{Boterman,BKM},
where it may be appropriate to neglect quark and gluon degrees of freedom,
and in the high energy domain directly relevant to a non-equilibrium
QGP~\cite{Geiger95,MH,ZW,neisel}.  One approach~\cite{smolann} to the
derivation of Vlasov-like transport equations can be described as a
relativistic generalisation of the Zubarev method~\cite{ZMR}.  Another is the
contour Green function technique~\cite{KBMSKel}, which has been applied
formally; e.g., in Refs.~\cite{MH,ZW,neisel}, and also in quantitative model
studies; e.g., Refs.~\cite{CMmeshustin,ole,aichelin,huefner,rehberg}.
The latter is widely used because it admits a systematic definition and
exploration of approximations, and will be our focus.

To be concrete, extant phenomenological applications begin with a model
Lagrangian density; e.g., that of the Friedberg-Lee or Nambu--Jona-Lasinio
model.  Such models are chosen because of their efficacy in describing DCSB,
and have been employed in studies of: the spacetime evolution of the
dressed-quark mass~\cite{CMmeshustin,ole,aichelin,huefner}, which
characterises the onset of DCSB; and hadronisation via the models'
fermion-antifermion--bound-state interaction terms~\cite{rehberg}.  In the
former case, the onset of DCSB is explored by solving a quantum Vlasov
equation {\it coupled} with a distribution-function-dependent gap equation,
which describes the evolution of the particles' mass and feeds this
information back into the system.  One important qualitative result is a
parton mass that increases as the system moves toward equilibrium; e.g.,
Ref.~\cite{ole}, which results in a dynamical softening of the EOS.  This
softening can have observable consequences; e.g., slowing the
[hydrodynamical] expansion of the plasma~\cite{rischke}.  These models,
however, do not incorporate confinement and hence cannot describe the {\it
replacement} of coloured degrees of freedom by hadronic matter through the
transition.

An indication that the onset of confinement might significantly affect the
evolution of the plasma was observed in Ref.~\cite{huefner}, wherein a
$T$-dependent dressed-quark mass was introduced by fitting the model's
quasi-particle energy density to that determined in lattice simulations.  A
self-consistent solution of the coupled Vlasov and gap equations then
exhibits confinement; i.e., the partons cannot leave the QGP volume because
their mass becomes infinite outside this region.  The model does not include
a parton-antiparton--hadron interaction and hence the confining effect also
stops the hydrodynamical expansion.  This study only admitted the importance
of the scalar piece of the dressed-quark self energy.  However, as we saw;
e.g., in connection with Fig.~\ref{pressurePic}, the vector self energy is
more important in QCD's deconfined domain.  The effect it has on plasma
evolution has not hitherto been explored.

The incorporation of such qualitatively robust features of DSE studies into
the Vlasov equation is an obvious next step~\cite{bastihirschegg}.  It is
complicated for [at least] two reasons: 1) The quasi-particle energy is a
functional of the scalar and vector dressed-quark self energies, which are
both nonzero and momentum dependent.  The scalar self energy is small in the
QGP.  [It vanishes in the chiral limit.]  However, the significant vector
self energy remains~\cite{bastiscm}; 2) In the confined domain the
dressed-parton $2$-point functions violate reflection positivity and hence a
single particle distribution function for these excitations cannot be
defined.  That is as it should be because hadrons are the relevant degrees of
freedom in this phase: a kinetic theory based on quarks and gluons is only
appropriate in the deconfined QGP phase.  As described in connection with
Fig.~\ref{confPic}, DSE models can adequately represent the transition from
confined excitation to propagating quasiparticle~\cite{axelT}, an attribute
exploited in Ref.~\cite{bastiscm} in calculating the thermodynamic functions
describing a QGP.  These observations suggest that a truly realistic
description of the quark and gluon distribution functions in a QGP will see
them vanish in the vicinity of the phase boundary.

We close this section with an illustration of the concepts introduced.  In a
QGP the evolution of the dressed-quark distribution function can be described
using a Vlasov equation with the quasi-particle energy
\begin{equation}
\label{v1}
E_\ast(\vec{p},\vec{x},t)=p_0\,[1+\Sigma_C(\vec{p},\vec{x},t)]= \sqrt{|\vec
p_\ast|^2 + B(\vec{p},\vec{x},t)^2}\,,
\end{equation}
where ${\vec p}_\ast= {\vec p}\,[1+\Sigma_A(\vec{p},\vec{x},t)]$ is the
rescaled three-momentum and $B(\vec{p},\vec{x},t) = [m +
\Sigma_B(\vec{p},\vec{x},t)]$.  Using Eqs.~(\ref{renself}), Eq.~(\ref{v1})
can be rewritten
\begin{equation}
E_\ast(\vec{p},\vec{x},t) = C(\vec{p},\vec{x},t)\, E(\vec{p},\vec{x},t)
=C(\vec{p},\vec{x},t)\,\sqrt{ |\vec p|^2 \, {\cal I}(\vec{p},\vec{x},t)^2 +
{\cal M}(\vec{p},\vec{x},t)^2}\,,
\end{equation}
where ${\cal I} = A/C$ and ${\cal M}=B/C$.  As our exemplar we follow
Ref.~\cite{bastihirschegg} and employ an instantaneous variant of the ${\cal
D}_{\rm A}$ model in Sec.~\ref{DAmodelsection}:
\begin{equation}
{\cal D}(p_{\Omega_k}) = 3 \pi^2\,\eta\,\delta^3(\vec{p})\,.
\end{equation}
With this interaction and using rainbow truncation, which is akin to a
mean-field approximation, the Matsubara sum can be evaluated algebraically
because the self energies are $p_0$-independent, and the gap equation assumes
the form
\begin{equation}
\label{sigBC}
B(\vec{p},\vec{x},t)= m +
\eta\,\frac{B(\vec{p},\vec{x},t)}
{E_\ast(\vec{p},\vec{x},t)}{f_\ast}(\vec{p},\vec{x},t)\,,\;\;
C(\vec{p},\vec{x},t) = 1 +
\sfrac{1}{2}\eta\,\frac{C(\vec{p},\vec{x},t)}{E_\ast(\vec{p},\vec{x},t)}{f_\ast}(\vec{p},\vec{x},t)\,,
\end{equation}
with $A=C$, so that ${\cal I} = 1$, and $f_\ast=f/C$.  [NB.  $f$ appears
explicitly here because, with a quasiparticle pole, the Matsurbara sum can be
evaluated algebraically.]  Again, in spite of the model's simplicity, the
solution exhibits qualitative features that are characteristic of a realistic
dressed-quark $2$-point function; e.g., momentum-dependent scalar and vector
self energies, and the persistence of this aspect of the solutions in the
deconfined domain~\cite{bastihirschegg}.  In this class of models, for a
collisionless plasma, $f_\ast$, satisfies~\cite{ZW}
\begin{eqnarray}
\label{sdevlasov}
\lefteqn{ 0 =\partial_t{f_\ast}(\vec{p},\vec{x},t) }\\
&& \nonumber + \frac{1}{E(\vec{p},\vec{x},t)}\, \left\{ \left[ {\vec p} +
{\cal M}(\vec{p},\vec{x},t) \vec{\nabla}_p\, {\cal M}(\vec{p},\vec{x},t)
\right]\cdot \vec{\nabla}_x{f_\ast}(\vec{p},\vec{x},t)
- {\cal M}(\vec{p},\vec{x},t) \vec{\nabla}_x {\cal M}(\vec{p},\vec{x},t) \cdot
                  \vec{\nabla}_p{f_\ast}(\vec{p},\vec{x},t) \right\}\,.
\end{eqnarray}
It is important to note that the dressed-quark mass function is
momentum-dependent so that in general
\begin{equation}
\vec{\nabla}_p\, {\cal M}(\vec{p},\vec{x},t) \cdot
                  \vec{\nabla}_x\,{f_\ast}(\vec{p},\vec{x},t)\neq  0\,.
\end{equation}
[NB.  The Vlasov equation in Ref.~\cite{huefner} is obtained by neglecting
this term, since $\vec{\nabla}_p M\equiv 0$ when the interaction is
momentum-independent, and setting $C=1$.]  Equations~(\ref{sigBC}) and
(\ref{sdevlasov}) provide a coupled system for the quark's distribution
function and self energies.

As remarked above, the derivation of a transport equation like
Eq.~(\ref{sdevlasov}) requires the existence of a quasi-particle mass shell.
The instantaneous model ensures that, with $p^0=E(\vec{p},\vec{x},t)$.  In
this case
$\frac{\partial f_\ast}{\partial p^0}=0$
and $\frac{\partial {\cal M}}{\partial p^0}=0$, hence Eq.~(\ref{sdevlasov})
can be written
\begin{equation}
p^\mu \,\partial^x_\mu f_\ast 
+ {\cal M} \left\{
\partial^x_\mu {\cal M}  \, \partial_p^\mu f_\ast
- \partial^p_\mu {\cal M}  \,\partial_x^\mu f_\ast \right\} = 0\,,
\end{equation}
using the Minkowski space metric conventions of Ref.~\cite{bd65}, which is
directly comparable with the equation described; e.g., in Ref.~\cite{neisel}.
However, the momentum dependence of ${\cal M}$, here driven explicitly via
Eqs.~(\ref{sigBC}), precludes a simple spherical space-volume scaling
solution~\cite{CMmeshustin,bozek}.

The coupled system, Eqs.~(\ref{sigBC}), (\ref{sdevlasov}), illustrates some
of the complexity to be anticipated in studying the re-emergence of DCSB and
confinement in an expanding QGP.  Even in this simple model, as in
Ref.~\cite{huefner}, proceeding further requires a numerical solution.  Such
studies are in their infancy but the qualitative feature is plain: the
nontrivial propagation characteristics of the dressed-partons will
significantly affect $f$, and hence QGP evolution, as $T\to T_c$ and the
distribution of partons begins to resemble a heat bath.

\sect{Epilogue} 
Continuum strong QCD is a broad field and we have only presented a snapshot.
Nevertheless, our discourse should serve to introduce many of the topics
currently occupying practitioners.  Since we employed the Dyson-Schwinger
equations as our primary medium, it should also provide an update of the
progress that has been made with this tool in the last decade and identify
the current challenges.  The text provides the details.  However, highlights
of the progress include: the application of a one-parameter model for the
dressed-gluon $2$-point function in a description of the masses, decays and
form factors of light pseudoscalar and vector mesons, and its simultaneous
prediction of the critical temperature for quark gluon plasma formation and
the properties of in-plasma correlations; significant progress in calculating
the baryon spectrum, and leptonic and nonleptonic interactions involving
baryons; a unified treatment of chemical potential and temperature, and their
effect on the equation of state for a quark gluon plasma and hadron
properties; and an incipient understanding of the evolution to equilibrium of
a quark gluon plasma, and the dynamical influence of the re-appearance of
confinement and dynamical chiral symmetry breaking on the plasma's expansion
and cooling.  On the other side, a primary challenge is to comprehend the
origin of the infrared enhancement in the kernel of the QCD gap equation that
is necessary to ensure dynamical chiral symmetry breaking.  In summarising we
cannot improve on Ref.~\cite{mikeCsQCD}: ``This programme has a long way to
go, but [we] hope you are convinced it has come far.''

While the Dyson-Schwinger equations have provided the backbone for our
discussion, we have made connections throughout with the results of other
methods.  Where there is agreement there can be little doubt that the
phenomena described are real.  That is the rationale underlying the
simultaneous pursuit of complementary methods.  Disagreement, in fact or
interpretation, provides a challenge, which should be met, but also an
opportunity for dialogue that should not be missed.

\parbox{185mm}{{\Large\bf\sc Acknowledgments}\\[\baselineskip]
We are grateful to
M.B.~Hecht,
P.~Maris
and P.C.~Tandy
for careful readings of the manuscript and useful comments.
We note too that, while there is no dearth of references, fallibility ensures
we have overlooked some contributions.  Such omissions are inadvertent.
This work was supported by the US Department of Energy, Nuclear Physics
Division, under contract no. W-31-109-ENG-38.  S.M.S. is grateful for
financial support from the A.v.~Humboldt foundation.}

%


\begin{thebibliography}{100}\vspace*{-\baselineskip}
\addcontentsline{toc}{section}{\protect\numberline{ }{\bf\sc References}}
\itemsep=0pt
%
\bibitem{mikeCsQCD} M.R. Pennington, ``Calculating hadronic properties in
strong QCD,'' hep-ph/9611242.
%
\bibitem{iz80} C. Itzykson and J.-B Zuber, Quantum Field Theory (McGraw Hill,
New York, 1980).
%
\bibitem{rivers} R.J. Rivers, Path integral methods in quantum field theory
(Cambridge University Press, Cambridge, 1987).
%
\bibitem{bd65} J.D. Bjorken and S.D. Drell, Relativistic Quantum Fields
(McGraw Hill, New York, 1965).
%
\bibitem{cdragw} C.D.~Roberts and A.G.~Williams, Prog.\ Part.\ Nucl.\ Phys.\
{\bf 33} (1994) 477.
%
\bibitem{gj81} J. Glimm and A. Jaffee, Quantum Physics. A Functional Point of
View (Springer-Verlag, New York, 1981).
%
\bibitem{seiler} E. Seiler, Gauge Theories as a Problem of Constructive
Quantum Theory and Statistical Mechanics (Springer-Verlag, New York, 1982). 
%
\bibitem{gross75} D.J.~Gross, ``Applications Of The Renormalization Group To
High-Energy Physics,'' in Proc. of Les Houches 1975,  Methods In
Field Theory (North Holland, Amsterdam, 1976) pp. 141-250.
%
\bibitem{zwanzigerold} D.~Zwanziger, Nucl.\ Phys.\ {\bf B 364} (1991) 127.
%
\bibitem{zwanziger} D.~Zwanziger, Nucl.\ Phys.\ {\bf B 412} (1994) 657.
%
\bibitem{revpeter} P.C.~Tandy, Prog.\ Part.\ Nucl.\ Phys.\ {\bf 39} (1997)
117.
%
\bibitem{vw91} U.~Vogl and W.~Weise, Prog.\ Part.\ Nucl.\ Phys.\  {\bf 27}
(1991) 195;
T.~Hatsuda and T.~Kunihiro, Phys.\ Rept.\ {\bf 247} (1994) 221.
%
\bibitem{njltwo} G.~Ripka and M.~Jaminon, Annals Phys.\ {\bf 218} (1992) 51;
D.~Ebert, H.~Reinhardt and M.K.~Volkov, Prog.\ Part.\ Nucl.\ Phys.\ {\bf 33}
(1994) 1.
%
\bibitem{sk92} S.P.~Klevansky, Rev.\ Mod.\ Phys.\ {\bf 64} (1992) 649.
%
\bibitem{gcm98} R.T.~Cahill and S.M.~Gunner, Fizika {\bf B 7} (1998) 171.
%
\bibitem{bjw64} K.~Johnson, M.~Baker and R.~Willey, Phys.\ Rev.\ {\bf 136}
(1964) B1111.
%
\bibitem{mrpAdelaide} M.R.~Pennington, ``Mass production requires precision
engineering,'' in Proc. of the Workshop on Nonperturbative Methods in Quantum
Field Theory, edited by A.W. Schreiber, A.G. Williams and A.W. Thomas (World
Scientific, Singapore, 1998) pp. 49-60.
%
\bibitem{fred94} F.T.~Hawes and A.G.~Williams, Phys.\ Rev.\ {\bf D 51} (1995)
3081.
%
\bibitem{manuel} M.~Reenders, Dynamical symmetry breaking in the gauged
Nambu-Jona-Lasinio model, PhD Thesis, University of Groningen (1999);
M.~Reenders, ``On the nontriviality of Abelian gauged Nambu-Jona-Lasinio
models in four dimensions,'' hep-th/9908158; and references therein.
%
\bibitem{mr97} P.~Maris and C.D.~Roberts, Phys.\ Rev.\  {\bf C 56} (1997)
3369. 
%
\bibitem{sizer} A.W.~Schreiber, T.~Sizer and A.G.~Williams, ``Dimensionally
regularized study of nonperturbative quenched QED,'' in Proc. of the Workshop
on Nonperturbative Methods in Quantum Field Theory, edited by A.W. Schreiber,
A.G. Williams and A.W. Thomas (World Scientific, Singapore, 1998)
pp. 299-307.
%
\bibitem{kusaka} K.~Kusaka, H.~Toki and S.~Umisedo, Phys.\ Rev.\ {\bf D 59}
(1999) 116010.
%
\bibitem{gastao} G.~Krein, C.D.~Roberts and A.G.~Williams, Int.\ J.\ Mod.\
Phys.\ {\bf A 7} (1992) 5607.
%
\bibitem{cp90} D.~C.~Curtis and M.R.~Pennington, Phys.\ Rev.\ {\bf D 42}
(1990) 4165.
%
\bibitem{dong} Z.~Dong, H.J.~Munczek and C.D.~Roberts, Phys.\ Lett.\ {\bf B
333} (1994) 536.
%
\bibitem{bashir} A.~Bashir, A.~Kizilersu and M.R.~Pennington, Phys.\ Rev.\
{\bf D 57} (1998) 1242.
%
\bibitem{ayse} A.~Bashir, A.~Kizilersu and M.R.~Pennington, ``Analytic form
of the one-loop vertex and of the two-loop fermion propagator in
3-dimensional massless QED,'' hep-ph/9907418.
%
\bibitem{adnan} A.~Bashir and M.R.~Pennington, Phys.\ Rev.\ {\bf D 50}
(1994) 7679.
%
\bibitem{pt84} P.~Pascual and R.~Tarrach, QCD: Renormalization For The
Practitioner (Springer-Verlag, Berlin, 1984).
%
\bibitem{bc80} J. S. Ball and T.-W. Chiu, Phys. Rev. {\bf D 22} (1980) 2542.
%
\bibitem{vertex} See also: S. K. Kim and M. Baker, Nucl. Phys. {\bf B 164}
(1980) 152; and J. S. Ball and T.-W. Chiu, Phys. Rev. D {\bf 22} (1980) 2550.
%
\bibitem{fredthesis} F.~T.~Hawes, Fermion Dyson-Schwinger studies in QED and
QCD: Comparisons of ansatz for boson propagator and vertex, PhD Thesis,
Florida State University (1994).
%
\bibitem{mn83} H.J.~Munczek and A.M.~Nemirovsky, Phys.\ Rev.\ {\bf D 28}
(1983) 181.
%
\bibitem{pdg98} C.~Caso {\it et al.}, Eur.\ Phys.\ J.\ {\bf C 3} (1998) 1.
%
\bibitem{fukuda} R.~Fukuda and T.~Kugo, Nucl.\ Phys.\ {\bf B 117} (1976) 250.
%
\bibitem{cahillQED} C.D.~Roberts and R.T.~Cahill, Phys.\ Rev.\ {\bf D 33}
(1986) 1755.
%
\bibitem{pieterrostock} P. Maris and C.D.~Roberts, ``Differences between
heavy and light quarks,'' in Proc. of the IVth International Workshop on on
Progress in Heavy Quark Physics, edited by M.~Beyer, T.~Mannel and
H.~Schr\"oder (University of Rostock, Rostock, 1998) pp. 159-162.
%
\bibitem{mishaSVY} M.A.~Ivanov, Yu.L.~Kalinovsky and C.D.~Roberts, Phys.\
Rev.\ {\bf D 60} (1999) 034018.
%
\bibitem{hdp76} K. Lane, Phys. Rev. {\bf 10} (1974) 2605;
H. D. Politzer, Nucl. Phys. {\bf B 117} (1976) 397.
%
\bibitem{higashijimaportermelbourne} K.~Higashijima, Phys.\ Rev.\ {\bf D 29}
(1984) 1228;
D.~Atkinson and P.W.~Johnson, Phys.\ Rev.\ {\bf D 37} (1988) 2296;
C.D.~Roberts and B.H.~McKellar, Phys.\ Rev.\ {\bf D 41} (1990) 672.
%
\bibitem{seattle} A.G.~Williams, G.Krein and C.D.~Roberts, Annals Phys.\ {\bf
210} (1991) 464.
%
\bibitem{mrt98} P.~Maris, C.D.~Roberts and P.C.~Tandy, Phys.\ Lett.\ {\bf B
420} (1998) 267.
%
\bibitem{diakonov} D.I.~Diakonov and V.Y.~Petrov, Nucl.\ Phys.\ {\bf B 272}
(1986) 457.
%
\bibitem{derek} D.B.~Leinweber, Annals Phys.\ {\bf 254} (1997) 328.
%
\bibitem{tonylatticequark} J.I.~Skullerud and A.G.~Williams, ``The quark
propagator in momentum space,'' hep-lat/9909142.
%
\bibitem{simon} S.~Capstick and B.D.~Keister, ``Baryon magnetic moments in a
relativistic quark model,'' \mbox{nucl-th/9611055}.
%
\bibitem{jain} P.~Jain and H.J.~Munczek,Phys.\ Rev.\ {\bf D 48} (1993) 5403.
%
\bibitem{fr96} M.R.~Frank and C.D.~Roberts, Phys.\ Rev.\ {\bf C 53} (1996)
390.
%
\bibitem{fm96} M.R.~Frank and T.~Meissner, Phys.\ Rev.\ {\bf C 53} (1996)
2410.
%
\bibitem{dubravko} D.~Klabu\v{c}ar and D.~Kekez, Phys.\ Rev.\ {\bf D 58} (1998)
096003.
%
\bibitem{regAdelaide} R.T.~Cahill and S.M.~Gunner, ``Low energy quark-gluon
processes from experimental data using the global colour model,'' in Proc. of
the Workshop on Nonperturbative Methods in Quantum Field Theory, edited by
A.W. Schreiber, A.G. Williams and A.W. Thomas (World Scientific, Singapore,
1998) pp. 261-270.
%
\bibitem{pieterVM} P.~Maris and P.C.~Tandy, Phys.\ Rev.\ {\bf C 60} (1999)
055214.
%
\bibitem{papa1} J.~Papavassiliou and J.M.~Cornwall, Phys.\ Rev.\ {\bf D 44}
(1991) 1285.
%
\bibitem{fredIR} F.T.~Hawes, C.D.~Roberts and A.G.~Williams, Phys.\ Rev.\
{\bf D 49} (1994) 4683.
%
\bibitem{axelIR} A.~Bender and R.~Alkofer, Phys.\ Rev.\ {\bf D 53} (1996)
446.
%
\bibitem{natale} A.A.~Natale and P.S.~Rodrigues da Silva, Phys.\ Lett.\ {\bf
B 392} (1997) 444.
%
\bibitem{fredIRnew} F.T.~Hawes, P.~Maris and C.D.~Roberts, Phys.\ Lett.\ {\bf
B 440} (1998) 353.
%
\bibitem{hauck} L.~v.~Smekal, A.~Hauck and R.~Alkofer, Annals Phys.\ {\bf
267} (1998) 1.
%
\bibitem{bbz} M.~Baker, J.S.~Ball and F.~Zachariasen, Nucl.\ Phys.\ {\bf B
186} (1981) 531; {\it ibid} 560.
%
\bibitem{west82} G.B.~West, Phys.\ Lett.\ {\bf B 115} (1982) 468.
%
\bibitem{atkinson83} D.~Atkinson, P.W.~Johnson, W.J.~Schoenmaker and
H.A.~Slim, Nuovo Cim.\ {\bf 77 A} (1983) 197.
%
\bibitem{west83} G.B.~West, Phys.\ Rev.\ {\bf D 27} (1983) 1878.
%
\bibitem{mandelstam} S.~Mandelstam, Phys.\ Rev.\ {\bf D 20} (1979) 3223.
%
\bibitem{bargadda} U.~Bar-Gadda, Nucl.\ Phys.\ {\bf B 163} (1980) 312.
%
\bibitem{brownpennington} N.~Brown and M.R.~Pennington, Phys.\ Lett.\ {\bf B
 202} (1988) 257.
%
\bibitem{brown88b} N.~Brown and M.R.~Pennington, Phys.\ Rev.\ {\bf D 38}
(1988) 2266.
%
\bibitem{brown89} N.~Brown and M.R.~Pennington, Phys.\ Rev.\ {\bf D 39}
(1989) 2723.
%
\bibitem{stingl12} U.~H\"abel, R.~K\"onning, H.G.~Reusch, M.~Stingl and
S.~Wigard, Z.\ Phys.\ {\bf A 336} (1990) 423; {\it ibid} 435.
%
\bibitem{stingl3} M.~Stingl, Z.\ Phys.\ {\bf A 353} (1996) 423.
%
\bibitem{hauckPRL} L.~v.~Smekal, R.~Alkofer and A.~Hauck, Phys.\ Rev.\
Lett.\ {\bf 79} (1997) 3591.
%
\bibitem{hauckAdelaide} A.~Hauck, R.~Alkofer and L.~v.~Smekal, ``A solution
to coupled Dyson-Schwinger equations for gluons and ghosts in Landau gauge,''
in Proc. of the Workshop on Nonperturbative Methods in Quantum Field Theory,
edited by A.W. Schreiber, A.G. Williams and A.W. Thomas (World Scientific,
Singapore, 1998) pp. 81-89.
%
\bibitem{bloch} D.~Atkinson and J.~C.~Bloch, Phys.\ Rev.\ {\bf D 58} (1998)
094036.
\bibitem{bloch2} D.~Atkinson and J.~C.~Bloch, Mod.\ Phys.\ Lett.\ {\bf A 13}
(1998) 1055.
%
\bibitem{davidAdelaide} D.~Atkinson, ``Infrared and ultraviolet coupling in
QCD,'' in Proc. of the Workshop on Nonperturbative Methods in Quantum Field
Theory, edited by A.W. Schreiber, A.G. Williams and A.W. Thomas (World
Scientific, Singapore, 1998) pp. 69-80.
%
\bibitem{jacquesAdelaide} J.C.R.~Bloch, ``Infrared fixed point of the running
coupling in QCD,'' in Proc. of the Workshop on Nonperturbative Methods in
Quantum Field Theory, edited by A.W. Schreiber, A.G. Williams and A.W. Thomas
(World Scientific, Singapore, 1998) pp. 90-96.
%
\bibitem{latticegluon} P.~Marenzoni, G.~Martinelli and N.~Stella, Nucl.\
Phys.\ {\bf B 455} (1995) 339;
D.B.~Leinweber, J.I.~Skullerud, A.G.~Williams and
C.~Parrinello [UKQCD Collaboration], Phys.\ Rev.\ {\bf D 60} (1999) 094507.
%
\bibitem{schilling} H.~Suman and K.~Schilling, Phys.\ Lett.\ {\bf B 373}
(1996) 314.
%
\bibitem{jacquesPrivate} J.C.R.~Bloch, unpublished.
%
\bibitem{instantonreview} E.~Shuryak and T.~Sch\"afer, Ann.\ Rev.\ Nucl.\
Part.\ Sci.\ {\bf 47} (1997) 359.
%
\bibitem{kevinpi0} M.R.~Frank, K.L.~Mitchell, C.D.~Roberts and P.C.~Tandy,
Phys.\ Lett.\ {\bf B 359} (1995) 17.
%
\bibitem{mikepomeron} M.A.~Pichowsky and T.S.~Lee, Phys.\ Rev.\ {\bf D 56}
(1997) 1644;
M.A.~Pichowsky, Nonperturbative quark dynamics in diffractive processes, PhD
Thesis, University of Pittsburgh (1996).
%
\bibitem{mrpion} P.~Maris and C.D.~Roberts, Phys.\ Rev.\ {\bf C 58} (1998)
3659.
%
\bibitem{klabucar} D.~Kekez and D.~Klabu\v{c}ar, Phys.\ Lett.\ {\bf B 457} (1999)
359;
D.~Klabu\v{c}ar and D.~Kekez, Fizika {\bf B 8} (1999) 303.
%
\bibitem{fizikaB} C.D.~Roberts, Fizika {\bf B 8} (1999) 285;
P.C.~Tandy, {\it ibid} 295.
%
\bibitem{klabucar2} B.~Bistrovic and D.~Klabu\v{c}ar, Phys.\ Lett.\ {\bf B
478} (2000) 127.
%
\bibitem{schwingermodel} J.~Schwinger, Phys.\ Rev.\ {\bf 128} (1962) 2425;
and, for example, W. Dittrich and M. Reuter, Lecture Notes in Physics, Vol.\
{\bf 244}: Selected Topics in Gauge Theories (Springer-Verlag, Berlin, 1985),
pp.~107-135; and J.~Zinn-Justin, Quantum Field Theory and Critical Phenomena
(Oxford University Press, Oxford, 1990), pp.~748-752.
%
\bibitem{hollenberg} L.C.L.~Hollenberg, C.D.~Roberts and B.H.J.~McKellar,
Phys.\ Rev.\ {\bf C 46} (1992) 2057.
%
\bibitem{pieterQED3} P.~Maris, Phys.\ Rev.\ {\bf D 52} (1995) 6087.
%
\bibitem{mack} M.~G\"opfert and G.~Mack, Commun.\ Math.\ Phys.\ {\bf 82}
(1981) 545.
%
\bibitem{appel} T.W.~Appelquist, M.~Bowick, D.~Karabali and
L.C.~Wijewardhana, Phys.\ Rev.\ {\bf D 33} (1986) 3704.
%
\bibitem{conrad} C.J.~Burden and C.D.~Roberts, Phys.\ Rev.\ {\bf D 44} (1991)
540.
%
\bibitem{justin} C.J.~Burden, J.~Praschifka and C.D.~Roberts, Phys.\ Rev.\
{\bf D 46} (1992) 2695.
%
\bibitem{einhorn} M.B.~Einhorn, Phys.\ Rev.\ {\bf D 14} (1976) 3451.
%
\bibitem{sw80} R.F.~Streater and A.S.~Wightman, PCT, Spin and Statistics, 3rd
edition (Addison-Wesley, Reading, Mass., 1980).
%
\bibitem{ericold} A.~Szczepaniak, E.S.~Swanson, C.~Ji and S.R.~Cotanch,
Phys.\ Rev.\ Lett.\ {\bf 76} (1996) 2011.
%
\bibitem{franz} \c{C}.~\c{S}avkli and F.~Gross, ``Quark-antiquark bound
states in the relativistic spectator formalism,'' hep-ph/9911319.
%
\bibitem{marshalldualreview} M.~Baker, J.S.~Ball and F.~Zachariasen, Phys.\
Rept.\ {\bf 209} (1991) 73.
%
\bibitem{robertreview} R.J.~Perry, ``Hamiltonian light front field theory and
Quantum Chromodynamics,'' in Proc. of Hadron Physics 94: Topics on the
Structure and Interaction of Hadronic Systems, edited by V.E.~Herscovitz,
C.A.Z.~Vasconcellos and E.~Ferreira (World Scientific, Singapore, 1995)
pp. 120-196.
%
\bibitem{tokilike} H.~Ichie and H.~Suganuma, ``Dual Higgs theory in the
confinement physics of QCD,'' in Proc. of the International Workshop on
Understanding Deconfinement in QCD, edited by D.~Blaschke, F.~Karsch and
C.D.~Roberts (World Scientific, Singapore, 2000) pp. 65-72.
%
\bibitem{nora} M.~Baker, J.S.~Ball, N.~Brambilla, G.M.~Prosperi and
F.~Zachariasen, Phys.\ Rev.\ {\bf D 54} (1996) 2829;
N.~Brambilla and A.~Vairo, Phys.\ Rev.\ {\bf D 56} (1997)
1445.
%
\bibitem{matthias} M.~Burkardt and B.~Klindworth, Phys.\ Rev.\ {\bf D 55}
(1997) 1001.
%
\bibitem{eric} A.P.~Szczepaniak and E.S.~Swanson, Phys.\ Rev.\ {\bf D 55}
(1997) 3987.
%
\bibitem{top} A.~Di Giacomo and B.~Lucini, ``What we do understand of colour
confinement,'' in Proc. of the International Workshop on Understanding
Deconfinement in QCD, edited by D.~Blaschke, F.~Karsch and C.D.~Roberts
(World Scientific, Singapore, 2000) pp.~55-64; 
K.~Langfeld, ``Vortex percolation and confinement,'' {\it ibid} pp.~73-78; 
H.~Reinhardt, ``Magnetic monopoles, vortices and the topology of gauge
fields,'' {\it ibid} pp.~86-93. 
%
\bibitem{pipi} C.D.~Roberts, R.T.~Cahill, M.E.~Sevior and N.~Iannella, Phys.\
Rev.\ {\bf D 49} (1994) 125.
%
\bibitem{echaya} C.D.~Roberts, Fiz. \'Elem. Chastits At. Yadra {\bf 30}
(1999) 537 (Phys. Part. Nucl. {\bf 30} (1999) 223).
%
\bibitem{serdar} C.D.~Roberts, ``Nonperturbative QCD with modern tools,'' in
Proc.  of the 11th Physics Summer School: Frontiers in Nuclear Physics,
edited by S.~Kuyucak (World Scientific, Singapore, 1999) pp. 212-261.
%
\bibitem{mtpion} P.~Maris and P.C.~Tandy, Phys. Rev. {\bf C 61} (2000) 45202.
%
\bibitem{truncscheme} A.~Bender, C.D.~Roberts and L.~v.~Smekal, Phys.\ Lett.\
{\bf B 380} (1996) 7.
%
\bibitem{alkoferbse} M.~Oettel and R.~Alkofer, ``A comparison between
relativistic and semi-relativistic treatment in the diquark-quark model,''
hep-ph/0001261.
%
\bibitem{rhopipiKLM} K.L.~Mitchell and P.C.~Tandy, Phys.\ Rev.\ {\bf C 55}
(1997) 1477.
%
\bibitem{rhopipiMAP} M.A.~Pichowsky, S.~Walawalkar and S.~Capstick, Phys.\
Rev.\ {\bf D 60} (1999) 054030.
%
\bibitem{miranskymunczek} V.A.~Miransky, Dynamical Symmetry Breaking in
Quantum Field Theories (World Scientific, Singapore, 1993) pp. 202-207;
%
H.J.~Munczek, Phys.\ Rev.\ {\bf D 52} (1995) 4736.
%
\bibitem{marisAdelaide} P.~Maris and C.D.~Roberts, ``QCD bound states and
their response to extremes of temperature and density,'' in Proc. of the
Workshop on Nonperturbative Methods in Quantum Field Theory, edited by
A.W. Schreiber, A.G. Williams and A.W. Thomas (World Scientific, Singapore,
1998) pp. 132-151.
%
\bibitem{conradsep} C.J.~Burden, L.~Qian, C.D.~Roberts, P.C.~Tandy and
M.J.~Thomson, Phys.\ Rev.\ {\bf C 55} (1997) 2649.
%
\bibitem{a1b1} J.C.R.~Bloch, Yu.L.~Kalinovsky, C.D.~Roberts and S.M.~Schmidt,
Phys.\ Rev.\ {\bf D 60} (1999) 111502.
%
\bibitem{cdrqcII} C.D.~Roberts, ``Confinement, diquarks and Goldstone's
theorem,'' in Proc. of the 2nd International Conference on Quark Confinement
and the Hadron Spectrum, edited by N.~Brambilla and G.M.~Prosperi (World
Scientific, Singapore, 1997) pp. 224-230.
%
\bibitem{mikescalars1} M.~Boglione and M.R.~Pennington, Eur.\ Phys.\ J.\ {\bf
C 9} (1999) 11.
%
\bibitem{mikescalars2} M.R.~Pennington, ``Riddle of the scalars: Where is the
$\sigma$?,'' in Proc. of the Workshop On Hadron Spectroscopy (WHS 99), edited
by T.~Bressani, A.~Feliciello and A.~Filippi (Frascati, INFN, 1999) pp.
95-114.
%
\bibitem{mikescalars3} M.R.~Pennington, ``Low Energy Hadron Physics,''
hep-ph/0001183.
%
\bibitem{etareinhard} M.R.~Frank and T.~Meissner, Phys.\ Rev.\ {\bf C 57}
(1998) 345;
L.~v.~Smekal, A.~Mecke and R.~Alkofer, ``A dynamical $\eta^\prime$ mass from
an infrared enhanced gluon exchange,'' hep-ph/9707210.
%
\bibitem{newsigmaT} P.~Maris, C.D.~Roberts, S.M.~Schmidt and P.C.~Tandy,
``T-dependence of pseudoscalar and scalar correlations,'' nucl-th/0001064.
%
\bibitem{carlshakin} L.S.~Celenza, S.~Gao, B.~Huang, H.~Wang and C.M.~Shakin,
Phys.\ Rev.\ {\bf C 61} (2000) 035201.
%
\bibitem{lewis} L.P.~Fulcher, Phys.\ Rev.\ {\bf D 57} (1998) 350.
%
\bibitem{noraheavy} N.~Brambilla, A.~Pineda, J.~Soto and A.~Vairo, Nucl.\
Phys.\ {\bf B 566} (2000) 275.
%
\bibitem{bpprivate} C.J.~Burden and M.A.~Pichowsky, unpublished.
%
\bibitem{perry} B.H.~Allen and R.J.~Perry, ``Glueballs in a Hamiltonian light
front approach to pure glue QCD,'' hep-th/9908124.
%
\bibitem{dualQCDglueball} Y.~Koma, H.~Suganuma and H.~Toki, Phys.\ Rev.\ {\bf
D 60} (1999) 074024.
%
\bibitem{hybrids} P.R.~Page, E.S.~Swanson and A.P.~Szczepaniak, Phys.\ Rev.\
{\bf D 59} (1999) 034016;
S.~Capstick and P.R.~Page, Phys.\ Rev.\ {\bf D 60} (1999)
111501.
%
\bibitem{grosspion} H.~Ito, W.W.~Buck and F.~Gross, Phys.\ Rev.\ {\bf C 45}
(1992) 1918.
%
\bibitem{cdrpion} C.D.~Roberts, Nucl.\ Phys.\ {\bf A 605} (1996) 475.
%
\bibitem{frankvertex} M.R.~Frank, Phys.\ Rev.\ {\bf C 51} (1995) 987.
%
\bibitem{pionffdat1} C.J.~Bebek {\it et al.}, Phys.\ Rev.\ {\bf D 13} (1976)
25.
%
\bibitem{pionffdat2} L.M.~Barkov {\it et al.}, Nucl.\ Phys.\ {\bf B 256}
(1985) 365.
%
\bibitem{pionffdat3} S.R.~Amendolia {\it et al.}  [NA7 Collaboration], Nucl.\
Phys.\ {\bf B 277} (1986) 168.
%
\bibitem{piloop} R.~Alkofer, A.~Bender and C.D.~Roberts, Int.\ J.\ Mod.\
Phys.\ {\bf A 10} (1995) 3319.
%
\bibitem{wwbuck} W.W.~Buck, R.A.~Williams and H.~Ito, Phys.\ Lett.\ {\bf B
351} (1995) 24.
%
\bibitem{conradkaon} C.J.~Burden, C.D.~Roberts and M.J.~Thomson, Phys.\
Lett.\ {\bf B 371} (1996) 163.
%
\bibitem{mtkaon} P.~Maris and P.C.~Tandy, ``The $\pi$, $K^+$, and $K^0$
electromagnetic form factors,'' nucl-th/0005015.
%
\bibitem{heath} H.B.~O'Connell, B.C.~Pearce, A.W.~Thomas and A.G.~Williams,
Prog.\ Part.\ Nucl.\ Phys.\  {\bf 39} (1997) 201;
M.~Benayoun, H.B.~O'Connell and A.G.~Williams, Phys.\ Rev.\ {\bf D 59} (1999)
074020.
%
\bibitem{fredFF} F.T.~Hawes and M.A.~Pichowsky, Phys.\ Rev.\ {\bf C 59}
(1999) 1743.
%
\bibitem{hecht} M.B.~Hecht and B.H.J.~McKellar, Phys.\ Rev.\ {\bf C 57} (1998)
2638;
{\it ibid} {\bf 60} (1999) 065202.
%
\bibitem{jpopiejustin1justin2} R.T.~Cahill, C.D.~Roberts and J.P.~Opie, ``A
Derivation Of Hadron Soliton Phenomenology From QCD,'' preprint
no. FIAS-R-158 (1985);
%
J.~Praschifka, C.D.~Roberts and R.T.~Cahill, Phys.\ Rev.\ {\bf D 36} (1987)
209;
%
C.D.~Roberts, R.T.~Cahill and J.~Praschifka, Annals Phys.\ {\bf 188} (1988)
20.
%
\bibitem{witten} E.~Witten, Nucl.\ Phys.\ {\bf B 223} (1983) 422.
%
\bibitem{bando} M.~Bando, M.~Harada and T.~Kugo, Prog.\ Theor.\ Phys.\ {\bf
91} (1994) 927.
%
\bibitem{reinhardA} R.~Alkofer and C.D.~Roberts,
Phys.\ Lett.\  {\bf B 369} (1996) 101.
%
\bibitem{serdarthomasfritzsimula} F.~Coester and D.O.~Riska, Few Body Syst.\
{\bf 25} (1998) 29;
A.W.~Thomas and S.~V.~Wright, ``Classical
quark models: An introduction,'' in Proc.  of the 11th Physics Summer School:
Frontiers in Nuclear Physics, edited by S.~Kuyucak (World Scientific,
Singapore, 1999) pp. 171-211;
F.~Cardarelli, E.~Pace, G.~Salm\'e and S.~Simula, Few Body Syst.\ Suppl.\
{\bf 11} (1999) 66.
%
\bibitem{regdq} R.T.~Cahill, J.~Praschifka and C.J.~Burden,
Austral. J. Phys. {\bf 42} (1989) 161;
{\it ibid} 171.
%
\bibitem{hugodq} H.~Reinhardt, Phys.\ Lett.\ {\bf B 244} (1990) 316.
%
\bibitem{regfe} R.T.~Cahill, C.D.~Roberts and J.~Praschifka, Austral.\ J.\
Phys.\ {\bf 42} (1989) 129.
%
\bibitem{justindq} R.T.~Cahill, C.D.~Roberts and J.~Praschifka, Phys.\ Rev.\
{\bf D 36} (1987) 2804.
%
\bibitem{jacquesdq} J.C.R.~Bloch, C.D.~Roberts and S.M.~Schmidt, Phys.\ Rev.\
{\bf C 60} (1999) 065208.
%
\bibitem{ishii} N.~Ishii, W.~Bentz and K.~Yazaki, Nucl.\ Phys.\ {\bf A 587}
(1995) 617.
%
\bibitem{raAB} M.~Oettel, G.~Hellstern, R.~Alkofer and H.~Reinhardt,
Phys. Rev. {\bf C 58} (1998) 2459 cf. Ref.~\cite{reinhard}.
%
\bibitem{reinhard} G.~Hellstern, R.~Alkofer, M.~Oettel and H.~Reinhardt,
Nucl. Phys. {\bf A 627} (1997) 679.
%
\bibitem{latticediquark} M.~Hess, F.~Karsch, E.~Laermann and I.~Wetzorke,
Phys.\ Rev.\ {\bf D 58} (1998) 111502.
%
\bibitem{anselmino} M.~Anselmino, E.~Predazzi, S.~Ekelin, S.~Fredriksson and
D.B.~Lichtenberg, Rev.\ Mod.\ Phys.\ {\bf 65} (1993) 1199;
P.~Kroll, Few Body Syst.\ Suppl.\ {\bf 11} (1999) 255.
%
\bibitem{piller} K.~Kusaka, G.~Piller, A.W.~Thomas and A.G.~Williams, Phys.\
Rev.\ {\bf D 55} (1997) 5299.
%
\bibitem{bentzFad} H.~Asami, N.~Ishii, W.~Bentz and K.~Yazaki, Phys.\ Rev.\
{\bf C 51} (1995) 3388.
%
\bibitem{keiner} V.~Keiner, Phys. Rev. {\bf C 54} (1996) 3232.
%
\bibitem{jacquesnucleon} J.C.R.~Bloch, C.D.~Roberts, S.M.~Schmidt, A.~Bender
and M.R.~Frank, Phys.\ Rev.\ {\bf C 60} (1999) 062201.
%
\bibitem{pichowskydiquark} M.~Oettel, M.~Pichowsky and L.~v.~Smekal,
``Current conservation in the covariant quark-diquark model of the nucleon,''
nucl-th/9909082; 
M.~Oettel, S.~Ahlig, R.~Alkofer and C.~Fischer, ``Form factors of baryons in
a confining and covariant diquark-quark model,'' nucl-th/9910079;
R.~Alkofer, S.~Ahlig, C.~Fischer and M.~Oettel, nucl-th/9911020, $\pi N$
Newslett.\ No.\ {\bf 15} (1999) 238.
%
\bibitem{myriad} J.C.R.~Bloch, C.D.~Roberts and S.M.~Schmidt, Phys. Rev. {\bf
C 61} (2000) 065207.
%
\bibitem{entire} H.J.~Munczek, Phys.\ Lett.\ {\bf B 175} (1986) 215.
%
\bibitem{entireCJB} C.J.~Burden, C.D.~Roberts and A.~G.~Williams, Phys.\
Lett.\ {\bf B 285} (1992) 347.
%
\bibitem{latticesigma} S.~Gusken {\it et al.} [SESAM Collaboration], Phys.\
Rev.\ {\bf D 59} (1999) 054504.
%
\bibitem{dereksigma} D.B.~Leinweber, A.W.~Thomas and S.V.~Wright,
``Lattice QCD calculations of the sigma commutator,'' hep-lat/0001007.
%
\bibitem{carlsigma} C.M.~Shakin, Phys.\ Rev.\ {\bf C 50} (1994) 1129.
%
\bibitem{knecht} M.~Knecht, hep-ph/9912443, $\pi N$ Newslett.\ No.\ {\bf 15}
(1999) 108; and references therein.
%
\bibitem{machleidt} R.~Machleidt, Adv.\ Nucl.\ Phys.\ {\bf 19} (1989) 189.
%
\bibitem{vincent} R.B.~Wiringa, V.G.~Stoks and R.~Schiavilla, Phys.\ Rev.\
{\bf C 51} (1995) 38.
%
\bibitem{harry} T.~Sato and T.-S.H.~Lee, Phys.\ Rev.\ {\bf C 54} (1996) 2660.
%
\bibitem{Friman} B.~Friman and M.~Soyeur, Nucl.\ Phys.\ {\bf A 600} (1996)
477.
%
\bibitem{ESFTtexts} J.I.~Kapusta, Finite-temperature field theory (Cambridge
University Press, Cambridge, 1989); M.\ le Bellac, Thermal Field Theory
(Cambridge University Press, Cambridge, 1996).
%
\bibitem{barbour} I.M.~Barbour, ``QCD at non-zero density,'' in Proc. of the
International Workshop on Understanding Deconfinement in QCD, edited by
D.~Blaschke, F.~Karsch and C.D.~Roberts (World Scientific, Singapore, 2000)
pp. 10-17.
%
\bibitem{edwinechaya} E.~Laermann, Fiz. \'Elem. Chastits At. Yadra {\bf 30}
(1999) 720 (Phys. Part. Nucl. {\bf 30} (1999) 304).
%
\bibitem{bastiscm} D.~Blaschke, C.D.~Roberts and S.M.~Schmidt, Phys.\ Lett.\
{\bf B 425} (1998) 232.
%
\bibitem{pressures} P.~Levai and U.~Heinz, Phys.\ Rev.\ {\bf C 57} (1998)
1879;
J.O.~Andersen, E.~Braaten and M.~Strickland, Phys.\ Rev.\
{\bf D 61} (2000) 074016;
R.~Baier and K.~Redlich, Phys.\ Rev.\ Lett.\ {\bf 84} (2000) 2100;
J.P.~Blaizot, E.~Iancu and A.~Rebhan, ``Approximately self-consistent
resummations for the thermodynamics of the quark-gluon plasma. I: Entropy and
density,'' hep-ph/0005003.
%
\bibitem{haymaker} R.W.~Haymaker, Riv.\ Nuovo Cim.\ {\bf 14} (1991) 1.
%
\bibitem{kalash} O.K.~Kalashnikov, Pis'ma Zh.\ Eksp.\ Teor.\ Fiz.\ {\bf 41}
(1985) 477 (JETP Lett.\ {\bf 41} (1985) 582).
%
\bibitem{bb95} R.D.~Bowler and M.C.~Birse, Nucl.\ Phys.\ {\bf A 582} (1995)
655.
%
\bibitem{axelT} A.~Bender, D.~Blaschke, Yu.L.~Kalinovsky and C.D.~Roberts,
Phys.\ Rev.\ Lett.\ {\bf 77} (1996) 3724.
%
\bibitem{cpbook} R.J.~Creswick, H.A.~Farach and C.P.~Poole, Introduction to
Renormalization Group Methods in Physics (John Wiley and Sons, Inc., New
York, 1992).
%
\bibitem{pisarski} R.D.~Pisarski and F.~Wilczek, Phys.\ Rev.\ {\bf D 29}
(1984) 338;
F.~Wilczek, Int.\ J.\ Mod.\ Phys.\ {\bf A 7} (1992) 3911.
%
\bibitem{neqfour} G.A.~Baker, B.G.~Nickel and D.I.~Meiron, Phys.\ Rev.\ {\bf
B 17} (1978) 1365;
``Compilation of 2-pt. and 4-pt. graphs for continuous spin models'',
University of Guelph report (1977), unpublished.
%
\bibitem{arne1} D.~Blaschke, A.~H\"oll, C.D.~Roberts and S.M.~Schmidt, Phys.\
Rev.\ {\bf C 58} (1998) 1758.
%
\bibitem{arne2} A.~H\"oll, P.~Maris and C.D.~Roberts, Phys.\ Rev.\ {\bf C 59}
(1999) 1751.
%
\bibitem{reinhardCoulomb} R.~Alkofer, P.A.~Amundsen and K.~Langfeld, Z.\
Phys.\ {\bf C 42} (1989) 199.
%
\bibitem{jackson} A.D.~Jackson and J.J.M.~Verbaarschot, Phys.\ Rev.\ {\bf D
53} (1996) 7223.
%
\bibitem{edwinTrento} E.~Laermann, ``Finite Temperature QCD on the Lattice,''
in Proc. of the International Workshop on Understanding Deconfinement in QCD,
edited by D.~Blaschke, F.~Karsch and C.D.~Roberts (World Scientific,
Singapore, 2000) pp. 3-9.
%
\bibitem{bastiO2} P.~Zhuang, J.H\"ufner and S.P.~Klevansky, Nucl.\ Phys.\
{\bf A 576} (1994) 525;
P.~Zhuang, Phys.\ Rev.\ {\bf C 51}, 2256 (1995);
D.~Blaschke, Yu.L.~Kalinovsky, G.~R\"opke, S.M.~Schmidt and M.K.~Volkov,
Phys.\ Rev.\ {\bf C 53} (1996) 2394.
%
\bibitem{bastiO1} D.~Blaschke, Yu.L.~Kalinovsky, V.N.~Pervushin, G.~R\"opke
and S.M.~Schmidt, Z.\ Phys.\ {\bf A 346} (1993) 85;
S.M.~Schmidt, D.~Blaschke and Yu.L.~Kalinovsky, Phys.\ Rev.\ {\bf C 50} (1994)
435;
D.~Blaschke, Yu.L.~Kalinovsky, L.~M\"unchow, V.N.~Pervushin, G.~R\"opke and
S.M.~Schmidt, Nucl.\ Phys.\ {\bf A 586} (1995) 711.
%
\bibitem{bruno} M.~Jaminon and B.~Van den Bossche, Z.\ Phys.\ {\bf C 64}
(1994) 339.
%
\bibitem{broniowski} B.~Szczerbinska and W.~Broniowski, Acta Phys.\ Polon.\
{\bf B 31} (2000) 835.
%
\bibitem{pctTrento} D.~Blaschke and P.C.~Tandy, ``Mesonic Correlations and
Quark Deconfinement,'' in Proc. of the International Workshop on
Understanding Deconfinement in QCD, edited by D.~Blaschke, F.~Karsch and
C.D.~Roberts (World Scientific, Singapore, 2000) pp. 218-230.
%
\bibitem{japan1} Y.~Taniguchi and Y.~Yoshida, Phys.\ Rev.\ {\bf D55} (1997)
2283.
%
\bibitem{stephanov1} J.B.~Kogut, M.A.~Stephanov and C.G.~Strouthos, Phys.\
Rev.\ {\bf D 58} (1998) 096001.
%
\bibitem{gregp} A.~Bender, G.I.~Poulis, C.D.~Roberts, S.M.~Schmidt and
A.W.~Thomas, Phys.\ Lett.\ {\bf B 431} (1998) 263.
%
\bibitem{japan2} O.~Kiriyama, M.~Maruyama and F.~Takagi, ``Chiral phase
transition at high temperature and density in the QCD-like theory,''
hep-ph/0001108.
%
\bibitem{cr85} R.T.~Cahill, C.D.~Roberts and A.G.~Williams, ``The Bag
Constant In QCD,'' preprint no. FIAS-R-122 (1994);
%
R.T.~Cahill and C.D.~Roberts, Phys.\ Rev.\ {\bf D 32} (1985) 2419.
%
\bibitem{blaschkebiel} D.~Blaschke and C.D.~Roberts, Nucl.\ Phys.\ {\bf A
642} (1998) 197c.
%
\bibitem{diakonov2} G.W.~Carter and D.~Diakonov, ``Chiral symmetry breaking
and color superconductivity in the Instanton picture,'' in Proc. of the
International Workshop on Understanding Deconfinement in QCD, edited by
D.~Blaschke, F.~Karsch and C.D.~Roberts (World Scientific, Singapore, 2000)
pp. 239-250.
%
\bibitem{stephanov2} M.A.~Stephanov, ``QCD critical point: What it takes to
discover,'' in Proc. of the International Workshop on Understanding
Deconfinement in QCD edited by D.~Blaschke, F.~Karsch and C.D.~Roberts (World
Scientific, Singapore, 2000) pp. 149-155.
%
\bibitem{cahillalone} R.T.~Cahill, Austral.\ J.\ Phys.\ {\bf 42} (1989) 171.
%
\bibitem{quarkstar1} D.~Blaschke, H.~Grigorian, G.~Poghosyan, C.D.~Roberts
and S.M.~Schmidt, Phys.\ Lett.\ {\bf B 450} (1999) 207.
%
\bibitem{solitongcm} C.W.~Johnson and G.~Fai, Phys.\ Rev.\ {\bf C 56} (1997)
3353.
%
\bibitem{drago} A.~Drago, ``Deconfinement signals in neutron stars and
supernova explosion,'' in Proc. of the International Workshop on
Understanding Deconfinement in QCD, edited by D.~Blaschke, F.~Karsch and
C.D.~Roberts (World Scientific, Singapore, 2000) pp. 342-348.
%
\bibitem{weber} F.~Weber, ``Signal of quark deconfinement in neutron stars,''
in Proc. of the International Workshop on Understanding Deconfinement in QCD,
edited by D.~Blaschke, F.~Karsch and C.D.~Roberts (World Scientific,
Singapore, 2000) pp. 334-341.
%
\bibitem{davidqqstar} D.~Blaschke, T.~Kl\"ahn and D.N.~Voskresensky,
Astrophys. J. {\bf 533}~(2000)~406.
%
\bibitem{kahana} D.~Kahana and U.~Vogl, Phys. Lett. {\bf B 244} (1990) 10.
%
\bibitem{bl84} D.~Bailin and A.~Love, Phys. Rept. {\bf 107} (1984) 325.
%
\bibitem{krishna} M.~Alford, K.~Rajagopal and F.~Wilczek, Phys.\ Lett.\ {\bf
B 422} (1998) 247;
R.~Rapp, T.~Sch\"afer, E.V.~Shuryak and M.~Velkovsky, Phys.\ Rev.\ Lett.\
{\bf 81} (1998) 53.
%
\bibitem{wilczekgossip} F.~Wilczek, Nucl. Phys. {\bf A 642} (1998) 1; and
references therein.
%
\bibitem{stephanov3} J.B.~Kogut, M.A.~Stephanov and D.~Toublan, Phys.\ Lett.\
{\bf B 464} (1999) 183.
%
\bibitem{bergesbiel} J.~Berges, Nucl.\ Phys.\ {\bf A 642} (1998) 51c.
%
\bibitem{krishna1} K.~Rajagopal, Nucl.\ Phys.\ {\bf A 642} (1998) 26c.
%
\bibitem{kisslinger} L.S.~Kisslinger, M.~Aw, A.~Harey and O.~Linsuain, Phys.\
Rev.\ {\bf C 60} (1999) 065204.
%
\bibitem{thoma} A.~Peshier and M.H.~Thoma, Phys.\ Rev.\ Lett.\ {\bf 84}
(2000) 841.
%
\bibitem{bastitrento} C.D.~Roberts and S.M.~Schmidt, ``Dyson-Schwinger
equations and the quark gluon plasma,'' in Proc. of the International
Workshop on Understanding Deconfinement in QCD, edited by D.~Blaschke,
F.~Karsch and C.D.~Roberts (World Scientific, Singapore, 2000) pp. 183-195.
%
\bibitem{schaeferbiel} T.~Sch\"afer, Nucl.\ Phys.\ {\bf A 642} (1998) 45c.
%
\bibitem{shuryakbiel} E.~Shuryak, Nucl.\ Phys.\ {\bf A 642} (1998) 14c.
%
\bibitem{jackson1} B. Vanderheyden and A.D. Jackson, ``Random matrix model
for chiral symmetry breaking and color superconductivity in QCD at finite
density,'' hep-ph/0003150.
%
\bibitem{birse} S.~Pepin, M.~C.~Birse, J.A.~McGovern and N.R.~Walet, Phys.\
Rev.\ {\bf C 61} (2000) 055209.
%
\bibitem{hands} S.~Hands and S.~Morrison, ``Diquark condensation in dense
matter: A Lattice perspective,'' in Proc. of the International Workshop on
Understanding Deconfinement in QCD, edited by D.~Blaschke, F.~Karsch and
C.D.~Roberts (World Scientific, Singapore, 2000) pp. 31-42.
%
\bibitem{jacquesscalar} J.C.R.~Bloch, M.A.~Ivanov, T.~Mizutani, C.D.~Roberts
and S.M.~Schmidt, ``$K \to \pi\pi$ and a light scalar meson,''
nucl-th/9910029, to appear in Phys. Rev. {\bf C}.
%
\bibitem{polemasses} N.P.~Landsman and C.G.~van Weert, Phys.\ Rept.\ {\bf
145} (1987) 141.
%
\bibitem{peterNew} D.~Blaschke, G.~Burau, Yu.L.~Kalinovsky, P.~Maris and
P.C.~Tandy, ``Finite-T meson correlations and quark deconfinement,''
nucl-th/0002024.
%
\bibitem{rhomuT} P.~Maris, C.D.~Roberts and S.M.~Schmidt, Phys.\ Rev.\ {\bf C
57} (1998) 2821.
%
\bibitem{yura2} S.M. Schmidt, D. Blaschke and Yu.L. Kalinovsky, Z.\ Phys.\
{\bf C 66} (1995) 485.
%
\bibitem{volkov} M.K.~Volkov, E.A.~Kuraev, D.~Blaschke, G.~R\"opke and
S.M.~Schmidt, Phys.\ Lett.\ {\bf B 424} (1998) 235.
%
\bibitem{cabo} A.~Cabo, O.K.~Kalashnikov and E.K.~Veliev, Nucl.\ Phys.\ {\bf
B 299} (1988) 367.
%
\bibitem{pisarski1} R.D.~Pisarski, Phys.\ Rev.\ Lett.\ {\bf 76} (1996) 3084.
%
\bibitem{pisarski2} R.D.~Pisarski, T.L.~Trueman and M.H.G.~Tytgat, Phys.\
Rev.\ {\bf D 56} (1997) 7077.
%
\bibitem{yura} D. Blaschke, Yu.L. Kalinovsky, P. Petrow, S.M. Schmidt,
M. Jaminon and B. Van den Bossche, Nucl.\ Phys.\ {\bf A 592} (1995) 561.
%
\bibitem{UA1yes} S.~Chandrasekharan, D.~Chen, N.~Christ, W.~Lee, R.~Mawhinney
and P.~Vranas, Phys.\ Rev.\ Lett.\ {\bf 82} (1999) 2463.
%
\bibitem{UA1no} J.B.~Kogut, J.F.~Lagae and D.K.~Sinclair, Phys.\ Rev.\ {\bf D
58} (1998) 054504.
%
\bibitem{wambachNc} M.~Oertel, M.~Buballa and J.~Wambach, ``Meson properties
in the $1/N_c$-corrected NJL model,'' hep-ph/0001239.
%
\bibitem{bastihirschegg} C.D.~Roberts and S.M.~Schmidt, ``Temperature,
chemical potential and the $\rho$-meson,'' in Proc. of the International
Workshop XXVIII on Gross Properties of Nuclei and Nuclear Excitations, edited
by M.~Buballa, W.~N\"orenberg, B.-J.~Schaefer and J.~Wambach (GSI mbH,
Darmstadt, 2000), pp.~185-191.
%
\bibitem{brown} G.E.~Brown and M.~Rho, Phys.\ Rev.\ Lett.\ {\bf 66} (1991)
2720.
%
\bibitem{wambach} R.~Rapp and J.~Wambach, ``Chiral symmetry restoration and
dileptons in relativistic heavy-ion collisions,'' hep-ph/9909229.
%
\bibitem{signatures} S.A.~Bass, M.~Gyulassy, H.~St\"ocker and W.~Greiner, J.\
Phys.\ G {\bf G 25} (1999) R1;
%
S.~Scherer {\it et al.}, Prog.\ Part.\ Nucl.\ Phys.\
{\bf 42} (1999) 279.
%
\bibitem{eskola} K.J.~Eskola, ``High energy nuclear collisions,'' hep-ph/9911350.
%
\bibitem{MS86} T.~Matsui and H.~Satz, Phys.\ Lett.\ {\bf B 178} (1986) 416.
%
\bibitem{JPsiTrento} L. Kluberg [for NA50 Collaboration] ``Results on
J$/\Psi$ Suppression,'' in Proc. of the International Workshop on
Understanding Deconfinement in QCD, edited by D.~Blaschke, F.~Karsch and
C.D.~Roberts (World Scientific, Singapore, 2000) pp. 291-302.
%
\bibitem{Gerschel} C.~Gerschel and J.~H\"ufner, Z.\ Phys.\ {\bf C 56} (1992)
171.
%
\bibitem{Nardi} M.~Nardi and H.~Satz, Phys.\ Lett.\  {\bf B 442} (1998) 14.
%
\bibitem{BlaTandy} D.B.~Blaschke, G.R.~Burau, M.A.~Ivanov, Yu.L.~Kalinovsky
and P.C.~Tandy, ``Dyson-Schwinger equation approach to the QCD deconfinement
transition and J$/\Psi$ dissociation,'' hep-ph/0002047.
%
\bibitem{heinz} U.~Heinz and M.~Jacob, ``Evidence for a new state of matter:
An assessment of the results from the CERN lead beam programme,''
nucl-th/0002042.
%
\bibitem{drees} A.~Drees, ``Recent results on low mass dileptons,'' in
Proc. of the International Workshop on Understanding Deconfinement in QCD,
edited by D.~Blaschke, F.~Karsch and C.D.~Roberts (World Scientific,
Singapore, 2000) pp. 285-290.
%
\bibitem{cassing} W.~Cassing and E.~L.~Bratkovskaya, Phys.\ Rept.\ {\bf 308}
(1999) 65.
%
\bibitem{brown2} G.Q.~Li, C.M.~Ko and G.E.~Brown, Phys.\ Rev.\ Lett.\ {\bf
75} (1995) 4007;
C.~Adami and G.E.~Brown, Phys.\ Rept.\ {\bf 234} (1993) 1.
%
\bibitem{georgefai} C.~Song and G.~Fai, Phys.\ Rev.\ {\bf C 58} (1998) 1689.
%
\bibitem{BECATTINI} F.~Becattini, M.~Ga\'zdzicki and J.~Sollfrank,
Eur. Phys. J. {\bf C 5} (1998) 143.
%
\bibitem{strangedat} T.~Alber {\it et al.}  [NA35 Collaboration],
Phys. Lett. {\bf B 366} (1996) 56;
P.G.~Jones {\it et al.}  [NA49 Collaboration], Nucl. Phys. {\bf A 610}
(1996) 188c.
%
\bibitem{strangedat2} F.~Antinori {\it et al}. [WA97 Collaboration],
Nucl. Phys. {\bf A 661} (1999) 130c; 
%
D.~Elia {\it et al}. [WA97 Collaboration], {\it ibid} 476c;
%
C.~H\"ohne {\it et al}. [NA 49 Collaboration], {\it ibid} 476c;
%
N.~Willis {\it et al}. [NA 50 Collaboration], {\it ibid} 534c.
%
\bibitem{bastithesis} S.M.~Schmidt, ``Nichtlokales Quarkmodell bei endlichen
Temperaturen und Dichten,'' PhD Thesis, University of Rostock (1995), in
German.
%
\bibitem{hydro} N.K.~Glendenning and T.~Matsui, Phys.\ Lett.\ {\bf B 141}
(1984) 419;
M.~Gyulassy and T.~Matsui, Phys.\ Rev.\ {\bf D 29} (1984) 419;
K.~Kajantie, R.~Raitio and P.V.~Ruuskanen, Nucl.\ Phys.\ {\bf B 222} (1983)
152;
G.~Baym, B.L.~Friman, J.P.~Blaizot, M.~Soyeur and W.~Czy\.z, Nucl.\ Phys.\
{\bf A 407} (1983) 541.
%
\bibitem{DCC} J.P.~Blaizot and A.~Krzywicki, Acta Phys.\ Polon.\ {\bf 27}
(1996) 1687;
T.C.~Brooks {\it et al.}  [MiniMax Collaboration], Phys.\ Rev.\ {\bf D 61}
(2000) 032003.
%
\bibitem{lerran} L.D.~McLerran and T.~Toimela, Phys.\ Rev.\ {\bf D 31} (1985)
545;
D.~Boyanovsky, H.J.~de Vega, R.~Holman and S.~Prem Kumar, Phys.\ Rev.\ {\bf D
56} (1997) 5233;
Y.~Kluger, V.~Koch, J.~Randrup and X.~Wang, Phys.\ Rev.\ {\bf C 57} (1998)
280.
%
\bibitem{Gribovpartons} L.V.~Gribov and M.G.~Ryskin, Phys.\ Rept.\ {\bf 189}
(1990) 29;
T.S.~Biro, C.~Gong, B.~M\"uller and A.~Trayanov, Int.\ J.\ Mod.\ Phys.\ {\bf
C 5} (1994) 113;
L. McLerran and R. Venugopalan, Phys.Rev. {\bf D 49} (1994) 2233; {\it ibid}
3352.
%
\bibitem{anders83} B.~Andersson, G.~Gustafson, G.~Ingelman and
T.~Sj\"ostrand, Phys.\ Rept.\ {\bf 97} (1983) 33.
%
\bibitem{partonMC} X.~Wang and M.~Gyulassy, Phys. Rev. {\bf D 44} (1991) 3501.
%
\bibitem{Geiger95} K.~Geiger, Phys.\ Rept.\ {\bf 258} (1995) 237.
%
\bibitem{stringMC} B.~Andersson, G.~Gustafson and B.~Nilsson-Almquist,
Nucl. Phys. {\bf B 281} (1987) 289;
K.~Werner, Phys.\ Rept.\ {\bf 232} (1993) 87;
S.A.~Bass {\it et al.}, Prog.\ Part.\ Nucl.\ Phys.\ {\bf 41} (1998) 225.
%
\bibitem{rwhay} R.W.~Haymaker, V.~Singh, Y.~Peng and J.~Wosiek, Phys.\ Rev.\
{\bf D 53} (1996) 389.
%
\bibitem{nuss} A.~Casher, H.~Neuberger and S.~Nussinov, Phys.\ Rev.\ {\bf D
20} (1979) 179.
%
\bibitem{SCH} F.~Sauter, Z.~Phys.~{\bf 69} (1931) 742; 
%
W.~Heisenberg and H.~Euler, Z.~Phys.~{\bf 98} (1936) 714;
J.~Schwinger, Phys.~Rev.~{\bf 82} (1951) 664.
%
\bibitem{greiner} W. Greiner, B. M\"uller, and J. Rafelski, Quantum
Electrodynamics of Strong Fields (Springer-Verlag, Berlin, 1985); A.A. Grib,
S.G. Mamaev and V.M. Mostepanenko, {\em Vacuum quantum effects in strong
external fields}, (Atomizdat, Moscow, 1988).
%
\bibitem{Back} Y.~Kluger, J.M.~Eisenberg, B.~Svetitsky, F.~Cooper and
E.~Mottola, Phys. Rev. Lett. {\bf 67} (1991) 2427;
F.~Cooper, J.M.~Eisenberg, Y.~Kluger, E.~Mottola and B.~Svetitsky,
Phys. Rev. {\bf D 48} (1993) 190.
%
\bibitem{juda} J.M.~Eisenberg, Phys.\ Rev.\ {\bf D 51} (1995) 1938.
%
\bibitem{pavelwong} H.-P.~Pavel and D.M.~Brink, Z.\ Phys.\ {\bf C 51} (1991)
119;
C.Y.~Wong and G.~Gatoff, Phys.\ Rept.\ {\bf 242} (1994) 489;
C.Y.~Wong, R.~Wang and J.~Wu, Phys.\ Rev.\ {\bf D 51} (1995) 3940.
%
\bibitem{ben} M.A.~Lampert and B.~Svetitsky, Phys.\ Rev.\ {\bf D 61} (2000)
034011.
%
\bibitem{bialas} A.~Bia\l as and W.~Czy\.z, Phys.~Rev.~{\bf D 30} (1984)
2371;
{\bf D 31} (1985) 198.
%
\bibitem{Elze89} H.~Elze and U.~Heinz, Phys.\ Rept.\ {\bf 183} (1989) 81.
%
\bibitem{bahl2} G.C.~Nayak and V.~Ravishankar, Phys. Rev. {\bf C 58} (1998)
356.
%
\bibitem{qcdwong} S.K.~Wong, Nuovo Cim.\ {\bf A 65} (1970) 689.
%
\bibitem{cooperboyaknollgies} F.~Cooper and E.~Mottola, Phys.\ Rev.\ {\bf D
36} (1987) 3114;
{\it ibid} {\bf D 40} (1989) 456;
T.S.~Biro, H.B.~Nielsen and J.~Knoll, Nucl.\ Phys.\ {\bf B 245} (1984) 449;
M.~Herrmann and J.~Knoll, Phys.\ Lett.\ {\bf B 234} (1990) 437;
D.~Boyanovsky, H.J.~de Vega, R.~Holman, D.S.~Lee and A.~Singh, Phys.\ Rev.\
{\bf D 51} (1995) 4419;
H.~Gies, Phys.\ Rev.\ {\bf D 61} (2000) 085021.
%
\bibitem{KM} K.~Kajantie and T.~Matsui, Phys. Lett. {\bf B 164} (1985) 373 .
%
\bibitem{GKM} G.~Gatoff, A.K.~Kerman, and T.~Matsui, Phys.~Rev.~{\bf D 36}
(1987) 114.
%
\bibitem{Rau} J.~Rau and B.~M\"uller, Phys.\ Rept.\ {\bf 272} (1996) 1.
%
\bibitem{gsi} S.A.~Smolyansky, G.~R\"opke, S.M.~Schmidt, D.~Blaschke,
V.D.~Toneev and A.V.~Prozorkevich, ``Dynamical derivation of a quantum
kinetic equation for particle production in the Schwinger mechanism,''
hep-ph/9712377;
S.M.~Schmidt D.~Blaschke, G.~R\"opke, S.A.~Smolyansky, A.V.~Prozorkevich and
V.D.~Toneev, Int.~J.~Mod.~Phys.~{\bf E 7} (1998) 709.
%
\bibitem{kme} Y.~Kluger, E.~Mottola, and J.M.~Eisenberg, Phys. Rev. {\bf D
58} (1998) 125015.
%
\bibitem{minijet} K.J.~Eskola, K.~Kajantie and J.~Lindford, Nucl.\ Phys.\
{\bf B 323} (1989) 37;
X.~Wang, Phys.\ Rev.\  {\bf D 43} (1991) 104;
K.~J.~Eskola and M.~Gyulassy, Phys.\ Rev.\ {\bf C 47} (1993) 2329.
%
\bibitem{nayak} R.S.~Bhalerao and G.C.~Nayak, Phys.\ Rev.\ {\bf C 61} (2000)
054907.
%
\bibitem{basti} S.M.~Schmidt, A.V.~Prozorkevich, and S.A.~Smolyansky,
``Creation of boson and fermion pairs in strong fields,'' in Proc. of the Vth
Workshop on Nonequilibrium Physics at Short Time Scales, edited by
K.~Morawetz, P.~Lipavsky', and V.~Spicka (University of Rostock, Rostock,
1998), pp. 142-145.
%
\bibitem{PRD} S.M.~Schmidt, D.~Blaschke, G.~R\"opke, A.V.~Prozorkevich,
S.A.~Smolyansky and V.D.~Toneev, Phys. Rev. {\bf D 59} (1999) 094005.
%
\bibitem{blochVE} J.C.R.~Bloch, V.A.~Mizerny, A.V.~Prozorkevich, C.D.~Roberts,
S.M.~Schmidt, S.A.~Smolyansky and D.V.~Vinnik, Phys.\ Rev.\ {\bf D 60} (1999)
116011.
%
\bibitem{flor} A.~Bialas, W.~Czy\.z, A.~Dyrek and W.~Florkowski,
Nucl. Phys. {\bf B 296} (1988) 611.
%
\bibitem{bhal} B.~Banerjee, R.S.~Bahlerao and V.~Ravishankar,
Phys. Lett. {\bf B 224} (1989) 16;
G.C.~Nayak and V.~Ravishankar, Phys. Rev. {\bf D 55} (1997) 6877.
%
\bibitem{eis} J.M.~Eisenberg, Found. Phys. {\bf 27} (1997) 1213.
%
\bibitem{memory} J.C.R.~Bloch, C.D.~Roberts and S.M.~Schmidt, Phys. Rev. {\bf
D 61} (2000) 117502.
%
\bibitem{nayak2} G.C.~Nayak, A.~Dumitru, L.~McLerran and W.~Greiner,
``Equilibration of the gluon-minijet plasma at RHIC and LHC,''
hep-ph/0001202.
%
\bibitem{Boterman} S.R.~de~Groot, W.A.~van~Leeuwen and Ch.G.~van Weert,
Relativistic Kinetic Theory (North-Holland, Amsterdam, 1980);
%
W.~Botermans and R.~Malfliet, Phys.\ Rept.\ {\bf 198} (1990) 115.
%
\bibitem{Cgreiner} C.~Greiner, K.~Wagner and P.G.~Reinhard, Phys.\ Rev.\ {\bf
C 49} (1994) 1693;
H.~Heiselberg and X.~Wang, Phys.\ Rev.\ {\bf C 53} (1996) 1892;
T.~S.~Biro and C.~Greiner, Phys.\ Rev.\ Lett.\ {\bf 79} (1997) 3138;
W.M.~Alberico, A.~Lavagno and P.~Quarati, Eur.\ Phys.\ J.\ {\bf C 12} (2000)
499;
O.V.~Utyuzh, G.~Wilk and Z.~Wlodarczyk, J.\ Phys.\ {\bf G 26} (2000) L39.
%
\bibitem{schuck} S.~Ayik, O.~Yilmaz, A.~Gokalp and P.~Schuck, Phys.\ Rev.\
{\bf C 58} (1998) 1594.
%
\bibitem{Colonna} M.~Colonna, M.~Di Toro and A.~Guarnera, Nucl.\ Phys.\ {\bf
A 580} (1994) 312.
%
\bibitem{fuhrman} U.~Fuhrmann, K.~Morawetz and R.~Walke, Phys.\ Rev.\ {\bf C
58} (1998) 1473.
%
\bibitem{Kluger:1993} Y.~Kluger, J.M.~Eisenberg and B.~Svetitsky, Int.\ J.\
Mod.\ Phys.\ {\bf E 2} (1993) 333.
%
\bibitem{wongHIC} C.Y.~Wong, Introduction to High-Energy Heavy-Ion
Collisions, (World Scientific, Singapore, 1994).
%
\bibitem{BKM} B.~Bl\"attel, V.~Koch and U.~Mosel, Rep. Prog. Phys. {\bf 56}
(1993) 1;
W.~Cassing and U.~Mosel, Prog. Part. Nucl. Phys. {\bf 25} (1990) 235;
P.A.~Henning, Phys. Rep. {\bf 253} (1995) 235.
%
\bibitem{MH} S.~Mrowczynski and U.~Heinz, Annals Phys.\ {\bf 229} (1994) 1.
%
\bibitem{ZW} W.-M.~Zhang and L.~Wilets, Phys. Rev. {\bf C 45} (1992) 1900.
%
\bibitem{neisel} S.P.~Klevansky, A.~Ogura and J.~H\"ufner, Annals Phys.\ {\bf
261} (1997) 37.
%
\bibitem{smolann} S.A.~Smolyansky, A.V.~Prozorkevich, S.M.~Schmidt,
D.~Blaschke, G.~R\"opke and V.D.~Toneev, Int.\ J.\ Mod.\ Phys.\ {\bf E 7}
(1998) 515;
S.A.~Smolyansky, A.V.~Prozorkevich, G. Maino and S.G.~Mashnik, Annals
Phys. {\bf 277} (1999) 193.
%
\bibitem{ZMR} D.N.~Zubarev, V.G.~Morozov and G.~R\"opke, Statistical
Mechanics of Nonequilibrium Processes, Vol.1 (Akademie Verlag, Berlin, 1996);
Vol. 2 (Wiley-VCH, Berlin, 1997).
%
\bibitem{KBMSKel} L.P.~Kadanoff and G.~Baym, Quantum Statistical Mechanics
(Benjamin, New York, 1962);
%
P.C.~Martin and J.~Schwinger, Phys. Rev. {\bf 115} (1959) 1342;
%
L.V.~Keldysh, Zh.\ Eksp.\ Teor.\ Fiz.\ {\bf 47} (1964) 1515 (Sov. Phys. JETP
{\bf 20} (1964) 1018).
%
\bibitem{CMmeshustin} L.P.~Csernai and I.N.~Mishustin, Phys.\ Rev.\ Lett.\
{\bf 74} (1995) 5005.
%
\bibitem{ole} I.N.~Mishustin, J.A.~Pedersen and O.~Scavenius, Heavy Ion
Phys.\ {\bf 5} (1997) 377.
%
\bibitem{aichelin} A.~Abada and J.~Aichelin, Phys. Rev. Lett. {\bf 74} (1995)
3130;
A.~Abada and M.C.~Birse, Phys.\ Rev.\ {\bf D 57} (1998) 292.
%
\bibitem{huefner} P.~Bo\v{z}ek, Y.B.~He and J.~H\"ufner, Phys.\ Rev.\ {\bf C 57}
(1998) 3263.
%
\bibitem{rehberg} D.S.~Isert, S.P.~Klevansky and P.~Rehberg, Nucl.\ Phys.\
{\bf A 643} (1998) 275;
P.~Rehberg and J.~Aichelin, Phys.\ Rev.\ {\bf C 60} (1999) 064905.
%
\bibitem{rischke} D.H.~Rischke, Nucl.\ Phys.\ {\bf A 610} (1996) 88.
%
\bibitem{bozek} P.~Bo\v{z}ek, Y.B.~He and J.~H\"ufner, ``Spinodal and
dynamical instabilities at the phase transition from the quark-gluon plasma
to hadrons,'' nucl-th/9806066.
%
\end{thebibliography}
\end{document}